\documentclass[onecolumn]{aastex631}

\usepackage{amsmath}
\usepackage{hyperref}
\usepackage{color,soul}
\usepackage{multirow}
\usepackage{longtable}
\usepackage[figuresright]{rotating}
\usepackage{colortbl}
\usepackage{hhline}
\usepackage{cellspace}
\usepackage{makecell}
\usepackage[version=4]{mhchem}

\begin{document}

\title{A Population Analysis of 20 Exoplanets Observed from the Optical to the Near-infrared Wavelengths with HST: Evidence for Widespread Stellar Contamination}

\author[0000-0002-1437-4228]{Arianna Saba}
\affiliation{Department of Physics and Astronomy, University College London, Gower Street, WC1E 6BT London, United Kingdom}
\author[0000-0003-4128-2270]{Alexandra Thompson}
\affiliation{Department of Physics and Astronomy, University College London, Gower Street, WC1E 6BT London, United Kingdom}
\author[0000-0002-9616-1524]{Kai Hou Yip}
\affiliation{Department of Physics and Astronomy, University College London, Gower Street, WC1E 6BT London, United Kingdom}
\author[0000-0001-9010-0539]{Sushuang Ma}
\affiliation{Department of Physics and Astronomy, University College London, Gower Street, WC1E 6BT London, United Kingdom}
\author[0000-0003-3840-1793]{Angelos Tsiaras}
\affiliation{Department of Physics and Astronomy, University College London, Gower Street, WC1E 6BT London, United Kingdom}
\author[0000-0003-2241-5330]{Ahmed Faris Al-Refaie}
\affiliation{Department of Physics and Astronomy, University College London, Gower Street, WC1E 6BT London, United Kingdom}
\author[0000-0001-6058-6654]{Giovanna Tinetti}
\affiliation{Department of Physics and Astronomy, University College London, Gower Street, WC1E 6BT London, United Kingdom}

\begin{abstract}
We present a population study of 20 exoplanets, ranging from Neptune-like to inflated hot-Jupiter planets, observed during transit with the STIS and WFC3 instruments aboard the \textit{Hubble Space Telescope}. To obtain spectral information from the near-UV to the near-infrared, we reanalysed sixteen WFC3 and over fifty STIS archival data sets with our dedicated HST pipeline. We also include twenty-four WFC3 data sets previously reduced with the same software. Across our target sample we observe significant divergence among multiple observations conducted with the same STIS grating at various epochs, whilst we do not detect variations in the WFC3 data sets. These results are suggestive of stellar contamination, which we have investigated further using known Bayesian tools and other tailored metrics, facilitating a more objective assessment of stellar activity intensity  within each system. Our findings reveal that stellar activity contaminates up to half of the studied exoplanet atmospheres, albeit at varying extents.
Accounting for stellar activity can significantly alter planetary atmospheric parameters like molecular abundances (up to 6 orders of magnitude) and temperature (up to 145\%), contrasting with the results of analyses that neglect activity. Our results emphasise the importance of considering the effects of stellar contamination in exoplanet transit studies; this issue is particularly true for data sets obtained with facilities that do not cover the optical and/or UV spectral range where the activity is expected to be more impactful but also more easily detectable. Our results also provide a catalogue of potentially active stars for further investigation and monitoring.
\end{abstract}

\section{Introduction}
\label{sec:intro}
We recently surpassed the mark of 5700 confirmed exoplanets, with more planets discovered by the day thanks to different and complementary detection techniques.
Studies involving large populations of planets and their stellar hosts have been pivotal to get a glimpse into the great diversity of planetary systems beyond our own. Notable examples of said studies focused on exoplanet demographics include e.g. \citet{2017AJ....154..109F,  2012ApJS..201...15H,  2011arXiv1109.2497M, 2012Natur.481..167C, 2019AJ....158...13N, 2022yCat..36740039G, 2022Sci...377.1211L}.
Exoplanet detection campaigns and dedicated missions, such as NASA \textit{Kepler}, \textit{TESS} and ESA \textit{Cheops}, have also significantly contributed to characterise their host stars by measuring key stellar parameters and monitoring their magnetic activity, see e.g.   \citet{2013Natur.500..427B,  2017ApJ...849...36Y,  2013ApJ...769...37B, 2015MNRAS.447.2714B, 2022arXiv221210776B, 2022A&A...663A.161M, 2023AJ....166...72S}.

Large-scale spectroscopic studies of exoplanetary atmospheres are today in their infancy. Space telescopes, such as NASA \textit{Spitzer} and NASA/ESA \textit{Hubble}, have delivered throughout their years of operation large enough data-sets to enable pioneering population studies for tens of exoplanet atmospheres, mainly hydrogen-rich (see e.g., \citet{2011ApJ...729...54C, 2016ApJ...823..109I, sing2016continuum, 2017ApJ...847L..22F, tsiaras2018population, 2018MNRAS.481.4698F, 2022ApJS..260....3C, edwards2022exploring}). However, the data quality provided by these space telescopes did not always match the ambition of the scientific community. To enable detailed comparative exoplanetology, population studies should ideally satisfy the following conditions:
\begin{itemize}
\item Exoplanet spectra should be simultaneously recorded across the broadest possible wavelength range to mitigate the hindrance of observations taken at different epochs \citep{2014ApJ...785...35D}. This issue is particularly acute if the stellar host is active and/or the atmosphere shows signs of temporal variability  due to changes in the cloud coverage, climate or weather patterns \citep{armstrong2016variability, 2023PSJ.....4...68C, 2024ApJS..270...34C}. In particular, optical-to-infrared atmospheric spectra may help to  address degeneracies among cloud parameters \citep{2019ApJ...883..144M} and to correct for stellar contamination \citep{2012A&A...539A.140B}. These mitigations allow to better constrain chemical abundances \citep{2023Natur.614..659R} and the thermal structure of the atmosphere.

    \item Planetary atmospheres should be observed by a single instrument or, if not possible, by combining multiple well calibrated instruments mounted on the same telescope. Numerous studies have shown how offsets among data sets caution against combining data from different instruments and/or observatories  \citep{yip2020compatibility, 2020MNRAS.497.5155W, 2020A&A...642A..50L, 2020A&A...642A..98Y, 2020A&A...641A.158M}. 

    \item A single pipeline should be used to analyse homogeneously all data sets, this approach minimising  biases caused by different data reduction methodologies \citep{2024MNRAS.531...35M}.
    Robust statistical tools should be adopted to  allow quantitative comparisons and estimate the reliability and degeneracy of the data interpretation \citep{2023RNAAS...7...54M, 2024ApJ...961...30Y}. 
\end{itemize}

The NASA/ESA/CSA \textit{James Webb Space Telescope} has enabled for the first time the observation of exoplanet atmospheres in spectral windows that where never sounded before. Given its high sensitivity, JWST is in a unique position for studying even the faintest exoplanet targets. A notable limitation is, however, its more limited optical coverage, starting at wavelengths longer than 0.6 $\mu$m, thus missing spectral coverage where  clouds, hazes, and stellar activity could leave significant imprints \citep{2024arXiv240307801F}. Moreover, given its relatively recent launch, most of the data collected so far are still publicly unavailable and therefore large population studies are not yet possible with \textit{Webb}.  
Upcoming exoplanet-dedicated satellites such as BSSL \textit{Twinkle} \citep{2019ExA....47...29E, stotesbury2022twinkle} and ESA \textit{Ariel} \citep{tinetti2018chemical, 2021arXiv210404824T}, will provide simultaneous optical to infrared exoplanet spectra through multi-year surveys, addressing the top two conditions mentioned above. In the meantime, population studies of exoplanet spectra spanning a broad spectral coverage, including optical wavelengths, must combine data sets from different observatories (e.g. JWST, HST, Spitzer, ground-based telescopes), despite the fact that this approach has its limitations, as explained above. 

To minimise the impact of potential offsets while maximising the wavelength coverage, we focus in this paper on data collected with two instruments onboard the \textit{Hubble Space Telescope}. In the past 20 years, \textit{Hubble} has allowed pioneering and ground-breaking  exoplanet atmospheric studies  from the far UV to the near-infrared, through a variety of instruments
\citep{2003Natur.422..143V, 2010ApJ...717.1291L, 2007ApJ...655..564K, 2009ApJ...690L.114S, 2013ApJ...774...95D}. The majority of these observations were obtained with the Space Telescope Imaging Spectrograph (STIS) and the Wide Field Camera 3 (WFC3) instruments, and now constitute the largest public database of space-based exoplanet transmission spectra. While WFC3 has been at the forefront of detecting and refining water vapour abundances  in the near-infrared \citep[e.g.][]{tsiaras2019water, anisman2020wasp, skaf2020ares, 2023ApJ...954L..52R}, STIS has often been employed to understand the nature of the scattering slopes in the visible wavelengths and to identify signatures of refractory compounds \citep[e.g.][]{2013MNRAS.436.2956S, 2018AJ....155...66L, 2021AJ....162...91E}. 

While significant strides have been made in characterising exoplanetary atmospheres, less emphasis has been placed so far on using the spectral information in the optical/near-UV regime to detect and correct for potential stellar contamination \citep[e.g.][]{2018MNRAS.480.5314P, 2021AJ....161...44E, 2022AJ....164...59L}. The difficulty of this task arises from the diverse ways in which an active star can influence the spectra of its companion planet(s). Stellar magnetic activity results in surface heterogeneities in the form of colder spots and hotter faculae that can cause the transit chord to differ from the observed disk-integrated stellar spectrum. On the stellar photosphere, these heterogeneities can either be occulted or unocculted during the planetary transit w.r.t. our line of sight. While occulted stellar signatures are comparatively easier to identify in transit light curves, unocculted features are trickier to recognise. The first imprint a bump (in the case of a spot) or a depression (for a facula) in the transit light curve, but can be masked or removed by inspection \citep{narita2013multi}. On the other hand, unocculted active regions affect the transit depth of the entire light curve, introducing a strong chromatic signal that deepens (decreases) the observed transit depth, overestimating (underestimating) the planetary radius compared to a scenario without spots (faculae) \citep{rackham2018transit}. Although stellar contamination effects are strongest in the optical regime, where they introduce characteristic, exponential slopes, they are also non-negligible in the infrared. In this region they are capable of introducing offsets, potentially mimicking or masking absorption features of interest in the planetary atmosphere, e.g. water \citep{2021AJ....162..300B} and/or resulting in erroneous retrieved chemical abundances \citep{2020ApJ...889...78I, 2024ApJ...960..107T}.

Leveraging the availability of many publicly available HST data sets, we combine here repeated observations belonging to 20 different exoplanets obtained with both STIS and WFC3. We aim to use these combined data sets to determine the origin of the distinct spectral signatures observed throughout the population. Subsequently, and largely due to the optical coverage allowed by STIS, we also find extensive evidence for stellar contamination. We apply a standardised data reduction across the entire sample using our data analysis pipeline \texttt{Iraclis} \citep{tsiaras2016new}, which is tailored specifically for HST data. We interpret the spectral results with the unified Bayesian retrieval code TauREx\,3.1 \citep{Al_Refaie_2021}, along with its stellar activity plugin \texttt{ASteRA} \citep{2024ApJ...960..107T}, which enables combined stellar-planetary retrievals. \texttt{ASteRA} allows for the consideration of potential stellar contamination effects during atmospheric retrievals, ensuring biases introduced by contamination are simultaneously accounted and corrected for in the retrieved planetary parameters. Overall, our retrieval framework is tailored to enable quantitative comparisons among atmospheric, planetary, and stellar parameters. 

We establish a set of novel stellar activity metrics which, by computing the departure from a spectrum unaffected by stellar influences, aid in both the qualitative and quantitative assessment of the significance of the stellar contamination on the planetary spectra. This survey effectively represents a systematic analysis of a population of 20 planets that benefit from a broad -- although not always continuous -- spectral coverage, uniform reduction, consistent atmospheric retrievals and the definition of new metrics to support or challenge previous findings related to stellar activity across a diverse set of targets.

The paper is organised as follows: in Section~\ref{sec:data_analysis} we describe the procedures we employed in our HST-dedicated data analysis pipeline to reduce and calibrate STIS and WFC3 data (Section~\ref{sec:stis_reduction} \& \ref{sec:wfc3_reduction}), the light curve detrending techniques applied to derive the final spectrum (Section~\ref{sec:lightcurve_fitting}) and a section on parameter estimation (Section~\ref{sec:parameters_estimation}) for completeness. The modelling methodology, including the variety of retrieval runs we performed on the data are described in Section~\ref{sec:spectral_models}. Section~\ref{sec:stellar_metrics} describes our use of the Bayesian evidence/Bayes factor, a well-established statistical tool (Section~\ref{sec:bayes_factor}) alongside two newly-developed metrics (Section~\ref{sec:sad}, \ref{sec:sat}) that we employed to describe the level of stellar contamination across our planetary sample. The results of our investigation are reported in Section~\ref{sec:results}, while population level findings together with three distinct case-studies are explored in more detail in Section~\ref{sec:discussion}. Finally, a summary highlighting our discoveries and their implications is presented in Section~\ref{sec:conclusions}.

\begin{figure*}[htp]
    \centering
    \includegraphics[width=\linewidth, height=0.9\textheight, keepaspectratio]{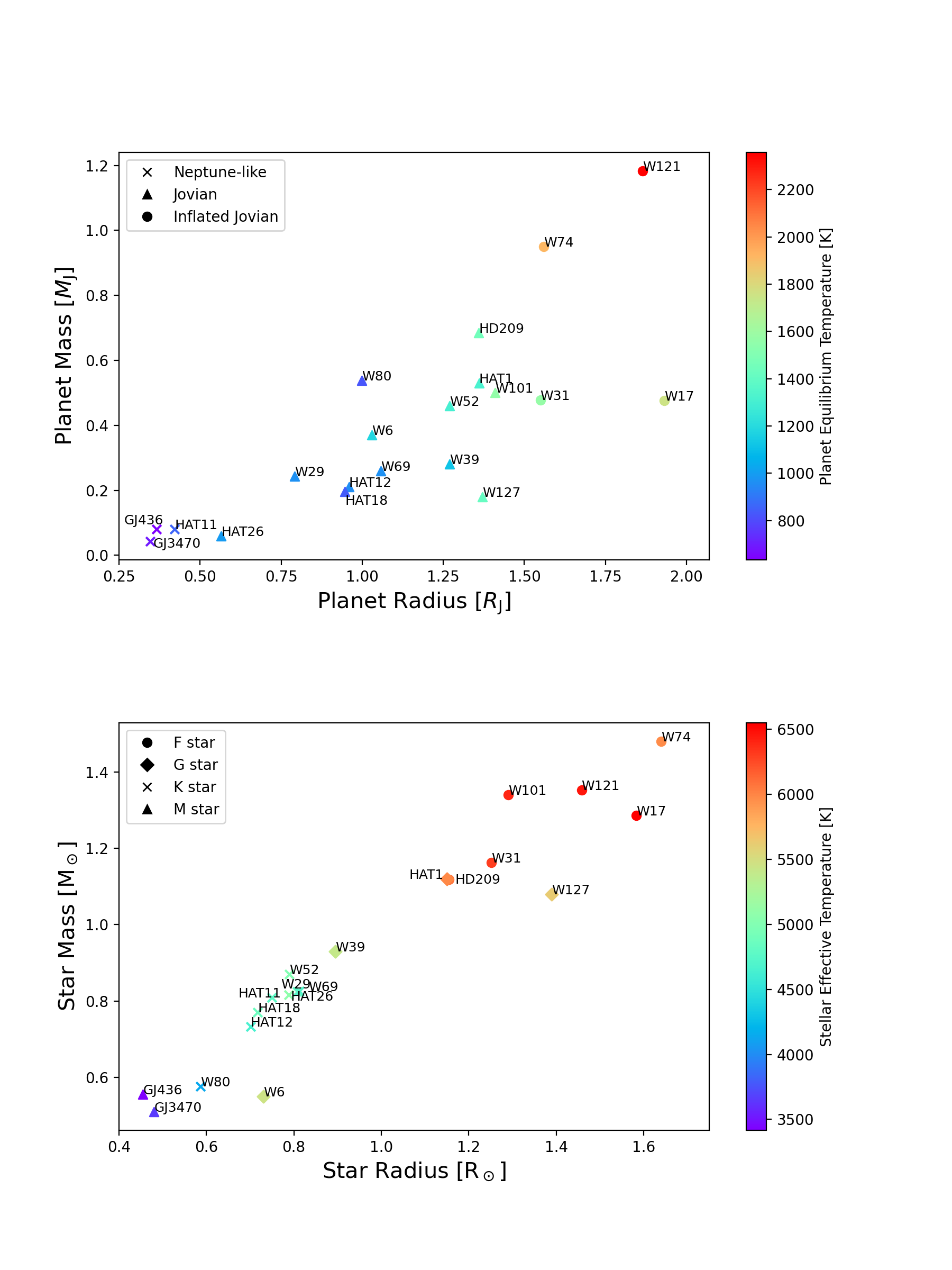}
    \caption{The sample of planets and stellar hosts investigated in this study. We defined a Neptune-like planet to have a radius between 0.31 and 0.54 R$_{\rm{J}}$, a Jovian to possess a radius between 0.54 and 1.42 R$_{\rm{J}}$ and an inflated Jovian to display an apparent radius at 10 bar of 1.42 R$_{\rm{J}}$ $<$ \textit{R$_{\rm{p}}$} $<$ 2.32 R$_{\rm{J}}$.}
    \label{fig:planets_sample}
\end{figure*}

\section{Data Analysis} 
\label{sec:data_analysis}
We searched the MAST archive for publicly available exoplanet transit data observed, with at least one grating/grism, by both the STIS and WFC3 instruments on HST. Although we were able to identify more than 30 planets that met this criteria, the data sets considered for this study refer to 20 planets only. To produce a consistent analysis and have a homogeneous data sample, we discarded HAT-P-70\,b and KELT-20\,b from the final selection as their STIS observations were taken with STIS/G430 at medium resolution. Similarly, we did not include WASP-43\,b as its G430L data set did not display a transit and its G750 observation was taken with the medium resolution grating (G750M). WASP-12\,b, despite being observed by both grisms on STIS and WFC3, was excluded from our sample due to the contamination by a nearby companion within the system. To obtain a reliable spectrum of WASP-12\,b, meticulous decontamination is essential, involving modelling of the companion star. Planetary spectra that displayed strong offsets among their STIS and WFC3 observations were also rejected. These include: HAT-P-17\,b, HAT-P-32\,b, HAT-P-41\,b, HD 189733\,b, KELT-7\,b, WASP-19\,b, WASP-62\,b, WASP-76\,b, WASP-79\,b. Extra processing steps are necessary to correct for the observed offsets and to explain their origin, therefore these planets will be the subject of a separate study. The final sample selection is displayed in Figure~\ref{fig:planets_sample}, while the observations' proposal ID and the names of their respective principal investigators are reported in Appendix~\ref{appendix:obs_info}. There, we also include the instrument-filter combination each planet was observed with and the date each observation was performed. We are aware that some of these exoplanets have also been observed with other instruments, including JWST, but for homogeneity reasons, here we restrict our analysis to data acquired with the two low resolution gratings on STIS and the infrared channel of WFC3. For the avoidance of doubt, each transit spectrum presented here was taken with all or a combination of the G430L and G750L grating on STIS, and the G102 and G141 grism on WFC3, specifically in spatial scanning mode.

While most of the STIS data sets we utilise have undergone prior analysis by different groups employing diverse data reduction pipelines, our study marks the first instance in the literature of analysing the G430L and G750L data sets of HAT-P-18\,b and the G430L data set of HAT-P-26\,b. 

In our wide-sample investigation, we introduce a novel data reduction framework for STIS data, previously presented in \cite{sabatransmission}, that enables us to uniformly process numerous data sets, with the goal of mitigating biases. WFC3-IR spatial scanned observations were also treated homogeneously across the target sample either by including in our study data sets previously analysed with the \texttt{Iraclis} pipeline or, where new observations were made available, by reducing them with \texttt{Iraclis} \citep{tsiaras2016new}. Such approach guarantees a standardised methodology for the selected sample and enables a comparative study. \texttt{Iraclis} is a standardised data reduction pipeline that has been extensively used to analyse exoplanet transit and eclipse data. The pipeline was featured in a recent comparison study with two other data analysis pipelines \citep{2024MNRAS.531...35M}. The authors of the study found that, despite using the same input data and parameters, the output spectra from these pipelines can differ substantially in some cases. An in-depth review of each processing step for each pipeline and their inter-comparison could help identify the origins of these differences and mitigate potential biases. Recently, a multi-pipeline analysis of the WASP-39\,b JWST ERS NIRISS data demonstrated that \texttt{Iraclis} produced results consistent with the other pipelines at every stage of the analysis \citep{2023Natur.614..670F}.

\subsection{STIS data processing}
\label{sec:stis_reduction}
Starting from the raw STIS data, we performed data reduction and calibration following the procedures highlighted in the STIS documentation \citep{bostroem2011stis}. The steps include: bias level subtraction (a.k.a. reference pixels correction), bias frame subtraction, dark image subtraction, background subtraction, flat-field correction, bad-pixel and cosmic-ray correction, wavelength calibration and light curves extraction. For the extraction we used 6-pixel-wide apertures along the cross-dispersion direction and smoothed aperture edges along the dispersion direction (the smoothing factors were 5 Å for the G430L grating and 10 Å for the G750L grating, corresponding to approximately 2 pixels in each case). The spectral bin sizes were chosen in such a way that roughly the same amount of flux would fall in each of them while keeping the S/N approximately constant across the spectrum. For each planetary spectra we extracted a broadband (white) light curve and a total of nine and seven spectral light curves for G430L and G750L respectively. For a number of planets, some of the chromatic light curves displayed strong outliers; we removed them by inspection before the light curve fitting step. 

\subsection{WFC3 data processing}
\label{sec:wfc3_reduction}
To ensure homogeneity in the data analysis process, we employed one single pipeline to analyse all transit data sets. \texttt{Iraclis} \citep{tsiaras2016new}, a specialised WFC3 spatial scanning mode data reduction pipeline and recently adapted to analyse STIS data as well, was used in this study. Most of the WFC3 data sets presented here were previously analysed with \texttt{Iraclis} and collected in two major population studies by \cite{tsiaras2018population} and \cite{edwards2022exploring}. However, these studies focused predominately on WFC3/G141 data, while here we aim to include WFC3/G102 as well. For this reason, we downloaded all publicly available WFC3/G102 data sets for the planets in our sample and processed them with \texttt{Iraclis}. We ensured that the planetary system parameters (Appendix~\ref{appendix:planet_params}) we employed were consistent with those used in previous WFC3/G141 studies and across the remaining data sets associated with the same planet. The planets observed with G102 in transmission include HAT-P-11\,b, HAT-P-26\,b, WASP-17\,b and WASP-39\,b. These are also the planets that possess a complete spectrum from 0.4 to 1.6 $\mu$m, i.e. were observed with STIS/G430L, STIS/G750L, WFC3/G102 and WFC3/G141, albeit at different epochs. 

As highlighted in \cite{tsiaras2016new}, \texttt{Iraclis} starts from the raw data and performs a number of steps (zero-read subtraction, reference pixels correction, non-linearity correction, dark current subtraction, gain conversion, sky background subtraction, calibration, flat-field correction, bad-pixel and cosmic-ray correction) before extracting the light curves. Both the white and spectral light curves are then fitted with the \texttt{PyLightcurve} package embedded into the pipeline (more details in Section \ref{sec:lightcurve_fitting}), to produce a final spectrum. Similarly to STIS, we chose the spectral bin sizes in such a way that roughly the same amount of flux would fall in each bin while keeping the S/N approximately constant across the whole spectrum.

\subsection{Light curve fitting}
\label{sec:lightcurve_fitting}
The light curve modelling parameters were pulled from the Exoplanet Characterisation Catalogue developed as part of the ExoClock project \citep{Kokori_2021, kokori2022exoclock, kokori2023exoclock}. In each of the light curve modelling steps we assumed a circular orbit and a fixed period for the planet, while the limb-darkening effect was modelled using the Claret 4-coefficient law \citep{Claret2000} and the \texttt{ExoTETHyS} \citep{morello2020exotethys} package, which takes into account the stellar parameters (Appendix~\ref{appendix:stellar_params}) and the response curve of the instrument. The white light curves were fitted with a function \textit{M(t)} that includes a transit model \textit{F(t)} and a systematics model \textit{R(t)} multiplied by a normalisation factor $n_w$, of the form:
\begin{equation}
    M(t) = n_w \cdot F(t) \cdot R(t) \ .
\end{equation}
The transit model, computed via the \texttt{PyLightcurve} software\footnote{\url{https://github.com/ucl-exoplanets/pylightcurve}}, follows the formalism of \cite{mandel2002analytic} and is dependent on the limb-darkening coefficients, the planet to star radii ratio and the orbital parameters $T_0$, $P$, $i$, $a/R_*$, $e$, $\omega$. For the systematics model \citep{kreidberg2015detection, tsiaras2016new}, we include a linear term to describe the long-term ramp and an exponential term to correct for the trend at the beginning of each HST orbit. The systematics function is implemented as follows:
\begin{equation}
    R(t) = (1 - r_a(t - T_0)) (1-r_{b1} e^{-r_{b2}(t-t_0)}) \ ,
\label{eq:systematics}
\end{equation} 
where $t$ is time, $T_0$ is the mid-transit time, $t_0$ is the time when each orbit starts, $r_a$ is the slope of the linear, long-term “ramp” and $r_{b1}$, $r_{b2}$ are the coefficients of the exponential short-term “ramp”. 

While the white light curves were fitted using the above technique for both WFC3 and STIS, we used two different fitting routines to treat WFC3 and STIS spectral light curves. After excluding the initial orbit from each WFC3 spectral light curve, as it exhibits more pronounced wavelength-dependent systematics compared to the subsequent orbits, we employed Equation~\ref{eq:systematics} to model the remaining orbits comprising the spectral light curves. The systematics-modelled light curves were then divided by the corresponding white light curve and further fitted via
\begin{equation}
    n_{\lambda}(1+\chi_{\lambda}(t-T_0))(F(\lambda, t)/F_{w}(t)) \ ,
\end{equation}
where $n_{\lambda}$ is a wavelength-dependent normalisation factor, $\chi_{\lambda}$ is the coefficient of the wavelength-dependent linear slope, $t$ is time, $T_0$ is the mid-transit time, $F(\lambda, t)$ is the wavelength dependent transit model, and $F_{w}(t)$ is the best-fit model on the white light curve.

Similarly to WFC3, we attempted to perform the detrending of the STIS spectral light curves with the divide-white method \citep{2015ApJ...814...66K} but, likely due to chromatic effects arising from the host star and the well-known wavelength-dependent thermal breathing of HST \citep{2001ApJ...552..699B, 2011MNRAS.416.1443S}, this approach was unsuccessful. A variety of spectroscopic STIS studies have highlighted how the systematics can be both time-dependent \citep[e.g.][]{2014MNRAS.437...46N, 2018AJ....155...66L} and wavelength-dependent during a single observation \citep{2013MNRAS.434.3252H}, therefore needing a specialised treatment case by case. However, we opted for a uniform systematics model throughout all data sets to maintain consistency: to correct for the combined instrumental and astrophysical trends characteristic of the STIS spectral light curves, we adopted the usual systematics model in Equation~\ref{eq:systematics} and applied it to each spectral light curve separately. Effectively, we modelled the spectral light curves in the same manner as the white light curve. This approach has been used extensively in past STIS studies and has been shown to perform better than employing the divide-white technique \citep[e.g.][]{2019AJ....158..244C}. It has consistently proven effective in detrending all data within our sample, thereby ensuring reliable and comparable results throughout.

In the literature there are many examples of how Gaussian Processes (GPs) are employed to fit exoplanet transit light curves \citep{2020AJ....160..240G}, especially at the optical wavelengths \citep{2017MNRAS.467.4591G}. Generally these algorithms are able to smooth out external features that do not belong to the phenomenon intended to be modelled, in this case by removing any potential spot-crossing signatures from the transit signal. Hence, we attempted to utilise them to detrend STIS light curves as reported in Appendix~\ref{appendix:GPs}. However, we found GPs to be ineffective in providing additional constraints on the light curve modelling. Therefore, all spectral data presented here and employed during the retrievals were produced by applying the standard linear times exponential detrending described in Equation~\ref{eq:systematics}.

\subsection{Parameters estimation: \texorpdfstring{$a/R_*$ and $i$}{a/R* and i}}
\label{sec:parameters_estimation}
During the white light curve fitting we fixed the $a/R_*$ and $i$ parameter values to those obtained from literature. Because of the gaps in the light curves, HST observations are not suited to constrain these parameters. A significant coverage in ingress and/or egress is necessary to constrain the semi-major axis and orbital inclination. To test whether combining many HST light curves could be a suitable approach to derive these parameters in the absence of complete light curves, we compared the values obtained by combining three STIS white light curves with those obtained by fitting a complete JWST/NIRISS light curve. We selected WASP-39\,b as our candidate planet, as at the time of writing this is the only planet in our sample for which there exist public JWST observations. We used again Equation~\ref{eq:systematics} to detrend the light curves, but leaving $a/R_*$ and $i$ as free parameters. We compared the results obtained using these two approaches with the values listed in ExoClock, and we conclude that the fitting of JWST light curves leads to the most constrained results, which are compatible well within 1$\sigma$ of the ExoClock parameters. The $a/R_*$ and $i$ values obtained from the combined STIS light curve fitting ($a/R_{*}$ = ${11.8} \pm 0.2$, $i$ = ${88.6}_{-0.3}^{+0.5}$) have a ten times larger uncertainty compared to the JWST fitted values ($a/R_{*}$ = 11.42$\pm$0.03, $i$ = 87.77$\pm$0.03, $b$ = 0.444$\pm$0.006) but they are comparable to the ExoClock uncertainties ($a/R_{*}$ = ${11.4} \pm 0.2$, $i$ = ${87.8}^{+0.3}_{-0.2}$). The semi-major axis and inclination parameters derived from the fitting of the three STIS light curves are 2$\sigma$ and 3$\sigma$ away respectively from both JWST and ExoClock, while JWST and ExoClock parameters agree within 1$\sigma$. Therefore, rather than trying to constrain these orbital parameters by fitting multiple incomplete light curves together, we suggest employing either tabulated parameters or fitted parameters from complete white light curves and fixing them during the spectral light curve fitting. For consistency we fitted all white and chromatic light curves using the ExoClock parameters as fixed values across the whole planetary sample, including on WASP-39\,b. 

\begin{figure*}[htp]
    \centering
    \includegraphics[width=\linewidth, height=0.8\textheight, keepaspectratio]{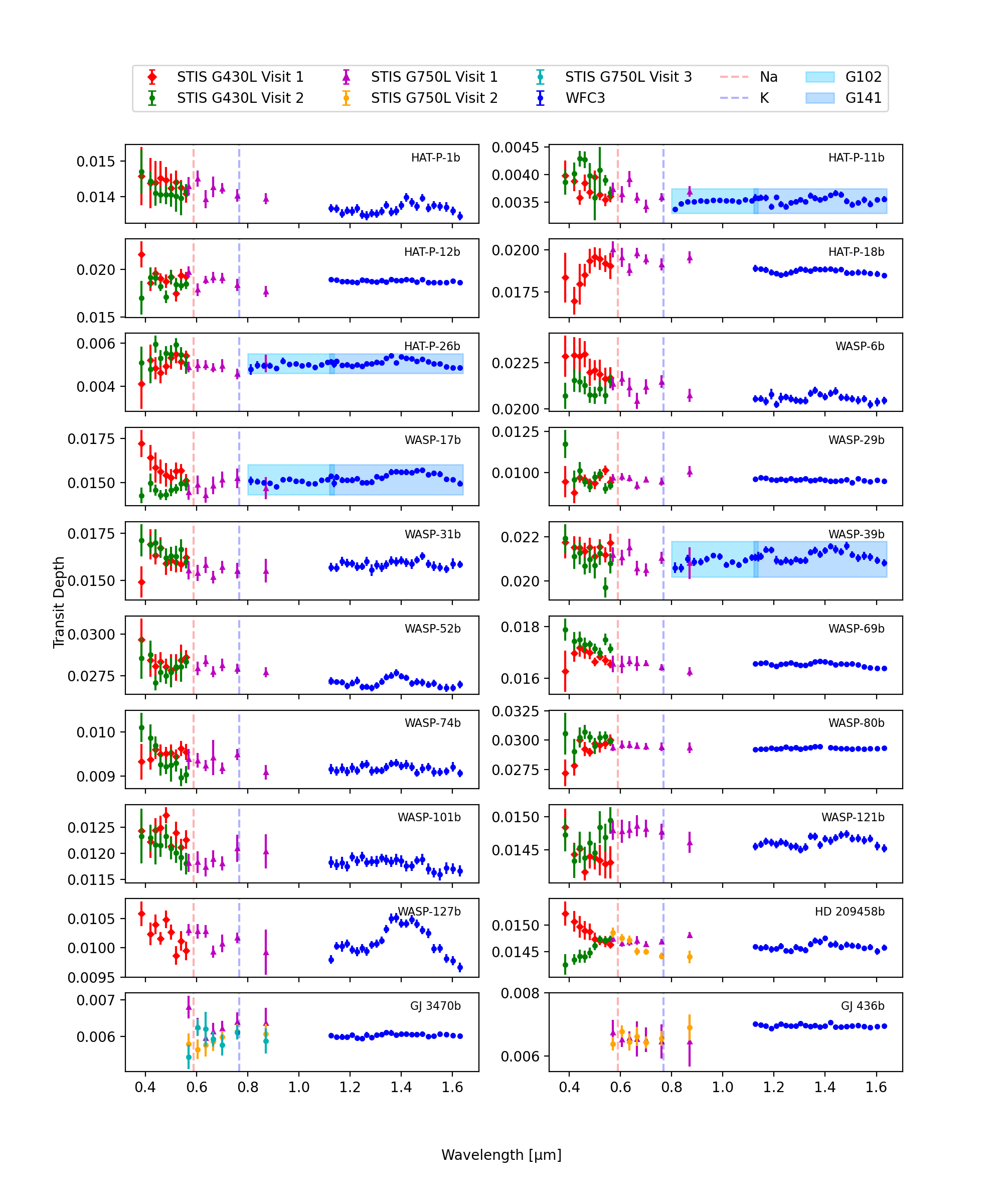}
    \caption{The spectral data resulting from our data analysis, colour-coded based on the observation timestamps of each STIS visit. ‘Visit 1’ denotes the first observation in chronological order conducted using the respective STIS grating, followed by ‘Visit 2’ etc.. For HAT-P-1\,b, HAT-P-26\,b, WASP-17\,b, and WASP-39\,b, we possess data from both the G102 and G141 grisms on WFC3. These data sets are highlighted to illustrate the corresponding data points and to visualise the overlap between the grisms. Conversely, observations for the remaining planets were exclusively conducted using the G141 grism on the WFC3 instrument. The dashed lines centred at 0.59 $\mu$m and 0.77 $\mu$m indicate where we expect the peaks of the absorption features of sodium and potassium to appear respectively.}
    \label{fig:all_spectra}
\end{figure*}

\section{Spectral Modelling} 
\label{sec:spectral_models}
Due to the often divergent STIS spectra resulting from their acquisition at varying epochs (Fig.~\ref{fig:all_spectra}), we decided to model the different combinations of data sets as separate spectral scenarios for each planet. As reported in Table~\ref{tab:cases}, each of the potential combinations of data sets is identified by a Case number. The spectral modelling is performed with the fully-Bayesian retrieval framework TauREx\,3.1 \citep{Al_Refaie_2021, 2022ApJ...932..123A}. The software has been utilised in a variety of modelling comparison projects \citep{2020MNRAS.493.4884B, 2022ExA....53..447B, 2024Natur.626..979P, 2024PSJ.....5...64V}, producing results that align well (within 1$\sigma$) with other retrieval tools. The robustness and consistency of TauREx 3.1 highlights its reliability as an atmospheric retrieval tool. Through the use of nested sampling techniques \citep{skilling2006nested}, the retrieval code maps the correlations among the atmospheric parameters in the defined parameter space. TauREx operates on an entirely open-source framework. Utilising the plug-in system, users have the flexibility to integrate their own models, allowing for the representation of various atmospheric conditions such as stellar contamination, cloud parameterisation, and temperature-pressure profiles of their choosing. To explore the parameter space, we employed the widely utilised nested sampling algorithm MultiNest \citep{feroz2009multinest, buchner2014x} with 250 live points and an evidence tolerance set to 0.5.

\begin{table}[hbp]
    \centering
        \caption{Data sets utilised to define each Case number. Each STIS visit indicated here is colour-coded in Fig.~\ref{fig:all_spectra}.}
    \begin{tabular}{|c|c|c|c|c|}
    \hline 
        Planet & Case 1 & Case 2 & Case 3 & Case 4\\
        \hline \hline
        \multirow{2}{6em}{GJ 3470\,b} & STIS G750L visit 1 & STIS G750L visit 2 & STIS G750L visit 3 & - \\
                                    & WFC3 & WFC3 & WFC3 & \\
        \hline
        \multirow{2}{6em}{GJ 436\,b} & STIS G750L visit 1 & STIS G750L visit 2 & - & - \\
                                    & WFC3 & WFC3 & & \\
        \hline                            
        \multirow{3}{6em}{HD 209458\,b} & STIS G430L visit 1 & STIS G430L visit 1 & STIS G430L visit 2 & STIS G430L visit 2 \\
        & STIS G750L visit 1 & STIS G750L visit 2 & STIS G750L visit 1 & STIS G750L visit 2 \\
        & WFC3 & WFC3 & WFC3 & WFC3 \\
        \hline
        \multirow{3}{6em}{Rest of the sample} & STIS G430L visit 1 & STIS G430L visit 2 & & \\
        & STIS G750L visit 1 & STIS G750L visit 1 & - & - \\
        & WFC3 & WFC3 & & \\
        \hline
    \end{tabular}
    \label{tab:cases}
\end{table}

\subsection{The \texttt{ASteRA} retrieval plugin}
\texttt{ASteRA} is a stellar activity plugin for TauREx 3.1, designed to account for photospheric heterogeneities in retrievals. This module allows for modelling the host star not only as a homogeneous body characterised by a single temperature and SED but also as an active star composed of multiple temperature components: the quiet photosphere, starspots, and faculae. The spectral emission of the heterogeneous star is modelled using BT-Settl models \citep{2012RSPTA.370.2765A}, while the stellar spectral grid is generated using the PHOENIX library \citep{husser2013new}. In essence, starspots and faculae are treated as separate cooler and hotter stars, respectively. The stellar emission densities (SEDs) from these three surface components (spots, faculae, and quiet photosphere) are combined based on their covering fractions to produce the observed disk-integrated stellar spectrum:
\begin{equation}
    S_{\text{star}, \lambda} = \left((1-F_{\text{spot}}-F_{\text{fac}}) \times S_{\text{phot}, \lambda}\right) + \left(F_{\text{spot}} \times S_{\text{spot}, \lambda}\right) + \left(F_{\text{fac}} \times S_{\text{fac}, \lambda}\right) \ .
    \label{eq:star}
\end{equation}
Here, \(S_{\text{star}, \lambda}\) is the flux of the heterogeneous star at a given wavelength, while \(S_{\text{phot}, \lambda}\), \(S_{\text{spot}, \lambda}\), and \(S_{\text{fac}, \lambda}\) represent the spectra of the quiet photosphere, starspots, and faculae, respectively. The parameters \(F_{\text{spot}}\) and \(F_{\text{fac}}\) denote the covering fractions of the spots and faculae relative to the observed stellar disk. According to Eq.~\ref{eq:star}, the heterogeneous star model defines the spot and facula temperatures (\(T_{\text{spot}}\) and \(T_{\text{fac}}\)), allowing for the calculation of their temperature contrasts relative to the quiescent photosphere. Additionally, the model derives the filling factors for spots and faculae, representing the fraction of the visible stellar disk covered by their projected areas (\(F_{\text{spot}}\) and \(F_{\text{fac}}\)). As a result, modelling an active star requires introducing two additional fitting parameters (for either spots or faculae) or four (when both are included) compared to a standard retrieval: the spot temperature (\(T_{\text{spot}}\)) and its filling factor (\(F_{\text{spot}}\)), as well as the facula temperature (\(T_{\text{fac}}\)) and its filling factor (\(F_{\text{fac}}\)).

\texttt{ASteRA} relies on the fundamental stellar parameters, effective temperature, metallicity, and \(\log(g)\), to select the corresponding PHOENIX spectra, which are fixed within the retrieval. The positional dependence of spots and their interaction with limb-darkening effects has been thoroughly investigated by \cite{2024ApJ...960..107T}, who found that these effects become significant only in cases of high stellar activity. Given the low-resolution nature of HST observations, we opted for the simplified two-parameter model (temperature and covering fraction) for both spots and faculae.

\subsection{Retrievals setup} \label{sec:retrieval_setup}
All planets in the sample were treated homogeneously: following a plane-parallel geometry, their atmospheres were split into 100 layers, uniformly distributed in log space between $10^{-5}$ and $10^{6}$ Pa. We assume that all planets possess a primary atmosphere, mainly constituted of atomic hydrogen and helium in a ratio of 1 to 0.17. For simplicity, we assumed the temperature structure to be constant with altitude for two main reasons: 1) the goal of this study is not to constrain the structure of the $T-p$ profile; 2) introducing a more complex $T-p$ profile would involve retrieving additional correlated parameters, potentially introducing degeneracies amongst themselves and other parameters of interest.

Taking advantage of the considerable wavelength coverage afforded by the combination of STIS and WFC3 observations, the high equilibrium temperature of some planets in the sample and for comparison with existing studies, we initially investigated the presence of various molecular opacities. These included TiH \citep{gharib2021exoplines}, ScH, CrH \citep{2022RASTI...1...43B}, TiO \citep{mckemmish2019exomol}, VO \citep{mckemmish2016exomol}, AlO \citep{patrascu2015exomol}, K \citep{allard2019new} Na \citep{allard2019new} and H$_{2}$O \citep{10.1093/mnras/sty1877}, employing linelists sourced from the ExoMol \citep{2016JMoSp.327...73T, chubb2021exomolop}, HITEMP \citep{rothman2010hitemp, rothman2014status}, HITRAN \citep{rothman1987hitran, gordon2022hitran2020} and NIST \citep{kramida2013critical} databases. Additionally, we explored different cloud parameterisations, encompassing wavelength-independent clouds, parametric Mie scattering clouds \citep{lee2013atmospheric}, cloud microphysics and radiative transfer models utilising \texttt{YunMa} \citep{2023ApJ...957..104M}, and accounted for the presence of stellar contamination signatures with \texttt{ASteRA} \citep{2024ApJ...960..107T}.

Upon realising the importance of accounting for potential stellar contamination and observing its predominant influence at a population level, surpassing that of clouds or refractory species, we opted against incorporating more intricate cloud microphysics models in this study. This decision was further influenced by the low resolution of the STIS observations and the expected parameter degeneracies arising from the interplay between cloud and stellar models which should instead be the focus of a separate, future study. Ultimately, we decided to consider in our retrievals potential stellar contamination by the host star as modelled by \texttt{ASteRA} \citep{2024ApJ...960..107T}, a simple gray cloud layer and H$_2$O, K and Na as trace gases. These minor species, while set to have a constant abundance at each atmospheric layer, were allowed to vary freely between -12
and -1 in volume mixing ratio (VMR). Additionally, collision-induced absorption (CIA) from H2-H2 \citep{PMID:21207941, Fletcher_2018} and H2-He \citep{PMID:22299883} and Rayleigh scattering for all molecules were included.

To assess the strength of the stellar contamination on each exoplanet spectrum and understand how it affects the retrieved results, we progressively made the retrieval model more complex. We organised the retrievals as follows:
\begin{itemize}
    \item \textbf{Base model}: Here we keep the model as simple as possible by only including water vapour absorption \citep{10.1093/mnras/sty1877}, CIA and Rayleigh scattering.
    \item \textbf{Gray clouds}: We expand the base model by including a simple gray cloud deck, for which we fit the top pressure in the range from $10^{-1}$ to $10^{6}$ Pa \citep{robinson2014common, kawashima2018theoretical, charnay2021formation}. 
    \item \textbf{Alkali metals}: Certain spectra exhibited pronounced features near the regions where sodium and potassium are expected to absorb. Where a potential detection of one or both of these elements was considered feasible, we included their opacities \citep{allard2019new} into the retrieval run, while retaining the presence of water absorption and the opaque cloud deck.
    \item \textbf{Stellar activity}: We conduct the three type of retrievals as previously outlined. However, in each of these retrievals, we now incorporate a stellar model capable of inferring both the temperature and filling factor of spots and faculae potentially present on the photosphere of the stellar host.
\end{itemize}

During the retrievals we fixed the planetary masses to their literature values (Table~\ref{tab:planet_params}). The planetary radii, temperatures and the \ce{H2O} volume mixing ratio were set as fitting parameters for which we only imposed wide, non-restrictive priors (0.1 to 3 for $R_{\text{p}}$; 100 to 5500 for $T_{\text{p}}$; $10^{-12}$ to $10^{-1}$ for molecular opacities) in order to avoid biasing the retrieval towards expected, a-priori values.
For comparison with our retrieved $T_{\text{p}}$ values, we calculated the planetary equilibrium temperature of each planet in our sample using the following equation:
\begin{equation}
    T_\text{p} = T_* \sqrt{\frac{R_*}{2a}} \left( \frac{1-A}{\epsilon} \right)^{\frac{1}{4}} \ ,
\label{eq:planet_temp}
\end{equation}
where $T_*$ is the host star’s effective temperature, $R_*$ is the stellar radius, $R_*/a$ is the inverse of the planet’s semi-major axis to stellar radius ratio. When calculating $T_{\text{p}}$, we set the albedo ($A$) and the emissivity ($\epsilon$) to be 0.2 and 0.8 respectively. The error bars associated with the equilibrium temperatures presented in Table~\ref{tab:planet_params} were computed under the assumption of different albedo and emissivity values. For the lower bound we adopted $A = 0.7$ and $\epsilon = 0.8$, indicating a planet with high reflectivity and radiative properties. For the upper bound, assuming a condition with low reflectivity but a high heat redistribution potentially resulting from the planet's formation process, we employed $A = 0.1$ and $\epsilon = 0.5$. In all of our retrievals, we modelled the stellar photosphere using the PHOENIX BT-Settl synthetic stellar atmosphere spectra \citep{2012RSPTA.370.2765A}. We fixed the fundamental stellar parameters, the stellar radii ($R_*$) and masses ($M_*$), photospheric effective temperatures ($T_*$) and metallicities to literature values, as reported in Table~\ref{tab:stellar_params}. To account for potential stellar contamination within the retrieval, four additional parameters relating to stellar activity are fit for alongside the planetary parameters of interest. These activity parameters are the spot and facula temperatures, $T_\text{spot}$ and $T_\text{fac}$ respectively (assuming that all spots/faculae are characterised by a single temperature), and their corresponding filling factors i.e. coverage as a fraction of the total projected stellar surface, which are denoted as $F_\text{spot}$ and $F_\text{fac}$. The prior bounds on the filling factors $F_\text{spot}$ and $F_\text{fac}$ were set between 0.0 and 0.9, where a filling factor of 0.9 would equate to 90\% of the photosphere being covered by either of these active regions, with the implicit condition that the sum of these two parameters cannot exceed 1. We set prior bounds for the temperatures of these active regions that are dependent on the effective (photospheric) temperature ($T_*$) of the host star and, as such, their absolute values vary on a case-by-case basis. For the spot temperature, $T_\text{spot}$, we set the lower bound to be equal to $T_*$ -- 1500K and the upper bound to be equal to $T_*$. For the facula temperature, $T_\text{fac}$, we set the lower bound to be equal to $T_*$ and the upper bound as $T_*$ + 500K. The reasoning behind imposing priors of different magnitudes on the faculae and spot temperatures is that we expect faculae to display a smaller positive temperature contrast ($\Delta T$) with respect to the photosphere as compared to the larger negative contrast of spots \citep[e.g.][]{berdyugina2005,gondoin2008,panja2020,norris2023}.

\section{Stellar Activity metrics}
\label{sec:stellar_metrics}

\subsection{Bayes factor}
\label{sec:bayes_factor}
To assess the strength of the model preference from a Bayesian perspective, we utilise the Bayes factor, defined as the ratio of two models' evidences given an observed data set. In our study we compare the evidence of the baseline model to that of the stellar baseline model for each planetary spectra, considering one spectral case at a time. The fitted parameters for the base model include water vapour, $R_{\text{p}}$, and $T$. Conversely, the stellar model incorporates these three parameters plus four additional activity parameters, as described in Section \ref{sec:retrieval_setup}. Thus, the dimensionality of the stellar model, i.e. the numbers of parameters in the model, is increased by four. All models used in the Bayes factor calculations are cloud-free and assume a uniform atmospheric temperature and water content with altitude. By employing the simplest atmospheric models we aim to distinguish at first order the signatures that stellar activity imprints on the planetary spectra. The Bayes factor serves as a fully Bayesian approach to model selection, relying solely on the evidence of the respective models. In log space it is expressed as follows:
\begin{equation}
\label{eq:bayes_factor}
    \ln(B) = \ln(E_\text{S}/E_\text{B}) =  \ln(E_\text{S}) - \ln(E_\text{B}) \ ,
\end{equation}
where $\ln(E_\text{S})$ represents the Bayesian evidence of the stellar model and $\ln(E_\text{B})$ corresponds to that of the base model. A negative ${\ln(B)}$ value suggests a preference for the base model, while a positive value indicates a preference for the model incorporating stellar activity, despite the additional dimensionality it requires. Following the Jeffreys' scale \citep{jeffreys1998theory, trotta2008bayes} we establish three thresholds to gauge the extent to which the stellar model is favoured over the baseline model (or vice versa). A Bayes factor 1 $\leq$ $\ln(B)$ $\leq$ 2.5 represents a weak evidence against the simpler base model; 2.5 $\leq$ $\ln(B)$ $\leq$ 5 is considered to be moderate evidence in favour of including stellar contamination and $\ln(B)$ $\geq$ 5 is considered strong evidence in favour of using a model that incorporates stellar activity. Similarly, the Jeffreys' scale can be used to explain the preference of the base model over the stellar model when the Bayes factor is negative. For instance, a $\ln(B)$ $\leq$ -5 signifies a strong preference for the base model over the active star model. A Bayes factor falling between -5 and -2.5 suggests a moderate preference for the baseline model, while a $\ln(B)$ ranging from -2.5 to -1 indicates weak evidence for it. The strength of the evidence becomes inconclusive for values 1 $\geq$ $\ln(B)$ $\geq$ -1, effectively implying that both models are equally capable of explaining the observed data. The Bayes factor is a robust indicator for goodness-of-fit and an essential criterion to discriminate against models. Arguably, the greatest strength of the Bayes factor is its tendency to favour simpler models over complex ones in order to prevent over-fitting. This is essential to make sure that our retrieval model is not biased towards finding stellar activity if there is not sufficient evidence for it. Due to the way Bayesian frameworks naturally interact with low-resolution spectra and their associated uncertainties, we explore additional, newly defined metrics. These metrics help provide a broader perspective on the activity environment by examining the contexts in which stellar activity plays a significant role. They are also meant to discern the extent of stellar activity on a qualitative scale, such as distinguishing between low and high activity regimes. A complete overview of these metrics will be given in Thompson et al. (in prep.), where a more extensive description will be presented alongside further applications.

\subsection{Stellar Activity Distance metric (SAD)}
\label{sec:sad}
The Bayesian approach to model selection tends to favour data sets with smaller uncertainties, prioritising precise data during the fitting routine. Consequently, STIS observations, which typically exhibit larger error bars compared to WFC3 due to their lower S/N, will carry less weight during the retrieval process, leading the model to assign less significance to the STIS data points. However, since stellar activity tends to imprint stronger signals within the wavelength range covered by STIS, it is crucial to evaluate how well the retrieval aligns with observations in the visible and to extract as much of the information content of this region as we can. Solely relying on STIS data sets for retrievals is not advisable; meaningful results require anchoring the optical observations to WFC3 data where the bulk of the planetary information content lies. Without the continuum provided by the near-infrared data, the task of constraining basic planetary parameters becomes exceedingly challenging, resulting in larger uncertainties, increased model and parameter degeneracies and a tendency for the retrieval to be pushed towards exploring unphysical territories. Hence, we introduce a metric that, focusing exclusively on the optical regime, aims at evaluating which retrieval model (with or without activity) better aligns with STIS observations without the need for running further retrievals or altering the input observation in any way.

We select solely the data points $<$1.0 $\mu$m since this is the wavelength range where stellar activity has the potential to induce the strongest chromatic effects. Due to the often contrasting STIS observations when multiple data sets are available for the same grating, we consider each Case on its own. For each Case we calculate the absolute difference of each observational data point from the stellar activity model and, separately, the absolute difference from the base model. We apply standard error propagation to calculate the uncertainty in the absolute difference, using the transit depth and model 1$\sigma$ errors. The calculations effectively return an absolute difference and an associated error per data point related to both the base model and the stellar model. As an example, if we take the spectrum of WASP-127\,b and we consider the 16 datapoints with wavelengths bluer than 1 $\mu$m (see Fig. \ref{fig:all_spectra}), we end up with sixteen absolute distances from the base model and sixteen absolute distances from the stellar model, for a total of thirty-two absolute distances. %$D_{i}$. 
By considering each spectral Case separately, we then calculate the average of all distances from each model and its uncertainty obtaining an average distance related to the base model ($\bar{D}_\text{B}$) and one referring to the stellar model ($\bar{D}_\text{S}$). Finally, the stellar activity distance metric (\textit{SAD}) is defined as the ratio between the two mean absolute distances
\begin{equation}
    SAD = \bar{D}_\text{B} / \bar{D}_\text{S}
\end{equation}
with an uncertainty of 
\begin{equation}
    \sigma_{SAD} = \left(\frac{\bar{D}_\text{B}}{\bar{D}_\text{S}}\right) \cdot \sqrt{\left(\frac{\sigma_{\bar{D}_\text{S}}}{\bar{D}_\text{S}}\right)^2 + \left(\frac{\sigma_{\bar{D}_\text{B}}}{\bar{D}_\text{B}}\right)^2} \ .
\end{equation}
The greater the \textit{SAD}, the more indicative it is of a strong preference for the stellar activity model, as it aligns more closely with the data. Conversely, a \textit{SAD} below 1 means that the base model best describes the data and a \textit{SAD} equal to 1 means that either model is good enough at explaining the observations. One of the unavoidable limitations of the SAD with STIS and WFC3 observations is that the associated uncertainties will always be fairly substantial, due to the large error bars on the observations themselves.

\subsection{Stellar Activity Temporal metric (SAT)}
\label{sec:sat}
To assess the repeatability and consistency of observations for each planet, we examine how the data sets acquired with the same STIS grating compare to each other. We note that stellar activity is not the only phenomenon which could be responsible for spectra taken at different epochs diverging from one another. Other temporally-variable processes e.g. heterogeneous, patchy clouds \citep{macdonald2017} or the influence of macro-scale planetary weather/storms \citep{skinner2021} could also contribute. However, stellar activity is likely the most plausible cause of such strong deviations observed for some of the planets in the sample over these relatively short timescales. This is because the activity modulation is synchronised primarily with the timescales of stellar rotation. This is particularly evident in cases where a reversal in the optical slope is observed between subsequent STIS visits, as seen, for example, in WASP-6\,b, WASP-17\,b, WASP-80\,b, and HD 209458\,b. Such reversals may indicate the host star transitioning between regimes dominated by starspots and those dominated by faculae. This metric cannot be used for WASP-127\,b and HAT-P-18\,b since we possess only one observation per grating for each of these planets. In this context, we disregard retrieval models and concentrate purely on observational data. Our aim is to compare two or more observations of the same planet normalised to a common factor. To maintain uniformity across the entire sample, we use the WFC3 data set of each planet as the normalisation factor for all its STIS observations considered for this metric. For each planet, the normalisation factor is determined as the weighted mean of its WFC3 data set, with weights assigned inversely proportional to the transit depth error bars, thereby giving less weight to larger error bars. Finally, we compute the average of the set of absolute normalised differences per pair of observations ($o_{1i}$ and $o_{2i}$), which we define as the \textit{SAT}:
\begin{equation}
\label{eq:sat}
    % SAT = \frac{1}{n} \sum_{i=1}^{n} x_i  = \frac{1}{n} \sum_{i=1}^{n} \frac{|o_{1i} - o_{2i}|}{\bar{n}_\text{WFC3}} \ ,
    SAT = \frac{1}{n} \sum_{i=1}^{n} \frac{|o_{1i} - o_{2i}|}{\bar{n}_\text{WFC3}} \ ,
\end{equation}
where ${\bar{n}_\text{WFC3}}$ is the normalisation factor. The uncertainty of the metric is given by
\begin{equation}
    \sigma_{SAT} = \frac{1}{n} \sqrt{\sum_{i=1}^{n} \sigma_{x_i}^2} \ ,
\end{equation}
where we define $x_i$ to be $|o_{1i} - o_{2i}|$ / ${\bar{n}_\text{WFC3}}$ for simplicity.
This methodology effectively addresses strong variations among data sets, particularly when the difference among the means are comparable but their extremes diverge significantly. 

For individual planets, determining the average deviation in transit depth in ppm between subsequent STIS visits would be informative on its own. However, to perform comparative planetology we need a method of normalising these differences across our entire population. As each of these planets have atmospheres characterised by different scale heights, for comparability we report the absolute differences as a percentage of the weighted average transit depth of the WFC3 observations, which are available for every planet in our sample. Our assumption here is that at first order, the weighted average transit depth taken from the WFC3 observation is a reasonable proxy for scale height effects, although it will of course be slightly influenced by any absorption features present in that region.

We calculate the \textit{SAT} for three different GJ 3470\,b combinations of its G750L data sets, while only one possible \textit{SAT} value is attainable for GJ 436\,b. For HD 209458\,b we calculate one \textit{SAT} for G430L and one for G750L, both normalised to WFC3. In total, these combinations produce three \textit{SATs} for GJ 3470\,b, two \textit{SATs} for HD 209458\,b, and one \textit{SAT} for GJ 436\,b.

\section{Results} 
\label{sec:results}
\subsection{Retrieved temperature and water abundance}
\label{sec:T_h2o_results}
As outlined in Section~\ref{sec:retrieval_setup}, multiple models incorporating various parameter combinations have been tested on each planetary spectrum. Since we considered multiple data set combinations per planet, we present the observations employed in each Case number in Table~\ref{tab:cases}. In Appendix~\ref{sec:retrieval_results}, Tables~\ref{tab:GJ 436b_retrieval_results} to \ref{tab:WASP-127_retrieval_results} we report for each retrieval model the setup used (for example, if we include stellar activity or not), the molecules considered (\ce{H2O}, K, Na or a combination of the three) and if a layer of opaque, grey clouds was added. Additionally, we present the retrieved values of $T$, $R_{\text{p}}$, $\log_{10}$(H$_2$O) and $\ln(E)$ with their respective uncertainties. When the retrieval process is able to fit multiple models for the same setup and spectral Case, we present both solutions for completeness. An example of multiple solutions on the same Case number and model setup is given in Table~\ref{tab:GJ 3470_retrieval_results}.

In the early stages of our investigations, we identified the need to integrate stellar contamination into our retrieval models. This was particularly necessary to explain the scenarios in which observations conducted at different epochs result in distinct spectral features; features that cannot reasonably be attributed to, for instance, refractory elements alone. We observed that the inclusion of stellar activity in its simplest model formalisation was already enough to alter the retrieval outcomes; on some of the targets we considered, stellar contamination is crucial for aligning the retrieved water abundances with those reported in the literature and ensuring temperature values consistent with planetary equilibrium temperatures. Clearly, when adjusting the complexity of the simplest models (base or stellar activity model) by including additional molecular absorbers, eventually we found parameter combinations that are better able to explain the observed spectral features. However, to ensure clarity, consistency, and uniformity across the entire sample, our discussion of results will specifically concentrate on the simplest baseline and stellar activity correction model. Furthermore, at a population level we analyse just the retrieved temperature and water abundance values. Individual planets are discussed separately in Appendix~\ref{appendix:individual_planets}, where we provide a brief overview of each exoplanet and frame our results in the context of existing literature studies.

\begin{figure*}[htp]
    \centering
    \includegraphics[width=\textwidth]{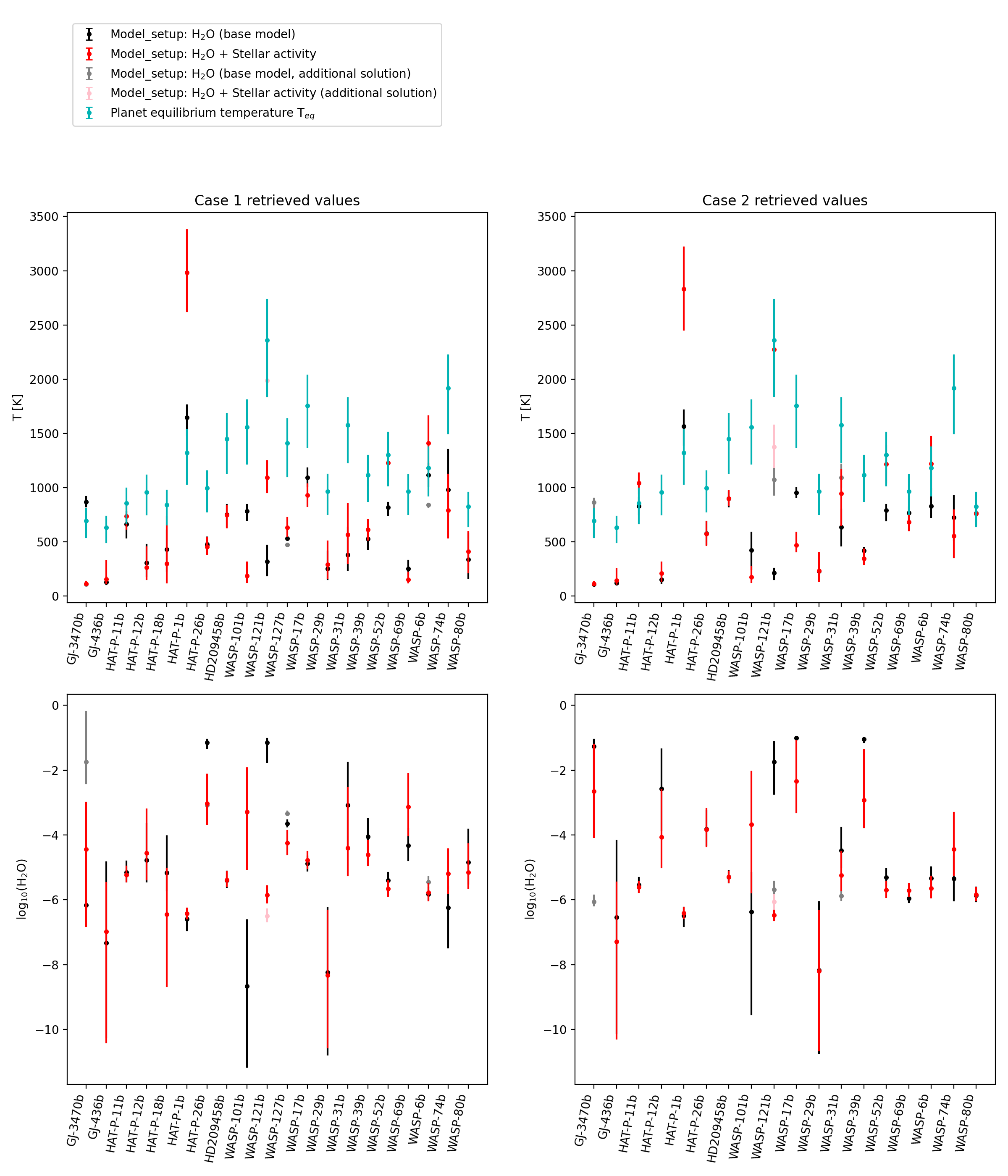}
    \caption{Retrieval results of the planetary equilibrium temperature and water vapour volume mixing ratio for each planet, divided by Case number. For an explanation of the data sets considered in each Case, we refer the reader to Table~\ref{tab:cases}. The results of additional Cases for GJ-3470\,b and HD 209458\,b are reported in Section~\ref{sec:retrieval_results}.}
    \label{fig:graph_results}
\end{figure*}

With this in mind, in Figure~\ref{fig:graph_results} we show the retrieved planetary temperatures and water abundances for each planet, arranged by Case number. Focusing specifically on the retrieved temperatures (upper two graphs in Fig.~\ref{fig:graph_results}), we can distinguish four scenarios that describe the planets in our sample:
\begin{enumerate}
    \item Planets for which the inclusion of stellar activity helps the model retrieve a temperature value close to the expected equilibrium temperature.
    \item Planets for which the retrieved temperature value is not affected whether stellar activity is included or not, with both models returning a temperature consistent with the equilibrium temperature.
    \item Planets for which both retrievals return similar results, yet they both fail to capture the equilibrium temperature.
    \item Planets for which each model returns differing results and both fail to find a temperature value consistent with the equilibrium temperature.
\end{enumerate} 
In group 1 we identify WASP-121\,b Case 1 and 2, WASP-52\,b Case 1 and 2, and WASP-6\,b Case 2, while group 2 comprises e.g. Case 1 and 2 for HAT-P-11\,b, WASP-6\,b Case 1, WASP-69\,b and WASP-80\,b Case 2. These scenarios highlight that employing a more complex model does not result in the unnecessary adjustment of retrieved parameters. On the other hand, it suggests that the retrieval process remains unbiased towards a more complex parameterisation, particularly in scenarios where stellar activity might not be present. Group 3 is where the majority of the planets lie. Here we place e.g. WASP-69\,b, WASP-39\,b, HAT-P-12\,b Case 1 and GJ-436\,b, WASP-29\,b, WASP-74\,b Case 2. The reason for the discrepancies is found behind model fine-tuning. In Fig.~\ref{fig:graph_results} we are just showing how the two simplest models compare, namely the base model with a quiescent star and the base model with an active star, neither of which are completely physically realistic. However, we find that the inclusion of clouds and/or alkali metals on the base model with and without stellar activity considerably improves the fit in most cases. Furthermore, it aids in retrieving temperature values closer to the ballpark provided by the calculated equilibrium temperatures. For instance, that is the case for HAT-P-26\,b and WASP-127\,b. By adding a layer or grey clouds on HAT-P-26\,b Case 2, we find a $T_{\text{p}}$ equal to 747 K in the base retrieval and $T_{\text{p}}$ = 1048 K when we consider stellar activity to influence the spectrum (Table~\ref{tab:HAT-P-26_retrieval_results}). Thus, while both the simplest models result in similar retrieved temperatures, the addition of opaque clouds is crucial to retrieve a more reasonable temperature value with the stellar contamination model w.r.t. to the one without. Likewise, for WASP-127\,b (Table~\ref{tab:WASP-127_retrieval_results}), the inclusion of clouds and alkali metal opacities leads to a notable shift in the retrieved temperatures. In the baseline setups, the retrieved temperatures range from 530 to 630 K, whereas in the more complex models--with and without the addition of stellar activity--the temperatures rise to 1100 K and 1520 K respectively. These findings align more closely with the calculated equilibrium temperature of the planet, which is $T_{\text{eq}}$ = ${1409}^{+223}_{-306}$ K.

An analysis of the retrieved water abundances reveals a variety of water volume mixing ratios (VMRs), which are closely related to the spectral slopes and the temperature values. On most planets, Case 1 and 2 spectra often display contrasting optical data, i.e. STIS visits obtained with the same grating at different epochs exhibit distinct slopes with differing gradients. The base model including Rayleigh scattering and water vapour can typically fit a spectrum with a positive slope in the visible wavelengths reasonably well. This is the case of WASP-17\,b, WASP-6\,b, WASP-69\,b and HD 209458\,b. The Case 1 spectra of these planets return a $\log_{10}(\ce{H2O})$ between -4 and -6, within the typical range seen for hot-Jupiters. However, an extreme optical slope, that can only be partially described by Rayleigh scattering, requires the inclusion of stellar contamination to be modelled optimally. This is the case of WASP-101\,b (Table~\ref{tab:WASP-101_retrieval_results}): its Case 1 base model returns a water VMR lower than -9 but the addition of stellar contamination brings the value to a nominal range ($\log_{10}(\ce{H2O})$ $\approx$ $-3$), hinting that stellar activity may be a relevant source of contamination on this spectrum.  Moreover, we find that the base model struggles to fit a spectrum displaying a negative slope in the optical wavelengths. Unocculted stellar faculae, however, provide a plausible explanation for the downward trend: when the planet crosses a transit chord that is characterised by a reduced flux compared to the surrounding stellar surface covered in high-temperature faculae, the resulting signal tends to underestimate the planetary radius. Consequently, this leads to a decrease in transit depth, generating a characteristic negative bluewards slope. In this group, we highlight the case of HAT-P-26\,b Case 1 (Table~\ref{tab:HAT-P-26_retrieval_results}). One of the base model solutions attempts to explain the slope by increasing the water content to a super-solar value ($\log_{10}(\ce{H2O})$ = $-1.15^{+0.09}_{-0.16}$) that is likely unrealistically high. When stellar contamination is included, a less extreme volume mixing ratio of $-3.0^{+0.9}_{-0.6}$ is obtained. Similarly for WASP-121\,b Case 1 (Table~\ref{tab:WASP-121_retrieval_results}), stellar activity reduces the retrieved water VMR from $-1.15$ to $\sim$ $-6$. Considering stellar contamination can therefore eliminate the need for resorting to super-solar metallicity models to explain elevated water values \citep{2019ApJ...887L..20W}.

It is important to note that the results reported above refer to the baseline models only, i.e. where water is the only active molecular absorber. Water content and planetary temperature are highly correlated. This explains why in a basic scenario with only water absorption, the retrieval can easily find two contrasting solutions: either an excessively high temperature being compensated by a lower water content, or a higher water abundance balanced out by a lower planetary temperature. The addition of clouds usually helps to disentangle these degeneracies, largely by reducing the water content and regulating the temperature to a more physically motivated value.

\subsection{Bayes factor results}
\label{sec:results_bayesF}

\begin{figure*}[htp]
    \centering
    \includegraphics[width=\textwidth]{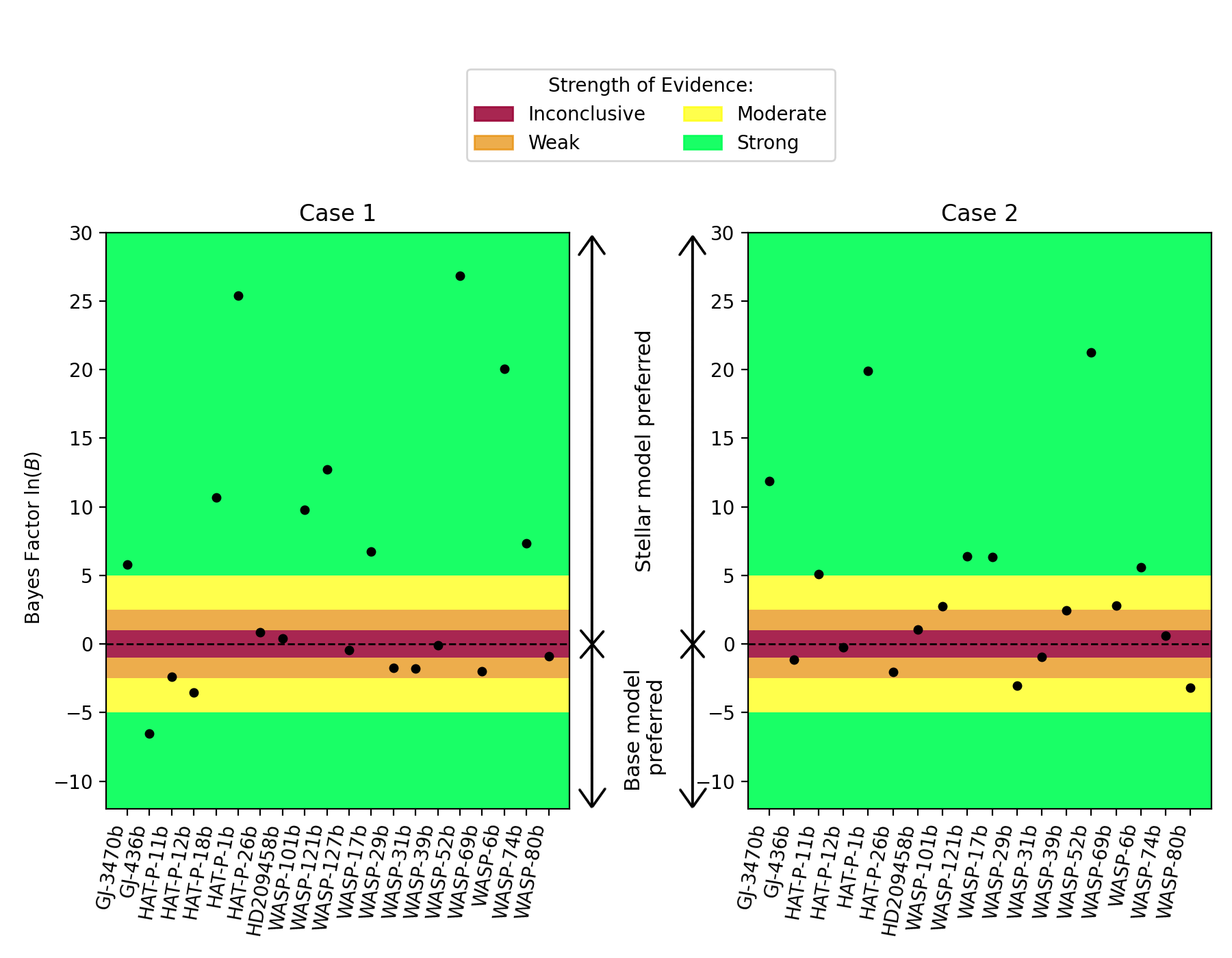}
    \caption{The Bayes factor resulting from the difference between the $\ln(E)$ of the stellar activity model minus that of the base model, divided by Case number. For an explanation of the data sets considered in each Case, we refer the reader to Table~\ref{tab:cases}. The coloured thresholds indicate the strength of the evidence, with negative values favouring the base model over the contamination model. Multiple solutions might be found by the retrieval in certain scenarios, however here we present either the solution with the highest Bayesian evidence if both models appear physically motivated, or the most physically motivated of the two solutions.}
    \label{fig:BF_results}
\end{figure*}

The Bayes factor, as described in Section~\ref{sec:bayes_factor} is a crucial criterion to compare how well different model parameterisations describe the data and is by no means specific to stellar activity. Here it provides us with precious insights into how important stellar contamination is to accurately fit a given spectrum. The results of the Bayes factor calculation, as defined in Equation~\ref{eq:bayes_factor}, are reported in Fig.~\ref{fig:BF_results} for all planets in the sample, considering, once again, each spectral case separately. In a few scenarios the retrieval outputs two possible solutions that fit the observations equally well from a statistical perspective. However in Fig.~\ref{fig:BF_results} we present either the solution with the highest Bayesian evidence if both models appear physically motivated (e.g. for WASP-17\,b Case 1), or the most physically motivated of the two solutions. To define a physically motivated model, we examine the retrieved water abundances and temperature values, comparing them with expectations from the literature. For instance, in the case of GJ 3470\,b, the baseline retrievals yield two possible solutions, one of which suggests a very low retrieved temperature ($\sim$100 K). However, we consider this unphysical due to its inconsistency with existing literature. Therefore, we exclude this solution from further analysis, even though the Bayesian evidence suggests it fits the data better. Another example is HAT-P-26 b. For this planet the baseline retrieval in Case 1 also provides two possible solutions that are generally consistent, except for the water abundance. In the second solution, the retrieved water abundance is nearly two orders of magnitude higher than expected, indicating a super-solar abundance. In this case, we opt for the first solution, as its retrieved water abundance aligns better with other retrievals, and it also has a higher Bayesian evidence.

The strength of the evidence in Fig.~\ref{fig:BF_results} has been colour-coded according to the Jeffreys' scale. The scale conventionally applies to the $\ln(B)$ results above zero, but it can be easily mirrored to explain the negative Bayes factor values. We remind the reader that here a negative $\ln(B)$ value is plausible, as the Bayes factor, defined as the difference between two logarithms, can yield negative results. For $\ln(B)$ values greater than zero, the stellar contamination model is preferred over the simpler base model. As it is evident from Fig.~\ref{fig:BF_results} - Case 1 results, the more complex model is strongly favoured ($\ln(B)$ $\geq$ 5) for nine out of the eleven spectra for which the stellar activity model is preferred overall. However, HAT-P-26\,b and HD 209458\,b exhibit no clear preference between the models. If we focus on the lower section of the left-hand side plot, where the Bayes factor is negative, we observe that among the nine planets for which the Bayesian evidence is in favour of the base model, only GJ 436\,b strongly prefers the base model ($\ln(B) = -6.6$). Notably this Bayes factor is still significantly smaller compared to those of the majority of planets showing strong preference for the activity model (the largest Bayes factor is seen for WASP-52b where $\ln(B) = 26.8$). HAT-P-12\,b moderately favours the base model, while the remaining spectra (HAT-P-11\,b, WASP-127\,b, WASP-29\,b, WASP-31\,b, WASP-39\,b, WASP-69\,b, WASP-80\,b) indicate a weak to inconclusive evidence in favour of the baseline model, suggesting that both the base and the stellar contamination model equivalently explain the data. 

Similarly for the Case 2 results, the Bayes factor of twelve planets lies above zero, ten of which indicate a moderate to strong evidence in favour of stellar activity. The remaining two indicate an inconclusive evidence or are at the border between a weak and an inconclusive evidence. Although the base model is preferred for six spectra, their evidence is mostly inconclusive or weak, with -2.5 $\eqslantless$ $\ln(B)$ $\eqslantless$ 0; even where a moderate evidence is identified for WASP-29\,b and WASP-80\,b, their values are closer to a weak strength of evidence (-2.5) rather than a strong one (-5). For Case 2, no strong preference for the base model is seen for any of the planets in our sample.

\begin{figure*}[ht]
    \centering
    \includegraphics[width=\textwidth]{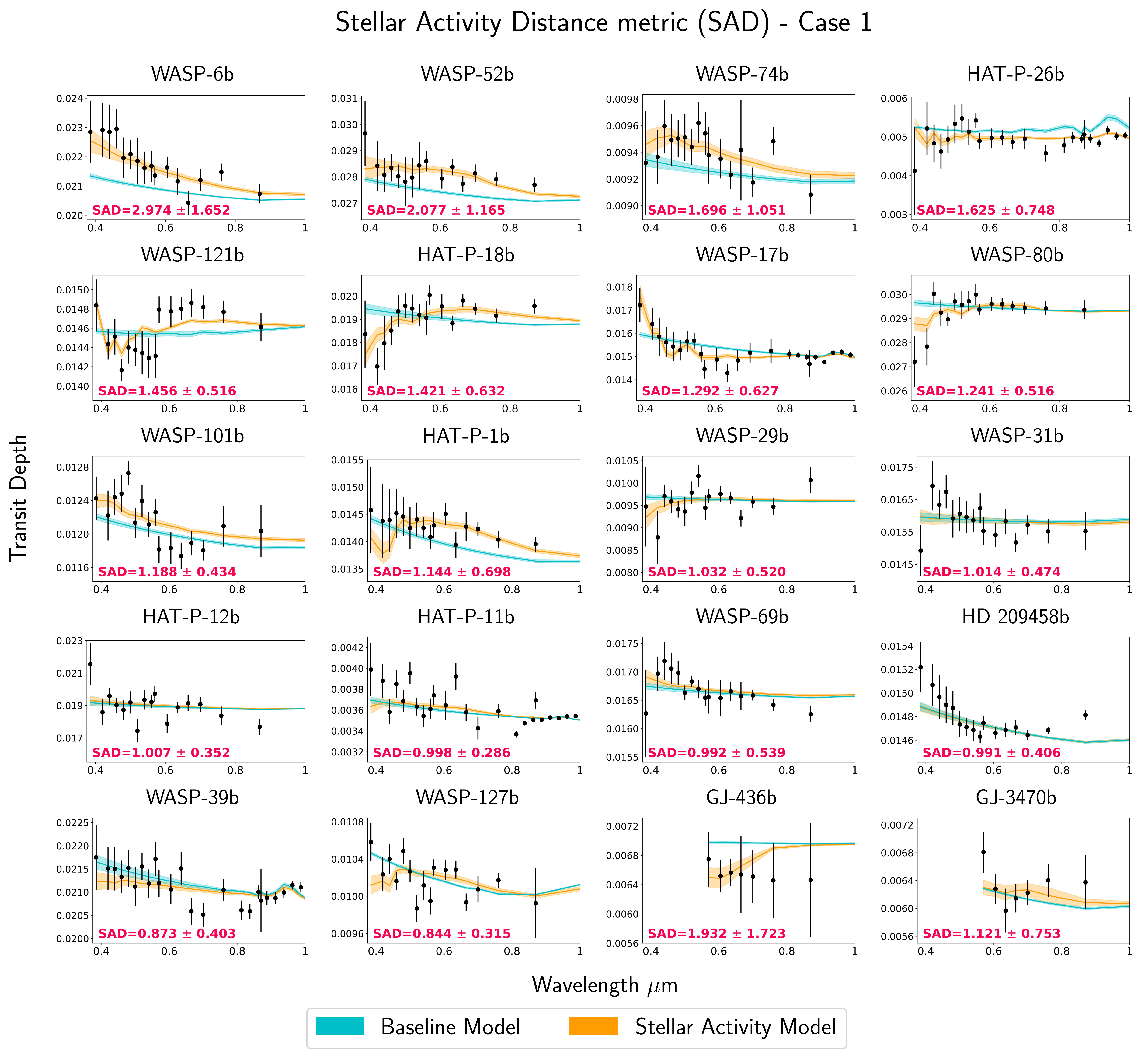}
    \caption{Case 1 planetary spectra ordered by increasing \textit{SAD}. For an explanation of the data sets considered in each Case, we refer the reader to Table~\ref{tab:cases}. A \textit{SAD} below 1 indicates that the spectral data modulations are better described by the base model, shown in cyan. Conversely, a \textit{SAD} above 1 indicates a preference for the stellar contamination model (orange) suggesting that some degree of stellar activity is required to explain the observation. This metric has been calculated for the data points bluewards of 1 $\mu$m, as this wavelength region typically experiences the most significant impact from stellar activity contamination. GJ 436\,b and GJ 3470\,b are added at the end of the list as their large error bars in the observed data and the restricted wavelength coverage in the optical hinder a proper comparison with the rest of the sample. If multiple solutions are found by the retrieval, we present here either the solution with the highest Bayesian evidence if both models appear physically motivated, or the most physically motivated of the two solutions.}
    \label{fig:sad_v1}
\end{figure*}

These results clearly indicate that in both Cases the preferences we obtain for the stellar contamination model are substantially stronger than those we obtain for the base model. This is in some ways to be expected, as a preference for the base model should predominantly result from the additional dimensionality penalty applied to the contamination model rather than from a goodness-of-fit perspective. In Case 1, only two planets, GJ 436\,b and HAT-P-12\,b, show a moderate to strong preference for the base model compared to nine planets that strongly or moderately prefer an active star model instead. This distinction is even more evident in the Case 2 results, where six planets strongly prefer the stellar model and none show a preference for the base model stronger than $\ln(B)=-3$. 
\begin{figure*}[ht]
    \centering
    \includegraphics[width=\textwidth]{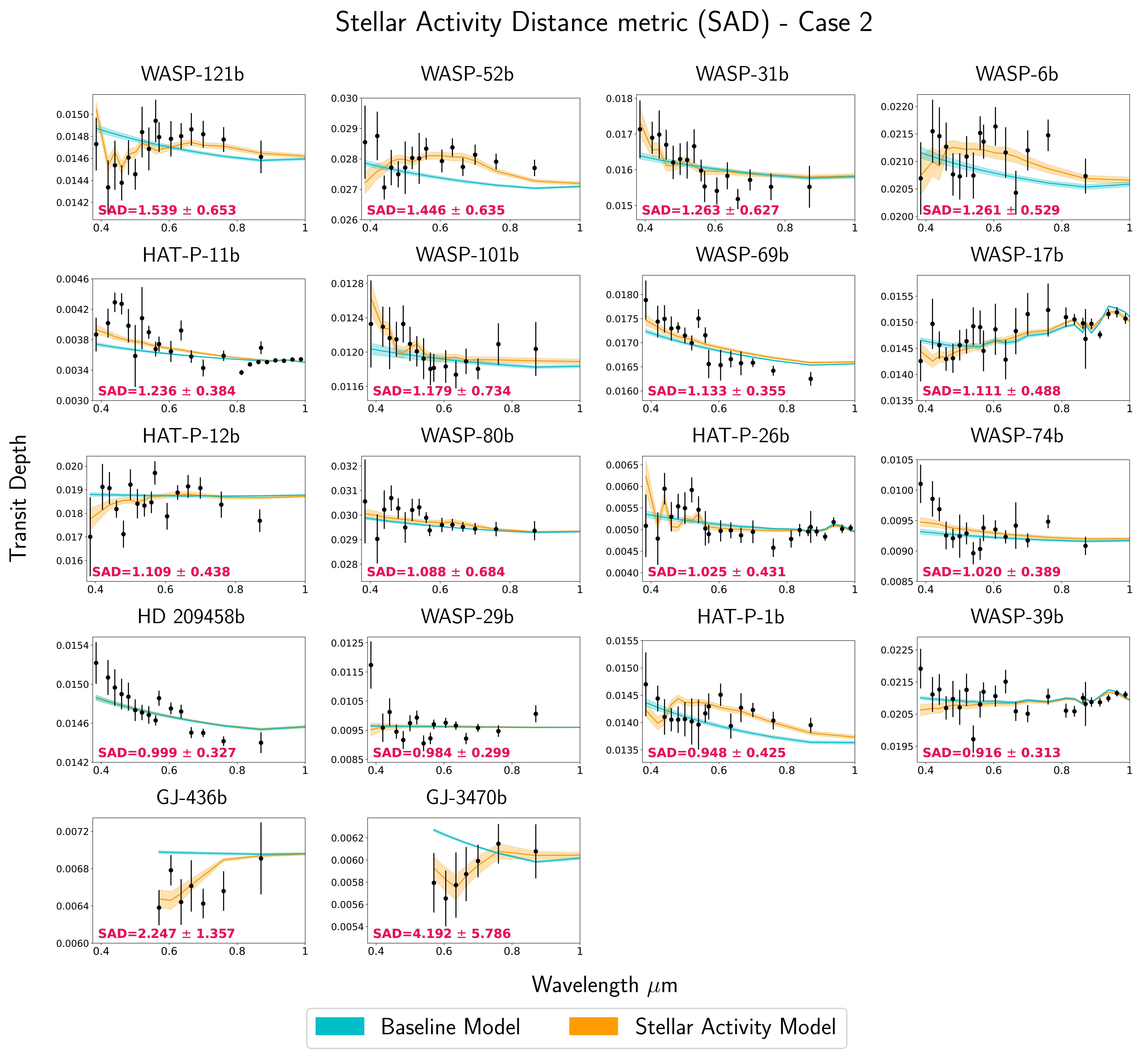}
    \caption{Same as Fig.~\ref{fig:sad_v1} but for the Case 2 spectra. For an explanation of the data sets considered in each Case, we refer the reader to Table~\ref{tab:cases}. HAT-P-18\,b and WASP-127\,b have not been included here as they have only been observed once by each STIS grating. GJ 436\,b and GJ 3470\,b are added at the end of the list as their large error bars in the observed data and the restricted wavelength coverage in the optical hinder a proper comparison with the rest of the sample. If multiple solutions are found by the retrieval, we present here either the solution with the highest Bayesian evidence if both models appear physically motivated, or the most physically motivated of the two solutions.}
    \label{fig:sad_v2}
\end{figure*}
The weighting of the error bars offers further evidence supporting the adequacy of the stellar contamination model in describing the data, particularly when a substantial Bayesian evidence favours it. During the retrieval calculations, the nested sampling algorithm assigns greater weight to smaller error bars, prioritising an optimal fit to the WFC3 data before addressing the STIS data. It is noteworthy that, even though the STIS data are weighted down, the stellar model (which carries the most relevance in the STIS bands) remains the preferred model for nearly half of the spectra in both Case 1 and Case 2 and can also visually improve the fit in the NIR region.

In both Case 1 and 2, a significant number of planets show either inconclusive or weak preferences for either model implying that both models give statistically equivalent fits. In this situation one can also apply Occam's Razor, therefore selecting the simplest model to avoid dealing with potential degeneracies or the risk of overfitting that is induced by the inclusion of more parameters than necessary.

\subsection{SAD metric results}
\label{sec:sad_results}

\begin{figure}[ht]
    \centering
    \includegraphics[width=0.55\textwidth, height=0.37\textheight]{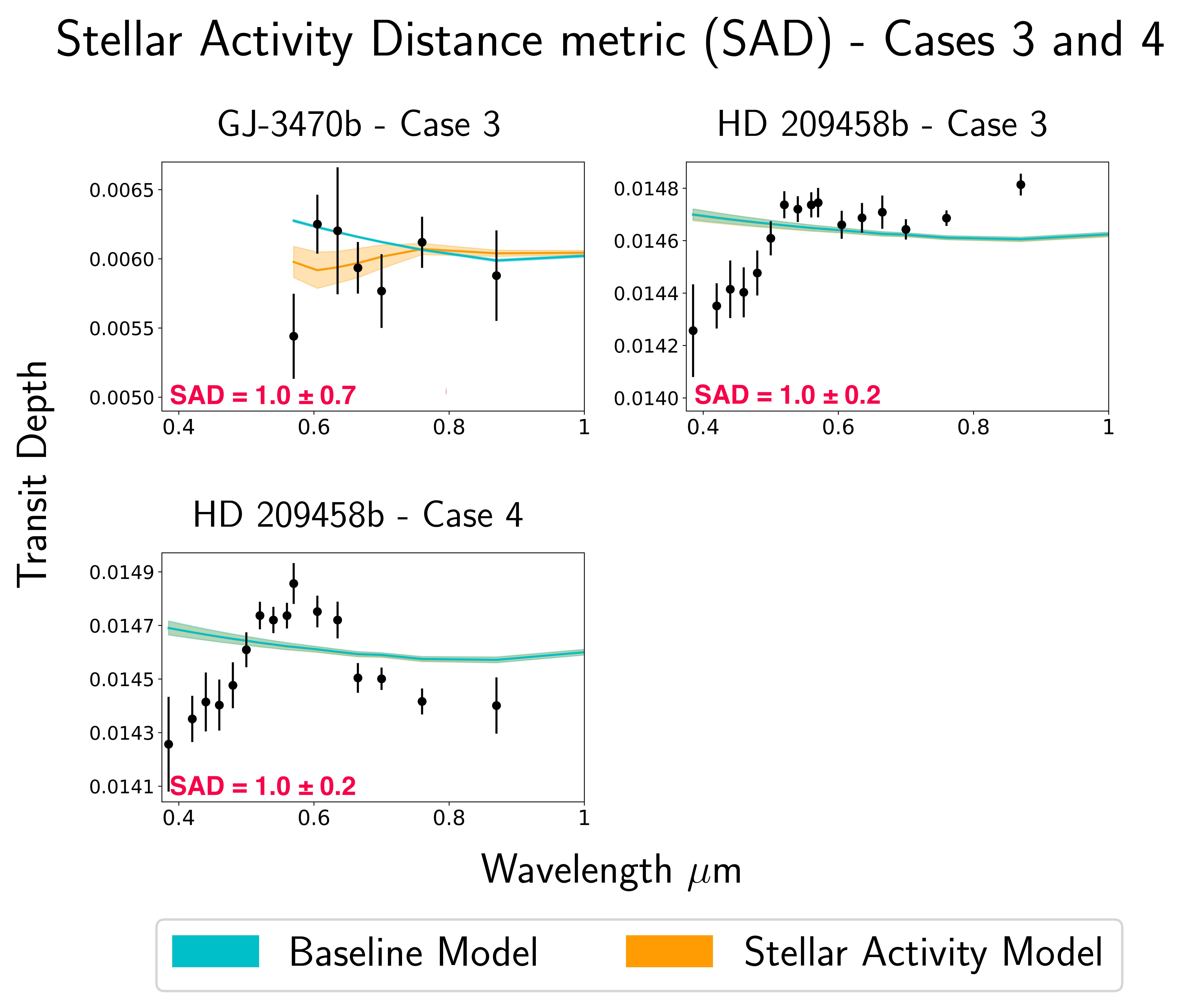}
    \caption{Same as Fig.~\ref{fig:sad_v1} but for the Case 3 and Case 4 spectra. For an explanation of the data sets considered in each Case, we refer the reader to Table~\ref{tab:cases}. If multiple solutions are found by the retrieval, we present here either the solution with the highest Bayesian evidence if both models appear physically motivated, or the most physically motivated of the two solutions.}
    \label{fig:sad_v3-v4}
\end{figure}

As discussed in Section~\ref{sec:results_bayesF}, the Bayesian approach favours precise data, giving less weight to data points with larger uncertainties. Attempting a retrieval on the spectral bands covered by STIS data only is not advisable, as the inclusion of WFC3 data is necessary to anchor the spectra to the continuum level. However, we still want to assess how well the retrieval captures potential stellar activity signals in the STIS data sets. Hence, we utilise the \textit{SAD} metric to assesses the goodness-of-fit in the optical wavelengths, which acts independently of Bayesian statistics. A description of the metric is given in Section~\ref{sec:sad} while the results are reported in Fig.~\ref{fig:sad_v1} and \ref{fig:sad_v2} for Case 1 and Case 2 respectively, in ascending \textit{SAD} order. The \textit{SAD} results for GJ 3470\,b Case 3 and HD 209458\,b Case 3 and 4 are displayed in Fig.~\ref{fig:sad_v3-v4}.

Based on the \textit{SAD} Case 1 results, it can be inferred that WASP-127\,b is the planet for which the stellar contamination model provides the least accurate fit to the optical data, likely due to the significant scatter in the data points. Conversely, the strong optical slope observed in WASP-6\,b's STIS data is much better explained by the activity model than by the base model. Between WASP-127\,b and WASP-6\,b, a wide range of \textit{SAD} values reflects the diversity in spectral shapes and  models’ ability to fit the data. Six planets (HD 209458\,b, WASP-69\,b, HAT-P-11\,b, HAT-P-12\,b, WASP-31\,b, and WASP-29\,b) have \textit{SAD} values near 1 (0.99$<$\textit{SAD}$<$1.10), indicating negligible/weak preference for either model. WASP-127\,b and WASP-39\,b mostly favour the base model (\textit{SAD} $<$ 0.99), while planets like GJ 3470\,b, WASP-80\,b, HAT-P-26\,b and others are better described by an active star model (\textit{SAD} $>$ 1.10).

\begin{figure*}[ht]
    \centering
    \includegraphics[width=0.7\textwidth, height=0.9\textheight, keepaspectratio]{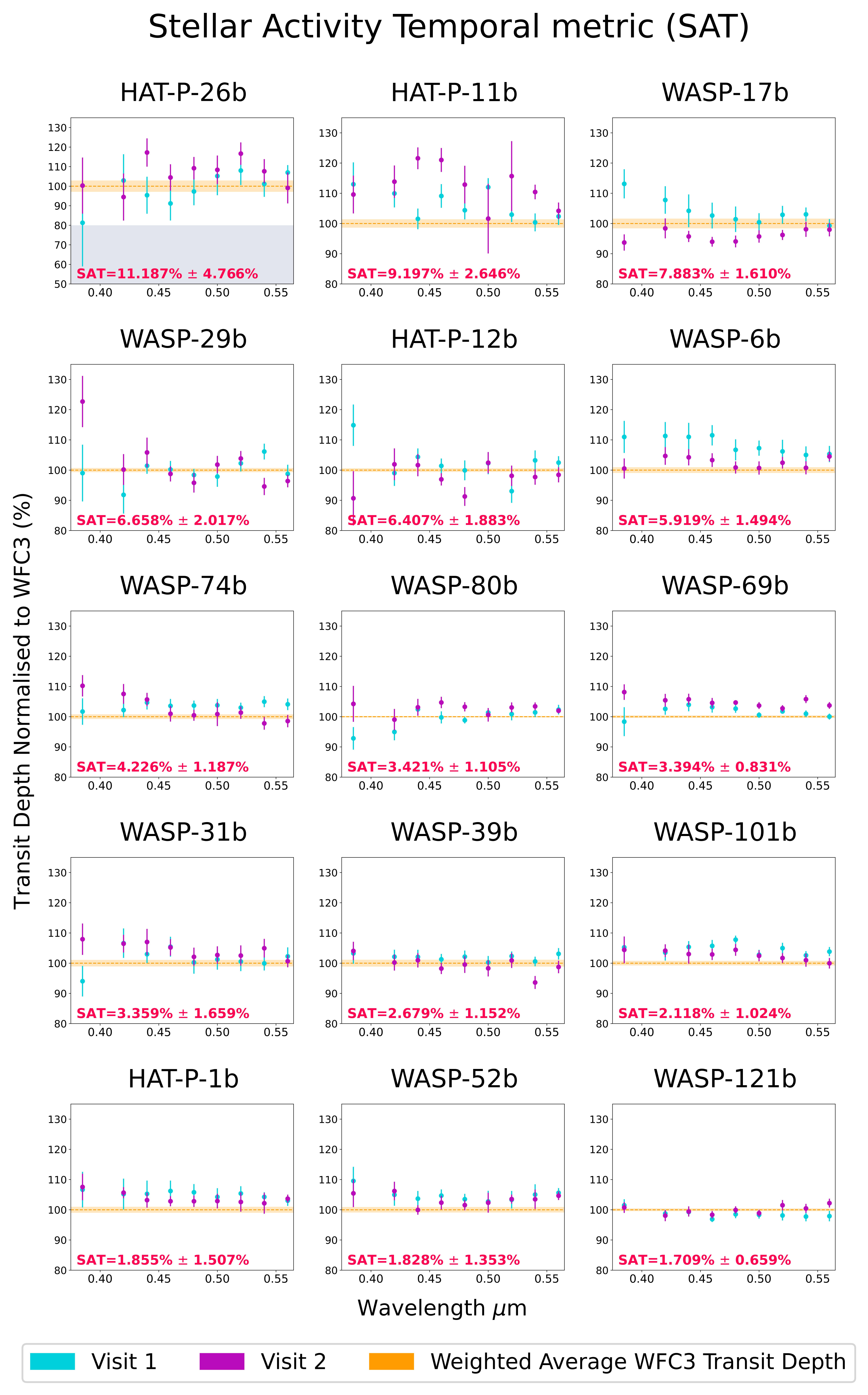}
    \caption{Results of the Stellar Activity Temporal (\textit{SAT}) metric organised by decreasing \textit{SAT} value i.e. from most likely to be active to least likely as inferred solely from this metric. HAT-P-18\,b and WASP-127\,b are excluded from this metric due to the absence of multiple observations taken with either STIS grating. Each subplot presents the two STIS G430L observations, normalised to the average WFC3 transit depth of the corresponding planet. To highlight the differences in transit depth between observations for the same planet and to enable comparisons within the population itself, the STIS transit depths have been converted into a percentage of the weighted average transit depth observed for the same planet with WFC3.}
    \label{fig:sat_results}
\end{figure*}
In Case 2, applying similar criteria, WASP-39\,b, HAT-P-1\,b, and WASP-29\,b show \textit{SAD} values below 0.99, while HD 209458\,b, WASP-74\,b, HAT-P-26\,b, and WASP-8\,b fall within the 0.99 to 1.10 range. The remaining eleven planets exceed a 1.10 value for the \textit{SAD}, with GJ 436\,b and GJ 3470\,b notably higher than 2. Case 2 exhibits a more uniform \textit{SAD} distribution compared to Case 1. Except for GJ 436\,b and GJ 3470\,b, which clearly prefer the stellar model but with large error bars, the rest of the spectra yield \textit{SAD} values between 0.9 and 1.4. \textit{SAD} values around 1.4 moderately favour the stellar model, but not decisively. For GJ 3470\,b Case 3 and HD 209458\,b Cases 3 and 4, displayed in Fig.~\ref{fig:sad_v3-v4}, \textit{SAD} values around 1.0 suggest both models fit the data comparably well.

The considerable uncertainties in \textit{SAD} values, often 30-40\% or more, warrant caution when favouring one model over the other. High \textit{SAD} errors result from large STIS data error bars coupled with poor model fits. This is particularly evident in GJ 3470\,b Case 2, where even though the stellar model reduces the average data-model distance, substantial uncertainties in both parameters lead to a considerable \textit{SAD} error. Given these large uncertainties, it is possible that any spectrum in the sample could be best fitted by either model.

\subsection{SAT metric results}
\label{sec:sat_results}

\begin{figure*}[htp]
    \centering
    \includegraphics[width=0.55\textwidth, height=0.4\textheight]{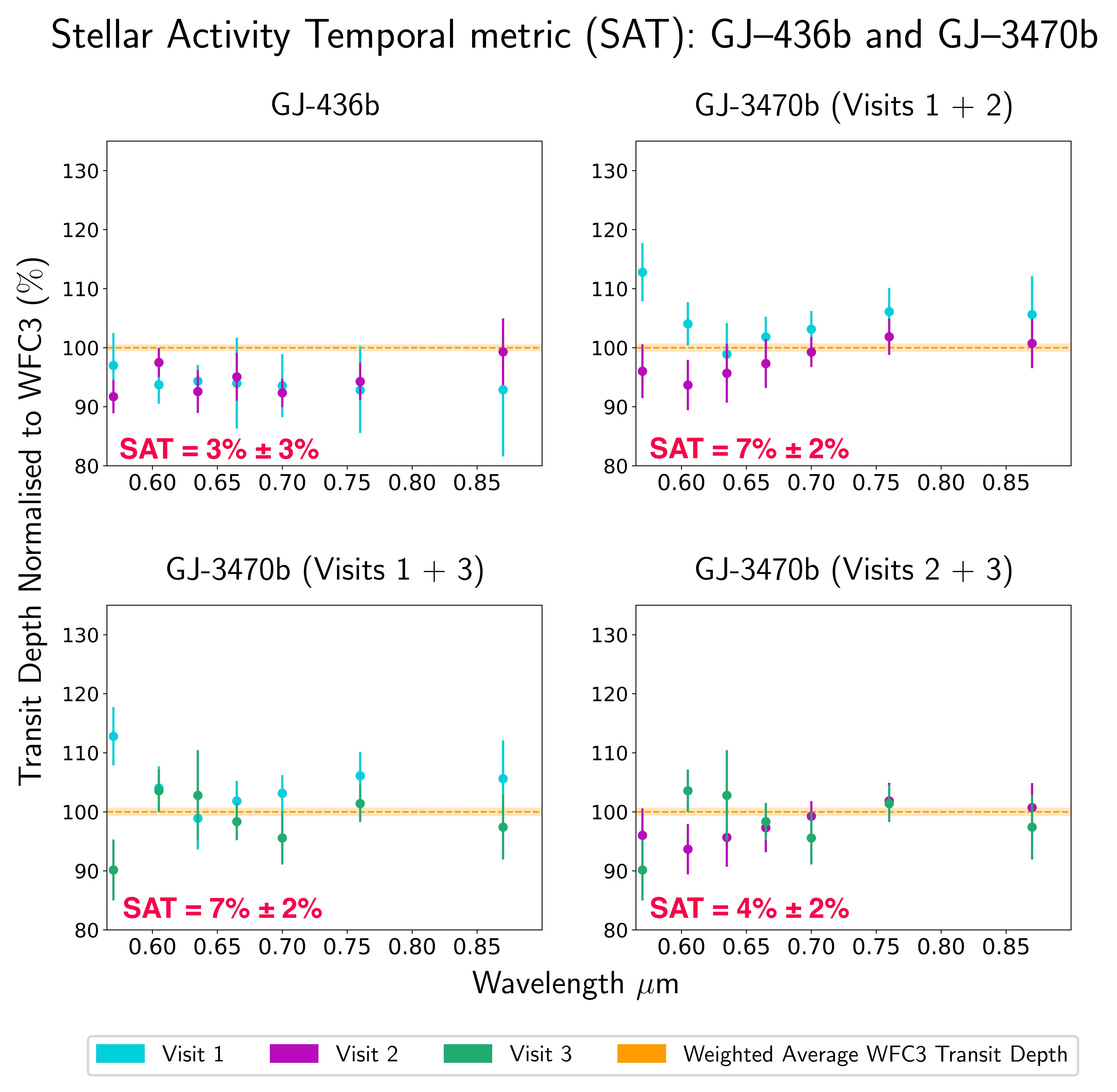}
    \caption{Results of the Stellar Activity Temporal (\textit{SAT}) metric for the Neptune-like planets GJ 436\,b and GJ 3470\,b. For GJ 3470\,b, three \textit{SAT} values were computed, resulting from the various combinations of its three G750L data sets. We have chosen to separate these planets from the rest of the population as, due to the absence of G430L observations and the large error bars on the G750L datapoints, the uncertainties in the computed \textit{SAT} values for these two planets are significantly higher than those seen for the rest of the population and as such should be treated with an element of caution.}
    \label{fig:sat_results_GJ}
\end{figure*}

\begin{figure}[htp]
    \centering
    \includegraphics[width=0.6\linewidth]{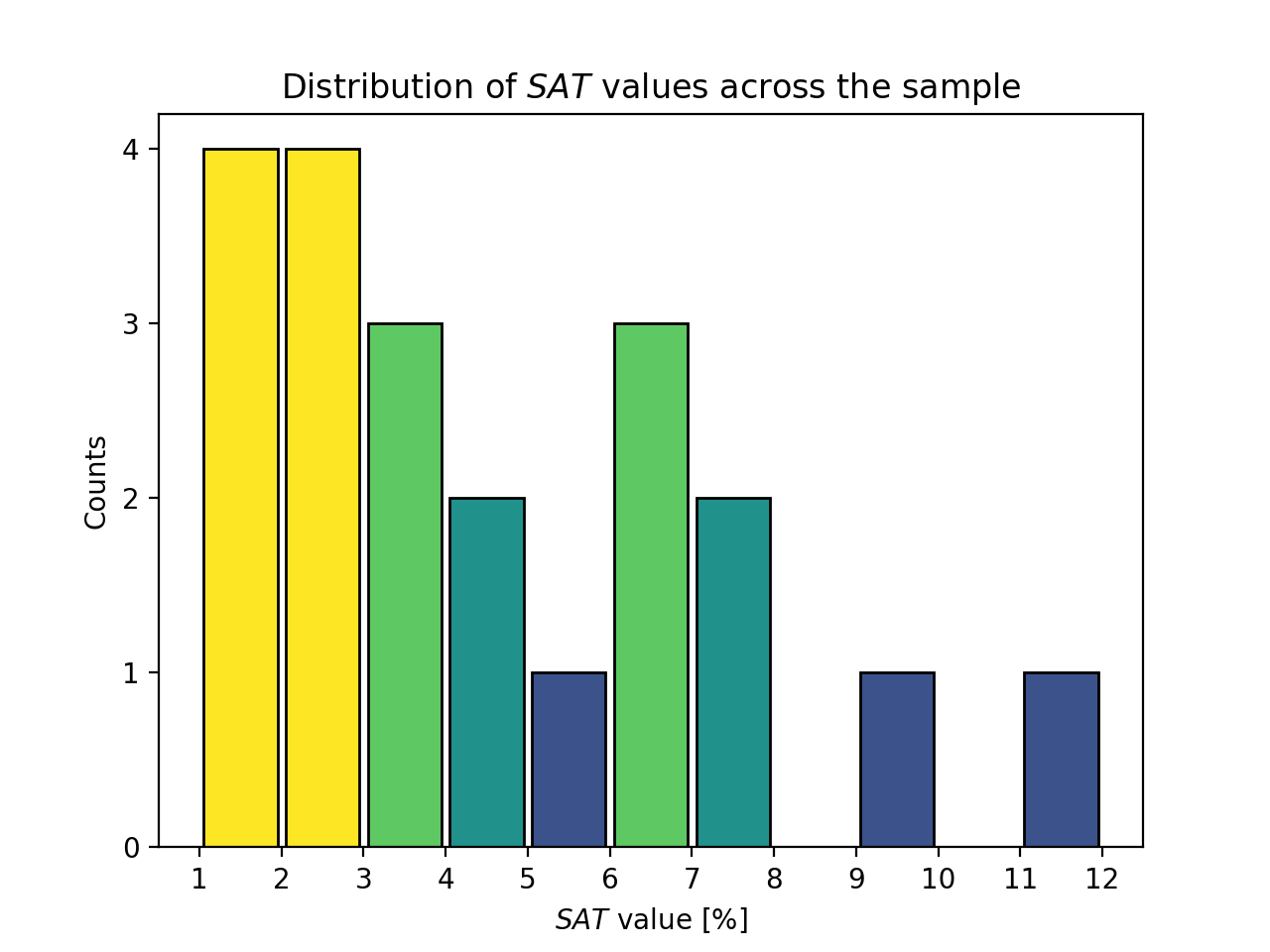}
    \caption{The distribution of \textit{SATs} across our target sample. The bars are colour-coded based on the total number of planets (or data set combinations per planet) falling into each bin; brighter colours indicate higher counts, while darker colours indicate lower counts. This colour scheme aids in visualising the binomial distribution of \textit{SATs}, revealing a prominent peak at 2\% and a smaller peak at 7\%.}
    \label{fig:sat_hist}
\end{figure}

\begin{figure*}[htp]
    \centering
    \includegraphics[width=0.55\textwidth, height=0.25\textheight]{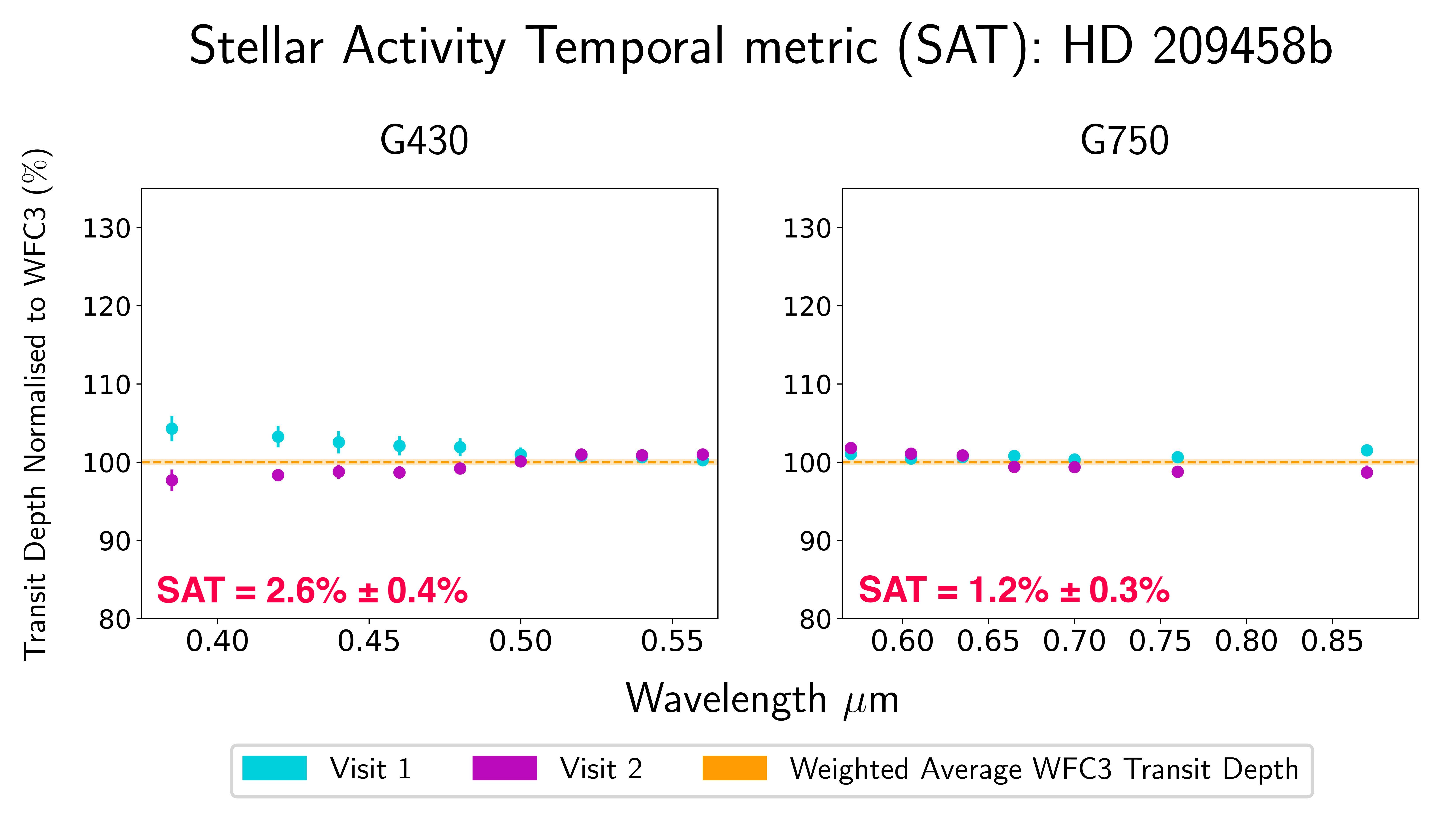}
    \caption{Results of the Stellar Activity Temporal (\textit{SAT}) metric for HD 209458\,b. We provide the \textit{SAT} derived from the two G430L observations (left) and the \textit{SAT} derived from the two G750L data sets (right).}
    \label{fig:sat_results_HD209}
\end{figure*}

The Stellar Activity Temporal metric, or \textit{SAT} in short, has been formulated to evaluate the consistency of observations, providing insights into the potential influence of stellar activity on spectral variations across different epochs. We report the \textit{SAT} results in Fig.~\ref{fig:sat_results}, where the transit depth has been converted into percentage points relative to the weighted average WFC3 data for each planet. The percentage \textit{SAT} enables direct comparisons among planets, facilitating visual identification of those exhibiting the greatest spectral variations across epochs with respect to their atmospheric extents.

Similarly to the \textit{SAD}, we observe a broad spectrum of \textit{SAT} values, each representing distinct amplitudes of spectral temporal variations. Specifically, among the twenty-one \textit{SATs} computed, eleven fall within the range of 1\% to 4\%. This indicates that over half of the planetary optical spectra exhibit largely consistent transit depth variations over time. On the other hand, about seven spectral combinations (involving six planets, as these include GJ 3470\,b visit 1-visit 2 and GJ 3470\,b visit 1-visit 3) out of twenty-one show a \textit{SAT} between 6\% and 12\%. This indicates that another substantial subset of planets is characterised by more significant spectral variabilities. Within the lower-to-moderate variability regime, i.e. with an \textit{SAT} falling within the 4-6\% range, we identify WASP-6\,b, GJ 3470\,b visit 2-visit 3 (shown in Fig.~\ref{fig:sat_results_GJ}), and WASP-74\,b. It is important to mention that this categorisation is entirely arbitrary, derived from a visualisation of the \textit{SAT} results in a histogram that reveals a bimodal distribution, roughly peaking at 2\% and 7\% (Fig.~\ref{fig:sat_hist}). HAT-P-26\,b is the planet that displays the greatest variation among STIS visits, boasting a remarkable \textit{SAT} variation of 11$\pm$5\%. By contrast, the planet with the smallest temporal variation is HD 209458\,b G750L, featuring a \textit{SAT} equal to 1.2$\pm$0.3\% (Fig.~\ref{fig:sat_results_HD209}). While the two visits' spectral features and overall modulations appear visually similar on HAT-P-26\,b, especially redder than 0.45 $\mu$m, the percentage scale reveals that the difference between data points at the same wavelength can reach up to 20\%. Such discrepancy could potentially be significant for the retrieval to favour the stellar activity model over the base model. On the contrary, the G750L visits on HD 209458\,b exhibit moderate disparities in spectral slopes between visits, yet the percentage discrepancy between any one pair of data points rarely exceeds 2\% of the absolute transit depth. 

As expected, the uncertainties are particularly pronounced for planets with the largest STIS error bars, like GJ 436\,b and HAT-P-26\,b. For GJ 436\,b, the substantial error suggests a potential variation ranging from close to 0\% to nearly 6\% across visits. On the other hand, even the lower \textit{SAT} bound for HAT-P-26\,b, indicates a considerable variation exceeding 6\% across visits, indicating that spectral modulations remain relevant despite the considerable uncertainty associated with the metric. 

For certain planets such as WASP-17\,b, GJ 3470\,b and WASP-74\,b a high \textit{SAT} value corresponds to large variations in spectral modulations across visits, as the metric is naturally most sensitive to scenarios where the optical slopes display a significant change in gradient. Nevertheless, the rationale behind this metric is not solely to emphasise the distinct spectral shapes. It is also intended to quantify the discrepancy between two visits and to flag temporal variations that warrant further monitoring and investigation, indicating potential activity within the system. Additionally, a low \textit{SAT} value does not necessarily indicate low levels of stellar activity, as the star could be similarly active at subsequent epochs. Therefore, while the metric is not suitable for ruling out stellar activity, it can be used to inform follow-up studies. For instance, if a planet previously considered inactive exhibits a high \textit{SAT}, we should not dismiss the possibility of activity altogether; rather, we should explore the system further.

\section{Discussion} 
\label{sec:discussion}

\begin{figure}[ht]
    \centering
    \includegraphics[width=\linewidth]{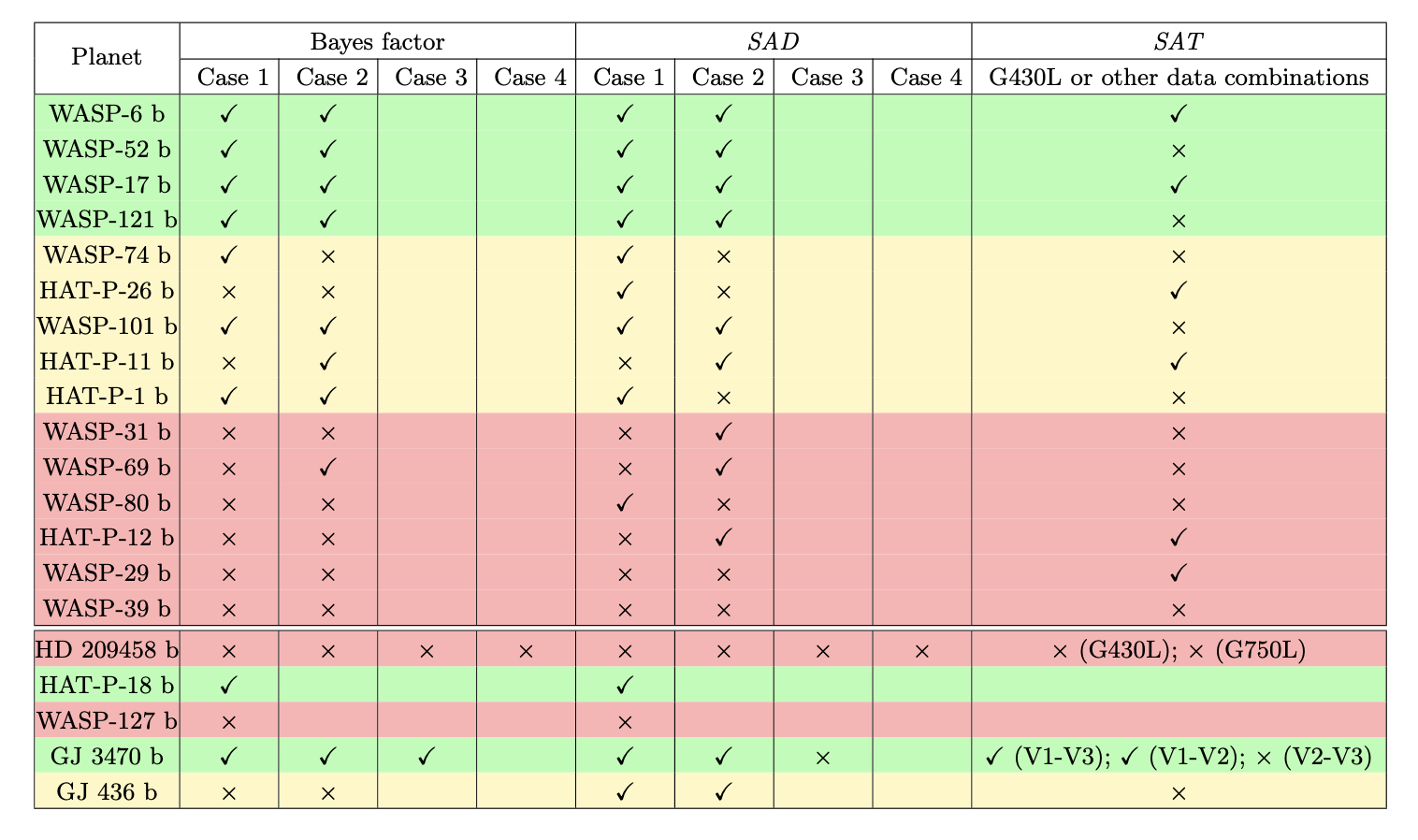}
    \caption{For each planet, we report the preference ($\checkmark$) or rejection ($\times$) for stellar contamination based on the metrics outlined in our investigation. The 15 planets with consistent sets of observations are ordered based on their ranking within the sample over all of the metrics from most likely to be contaminated (WASP-6\,b) to least likely (WASP-39\,b). Colour coding reflects the consensus among metrics: green signifies unanimous or nearly unanimous indications of significant activity; yellow denotes uncertain outcomes, with some metrics leaning towards stellar activity and others against it; red indicates a predominant preference for an inactive star. The 5 planets that do not have consistent data sets are separated by the horizontal line. They are not included in the ranking but are still colour coded in the same way.}
    \label{fig:colour-coded_results}
\end{figure}
This study aims to characterise a variety of planets in a fully consistent way, both from a data analysis and atmospheric retrieval perspective. The comprehensive coverage provided by the combination of STIS and WFC3 data enabled us to explore the presence of various optical absorbers, cloud parameterisations, and the influence of stellar surface heterogeneities on planetary spectra. However, the quality and spectral coverage of the HST data available to us posed limitations on confidently identifying molecules beyond water. As the study progressed, it rapidly became evident that, at the population level, the spectra of these planets are predominantly affected by stellar activity. Stellar photospheric structures are capable of imprinting much stronger signatures on the planetary spectra compared to other atmospheric phenomena. Consequently, we redirected our focus towards assessing stellar contamination across the planetary sample, examining whether and to what extent the spectra of these planets are contaminated by the presence time-varying heterogeneities on the stellar photosphere. In Fig.~\ref{fig:colour-coded_results} we present the collated stellar activity metric results for all of the planets in our sample. For the 15 planets whose observations are entirely consistent with one another we used standard competition-style ranking in an attempt to order them from most likely to be affected by stellar contamination to least likely. We stress that this is not a definitive assessment of stellar contamination; instead, it is merely a way of visually summarising the results of our study and highlighting the systems for which follow up, activity-oriented observations e.g. photometric monitoring surveys should be of highest priority. We also still include within Fig.~\ref{fig:colour-coded_results} the 5 planets whose data sets are not consistent with the rest of the sample. However these are not included within the overall ranking as they are not directly comparable.

Furthermore, in the sections below, we explore three of the planets in our population as in-depth case studies, putting the outcomes of our stellar activity metrics into the context of prior investigations, including supplementary stellar activity indicators, such as Ca H\&K lines (S-indexes and $\log(R'_\text{HK})$ values) and spot/facula crossing events identified in light curves. Our planet selection includes WASP-6\,b, characterised by high levels of activity across all indicators; WASP-39\,b, whose observations suggest a quiet stellar profile according to our activity metrics; and HAT-P-11\,b, orbiting a notably active star, yet our findings consistently suggest that the host star may have been less active at the time of the STIS observations. The remaining planets included in our study undergo a more condensed analysis. We compile our findings comprehensively and present them in Appendix~\ref{appendix:individual_planets} for completeness.

\subsection{WASP-6 b: exploring a renowned active exoplanetary system}
\label{sec:discussion_wasp-6}
\begin{figure*}[ht]
    \centering
    \includegraphics[width=\textwidth]{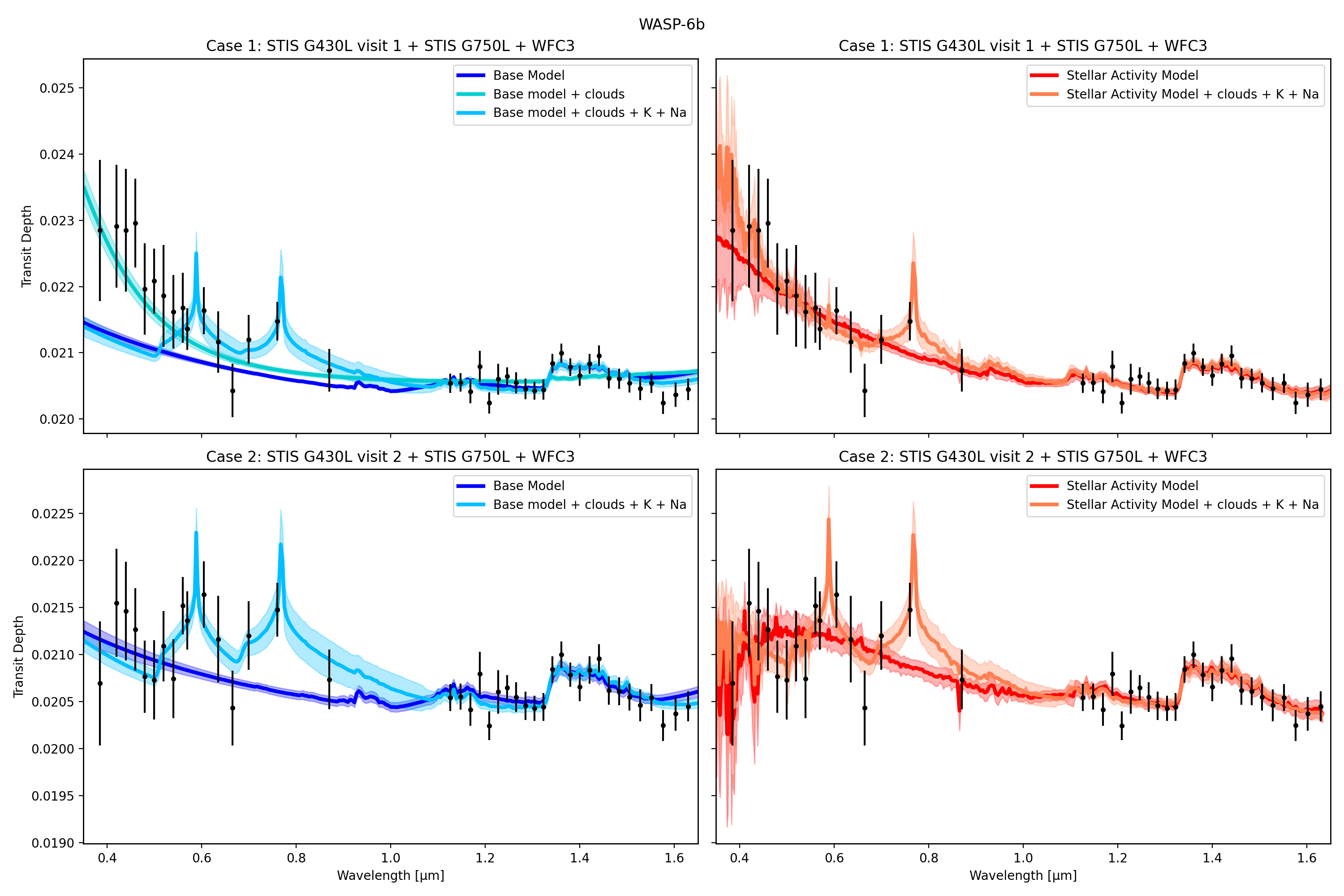}
    \caption{Left subplots: best-fit model retrievals of different combinations of the base model on the Case 1 (upper left plot) and Case 2 spectra (lower left plot). Right subplots: best-fit model retrievals of different combinations of the stellar activity model on the Case 1 (upper right plot) and Case 2 spectra (lower right plot). It is evident that a quiescent star model struggles to explain the Case 1 data adequately, unless it is forced to adopt nonphysical temperature values, like in the base + clouds model. Conversely, while the modulations in the Case 2 spectrum can be explained by either an active or non-active star model, the Bayesian evidence favours models that incorporate an active star.}
    \label{fig:wasp-6_models_plot}
\end{figure*}

\begin{figure*}[htp]
    \centering
    \includegraphics[width=\textwidth]{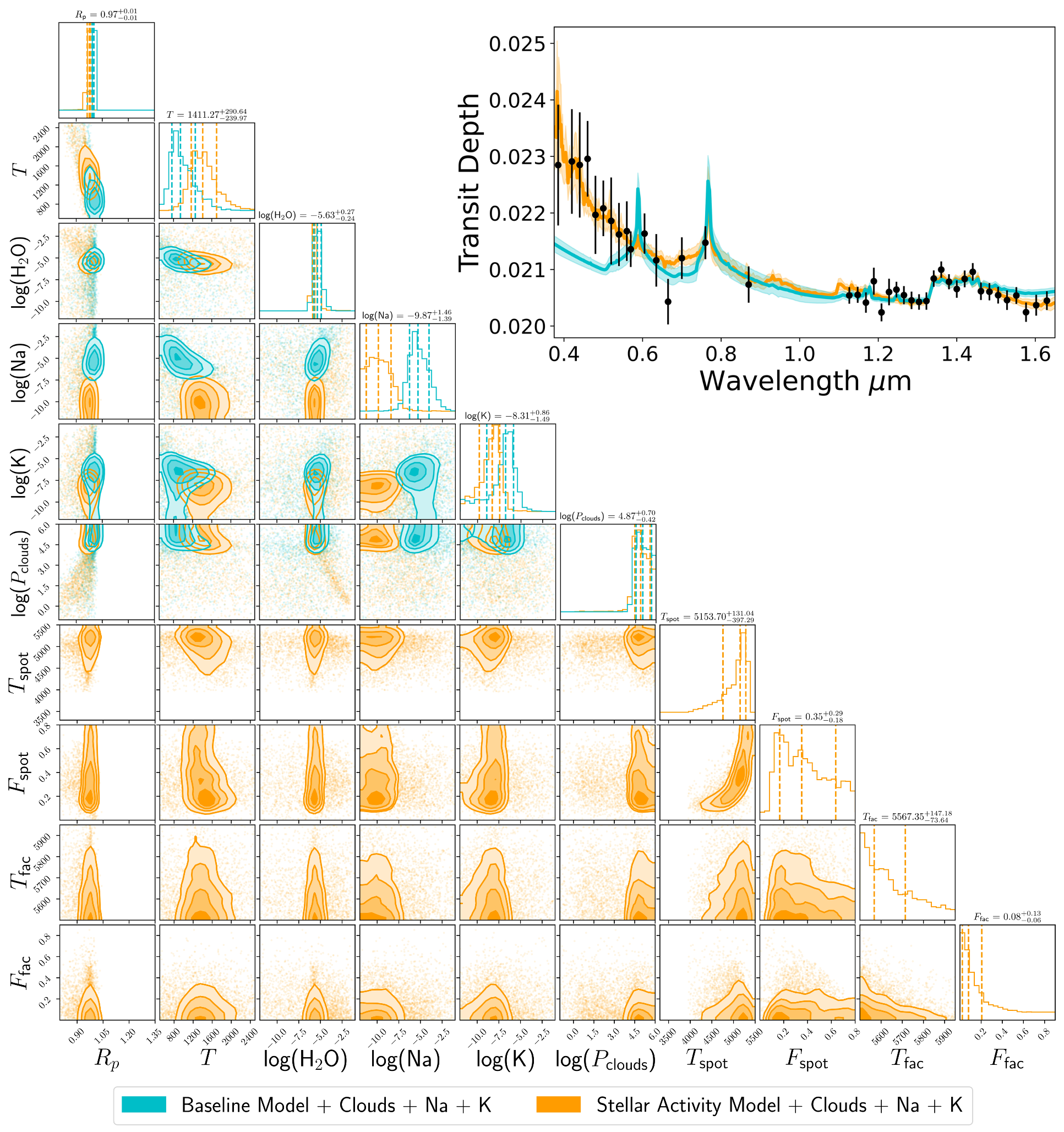}
    \caption{Posterior distributions for the transmission spectrum of WASP-6\,b accounting for a quiet star (cyan) and an active star (orange) retrieved with the Case 1 data set. The dashed vertical lines in each histogram refer to the median value, the first quantile (lower bound error), and the third quantile (upper bound error) of each parameter. Similarly, the reported values above each histogram denote the retrieved median value for that parameter and its corresponding uncertainty. All retrieved values shown are those obtained using the active star model for consistency. Inset: Retrieved best-fit transmission spectra with shaded regions corresponding to their 1$\sigma$ uncertainties.}
    \label{fig:wasp-6_v1}
\end{figure*}

\begin{figure*}[htp]
    \centering
    \includegraphics[width=\textwidth]{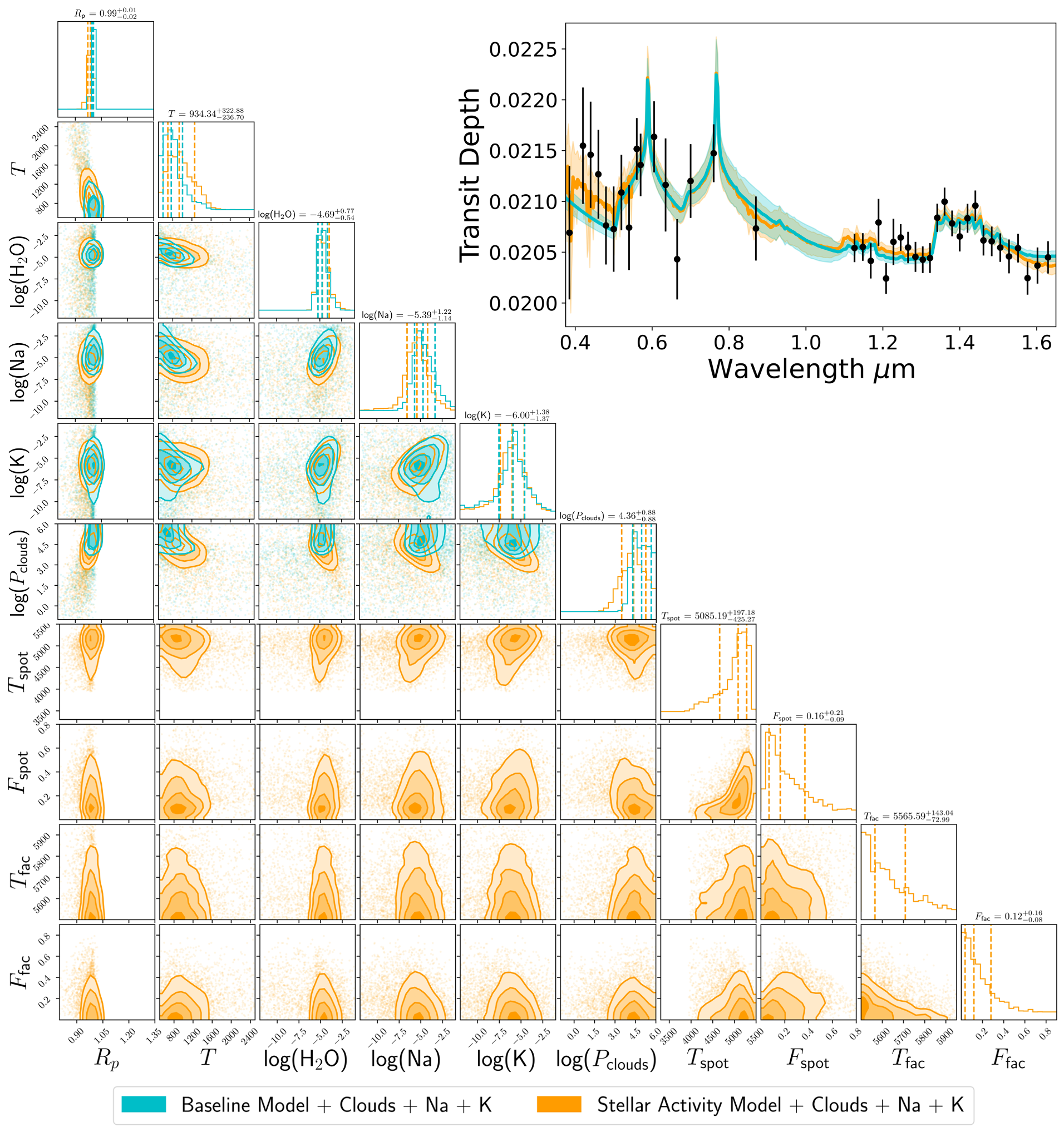}
    \caption{Posterior distributions for the transmission spectrum of WASP-6\,b accounting for a quiet star (cyan) and an active star (orange) retrieved with the Case 2 data set. Figure elements are the same as those in Fig.~\ref{fig:wasp-6_v1}.}
    \label{fig:wasp-6_v2}
\end{figure*}

\begin{sidewaystable}[ht]
\centering
\caption{Results of the retrieval models tested on WASP-6\,b, split by Case number.} 
\begin{tabular}{|c|c|c|c|c|c|c|c|}
\multicolumn{8}{c} {WASP-6\,b } \\ \hline \hline 
\multicolumn{8}{c} {Case 1: STIS G430L visit 1 + STIS G750L + WFC3} \\ \hline 
Model setup & $T$ [K] & $R_\text{p}$ [R$_\text{J}$] & $\log_{10}$(H$_2$O) & $\log_{10}$($P_{\text{clouds}}$) [Pa] & $\log_{10}$(K) & $\log_{10}$(Na) & $\ln$(E) \\
\hline
H$_2$O (base model) (sol 0) & $1114^{+106}_{-101}$ & $0.997^{+0.002}_{-0.002}$ & $-5.8^{+0.2}_{-0.2}$ & - & - & - & 253.6$\pm$0.2 \\
H$_2$O (base model) (sol 1) & $839^{+14}_{-17}$ & ${1.0026}^{+0.0005}_{-0.0005}$ & $-5.5^{+0.2}_{-0.2}$ & - & - & - & 251.7$\pm$0.2 \\
H$_2$O + clouds (sol 0) & $5255^{+171}_{-432}$ & $0.914^{+0.008}_{-0.006}$ & $-9^{+2}_{-2}$ & $4.29^{+0.07}_{-0.07}$ & - & - &  256.7$\pm$0.2 \\ 
H$_2$O + clouds (sol 1) & $1201^{+293}_{-177}$ & $0.995^{+0.003}_{-0.008}$ & $-5.9^{+0.2}_{-0.3}$ & $4.9^{+0.7}_{-0.3}$ & - & - &  253.1$\pm$0.2 \\
H$_2$O + clouds (sol 2) & $755^{+337}_{-165}$ & $0.98^{+0.01}_{-0.02}$ & $-2.6^{+0.9}_{-0.9}$ & ${1.8}^{+0.4}_{-0.3}$ & - & - & 234.0$\pm$0.2 \\
H$_2$O + K + Na & $1121^{+218}_{-176}$ & $0.978^{+0.008}_{-0.009}$ & $-4.0^{+0.4}_{-0.4}$ & - & $-4.1^{+0.9}_{-4.1}$ & $-2.4^{+0.7}_{-0.7}$ & 253.9$\pm$0.2 \\
H$_2$O + clouds + K + Na (sol 0) & $940^{+330}_{-170}$ & $0.998^{+0.004}_{-0.008}$ & $-5.3^{+0.4}_{-0.4}$ & $5.0^{+0.7}_{-0.5}$ & $-6.7^{+0.9}_{-2}$ & $-5^{+1}_{-1}$ & 262.0$\pm$0.2 \\
H$_2$O + clouds + K + Na (sol 1) & $5301^{+141}_{-284}$ & $0.911^{+0.005}_{-0.005}$ & $-7.7^{+0.7}_{-1.2}$ & $4.26^{+0.08}_{-0.06}$ & $-10.8^{+0.9}_{-0.8}$ & $-10.0^{+1.0}_{-0.9}$ & 255.4$\pm$0.2 \\ \cline{1-1}

H$_2$O + Stellar activity & $1410^{+247}_{-202}$ & $0.973^{+0.007}_{-0.010}$ & $-5.8^{+0.3}_{-0.2}$ & - & - & - & 273.7$\pm$0.2 \\
H$_2$O + Stellar activity + clouds & $1454^{+356}_{-242}$ & $0.972^{+0.008}_{-0.01}$ & $-5.7^{+0.3}_{-0.2}$ & $4.8^{+0.7}_{-0.4}$ & - & - &  273.4$\pm$0.2 \\
H$_2$O + Stellar activity + K + Na & $1259^{+218}_{-184}$ & $0.979^{+0.005}_{-0.007}$ & $-5.6^{+0.3}_{-0.3}$ & -& $-8.5^{+0.9}_{-1.6}$ & $-10^{+1}_{-1}$ & 273.9$\pm$0.2 \\
H$_2$O + Stellar activity + clouds + K + Na & $1411^{+291}_{-239}$ & $0.975^{+0.006}_{-0.01}$ & $-5.6^{+0.3}_{-0.2}$ & $4.9^{+0.7}_{-0.4}$ & $-8.3^{+0.9}_{-1.5}$ & $-10^{+1}_{-1}$ & 273.2$\pm$0.3 \\
\hline
\multicolumn{8}{c} {Case 2: STIS G430L visit 2 + STIS G750L + WFC3} \\
\hline
Model setup & $T$ [K] & $R_\text{p}$ [R$_\text{J}$] & $\log_{10}$(H$_2$O) & $\log_{10}$($P_{\text{clouds}}$) [Pa] & $\log_{10}$(K) & $\log_{10}$(Na) & $\ln$(E) \\
\hline
H$_2$O (base model) & $830^{+158}_{-103}$ & $1.002^{+0.002}_{-0.003}$ & $-5.3^{+0.3}_{-0.3}$ & - & - & - & 269.4$\pm$0.2 \\
H$_2$O + clouds & $840^{+233}_{-110}$ & $1.001^{+0.003}_{-0.005}$ & $-5.4^{+0.4}_{-0.3}$ & $5.2^{+0.5}_{-0.5}$ & - & - &  268.8$\pm$0.2 \\ 
H$_2$O + K + Na & $1083^{+247}_{-196}$ & $0.977^{+0.009}_{-0.010}$ & $-4.0^{+0.5}_{-0.4}$ & - & $-3.7^{+0.7}_{-1.3}$ & $-2.9^{+0.7}_{-0.6}$ & 276.2$\pm$0.2 \\
H$_2$O + clouds + K + Na & $772^{+220}_{-176}$ & $1.001^{+0.004}_{-0.007}$ & $-4.7^{+0.6}_{-0.5}$ & $4.9^{+0.8}_{-0.6}$ & $-6^{+1}_{-2}$ & $-5^{+1}_{-1}$ & 280.0$\pm$0.2 \\ \cline{1-1}
H$_2$O + Stellar activity & $1222^{+248}_{-241}$ & $0.98^{+0.01}_{-0.01}$ & $-5.6^{+0.4}_{-0.3}$ & - & - & - & 274.9$\pm$0.2 \\
H$_2$O + Stellar activity + clouds & $1355^{+368}_{-302}$ & $0.97^{+0.01}_{-0.01}$ & $-5.6^{+0.6}_{-0.3}$ & $4.9^{+0.7}_{-0.8}$ & - & - & 275.3$\pm$0.2 \\
H$_2$O + Stellar activity + K + Na & $801^{+187}_{-175}$ & $0.996^{+0.005}_{-0.008}$ & $-4.9^{+0.6}_{-0.5}$ & - & $-7^{+1}_{-1}$ & $-6^{+1}_{-1}$ & 278.2$\pm$0.2 \\
H$_2$O + Stellar activity + clouds + K + Na & $934^{+322}_{-237}$ & $0.99^{+0.01}_{-0.02}$ & $-4.7^{+0.8}_{-0.5}$ & $4.4^{+0.9}_{-0.9}$ & $-6^{+1}_{-1}$ & $-5^{+1}_{-1}$ & 278.3$\pm$0.2 \\
\hline \hline   
\end{tabular}
\label{tab:WASP-6_retrieval_results}
\end{sidewaystable}

Discovered in 2009, WASP-6\,b is a sub-Jupiter mass planet with an inflated atmosphere, orbiting a metal-poor solar-type star \citep{2009Gillon}. Its initial atmospheric characterisation utilised the ground-based facility at Las Campanas Observatory in Chile, covering the wavelengths from 480 nm to 860 nm. The resulting transmission spectrum exhibited a trend of decreasing apparent planetary size with wavelength, lacking evidence for the broad spectral features of Na and K predicted by clear atmosphere models. Hence, the blue part of the spectrum was deemed to indicate the presence of scattering caused by hazes or condensates in its atmosphere \citep{2013ApJ...778..184J}. Subsequent studies using HST/STIS and Spitzer's InfraRed Array Camera (IRAC), also revealed an increasing transit depth with decreasing wavelength. The overall spectrum was well-described by models incorporating significant optical opacity from aerosols, including Fe-poor \ce{Mg2SiO4}, \ce{MgSiO3}, KCl, and \ce{Na2S} dust condensates \citep{2015MNRAS.447..463N}. 

Our analysis unveils distinct spectral slopes in the two STIS G430L observations. The optical to NIR spectrum including G430L visit 1, depicts a predominantly cloud-free terminator, confirmed by the presence of a substantial water feature at 1.4 $\mu$m measuring 1.8$\pm$0.4 times the atmospheric scale height \citep{edwards2022exploring}, along with subtle alkali metal signatures. Our base model retrieval struggles to reconcile the steep optical slope with a simple Rayleigh scattering slope. Attempts to accommodate this slope by adding other possible atmospheric optical opacity sources lead the retrieval to drastically inflate the retrieved temperature to nonphysical values, while fitting the NIR/WFC3 data with a flat line. Consequently, the base model is inadequate to explain the observed data. The steep visible slope observed in Case 1 is better described by modelling the host star as displaying an heterogeneous photosphere. Adding more realistic complexity to the planet improves the fits of both the base and the activity models as shown in Fig.~\ref{fig:wasp-6_models_plot}. Incorporating a heterogeneous star with contributions from water, alkali metals and grey clouds in the atmosphere of the planet provide a much better fit to the observed spectrum than the model in which stellar activity is neglected. This is particularly evident in the optical (STIS) but a better fit is also obtained in the NIR (WFC3). The posterior distribution (Fig.~\ref{fig:wasp-6_v1}) yields a temperature of ${1411}^{+291}_{-234}$, which is consistent with the expected equilibrium temperature of ${1183}^{+187}_{-257}$ (within 1$\sigma$) and a water abundance in the range of $\log_{10}({\ce{H2O}})$ = ${-5.6}^{+0.3}_{-0.2}$. The sodium abundance is not well constrained, whereas potassium is better constrained with a retrieved abundance of $\log_{10}({\ce{K}})$ = ${-8.3}^{+0.9}_{-1.5}$. We observe a substantial preference for the presence of spots, with a best fit filling factor of $F_\text{spot}$ = ${0.4}^{+0.3}_{-0.2}$, indicating that between 20-70\% of the stellar photosphere may plausibly be covered by spots with an associated temperature of $T_\text{spot}$ = ${5219}^{+97}_{-347}$ K. The retrieval is less conclusive regarding the presence of faculae or lack thereof, as both the faculae temperature ($T_\text{fac}$) and filling factor ($F_\text{fac}$) converge towards their lower bounds, indicative that their presence is not necessarily required. We note that for WASP-6, the retrieved spot temperatures for all retrievals including stellar activity are self-consistent, ranging from 4900 K to 5218 K (Table~\ref{tab:WASP-6_retrieval_results}). This is significantly higher than the anticipated value of $\sim$3900 K determined in \citet{rackham2019transit} with their Eq.~1. This is explainable as, especially due to the large error bars on the STIS data points, the spot temperature and filling factor are highly degenerate with respect to one another as seen in the corresponding posterior (Fig.~\ref{fig:wasp-6_v1}). The result is that the retrieval is not capable of distinguishing between two broad scenarios, one being a higher filling factor of hotter, lower contrast spots and the other being a lower coverage of cooler, higher contrast spots. The significant implied spot coverage and significantly lower faculae filling factor aligns with the pronounced positive optical slope observed by STIS, which is widely accepted as a distinctive signature of stellar contamination from starspots. Our findings align with \cite{2020MNRAS.494.5449C}, who, by combining Very Large Telescope FORS2 ground-based observations with space-based observations from TESS, Spitzer, and the \textit{Hubble Space Telescope} WFC3 and STIS, detected Na, K, and \ce{H2O}. Their findings also suggest that stellar heterogeneities are required over enhanced hazes in order to fully explain the optical slope observed in the atmosphere of WASP-6\,b. 

On the other hand, the Case 2 spectrum comprising STIS G430L visit 2 displays a less steep optical slope. Due to the shallower slope and increased modulation in the optical data, a model incorporating clouds and alkali metals aligns most closely with the data by exhibiting the highest evidence among all base model combinations. Both alkali metals abundances are well constrained and their distinct signatures are well identified in the spectrum by the retrieval. However, this model yields an isothermal temperature approximately 430 K lower than anticipated. In this context, introducing stellar activity to the water-only model shifts the retrieved temperature upwards, closer to the expected value given by the calculated equilibrium temperature. The addition of a layer of gray clouds to the base model and to the stellar activity model doesn't significantly affect the retrieved parameters, as they remain consistent within a 1$\sigma$ range. From the posterior distributions of the stellar contamination model (Fig.~\ref{fig:wasp-6_v2}), the retrieved cloud pressure of $10^{4.86}$ Pa suggests that, if present, the cloud deck tends towards higher pressures. Incorporating alkali metals enhances the model's Bayesian evidence due to the presence of potential alkali features, yet their inclusion simultaneously results in a temperature decrease. It is plausible that the model interprets the modulation within the data, coupled with the limited precision, as pronounced alkali features. Alternatively these broader modulations could feasibly also stem from the competing contamination effects of stellar spots and faculae if both were present on the photosphere. Discriminating between these two signatures with the current data quality poses a challenge. Ultimately, higher resolution data could aid in distinguishing between models more effectively. 

At first order the baseline stellar model improves the base model, as evidenced by the higher Bayesian evidence ($\ln(E)$) associated with it. Compared to Case 1 and as shown in the posterior distribution in Fig.~\ref{fig:wasp-6_v2}, the star in this scenario is expected to exhibit a slightly greater concentration of faculae ($F_\text{fac}$ = ${0.12}^{+0.16}_{-0.08}$) and fewer spots ($F_\text{spot}$ = ${0.16}^{+0.21}_{-0.09}$). The former are necessary to slightly inhibit the strong positive bluewards optical slope imparted by spots. From the STIS observations, which were taken 6 days apart, the star appears to have been consistently active over that timescale, albeit with varying degrees of activity. The activity level appearing to remain relatively high over this period would be consistent with the stellar rotation period of 23.8$\pm$0.2 days determined in \citet{2017A&A...602A.107B}. Our analysis of the stellar activity indicators reveals that WASP-6\,b exhibits a Bayes factor of approximately 20 for Case 1 and above 5 for Case 2, strongly favouring the inclusion of stellar contamination in both cases. Similarly, with a \textit{SAD} = 3$\pm$2 for Case 1 and 1.3$\pm$0.5 for Case 2, WASP-6 emerges as the most active star among all Case 1 scenarios and amongst the most active for Case 2. The \textit{SAT} is also notably high at 6\%$\pm$1\%, emphasising the diversities amongst the two G430L visits, which likely requires a degree of active region evolution or decay to fully explain. A visual inspection of the STIS white light curves doesn't definitively reveal any instances of stellar spot or faculae crossings, although it's challenging to ascertain their presence due to gaps in the light curves. The active nature of the star has been confirmed extensively in the literature. \cite{2018MNRAS.480.5314P} mention that WASP-6\,b is the planet with the highest level of stellar activity contamination among their whole sample. High-resolution spectra of the stellar host obtained with HARPS, revealed indications of chromospheric activity. Within the wavelength region encompassing the Ca{\sc ii} H\&K lines, all HARPS spectra exhibit evidence of emission, suggesting a moderate level of stellar activity \citep{2015MNRAS.447..463N}. Interestingly, despite the presence of such activity, the three \textit{Hubble Space Telescope} visits analysed by \cite{2015MNRAS.447..463N} demonstrated consistent stellar flux levels, and the corresponding light curves displayed no discernible evidence of spot-crossing events. Furthermore, literature values of $\log(R'_\text{HK})$ corroborate our suggestion that WASP-6 is moderate-to-highly active: \citet{sing2016continuum} report a value of $\log(R'_\text{HK})$ = -4.741 while \citet{2020MNRAS.494.5449C} find a value of $\log(R'_\text{HK})$ = -4.51$\pm$0.04, which was derived from the HARPS spectra presented in \citet{2009Gillon}. In summary, WASP-6 appears to be a star with moderate activity when compared to other stars of similar spectral type \citep{2018A&A...616A.108B, 2020MNRAS.494.5449C}.

\subsection{An inspection of the quiet star hosting WASP-39 b}

\begin{figure*}[htp]
    \centering
    \includegraphics[width=\textwidth]{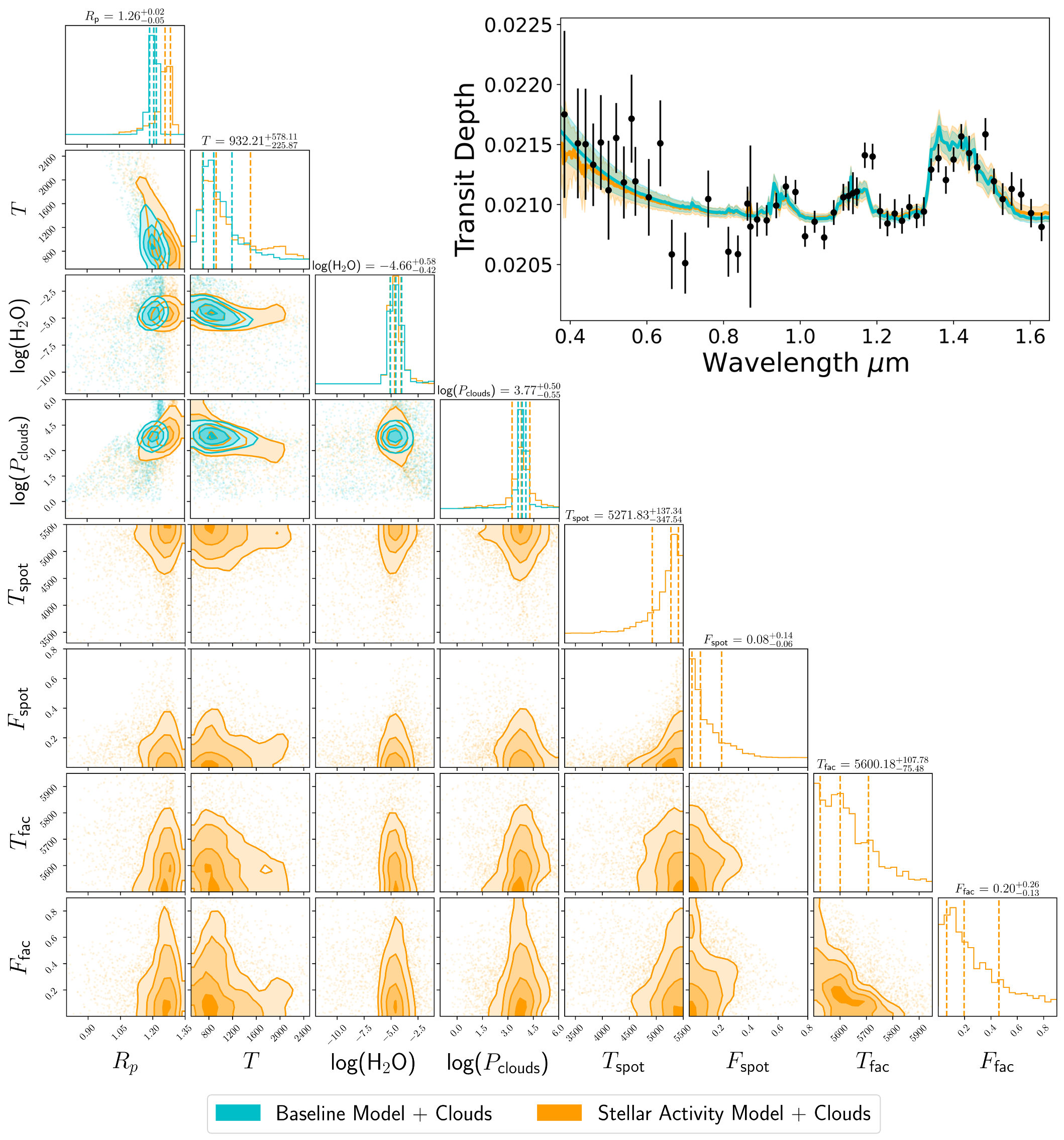}
    \caption{Posterior distributions for the transmission spectrum of WASP-39\,b accounting for a quiet star (cyan) and an active star (orange) retrieved with the Case 1 data set. Figure elements are the same as those in Fig.~\ref{fig:wasp-6_v1}.}
    \label{fig:wasp-39_v1}
\end{figure*}

\begin{figure*}[htp]
    \centering
    \includegraphics[width=\textwidth]{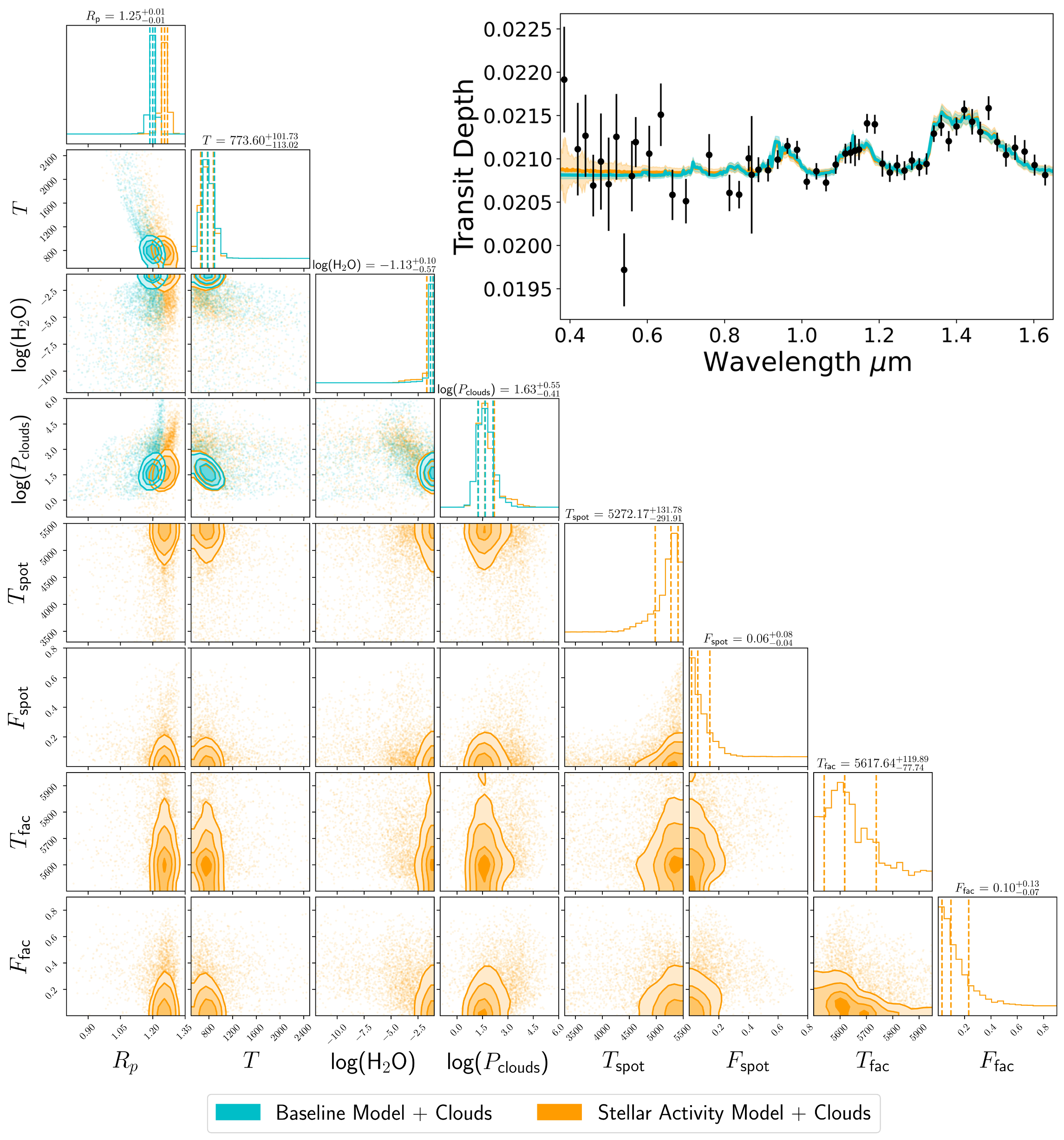}
    \caption{Posterior distributions for the transmission spectrum of WASP-39\,b accounting for a quiet star (cyan) and an active star (orange) retrieved with the Case 2 data set. Figure elements are the same as those in Fig.~\ref{fig:wasp-6_v1}.}
    \label{fig:wasp-39_v2}
\end{figure*}

\begin{table}[ht]
\centering
\caption{Results of the retrieval models tested on WASP-39\,b, split by Case number.} 
\begin{tabular}{|c|c|c|c|c|c|c|}
\multicolumn{7}{c} {WASP-39\,b} \\ \hline \hline
\multicolumn{7}{c} {Case 1: STIS G430L visit 1 + STIS G750L + WFC3} \\ \hline
Model setup & $T$ [K] & $R_\text{p}$ [R$_\text{J}$] & $\log_{10}$(H$_2$O) & $\log_{10}$($P_{\text{clouds}}$) [Pa] & $\log_{10}$(K) & $\ln$(E) \\
\hline
H$_2$O (base model) & $528^{+95}_{-96}$ & ${1.295}^{+0.004}_{-0.004}$ & $-4.1^{+0.5}_{-0.4}$ & - & - & 367.3$\pm$0.2 \\
H$_2$O + clouds & $892^{+306}_{-178}$ & $1.21^{+0.01}_{-0.02}$ & $-4.6^{+0.5}_{-0.5}$ & $3.8^{+0.2}_{-0.2}$ & - & 371.5$\pm$0.2 \\
H$_2$O + clouds + K & $866^{+286}_{-164}$ & $1.21^{+0.01}_{-0.02}$ & $-4.5^{+0.5}_{-0.5}$ & $3.8^{+0.2}_{-0.3}$ & $-10^{+1}_{-1}$ & 370.9$\pm$0.2 \\ \cline{1-1}
H$_2$O + Stellar activity & $611^{+89}_{-92}$ & $1.296^{+0.004}_{-0.005}$ & $-4.6^{+0.4}_{-0.3}$ & - & - & 367.2$\pm$0.3 \\
H$_2$O + Stellar activity + clouds & $932^{+577}_{-226}$ & $1.26^{+0.02}_{-0.05}$ & $-4.7^{+0.6}_{-0.4}$ & $3.8^{+0.5}_{-0.6}$ & - & 368.5$\pm$0.2 \\
H$_2$O + Stellar activity + clouds + K & $949^{+363}_{-200}$ & $1.25^{+0.03}_{-0.07}$ & $-4.3^{+2.1}_{-0.5}$ & $3.6^{+0.5}_{-2.4}$ & $-9^{+3}_{-2}$ & 367.6$\pm$0.3 \\
\hline
\multicolumn{7}{c} {Case 2: STIS G430L visit 2 + STIS G750L + WFC3} \\ \hline
Model setup & $T$ [K] & $R_\text{p}$ [R$_\text{J}$] & $\log_{10}$(H$_2$O) & $\log_{10}$($P_{\text{clouds}}$) [Pa] & $\log_{10}$(K) & $\ln$(E) \\
\hline
H$_2$O (base model) & $418^{+23}_{-63}$ & $1.301^{+0.004}_{-0.002}$ & $-1.04^{+0.03}_{-0.10}$ & - & - & 362.0$\pm$0.2 \\
H$_2$O + clouds & $780^{+112}_{-97}$ & $1.20^{+0.01}_{-0.01}$ & $-1.11^{+0.08}_{-0.25}$ & $1.7^{+0.5}_{-0.4}$ & - & 372.7$\pm$0.2 \\
H$_2$O + clouds + K & $754^{+66}_{-71}$ & $1.202^{+0.009}_{-0.009}$ & $-1.12^{+0.09}_{-0.41}$ & $1.8^{+0.3}_{-0.3}$ & $-7^{+3}_{-3}$ & 372.0$\pm$0.2 \\ \cline{1-1}
H$_2$O + Stellar activity & $345^{+84}_{-49}$ & $1.310^{+0.004}_{-0.004}$ & $-2.9^{+1.5}_{-0.9}$ & - & - & 364.4$\pm$0.2 \\
H$_2$O + Stellar activity + clouds & $774^{+101}_{-113}$ & $1.25^{+0.01}_{-0.01}$ & $-1.1^{+0.1}_{-0.6}$ & $1.6^{+0.6}_{-0.4}$ & - & 367.9$\pm$0.3 \\
H$_2$O + Stellar activity + clouds + K & $835^{+117}_{-104}$ & $1.25^{+0.01}_{-0.02}$ & $-1.09^{+0.06}_{-0.17}$ & $1.4^{+0.4}_{-0.4}$ & $-8^{+3}_{-3}$ & 369.6$\pm$0.3 \\
\hline \hline   
\end{tabular}
\label{tab:WASP-39_retrieval_results}
\end{table}

WASP-39\,b is a Jupiter-size exoplanet that transits a mature G-type dwarf star \citep{faedi2011w39}. Observations using various telescopes have provided insights into WASP-39\,b's atmospheric composition and properties. Using the \textit{Hubble Space Telescope} and the Very Large Telescope FORS2 instrument, studies led by \cite{2016ApJ...827...19F} and \cite{2016ApJ...832..191N} conducted transmission spectroscopy on WASP-39\,b. They revealed Rayleigh scattering, sodium and potassium absorption, and a predominantly clear atmosphere, consistent across both ground-based and space-based observations. Further analyses with WFC3 G102 and G141 data exhibited prominent water absorption features in the mid-infrared \citep{2018AJ....155...29W}, while ground-based optical data suggested super-solar atmospheric metallicities \citep{2019AJ....158..144K}. Our findings are largely consistent with previous studies. In Case 1 (Fig.~\ref{fig:wasp-39_v1}), we observe a haze-free spectrum with a distinct Rayleigh scattering slope. Contrary to previous analyses \citep{sing2016continuum}, our best-fit model contains a layer of opaque clouds at $10^{3.8}$ Pa, alongside water vapour at a volume mixing ratio of ${10}^{-4.6}$ (Table~\ref{tab:WASP-39_retrieval_results}). Similarly, Case 2 also prefers the presence of clouds to a clear model, as in \cite{2018MNRAS.480.5314P}. However, in this scenario we retrieve a water abundance exceeding solar levels, likely influenced by a potential small offset in the STIS G430L visit 2 data compared to the rest of the observations (Fig.~\ref{fig:wasp-39_v2}). The less pronounced optical slope in this STIS visit prompts the retrieval to increase the water content to counteract the nearly flat spectrum in the optical range. Confirmation of wavelength-independent clouds, recently discovered through JWST NIRSpec/PRISM \citep{2023Natur.614..659R}, supports our findings, and the absence of sharp alkali signatures in WASP-39\,b's spectrum reinforces this conclusion. Although alkali metals were previously detected by FORS2 and more recently by the \textit{James Webb Space Telescope}, including in our independent analysis of the WASP-39\,b NIRISS transit \citep{2023Natur.614..670F}, the STIS data quality and its limited resolution hinder the identification of these distinct signatures.

There is consensus across previous studies, except from \cite{2018MNRAS.480.5314P} who report substantial evidence for stellar heterogeneities, that the star appears to be magnetically quiet: \cite{sing2016continuum} report a value of $\log(R'_\text{HK})$ = -4.994, while \cite{Mancini2018} determine an S-index of $S_\text{HK}$ = 0.18$\pm$0.02 and a corresponding $\log(R'_\text{HK})$ = -4.97$\pm$0.06. \cite{Claudi2024} report an S-index of $S_\text{HK}$ = 0.17$\pm$0.04 and a corresponding $\log(R'_\text{HK})$ = -5.0$\pm$0.2, or $\log(R'_\text{HK})$ = -4.8$\pm$0.6 when the correction detailed in \citet{Mittag2013} is applied, all of which suggest relative chromospheric inactivity. The Bayesian evidence for most of our active-star scenarios are consistently lower than those that treat the star as being inactive. Specifically, introducing stellar contamination to the cloudy base model in both spectral cases fails to improve the fit. In Figure~\ref{fig:BF_results}, we observe a Bayes factor of approximately 0 for Case 1 and around 2.5 for Case 2. These values suggest inconclusive evidence for Case 1 and a weak preference for the stellar model in Case 2. This reinforces the fact that stellar activity is likely not affecting the planetary spectrum. The higher evidence for Case 2 may be attributed to a flatter optical region from 0.4 to $\sim$0.6 $\mu$m, which the retrieval attempts to explain with a high concentration of faculae (with a retrieved filling factor of $0.3 \pm 0.2$). However, as mentioned above, the inclusion of clouds further favours the base model over the stellar one, resulting in a Bayes factor of 4.79 in moderate preference of the base model. This serves as another indication supporting the robustness of our stellar contamination model, which appears to correctly identify the absence/non-necessity of activity where it is not playing a critical role.

The \textit{SAD} for this planet corroborates the Bayes factor results. In Case 2, we identify an \textit{SAD} of 0.9$\pm$0.3, the lowest among all Case 2 spectra, while for Case 1, the \textit{SAD} is the second lowest at 0.9$\pm$0.4. Both \textit{SAD} values being less than 1 reaffirm the star's inactivity. Additionally, WASP-39\,b exhibits a low \textit{SAT} of 3$\pm$1\%, placing it among the planets that do not display strong spectral modulations across the two observational epochs available to us. Notably, no clear spot- or faculae-crossing events are evident in the STIS white light curves, further supporting the star's quiet nature. The absence of lithium in the star's spectrum, the relatively slow rotation rate of 44$\pm$4 days \citep{Mancini2018}, and the lack of stellar activity, as evidenced by the absence of Ca{\sc ii}  H \& K emission, collectively suggest that the host star is relatively old, consistent with the age of 7$\pm$1 Gyrs determined in \citet{Mancini2018}. 

Although cloudy, WASP-39\,b's broad spectral features combined with the quiet nature of the host star, led to its selection as one of the targets for the \textit{James Webb Space Telescope} Early Release Science Programme Observations \citep{2023Natur.614..649J}. The transit observations acquired with all four JWST instruments confirmed a high atmospheric metallicity and identified various molecular components including \ce{H2O} \citep{2023Natur.614..653A}, \ce{CO2} \citep{2023Natur.614..664A}, \ce{CO} \citep{2023ApJ...949L..15G}, \ce{Na} and \ce{K} \citep{2023Natur.614..659R}. An unexpected detection of \ce{SO2} \citep{2023Natur.617..483T} was linked to the oxidation of sulphur radicals released during the breakdown of hydrogen sulphide (\ce{H2S}), signifying that active photochemistry is occurring in this exo-atmosphere. 

\subsection{A notoriously active star observed whilst sleeping: the curious apparent quiescence of HAT-P-11}
%     \item section on how shifting all stis data by a fixed value affects the results.

HAT-P-11\,b, the first hot-Neptune discovered through transit searches, orbits a bright and metal-rich K4 dwarf star \citep{bakos2010hat}, displaying high levels of chromospheric activity. The Neptune-sized exoplanet's orbit is found to be highly inclined relative to the equatorial plane of its host star, with a spin-orbit misalignment of 103 degrees \citep{2010ApJ...723L.223W, 2011PASJ...63S.531H}. More recently, a second planetary companion, HAT-P-11\,c, has been discovered in the system, being similar to Jupiter in mass but with a more eccentric orbit \citep{2018AJ....155..255Y}.

\begin{figure*}[htp]
    \centering
    \includegraphics[width=\textwidth]{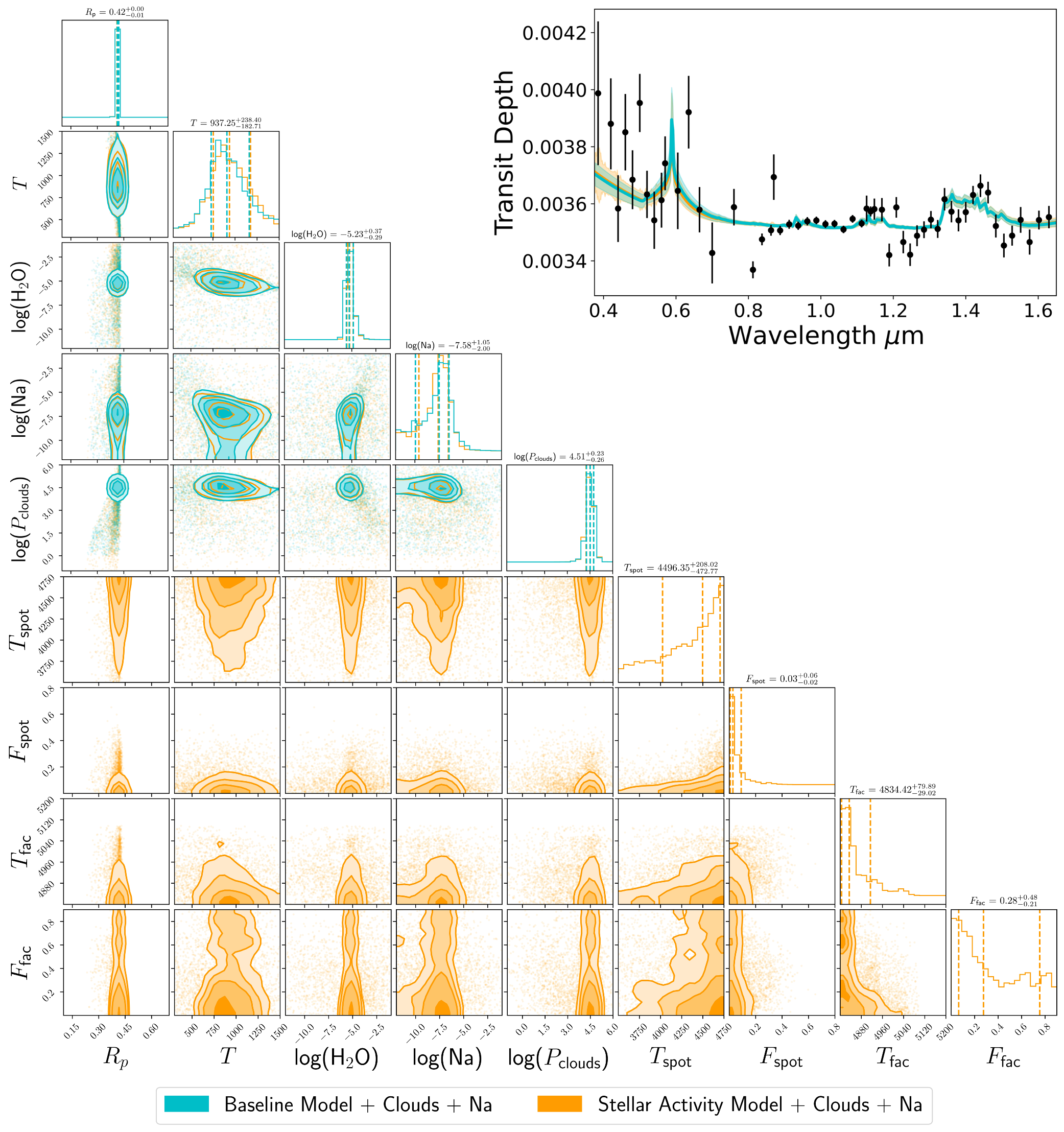}
    \caption{Posterior distributions for the transmission spectrum of HAT-P-11\,b accounting for a quiet star (cyan) and an active star (orange) retrieved with the Case 1 data set. Figure elements are the same as those in Fig.~\ref{fig:wasp-6_v1}.}
    \label{fig:hat-p-11b_v1}
\end{figure*}

\begin{figure*}[htp]
    \centering
    \includegraphics[width=\textwidth]{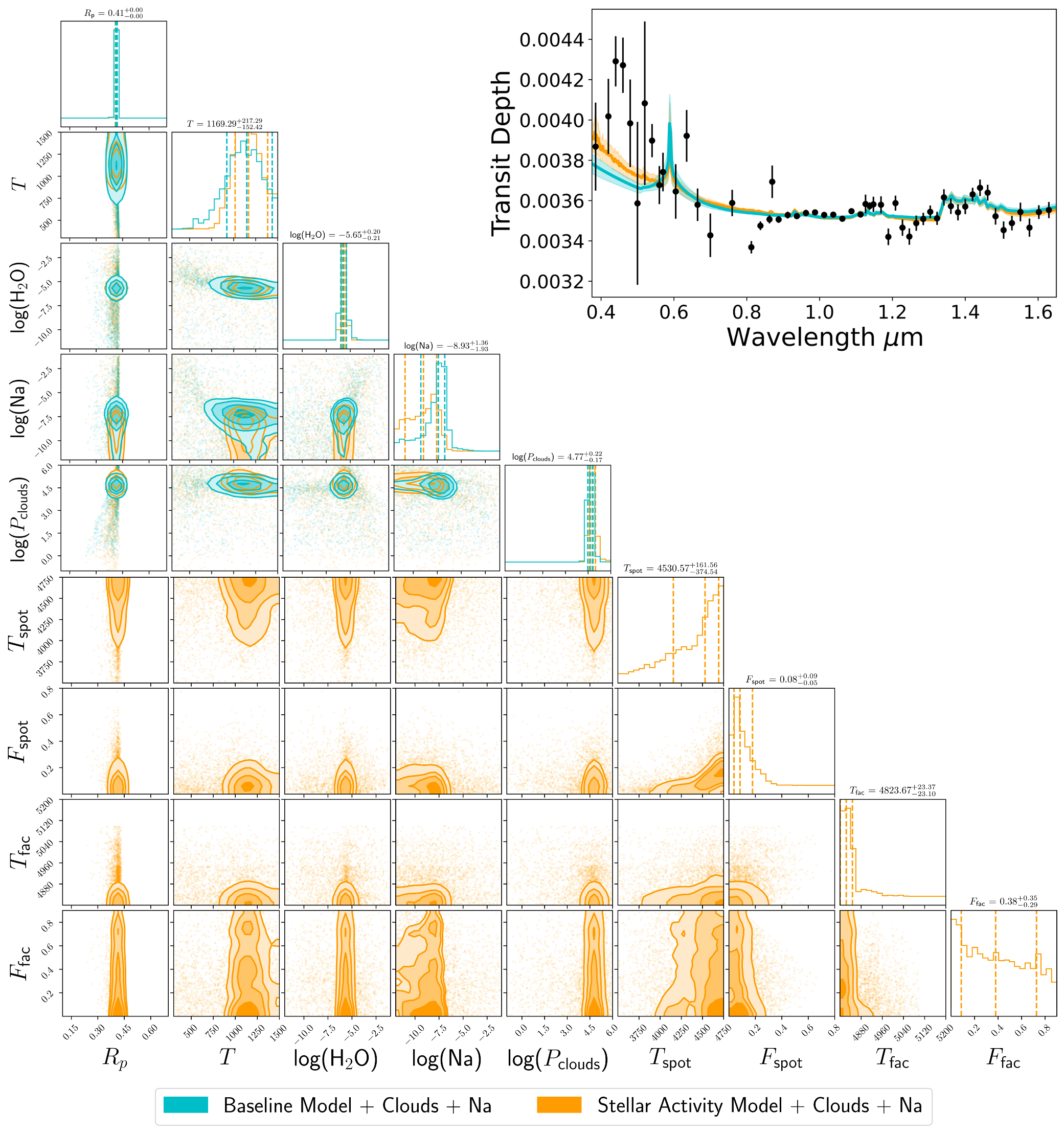}
    \caption{Posterior distributions for the transmission spectrum of HAT-P-11\,b accounting for a quiet star (cyan) and an active star (orange) retrieved with the Case 2 data set. Figure elements are the same as those in Fig.~\ref{fig:wasp-6_v1}.}
    \label{fig:hat-p-11b_v2}
\end{figure*}

Observations of HAT-P-11\,b's transmission spectrum using data from the HST, Spitzer and Giant Meterwave Radio Telescope (GMRT) telescopes have shown water vapour absorption \citep{2014Natur.513..526F}, helium \citep{2018ApJ...868L..34M, 2018Sci...362.1384A}, and hints of radio emission \citep{2013A&A...552A..65L,2016MNRAS.461.1222H}. The planet's atmosphere has been found to be predominantly flat in the optical and with weakened features in the NIR, suggesting a high-altitude cloud layer \citep{2019AJ....158..244C}. Our findings (Table~\ref{tab:HAT-P-11_retrieval_results}) are consistent with a cloudy atmosphere in both spectral scenarios (Fig.~\ref{fig:hat-p-11b_v1} \& \ref{fig:hat-p-11b_v2}), that attenuates the water feature at 1.4 $\mu$m, and masks the potassium feature entirely. Our retrievals detect a faint sodium signature, also attenuated by the presence of a high-altitude cloud layer. The cloudiness in HAT-P-11\,b's atmosphere is supported by various studies indicating that Neptune-like planets are predominantly cloudy \citep{2020NatAs...4..951G, 2024ApJ...961L..23B}.

\begin{table}[hbp]
\centering
\caption{Results of the retrieval models tested on HAT-P-11\,b, split by Case number.} 
\begin{tabular}{|c|c|c|c|c|c|c|}
\multicolumn{7}{c} {HAT-P-11\,b} \\ \hline \hline
\multicolumn{7}{c} {Case 1: STIS G430L visit 1 + STIS G750L + WFC3} \\ \hline
Model setup & $T$ [K] & $R_\text{p}$ [R$_\text{J}$] & $\log_{10}$(H$_2$O) & $\log_{10}$($P_{\text{clouds}}$) [Pa] & $\log_{10}$(Na) & $\ln$(E) \\
\hline
H$_2$O (base model) & $663^{+112}_{-124}$ & $0.424^{+0.001}_{-0.001}$ & $-5.2^{+0.4}_{-0.3}$ & - & - & 412.0$\pm$0.2 \\
H$_2$O + Na & $600^{+128}_{-137}$ & $0.425^{+0.002}_{-0.002}$ & $-5.0^{+0.4}_{-0.3}$ & - &  ${-8}^{+2}_{-3}$ & 412.5$\pm$0.2 \\
%H$_2$O + clouds & $923.45^{+248.52}_{-151.22}$ & $0.4184^{+0.0031}_{-0.0049}$ & $-5.36^{+0.29}_{-0.29}$ & ${4.60}^{+0.22}_{-0.23}$ & - & 416.72$\pm$0.23 \\ 
H$_2$O + clouds + Na & $908^{+252}_{-179}$ & $0.418^{+0.003}_{-0.005}$ & $-5.3^{+0.4}_{-0.3}$ & ${4.5}^{+0.2}_{-0.2}$ & ${-8}^{+1}_{-2}$ & 416.7$\pm$0.2 \\
H$_2$O (STIS shifted down) & $174^{+54}_{-32}$ & $0.4302^{+0.0006}_{-0.0010}$ & $-1.9^{+0.7}_{-1.2}$ & - & - & 415.1$\pm$0.2 \\ \cline{1-1}
H$_2$O + Stellar activity & $736^{+100}_{-114}$ & $0.422^{+0.002}_{-0.003}$ & $-5.2^{+0.3}_{-0.2}$ & - & - & 409.6$\pm$0.2 \\
H$_2$O + Stellar activity + Na & $679^{+121}_{-114}$ & $0.423^{+0.003}_{-0.03}$ & $-5.1^{+0.4}_{-0.3}$ & - & ${-8}^{+2}_{-2}$ & 409.7$\pm$0.2 \\
%\textcolor{red}{H$_2$O + Stellar activity + clouds} & $936.49^{+238.97}_{-182.07}$ & $0.4165^{+0.0042}_{-0.0054}$ & $-5.22^{+0.37}_{-0.27}$ & & - & 413.817$\pm$0.26 \\
H$_2$O + Stellar activity + clouds + Na & $936^{+239}_{-182}$ & $0.417^{+0.004}_{-0.005}$ & $-5.2^{+0.4}_{-0.3}$ & ${4.5}^{+0.2}_{-0.3}$ & ${-8}^{+1}_{-2}$ & 413.8$\pm$0.3 \\
\hline
\multicolumn{7}{c} {Case 2: STIS G430L visit 2 + STIS G750L + WFC3} \\
\hline
Model setup & $T$ [K] & $R_\text{p}$ [R$_\text{J}$] & $\log_{10}$(H$_2$O) & $\log_{10}$($P_{\text{clouds}}$) [Pa] & $\log_{10}$(Na) & $\ln$(E) \\
\hline
H$_2$O (base model) & $827^{+74}_{-75}$ & ${0.4223}^{+0.0009}_{-0.0009}$ & $-5.5^{+0.2}_{-0.2}$ & - & - & 383.2$\pm$0.2 \\
H$_2$O + Na & $703^{+135}_{-160}$ & $0.424^{+0.002}_{-0.002}$ & $-5.3^{+0.4}_{-0.3}$ & - & ${-7}^{+2}_{-2}$ & 384.2$\pm$0.2 \\
%H$_2$O + clouds (sol 0) & $1215.72^{+382.01}_{-208.28}$ & $0.4150^{+0.0038}_{-0.0065}$ & $-5.86^{+0.27}_{-0.26}$ & ${4.68}^{+0.16}_{-0.15}$ & - & 388.28$\pm$0.24 \\ 
%H$_2$O + clouds (sol 1) & $4398.61^{+179.62}_{-265.63}$ & $0.3646^{+0.0047}_{-0.0041}$ & $-6.60^{+0.10}_{-0.15}$ & ${4.27}^{+0.08}_{-0.06}$ & - & 386.58$\pm$0.25 \\
H$_2$O + clouds + Na & $1149^{+288}_{-227}$ & $0.416^{+0.004}_{-0.005}$ & $-5.7^{+0.3}_{-0.3}$ & ${4.5}^{+0.2}_{-0.2}$ & ${-7}^{+1}_{-2}$ & 389.5$\pm$0.2 \\
H$_2$O (STIS shifted down) & $589^{+130}_{-138}$ & $0.424^{+0.002}_{-0.002}$ & $-4.9^{+0.5}_{-0.3}$ & - & - & 399.8$\pm$0.2 \\ \cline{1-1}
H$_2$O + Stellar activity & $1042^{+90}_{-95}$ & $0.415^{+0.002}_{-0.002}$ & $-5.6^{+0.2}_{-0.2}$ & - & - & 388.3$\pm$0.3 \\
H$_2$O + Stellar activity + Na & $1032^{+82}_{-96}$ & ${0.415}^{+0.002}_{-0.002}$ & $-5.6^{+0.2}_{-0.2}$ & - & ${-10}^{+1}_{-1}$ & 387.7$\pm$0.3 \\
%\textcolor{red}{H$_2$O + Stellar activity + clouds} & $1168.31^{+218.03}_{-151.55}$ & $0.4130^{+0.0029}_{-0.0041}$ & $-5.65^{+0.20}_{-0.21}$ & & - & 388.90$\pm$0.26 \\
H$_2$O + Stellar activity + clouds + Na & $1168^{+218}_{-152}$ & $0.413^{+0.003}_{-0.004}$ & $-5.7^{+0.2}_{-0.2}$ & ${4.8}^{+0.2}_{-0.2}$ & ${-9}^{+1}_{-2}$ & 388.9$\pm$0.3 \\
\hline \hline   
\end{tabular}
\label{tab:HAT-P-11_retrieval_results}
\end{table}

Numerous studies have reported HAT-P-11 to be one of the most active exoplanet hosts. The star exhibits a high level of chromospheric activity, as evidenced by observations of the Ca{\sc ii} H \& K lines \citep{2017ApJ...848...58M}. The reported activity indexes are consistent with a high level of activity for HAT-P-11, with a S-index ($S_\text{HK}$) equal to 0.580 and a $\log(R'_\text{HK})$ = -4.567 respectively \citep{Knutson2010}. More recently, values of  $S_\text{HK}$ = 0.54$\pm$0.01 and $\log(R'_\text{HK})$ = -4.68$\pm$0.08 or $\log(R'_\text{HK})$ = -4.302$\pm$0.009 when the correction detailed in \citet{Mittag2013} is applied are obtained in \citet{Claudi2024}. \textit{Kepler} transit data further reveals the occurrence of frequent star spot crossing events, concentrated primarily at phases -0.002 and +0.006, indicating a transit path nearly perpendicular to the stellar equator \citep{2011ApJ...740...33D}. Studies by \citet{2015ApJ...807..170H} find that spots on HAT-P-11 are predominantly located at low stellar latitudes. The distribution of spots on HAT-P-11 mirrors that of sunspots near solar activity maximum, with a mean spot latitude of approximately 16 degrees \citep{2017ApJ...846...99M}. Moreover, the majority of starspots on HAT-P-11 closely resemble the sizes of sunspots at solar maximum, with an estimated mean spotted area coverage two orders of magnitude greater than typical solar spotted areas \citep{2017ApJ...846...99M}. Observations acquired with the Gran Telescopio Canarias suggest a spot filling factor of $T_\text{spot}$ = ${0.6} \pm 0.2$ and a spot-to-photosphere temperature difference of $\Delta$T = ${429}_{-299}^{+184}$ K \citep{2019A&A...622A.172M}. However, our retrieval results do not present compelling evidence for stellar activity contamination. In Case 1, the Bayes factor indicates a weak preference for the base model, supported by a \textit{SAD} value of 1.0$\pm$0.3. For Case 2, our retrievals report a stronger inclination towards stellar contamination, with a $\ln(B)$ slightly above 5. Nevertheless, a \textit{SAD} of 1.2$\pm$0.4 still suggests a weak preference for stellar activity, the possibility of a quiet star remaining plausible considering the measurement uncertainty of the metric. The system's high temporal variability is highlighted by its Stellar Activity Temporal (\textit{SAT}) value, which exceeds 9\%. The exoplanet was observed by the STIS instrument at three different times over the course of three months, namely on 22 February 2017, 12 April 2017 and 26 May 2017. The substantial variations among the optical data are unlikely to arise from photospheric activity within this brief 3-month period. This is especially noteworthy as the star exhibits an unexpectedly long activity cycle, known to last for over 10 years.
Moreover, its unusually strong chromospheric emission, compared to planet-hosting stars of similar effective temperature and rotation period, may be influenced by tides raised by its planetary companion \citep{2017ApJ...848...58M}. What we observed is likely due to the interplay between an active photosphere, a disruptive flaring event and perhaps more complex cloud parameterisations such as hazes made of tholins or soots. Our stellar activity characterisation is also likely hindered to an extent by the WFC3 observations, particularly that of G102 which exhibit a downward trend caused by the reduced throughput at the grism's edge. This effect,  coupled with small error bars achievable with the combination of seven different G102 observations, causes the Bayesian approach of the retrieval to prioritise WFC3 obtaining the best possible fit at these longer wavelengths. This may prevent the retrieval from extracting the full information content afforded to us by the STIS observations, where most of the information surrounding stellar contamination lies.

\section{Conclusions}
\label{sec:conclusions}
We have presented here a population study of 20 transiting hydrogen-rich exoplanets, currently the largest sample of exoplanet atmospheres observed to date in the spectral range near-UV to near-IR. Archive data obtained with \textit{Hubble} STIS and WFC3 instruments were uniformly analysed using the \texttt{Iraclis} pipeline and a suite of spectral retrieval plugins integrated in TauREx\,3.1, including the stellar activity plugin \texttt{ASteRA} and the cloud microphysics plugin \texttt{YunMa}. Most exoplanets included in the sample were observed multiple times with the STIS G430L and/or G750L gratings, allowing for the study of temporal variability caused by changes in the exoplanet atmosphere (e.g. due to cloud/haze variability) and/or stellar magnetic activity.

Our analysis revealed significant differences among the observations obtained with STIS at various epochs for about half of the planets included in the sample, with amplitudes and patterns that are suggestive of time-varying stellar surface heterogeneities imprinted in the exoplanetary spectra. We used the \texttt{ASteRA} model, integrated into TauREx\,3.1, to assess the potential impact of stellar contamination in the analysis of the exoplanet spectra. At a population level, our analysis reveals that for planets on which stellar activity exerts the most significant influence on their spectrum, even employing the simplest stellar contamination model is sufficient to align water abundance values to be more consistent with the average values in the sample. This removes the necessity of employing super-solar metallicity models to account for increased water abundances. 

To statistically assess the preference for an active model over a simpler inactive star model, we computed the Bayes factor resulting from the difference in Bayesian evidence between two retrieval models: one model representing a primary atmosphere with water vapour as the sole opacity source, assuming a uniform stellar photosphere, and the other incorporating the same atmospheric model but considering the star to have a heterogeneous surface. Despite the active star model being statistically penalised for its increased complexity compared to the simpler model, it remains significantly favoured across many spectral cases, including those incorporating more realistic atmospheric physics like clouds and alkali metals. This result emphasises the critical nature of accounting for the potential presence of unocculted stellar spots and faculae when performing spectral retrievals in the optical regime. Among the total number of targets investigated, nearly half of them exhibit a strong preference towards the model incorporating an active star, while only one planet strongly favours a quiet star. 

Furthermore, we established new, complementary metrics to assess the extent of stellar activity contamination; the Stellar Activity Distance metric (\textit{SAD}) and the Stellar Activity Temporal metric (\textit{SAT}). The \textit{SAD}, by corroborating the findings of the Bayes factor analysis, indicates that an active star model better describes the observations in approximately half of the planetary spectra. Similarly, the \textit{SAT} results reveal a bimodal distribution, indicating that spectral variations in the visible regime within this limited planet sample are either minimal or moderately significant.

We analyse in more detail three case studies, WASP-6\,b, WASP-39\,b and HAT-P-11\,b. Our findings align with previous literature results in the case of WASP-6\,b and WASP-39\,b, for which we find an active stellar environment and a quiet star respectively. However, HAT-P-11, expected to be active, is found to show no significant signs of stellar contamination. 

Our findings demonstrate the significant role that stellar contamination may have in all exoplanet spectra observations. Therefore, comprehending, modelling, and correcting for the impact of stellar activity is important for a complete characterisation of exoplanet atmospheres. This issue becomes particularly relevant with the ongoing observations of small planets, which are statistically more likely to orbit active stars, performed by JWST. Additionally, addressing stellar activity becomes even more pertinent in anticipation of the extensive chemical surveys to be conducted by space telescopes such as \textit{Ariel} and \textit{Twinkle} in the coming decade. 

Ideally, the community should aim to have simultaneous photometry alongside all of the future exoplanetary observations. This is to place exoplanets in the correct stellar context of both localised photospheric activity and proximity to stellar minima or maxima. As stellar variability operates over a wide range of timescales, two different observing strategies would be recommended. The first would involve a finely sampled, short cadence survey to capture the stellar rotational modulation, typically on the order of days to weeks. The second should use a longer cadence survey conducted over many years to constrain long-term magnetic cycles analogous to the 11-year solar cycle.

For both approaches we would recommend the use of continuous photospheric monitoring, preferably in the B or V bands where stellar activity effects are the most enhanced. Existing ground-based networks of moderately-sized telescopes ($\geq$ 40 cm) such as the Las Cumbres Observatory (LCO) or those used by the MEarth survey could be ideal for this purpose. In the meantime, robust homogeneous analyses of TESS light curves will provide invaluable insights into a star's activity levels. Particularly for the targets that have already been observed over multiple epochs, we could extrapolate activity information on the timescale of their rotation period and potentially on an epoch-to-epoch basis.

Spectroscopic observations of e.g. Ca H\&K lines would also be advantageous. However, the chromospheric activity measured may not be a direct representation of the photospheric contamination observed. Equally, the time and resource intensive nature of such observations may make this approach less plausible for the large homogeneous survey we recommend.

Additionally, we emphasise the importance of data reduction homogeneity, to prevent the introduction of spurious signals through systematics and/or calibration errors. This could perhaps make the exploitation of pre-existing archival data difficult but not impossible.

The observational strategies to monitor stellar activity highlighted above would significantly contribute to the success of the exoplanet field.

\section*{Acknowledgements}
We would like to express our gratitude to the anonymous reviewer for their invaluable feedback, which significantly enhanced the quality of our work.
\vspace{0.7mm}
\newline

\noindent \textbf{Funding:} This project has received funding from the European Research Council under the European Union Horizon 2020 research and innovation program (grant agreement No. 758892, ExoAI) and the European Union Horizon 2020 COMPET program (grant agreement No. 776403, ExoplANETS A). Furthermore, we acknowledge funding by the UK Space Agency and Science and Technology Funding Council grants ST/K502406/1, ST/P000282/1, ST/P002153/1, ST/ S002634/1, ST/T001836/1, ST/V003380/1, and ST/W00254X/1. This project has received funding from the European Research Council (ERC) under the European Union Horizon 2020 research and innovation programme through Advanced Investigator grant number 883830 (ExoMolHD).
\vspace{0.7mm}
\newline

\noindent \textbf{Computing:} We acknowledge the availability and support from the High Performance Computing platforms (HPC) from University College London and DIRAC, both of whom provided the computing resources necessary to perform this work. This work utilised the Cambridge Service for Data Driven Discovery (CSD3), part of which is operated by the University of Cambridge Research Computing on behalf of the STFC DiRAC HPC Facility (www.dirac.ac.uk). The DiRAC component of CSD3 was funded by BEIS capital funding via STFC capital grants ST/P002307/1 and ST/R002452/1 and STFC operations grant ST/R00689X/1. DiRAC is part of the National e-Infrastructure. Additionally, the authors acknowledge the use of the UCL Kathleen High Performance Computing Facility (Kathleen@UCL) and the UCL Myriad High Performance Computing Facility (Myriad@UCL), and associated support services, in the completion of this work.
\vspace{0.7mm}
\newline

\noindent \textbf{Software:} astropy \citep{2013A&A...558A..33A,2018AJ....156..123A, price2018astropy}, Iraclis \citep{tsiaras2016detection}, PyLightcurve \citep{tsiaras2016new}, ExoTETHyS \citep{morello2020exotethys}, TauREx\,3.1 \citep{Al_Refaie_2021}, ASteRA \citep{2024ApJ...960..107T}, YunMa \citep{2023ApJ...957..104M}, h5py \citep{collette2013python}, emcee \citep{foreman2013emcee}, Matplotlib \citep{hunter2007matplotlib}, Multinest \citep{feroz2009multinest}, PyMultinest \citep{buchner2014x}, Numpy \citep{oliphant2006guide}, corner \citep{foreman2016corner}.
\vspace{0.7mm}
\newline

\noindent \textbf{Facilities:} HST (STIS and WFC3).
\vspace{0.7mm}
\newline

\noindent \textbf{Data:} This work is based on observations with the NASA/ESA \textit{Hubble Space Telescope}, obtained at the Space Telescope Science Institute (STScI), operated by AURA, Inc under NASA contract NAS 5-26555. The publicly available HST observations presented here (Table~\ref{tab:transit_data}) were obtained from the Hubble Archive, which is part of the Mikulski Archive for Space Telescopes. We are thankful to those who operate this archive, the public nature of which increases scientific productivity and accessibility \citep{peek2019robust}. Additionally, we express gratitude to the Principal Investigators (PIs) of the observations used in study, as well as their respective teams, for requesting the observations that have and will continue to expand the characterisation of exoplanetary atmospheres.
The observations analysed in this study can be accessed via\dataset[10.17909/7nt2-mk29]{https://doi.org/10.17909/7nt2-mk29}.
\vspace{0.7mm}
\newline

\noindent \textbf{Author contributions:} 
AS and ATs contributed to the development of the data-analysis pipeline used in this work. AS reduced the data, modelled the light curves and produced the planetary spectra. AFA developed the retrieval code while AS and ATh produced the atmospheric retrieval models and generated figures and tables for this manuscript. ATh conceived the stellar activity metrics employed to assess the data analysis and retrieval results. KHY and GT provided support on statistical theory and tools. SM ran the exploratory cloud microphysics retrievals. AS wrote the manuscript along with ATh. All authors discussed the results and commented on the draft.

%\clearpage

\appendix
\renewcommand{\thefigure}{\thesection.\arabic{figure}}
\renewcommand{\thetable}{\thesection.\arabic{table}}

\section{Observations information}
\setcounter{table}{0} 
\label{appendix:obs_info}
\begin{longtable}{|c|c|c|c|c|c|}
\caption{Transit observations analysed in this study, highlighting the instrument with the corresponding grism or grating employed, the date the observation was conducted and the proposal ID with its principal investigator.}
\label{tab:transit_data} \\
\hline
Planet & Instrument & Grism/Grating & Date of observation & Proposal ID & Proposal PI \\ 
\hline \hline
\multirow{6}{6em}{GJ 436\,b} & STIS & G750 & 2015-06-10 & \multirow{2}{2em}{13665} & \multirow{2}{3em}{Benneke} \\
                            & STIS & G750 & 2016-06-14 & & \\\cline{2-6}
                            & WFC3 & G141 & 2012-10-26 & \multirow{4}{2em}{11622} & \multirow{4}{3em}{Knutson} \\
                            & WFC3 & G141 & 2012-11-29 & & \\
                            & WFC3 & G141 & 2012-12-09 & & \\
                            & WFC3 & G141 & 2013-01-02 & & \\
\hline
\multirow{5}{6em}{GJ 3470\,b} & STIS & G750 & 2015-02-07 & \multirow{5}{2em}{13665} & \multirow{5}{3em}{Benneke} \\
                              & STIS & G750 & 2015-05-12 & & \\
                              & STIS & G750 & 2016-04-03 & & \\
                              & WFC3 & G141 & 2015-01-29 & & \\
                              & WFC3 & G141 & 2015-03-13 & & \\
\hline
\multirow{4}{6em}{HAT-P-1\,b} & STIS & G430 & 2012-05-26 & \multirow{4}{2em}{12473} & \multirow{4}{2em}{Sing} \\
                              & STIS & G430 & 2012-09-19 & & \\
                              & STIS & G750 & 2012-05-31 & & \\
                              & WFC3 & G141 & 2012-07-05 & & \\
\hline
\multirow{11}{6em}{HAT-P-11\,b} & STIS & G430 & 2017-02-22 & \multirow{3}{2em}{14767} & \multirow{3}{2em}{Sing} \\
                                & STIS & G430 & 2017-05-26 & & \\
                                & STIS & G750 & 2017-04-12 & & \\\cline{2-6}
                                & WFC3 & G102 & 2016-09-14 & \multirow{7}{2em}{14793} & \multirow{7}{2em}{Bean} \\ 
                                & WFC3 & G102 & 2016-10-13 & & \\
                                & WFC3 & G102 & 2016-11-02 & & \\  
                                & WFC3 & G102 & 2016-11-07 & & \\
                                & WFC3 & G102 & 2016-11-26 & & \\
                                & WFC3 & G102 & 2016-12-26 & & \\
                                & WFC3 & G102 & 2016-09-14 & & \\\cline{3-6}
                                & WFC3 & G141 & 2012-10-18 & 12449 & Deming \\
\hline
\multirow{4}{6em}{HAT-P-12\,b}  & STIS & G430 & 2012-04-11 & \multirow{3}{2em}{12473} & \multirow{3}{2em}{Sing} \\
                                & STIS & G430 & 2012-04-30 & & \\
                                & STIS & G750 & 2013-02-04 & & \\\cline{2-6}
                                & WFC3 & G141 & 2012-10-18 & 14260 & Deming \\
\hline                    
\multirow{5}{6em}{HAT-P-18\,b} & STIS & G430 & 2015-11-15 & \multirow{2}{2em}{14099} & \multirow{2}{6em}{Mikal-Evans} \\
                               & STIS & G750 & 2015-11-21 & & \\\cline{2-6}
                               & WFC3 & G141 & 2016-02-11 & \multirow{3}{2em}{14260} & \multirow{3}{3em}{Deming} \\
                               & WFC3 & G141 & 2017-01-12 & & \\
                               & WFC3 & G141 & 2015-10-30 & & \\ 
\hline
\multirow{6}{6em}{HAT-P-26\,b} & STIS & G430 & 2018-03-27 & \multirow{2}{2em}{14767} & \multirow{2}{2em}{Sing} \\
                               & STIS & G430 & 2019-03-14 & & \\\cline{3-6}
                               & STIS & G750 & 2016-01-26 & \multirow{2}{2em}{14110} & \multirow{2}{2em}{Sing} \\\cline{2-2}
                               & WFC3 & G102 & 2016-08-16 & & \\\cline{3-6}
                               & WFC3 & G141 & 2016-05-02 & \multirow{2}{2em}{14260} & \multirow{2}{3em}{Deming} \\
                               & WFC3 & G141 & 2016-03-12 & & \\
\hline
\multirow{5}{6em}{HD 209458\,b} & STIS & G430 & 2003-05-03 & \multirow{4}{2em}{9447} & \multirow{4}{6em}{Charbonneau} \\
                               & STIS & G430 & 2003-06-25 & & \\
                               & STIS & G750 & 2003-05-31 & & \\
                               & STIS & G750 & 2003-07-05 & & \\\cline{2-6}
                               & WFC3 & G141 & 2013-09-26 & 12181 & Deming \\                           
\hline
\multirow{4}{6em}{WASP-6\,b} & STIS & G430 & 2012-06-10 & \multirow{3}{2em}{12473} & \multirow{3}{2em}{Sing} \\
                             & STIS & G430 & 2012-06-16 & & \\
                             & STIS & G750 & 2012-07-23 & & \\\cline{2-6}
                             & WFC3 & G141 & 2017-05-06 & 14767 & Sing \\ 
\hline
\multirow{5}{6em}{WASP-17\,b} & STIS & G430 & 2012-06-08 & \multirow{3}{2em}{12473} & \multirow{3}{2em}{Sing} \\
                              & STIS & G430 & 2013-03-15 & & \\
                              & STIS & G750 & 2013-03-19 & & \\\cline{2-6}
                              & WFC3 & G102 & 2017-09-25 & \multirow{2}{2em}{14918} & \multirow{2}{6em}{Wakeford} \\ 
                              & WFC3 & G141 & 2017-07-23 & & \\
\hline
\multirow{4}{6em}{WASP-29\,b} & STIS & G430 & 2017-07-02 & \multirow{3}{2em}{14767} & \multirow{3}{2em}{Sing} \\
                              & STIS & G430 & 2017-07-09 & & \\
                              & STIS & G750 & 2017-08-29 & & \\\cline{2-6}
                              & WFC3 & G141 & 2016-04-14 & 14260 & Deming \\
\hline
\multirow{4}{6em}{WASP-31\,b} & STIS & G430 & 2012-06-13 & \multirow{4}{2em}{12473} & \multirow{4}{2em}{Sing} \\
                              & STIS & G430 & 2012-06-26 & & \\
                              & STIS & G750 & 2012-07-10 & & \\
                              & WFC3 & G141 & 2012-05-13 & & \\
\hline
\multirow{6}{6em}{WASP-39\,b} & STIS & G430 & 2013-02-09 & \multirow{3}{2em}{12473} & \multirow{3}{2em}{Sing} \\
                              & STIS & G430 & 2013-02-13 & & \\
                              & STIS & G750 & 2013-03-17 & & \\\cline{2-6}
                              & WFC3 & G102 & 2016-07-07 & 14169 & Wakeford \\\cline{3-6}
                              & WFC3 & G141 & 2016-08-29 & \multirow{2}{2em}{14260} & \multirow{2}{3em}{Deming} \\
                              & WFC3 & G141 & 2017-02-07 & & \\\cline{2-6}
\hline
\multirow{4}{6em}{WASP-52\,b} & STIS & G430 & 2016-11-01 & \multirow{3}{2em}{14767} & \multirow{3}{2em}{Sing} \\
                              & STIS & G430 & 2016-11-29 & & \\
                              & STIS & G750 & 2017-05-11 & & \\\cline{2-6}
                              & WFC3 & G141 & 2016-08-28 & 14260 & Deming \\
\hline
\multirow{4}{6em}{WASP-69\,b} & STIS & G430 & 2017-05-14 &\multirow{3}{2em}{14767} & \multirow{3}{2em}{Sing} \\
                              & STIS & G430 & 2017-05-25 & & \\
                              & STIS & G750 & 2017-10-04 & & \\\cline{2-6}
                              & WFC3 & G141 & 2016-08-16 & 14260 & Deming \\
\hline
\multirow{4}{6em}{WASP-74\,b} & STIS & G430 & 2017-05-03 & \multirow{4}{2em}{14767} & \multirow{4}{2em}{Sing} \\
                              & STIS & G430 & 2017-07-19 & & \\
                              & STIS & G750 & 2017-06-19 & & \\
                              & WFC3 & G141 & 2016-10-06 & & \\                      
\hline
\multirow{4}{6em}{WASP-80\,b} & STIS & G430 & 2017-05-15 & \multirow{3}{2em}{14767} & \multirow{3}{2em}{Sing} \\
                              & STIS & G430 & 2017-07-13 & & \\
                              & STIS & G750 & 2017-10-25 & & \\\cline{2-6}
                              & WFC3 & G141 & 2016-06-21 & 14260 & Deming \\
\hline 
\multirow{4}{6em}{WASP-101\,b} & STIS & G430 & 2017-04-24 & \multirow{4}{2em}{14767} & \multirow{4}{2em}{Sing} \\
                               & STIS & G430 & 2017-08-24 & & \\
                               & STIS & G750 & 2017-11-07 & & \\
                               & WFC3 & G141 & 2016-10-02 & & \\ 
\hline
\multirow{4}{6em}{WASP-121\,b} & STIS & G430 & 2016-10-24 & \multirow{3}{2em}{14767} & \multirow{3}{2em}{Sing} \\
                               & STIS & G430 & 2016-11-06 & & \\ 
                               & STIS & G750 & 2016-11-12 & & \\\cline{2-6}
                               & WFC3 & G141 & 2016-02-06 & 14468 & \multirow{3}{6em}{Mikal-Evans} \\\cline{4-5}
                               & WFC3 & G141 & 2018-03-12 & \multirow{3}{2em}{15134} &  \\
                               & WFC3 & G141 & 2019-02-03 & &  \\
\hline
\multirow{3}{6em}{WASP-127\,b} & STIS & G430 & 2018-06-23 & \multirow{3}{2em}{14619} & \multirow{3}{2em}{Spake} \\
                               & STIS & G750 & 2018-02-18 & & \\
                               & WFC3 & G141 & 2018-04-09 & & \\
\hline \hline                           
\end{longtable}

\clearpage

\begin{sidewaystable}
\section{Planetary Parameters}
\label{appendix:planet_params}
\setcounter{table}{0} 
\centering
\caption{Planetary parameters for each system examined in this study. In instances where certain parameters were not provided in the main study referenced in the column, data were sourced from other studies noted as footnotes.}
\label{tab:planet_params}
\begin{tabular}{cccccccccc}
\multicolumn{10}{c} {} \\ \hline \hline
Planet & T$_\text{eq}$ [K] & Mass [M$_\text{J}$] & Radius [R$_\text{J}$] & Period [days] & i [$^{\circ}$] & a/R$_*$ & e & $\omega$ [$^{\circ}$] & Reference \\
\hline \hline   
GJ 436\,b & ${633}^{+100}_{-138}$ & ${0.080}^{+0.007}_{-0.006}$ & 0.366$\pm$0.014 & ${2.64389803}^{+0.00000027}_{-0.00000025}$ & ${86.858}^{+0.049}_{-0.052}$ & ${14.54}^{+0.14}_{-0.15}$ & ${0.1616}^{+0.0041}_{-0.0032}$ & ${327.2}^{+1.8}_{-2.2}$ & \cite{Lanotte2014} \\
\hline

GJ 3470\,b & ${692}^{+109}_{-150}$ & 0.043$\pm$0.005 & 0.346$\pm$0.029 & ${3.3366487}^{+0.0000043}_{-0.0000033}$ & ${88.88}^{+0.62}_{-0.45}$ & ${13.94}^{+0.44}_{-0.49}$ & ${0.017}^{+0.016}_{-0.012}$ & ${1.70}^{+0.96}_{-1.20}$ & \cite{2014MNRAS.443.1810B} \\
\hline

HAT-P-1\,b & ${1320}^{+209}_{-287}$ & 0.53$\pm$0.04 & ${1.36}^{+0.11}_{-0.09}$ & 4.46529$\pm$0.00009 & 85.9$\pm$0.8 & 10.246 %but couldn't find it in the paper
& 0.0 & 0.0 & \cite{bakos2007hat} \\
\hline

HAT-P-11\,b & ${856}^{+136}_{-186}$ & 0.081$\pm$0.009 & 0.422$\pm$0.014 & 4.8878162$\pm$0.0000071 & 88.5$\pm$0.6 & $15.58^{+0.17}_{-0.82}$ & 0.198$\pm$0.046 & 355.2$\pm$17.3 & \cite{bakos2010hat} \\
\hline

HAT-P-12\,b & ${958}^{+152}_{-208}$ & 0.211$\pm$0.012 & ${0.959}^{+0.029}_{-0.021}$ & 3.2130598$\pm$0.0000021 & 89.0$\pm$0.4 & ${11.77}^{+0.15}_{-0.21}$ & 0.0 & - & \cite{2009ApJ...706..785H} \\
\hline

HAT-P-18\,b & ${841}^{+133}_{-183}$ & 0.196$\pm$0.008 & 0.947$\pm$0.044 & 5.507978$\pm$0.000043 & 88.79$\pm$0.25 & 16.77$\pm$0.04 $^{*}$ & ${0.0090}^{+0.0300}_{-0.0090}$ & 104$\pm$50 & \cite{esposito2014gaps}\\
\hline

HAT-P-26\,b & ${994}^{+157}_{-216}$ & 0.059$\pm$0.007 & 
${0.565}^{+0.072}_{-0.032}$ & 4.234516$\pm$0.000015 & ${88.6}^{+0.5}_{-0.9}$ & 13.06$\pm$0.83 & 0.124$\pm$0.060 & 54$\pm$165 & \cite{2011ApJ...728..138H} \\
\hline

HD 209458\,b & ${1449}^{+229}_{-315}$ & ${0.685}^{+0.015}_{-0.014}$ & ${1.359}^{+0.016}_{-0.019}$ & 3.524746 & 86.71$\pm$0.05 & 8.76$\pm$0.04 & - & - & \cite{2008ApJ...677.1324T} \\
\hline

WASP-6\,b & ${1183}^{+187}_{-257}$ & 0.37$\pm$0.08 & 1.03$\pm$0.10 & 3.3610100$\pm$0.0000028 & 88.47$\pm$0.56 & 10.62$\pm$0.41 & 0.05$\pm$0.02 & - & \cite{2017AJ....153..136S} \\
\hline

WASP-17\,b & ${1755}^{+278}_{-382}$ & 0.477$\pm$0.033 & 1.932$\pm$0.053 & 3.7354845$\pm$0.0000019 & 86.71$\pm$0.30 & ${6.9638}^{+0.0884}_{-0.0862}$ & 0.0 & - & \cite{southworth2012w17}\\
\hline

WASP-29\,b & ${965}^{+153}_{-210}$ $^{\diamond}$ & 0.244$\pm$0.020 & ${0.792}^{+0.056}_{-0.035}$ & 3.922727$\pm$0.000004 & 88.8$\pm$0.7 & ${12.36}^{+0.13}_{-0.22}$ $^{\diamond}$ & ${0.03}^{+0.05}_{-0.03}$ & - & \cite{2010ApJ...723L..60H}\\
\hline

WASP-31\,b & ${1575}^{+249}_{-343}$ & 0.478$\pm$0.029 & 1.549$\pm$0.050 & 3.4059096$\pm$0.000005 & 84.41$\pm$0.22 & 8.00$\pm$0.03 $^{*}$ & 0.0 & - & \cite{anderson2011wasp} \\
\hline

WASP-39\,b & ${1117}^{+177}_{-243}$ & 0.28$\pm$0.03 & 1.27$\pm$0.04 & 4.055259$\pm$0.000009 & ${87.83}^{+0.25}_{-0.22}$ & 11.68$\pm$0.03 $^{*}$ & 0.0 & - & \cite{faedi2011w39} \\
\hline

WASP-52\,b & ${1301}^{+206}_{-283}$ & 0.46$\pm$0.02 & 1.27$\pm$0.03 & 1.7497798$\pm$0.0000012 & 85.35$\pm$0.20 & ${7.3801}^{+0.1106}_{-0.1073}$ & 0.0 & - & \cite{hebrard2013wasp} \\
\hline

WASP-69\,b & ${964}^{+153}_{-210}$ & 0.260$\pm$0.017 & 1.057$\pm$0.047 & 3.8681382$\pm$0.0000017 & 86.71$\pm$0.20 & 11.97$\pm$0.04 $^{*}$ & 0.0 & - & \cite{2014MNRAS.445.1114A} \\
\hline

WASP-74\,b & ${1917}^{+303}_{-417}$ & 0.950$\pm$0.060 & 1.56$\pm$0.06 & 2.137750$\pm$0.000001 & 79.81$\pm$0.24 & 4.85$\pm$0.04 $^{*}$ & 0.0 & - & \cite{2015AJ....150...18H} \\
\hline

WASP-80\,b & ${824}^{+130}_{-179}$ & ${0.538}^{+0.035}_{-0.036}$ & ${0.999}^{+0.030}_{-0.031}$ & ${3.06785234}^{+0.00000083}_{-0.00000079}$ & ${89.02}^{+0.11}_{-0.10}$ & ${12.63}^{+0.08}_{-0.13}$ & ${0.0020}^{+0.0100}_{-0.0020}$ & ${94.0}^{+120.0}_{-21.0}$ & \cite{triaud2015dayside} \\
\hline

WASP-101\,b & ${1558}^{+247}_{-339}$ & 0.50$\pm$0.04 & 1.41$\pm$0.05 & 3.585722$\pm$0.000004 & 85.0$\pm$0.2 & 8.44$\pm$0.04 $^{*}$ & 0.0 & - & \cite{2014MNRAS.440.1982H} \\
\hline

WASP-121\,b & ${2357}^{+373}_{-513}$ & $1.183^{+0.064}_{-0.062}$ & 1.865$\pm$0.044 & ${1.27492550}^{+0.00000020}_{-0.00000025}$ & 87.6$\pm$0.6 & ${3.754}^{+0.023}_{-0.028}$ & 0.0 & - & \cite{delrez2016w121}\\
\hline

WASP-127\,b & ${1409}^{+223}_{-306}$ & 0.18$\pm$0.02 & 1.37$\pm$0.04 & 4.178062$\pm$0.000002 & ${88.7}^{+0.8}_{-0.6}$ & ${7.95}^{+0.19}_{-0.27}$ $^{\triangleright}$ & - & - & \cite{lam2017} \\
\hline \hline \\
\end{tabular}
\vspace{1ex}
    {\raggedright $^{\diamond}$ \cite{2013MNRAS.428.3680G} \par}
    {\raggedright $^{\triangleright}$ \cite{2017A&A...602L..15P} \par}
    {\raggedright $^{*}$ derived parameter \par}
\end{sidewaystable}

\begin{sidewaystable}
\section{Stellar Parameters}
\label{appendix:stellar_params}
\setcounter{table}{0} 
\caption{Stellar parameters for each system examined in this study. In cases where certain parameters were not provided in the main study referenced in the column, information was sourced from other studies noted as footnotes. Spectral types were primarily obtained from the main reference; however, if unavailable, they were derived from the study referenced in brackets next to the spectral type.}
\label{tab:stellar_params}
\centering
\begin{tabular}{cccccccc}

\multicolumn{8}{c} {} \\ \hline \hline
Star & Fe/H & Temperature [K] & log(g) & Radius [R$_\odot$] & Mass [M$_\odot$] & Spectral Type & Reference \\
\hline \hline   

GJ 436\,b & 0.02$\pm$0.20 & 3416$\pm$100 & 4.843$\pm$0.018 & 0.455$\pm$0.018 & ${0.556}^{+0.071}_{-0.065}$ & M2.5 \citep{2015Salz} & \cite{Lanotte2014} \\
\hline

GJ 3470\,b & 0.18$\pm$0.08 & 3652$\pm$50 & 4.78$\pm$0.12 & 0.48$\pm$0.04 & 0.51$\pm$0.06 & M1.5 \citep{2019AJ....157...97K} & \cite{2014MNRAS.443.1810B} \\
\hline

HAT-P-1\,b & 0.13$\pm$0.02 & 5975$\pm$45 & 4.45$\pm$0.06 & ${1.15}^{+0.10}_{-0.07}$ & 1.12$\pm$0.09 & G0V \citep{johnson2008} &  \cite{bakos2007hat} \\
\hline

HAT-P-11\,b & 0.31$\pm$0.05 & 4780$\pm$50 & 4.59$\pm$0.03 
& 0.75$\pm$0.02 & ${0.81}^{+0.02}_{-0.03}$ & K4 & \cite{bakos2010hat} \\
\hline

HAT-P-12\,b & -0.29$\pm$0.05 & 4650$\pm$60 & 4.61$\pm$0.01 & ${0.701}^{+0.017}_{-0.012}$ & 0.733$\pm$0.018 & K4 & \cite{2009ApJ...706..785H} \\
\hline

HAT-P-18\,b & 0.10$\pm$0.06 & 4870$\pm$50 & 4.613$\pm$0.031 & 0.717$\pm$0.026 & 0.770$\pm$0.027 & K2 \citep{2011ApJ...726...52H} & \cite{esposito2014gaps} \\
\hline

HAT-P-26\,b & -0.04$\pm$0.08 & 5079$\pm$88 & 4.56$\pm$0.06 & ${0.788}^{+0.098}_{-0.043}$ & 0.816$\pm$0.033 & K1 & \cite{2011ApJ...728..138H}\\
\hline

HD 209458\,b & 0.00$\pm$0.05 & 6065$\pm$50 & ${4.361}^{+0.007}_{-0.008}$ & ${1.155}^{+0.014}_{-0.016}$ & 1.119$\pm$0.033 & F9 \citep{gray2001physical} & \cite{2008ApJ...677.1324T} \\
\hline

WASP-6\,b & -0.20 & 5450$\pm$100 & 4.60$\pm$0.20 & 0.73$\pm$0.07 & 0.55$\pm$0.17 & G8 V \citep{2009Gillon} & \cite{2017AJ....153..136S} \\
\hline

WASP-17\,b & -0.25 $^{\dagger}$ & 6550$\pm$100 $^{\dagger}$ & 4.149$\pm$0.014 & 1.583$\pm$0.041 & 1.286$\pm$0.079 & F4 \citep{triaud2010} & \cite{southworth2012w17}\\
\hline

WASP-29\,b & 0.11 $^{\dagger}$ & 4800$\pm$150 & 4.50$\pm$0.20 $^{\dagger}$ & 0.808$\pm$0.044 & 0.825$\pm$0.033 & K4 & \cite{2010ApJ...723L..60H} \\
\hline

WASP-31\,b & -0.200$\pm$0.090 $^{\ddagger}$ & 6302$\pm$102 & ${4.31}^{+0.07}_{-0.08}$ $^{\triangleleft}$ & 1.252$\pm$0.030 & 1.163$\pm$0.026 & F7-8 & \cite{anderson2011wasp} \\
\hline

WASP-39\,b & -0.12$\pm$0.10 $^{\ddagger}$ & 5400$\pm$150 & ${4.480}^{+0.029}_{-0.025}$ $^{\triangle}$ & 0.895$\pm$0.023 & 0.93$\pm$0.03 & G7 & \cite{faedi2011w39} \\
\hline

WASP-52\,b & 0.03$\pm$0.12 & 5000$\pm$100 & 4.582$\pm$0.014 & 0.79$\pm$0.02 & 0.87$\pm$0.03 & K2 V & \cite{hebrard2013wasp} \\
\hline

WASP-69\,b & 0.144$\pm$0.077 & 4715$\pm$50 & 4.535$\pm$0.023 & 0.813$\pm$0.028 & 0.826$\pm$0.029 & K5 & \cite{2014MNRAS.445.1114A} \\
\hline

WASP-74\,b & 0.39$\pm$0.13 & 5970$\pm$110 & 4.180$\pm$0.018 & 1.64$\pm$0.05 & 1.48$\pm$0.12 & F9 & \cite{2015AJ....150...18H} \\
\hline

WASP-80\,b & ${-0.13}^{+0.15}_{-0.17}$ & ${4143}^{+92}_{-94}$ & ${4.663}^{+0.015}_{-0.016}$ & ${0.586}^{+0.017}_{-0.018}$ & ${0.577}^{+0.051}_{-0.054}$ & K7 V \cite{triaud2013} & \cite{triaud2015dayside} \\
\hline

WASP-101\,b & 0.20$\pm$0.12 & 6400$\pm$110 & 4.345$\pm$0.019 & 1.29$\pm$0.04 & 1.34$\pm$0.07 & F6 & \cite{2014MNRAS.440.1982H} \\
\hline

WASP-121\,b & 0.13$\pm$0.09 & 6459$\pm$140 & ${4.242}^{+0.011}_{-0.012}$ & 1.458$\pm$0.030 & ${1.353}^{+0.080}_{-0.079}$ & F6 V & \cite{delrez2016w121} \\
\hline

WASP-127\,b & -0.18$\pm$0.06 & 5620$\pm$85 & 4.18$\pm$0.01 & 1.39$\pm$0.03 & 1.08$\pm$0.03 & G5 & \cite{lam2017} \\
\hline \hline \\
\end{tabular}
\vspace{1ex}
    {\raggedright  \par}
    {\raggedright $^{\dagger}$ \cite{2017AJ....153..136S} \par}
    {\raggedright $^{\ddagger}$ \cite{2017A&A...602A.107B}  \par}
    {\raggedright $^{\triangleleft}$ \cite{2019AJ....158..138S} \par}
    {\raggedright $^{\triangle}$ \cite{2016AcA....66...55M} \par}
\end{sidewaystable}

\clearpage

\section{Additional test on light curves: Gaussian Processes (GP) detrending}
\label{appendix:GPs}
We employed Juliet \citep{2019MNRAS.490.2262E}, in an attempt to fit the STIS broadband and spectral light curves and to assess its performance on STIS data. The software is specialised on the fitting of transit and radial velocity data, making use of the george \citep{2015ITPAM..38..252A} and celerite \citep{celerite} GP codes to perform Bayesian inference through nested sampling algorithms, including pymultinest \citep{2014A&A...564A.125B}, dynesty \citep{2020MNRAS.493.3132S} and ultranest \citep{2016S&C....26..383B, 2019PASP..131j8005B, 2021JOSS....6.3001B}. We initially employed Juliet to perform the fitting of the white and spectral light curves of HD189733\,b and KELT-7\,b. 
Given the strong instrumental offsets and challenging spectral features exhibited by these planets, we explored whether employing a different fitting method could potentially improve the results. The fitting model we use includes a variety of parameters, such as transit-only parameters, limb-darkening coefficients (only linear and quadratic laws are supported), instrumental-specific parameters and GPs hyper-parameters (which set the Kernel used). Gaussian processes are indeed powerful to model strong systematics and repeated trends in the light curves. However, they are also prone to strong overfitting, while Kernel selection can potentially strongly affect the results. We find that the algorithm does not improve the results obtained with the standard detrending technique presented in Section~\ref{sec:lightcurve_fitting}. Furthermore, to compensate for the strong chromatic stellar signal, the error bars were often inflated. Overall, while the transit depth mean values obtained with Juliet matched those resulting from the wavelength-dependent detrending, we report a two orders of magnitude increase in the average error bar as calculated by Gaussian Process techniques. 

\section{Retrieval results}
\label{sec:retrieval_results}
\setcounter{table}{0} 

\begin{table}[hbp]
\centering
\caption{Results of the retrieval models tested on GJ 436\,b, split by Case number.} 
\begin{tabular}{|c|c|c|c|c|}
\multicolumn{5}{c} {GJ 436\,b} \\ \hline \hline
\multicolumn{5}{c} {Case 1: STIS G750L visit 1 + WFC3} \\ \hline
Model setup & $T$ [K] & $R_\text{p}$ [R$_\text{J}$] & $\log_{10}$(H$_2$O) & ln(E) \\
\hline
H$_2$O (base model) & $130^{+49}_{-22}$ & $0.3678^{+0.0004}_{-0.0004}$ & $-7^{+2}_{-3}$ & 247.3$\pm$0.2 \\
%H$_2$O + Na & \\
H$_2$O + clouds & $1266^{+2540}_{-817}$ & $0.33^{+0.03}_{-0.05}$ & $-8^{+2}_{-3}$ & 252.3$\pm$0.2 \\ 
 \cline{1-1}
H$_2$O + Stellar activity & $155^{+166}_{-41}$ & $0.3680^{+0.0004}_{-0.0015}$ & $-7^{+2}_{-3}$ & 240.8$\pm$0.3 \\
%H$_2$O + Stellar activity + Na & $679.08^{+120.85}_{-114.20}$ & $0.4227^{+0.0025}_{-0.0027}$ & $-5.11^{+0.36}_{-0.26}$ & 409.70$\pm$0.24 \\
H$_2$O + Stellar activity + clouds & $1098^{+1021}_{-615}$ & $0.34^{+0.02}_{-0.03}$ & $-8^{+3}_{-3}$ & 243.9$\pm$0.3 \\
\hline
\multicolumn{5}{c} {Case 1: STIS G750L visit 2 + WFC3} \\
\hline
Model setup & $T$ [K] & $R_\text{p}$ [R$_\text{J}$] & $\log_{10}$(H$_2$O) & ln(E) \\
\hline
H$_2$O (base model) & $120^{+30}_{-15}$ & $0.3678^{+0.0003}_{-0.0004}$ & $-7^{+2}_{-4}$ & 240.7$\pm$0.2 \\
%H$_2$O + Na & $703.33^{+134.91}_{-160.09}$ & $0.4236^{+0.0018}_{-0.0015}$ & $-5.26^{+0.44}_{-0.31}$ & 384.21$\pm$0.23 \\
H$_2$O + clouds & $1293^{+1996}_{-788}$ & $0.33^{+0.02}_{-0.05}$ & $-8^{+3}_{-2}$ & 246.4$\pm$0.2 \\ 
\cline{1-1}
H$_2$O + Stellar activity & $144^{+103}_{-33}$ & $0.3681^{+0.0004}_{-0.0009}$ & $-7^{+2}_{-3}$ & 239.6$\pm$0.3 \\
%H$_2$O + Stellar activity + Na & $1031.81^{+82.46}_{-96.16}$ & 0.4148$\pm$0.0022 & $-5.60^{+0.17}_{-0.16}$ & 387.72$\pm$0.25 \\
H$_2$O + Stellar activity + clouds & $1841^{+2118}_{-1173}$ & $0.31^{+0.04}_{-0.06}$ & $-8^{+3}_{-2}$ & 243.6$\pm$0.3 \\
\hline \hline   
\end{tabular}
\label{tab:GJ 436b_retrieval_results}
\end{table}

\begin{table}[htp]
\centering
\caption{Results of the retrieval models tested on GJ 3470\,b, split by Case number.} 
\begin{tabular}{|c|c|c|c|c|}
\multicolumn{5}{c} {GJ 3470\,b} \\ \hline \hline
\multicolumn{5}{c} {Case 1: STIS G750L visit 1 + WFC3} \\ \hline
Model setup & $T$ [K] & $R_\text{p}$ [R$_\text{J}$] & $\log_{10}$(H$_2$O) & ln(E) \\
\hline
H$_2$O (base model) (sol 0) & $867^{+46}_{-41}$ & ${0.347}^{+0.001}_{-0.001}$ & $-6.2^{+0.2}_{-0.1}$ & 243.7$\pm$0.2 \\
H$_2$O (base model) (sol 1) & $110^{+15}_{-7}$ & $0.3601^{+0.0004}_{-0.0005}$ & $-1.8^{+0.7}_{-1.5}$ & 247.6$\pm$0.2 \\
H$_2$O + clouds & $1613^{+629}_{-624}$ & $0.32^{+0.02}_{-0.02}$ & $-6.4^{+2.0}_{-0.4}$ & 255.59$\pm$0.2 \\
H$_2$O + clouds + Na + K & $1072^{+933}_{-650}$ & $0.33^{+0.02}_{-0.03}$ & $-5.7^{+1.6}_{-0.8}$ & 257.12$\pm$0.2 \\ \cline{1-1}
H$_2$O + Stellar activity & $113^{+19}_{-9}$ & $0.356^{+0.002}_{-0.002}$ & $-4^{+1}_{-2}$ & 249.46$\pm$0.2 \\
H$_2$O + Stellar activity + clouds & $1200^{+1190}_{-632}$ & $0.31^{+0.03}_{-0.05}$ & $-7^{+2}_{-3}$ & 256.8$\pm$0.2 \\
H$_2$O + Stellar activity + clouds + Na + K & $746^{+566}_{-372}$ & $0.32^{+0.02}_{-0.03}$ & $-6^{+3}_{-3}$ & 256.77$\pm$0.2 \\
\hline
\multicolumn{5}{c} {Case 2: STIS G750L visit 2 + WFC3} \\
\hline
Model setup & $T$ [K] & $R_\text{p}$ [R$_\text{J}$] & $\log_{10}$(H$_2$O) & ln(E) \\
\hline
H$_2$O (base model) (sol 0) & $109^{+13}_{-7}$ & $0.3601^{+0.0005}_{-0.0004}$ & $-1.27^{+0.21}_{-1.01}$ & 252.5$\pm$0.2 \\
H$_2$O (base model) (sol 1) & $866^{+33}_{-47}$ & $0.3462^{+0.0012}_{-0.0006}$ & $-6.1^{+0.1}_{-0.2}$ & 241.9$\pm$0.2 \\
H$_2$O + clouds & $1702^{+912}_{-866}$ & $0.29^{+0.04}_{-0.03}$ & $-4^{+1}_{-1}$ & 260.4$\pm$0.2 \\
H$_2$O + clouds + Na + K & $1209^{+764}_{-656}$ & $0.30^{+0.03}_{-0.03}$ & $-4^{+1}_{-1}$ & 259.4$\pm$0.2 \\ \cline{1-1}
H$_2$O + Stellar activity &  $112^{+18}_{-8}$ & $0.360^{+0.001}_{-0.002}$ & $-3^{+1}_{-1}$ & 253.8$\pm$0.3 \\
H$_2$O + Stellar activity + clouds & $1049^{+966}_{-506}$ & $0.31^{+0.02}_{-0.05}$ & $-5^{+2}_{-3}$ & 259.3$\pm$0.2 \\
H$_2$O + Stellar activity + clouds + Na + K & $607^{+494}_{-298}$ & $0.33^{+0.02}_{-0.03}$ & $-5^{+2}_{-4}$ & 258.6$\pm$0.2 \\
\hline
\multicolumn{5}{c} {Case 3: STIS G750L visit 3 + WFC3} \\
\hline
Model setup & $T$ [K] & $R_\text{p}$ [R$_\text{J}$] & $\log_{10}$(H$_2$O) & ln(E) \\
\hline
H$_2$O (base model) (sol 0) & $110^{+15}_{-7}$ & $0.3601^{+0.0004}_{-0.0005}$ & $-1.3^{+0.2}_{-1.4}$ & 250.6$\pm$0.2 \\
H$_2$O (base model) (sol 1) & $867^{+32}_{-37}$ & $0.3463^{+0.0010}_{-0.0006}$ & $-6.1^{+0.1}_{-0.2}$ & 241.3$\pm$0.2 \\
H$_2$O + clouds & $1590^{+629}_{-790}$ & $0.30^{+0.03}_{-0.03}$ & $-4^{+1}_{-1}$ & 258.1$\pm$0.2 \\
H$_2$O + clouds + Na + K & $1257^{+621}_{-816}$ & $0.30^{+0.04}_{-0.03}$ & $-4^{+1}_{-1}$ & 257.3$\pm$0.2 \\ \cline{1-1}
H$_2$O + Stellar activity & $109^{+15}_{-7}$ & $0.358^{+0.002}_{-0.002}$ & $-2^{+1}_{-2}$ & 248.6$\pm$0.3 \\ 
H$_2$O + Stellar activity + clouds & $900^{+829}_{-486}$ & $0.31^{+0.02}_{-0.04}$ & $-5^{+2}_{-3}$ & 255.3$\pm$0.2 \\
H$_2$O + Stellar activity + clouds + Na + K & $777^{+586}_{-453}$ & $0.31^{+0.03}_{-0.03}$ & $-4^{+2}_{-4}$ & 255.3$\pm$0.2 \\
\hline \hline                           
\end{tabular}
\label{tab:GJ 3470_retrieval_results}
\end{table}

\begin{table}[htp]
\centering
\caption{Results of the retrieval models tested on HAT-P-1\,b, split by Case number.} 
\begin{tabular}{|c|c|c|c|c|}
\multicolumn{5}{c} {HAT-P-1\,b} \\ \hline \hline
\multicolumn{5}{c} {Case 1: STIS G430L visit 1 + STIS G750L + WFC3} \\ \hline
Model setup & $T$ [K] & $R_\text{p}$ [R$_\text{J}$] & $\log_{10}$(H$_2$O) & ln(E) \\
\hline
H$_2$O (base model) & $1645^{+113}_{-162}$ & $1.273^{+0.004}_{-0.003}$ & $-6.6^{+0.2}_{-0.4}$ & 269.7$\pm$0.2 \\
H$_2$O + Na (sol 0) & $1713^{+302}_{-220}$ & $1.24^{+0.01}_{-0.02}$ & $-4.3^{+0.2}_{-0.3}$ & 282.9$\pm$0.2 \\
H$_2$O + clouds & $5127^{+274}_{-490}$ & $1.188^{+0.011}_{-0.007}$ & $-6.9^{+0.2}_{-1.0}$ & 272.8$\pm$0.2 \\
H$_2$O + clouds + Na (sol 0) & $922^{+68}_{-73}$ & $1.289^{+0.003}_{-0.003}$ & $-4.5^{+0.2}_{-0.2}$ & 283.1$\pm$0.2 \\
H$_2$O + clouds + Na (sol 1) & $1491^{+405}_{-288}$ & $1.270^{+0.008}_{-0.015}$ & $-5.8^{+0.3}_{-0.2}$ & 286.4$\pm$0.2 \\ \cline{1-1}
H$_2$O + Stellar activity (spots+faculae) & $2981^{+391}_{-353}$ & $1.19^{+0.01}_{-0.02}$ & $-6.4^{+0.1}_{-0.2}$ & 295.2$\pm$0.3 \\
H$_2$O + Stellar activity (spots only) & $2056^{+232}_{-238}$ & $1.24^{+0.01}_{-0.01}$ & ${-6.2}^{+0.2}_{-0.2}$ & 288.2$\pm$0.2 \\
H$_2$O + Stellar activity (spots+faculae) + Na & $3011^{+447}_{-381}$ & ${1.19}^{+0.02}_{-0.02}$ & $-6.4^{+0.2}_{-0.2}$ & 294.8$\pm$0.3 \\
H$_2$O + Stellar activity (spots only) + Na & $1938^{+271}_{-296}$ & $1.24^{+0.01}_{-0.01}$ & $-6.1^{+0.2}_{-0.2}$ & 288.1$\pm$0.2 \\
H$_2$O + Stellar activity (spots+faculae) + clouds & $3728^{+954}_{-726}$ & $1.17^{+0.03}_{-0.07}$ & $-6.2^{+1.8}_{-0.2}$ & 295.8$\pm$0.3 \\
H$_2$O + Stellar activity (spots only) + clouds & $4383^{+677}_{-726}$ & $1.07^{+0.06}_{-0.04}$ & $-4^{+1}_{-1}$ & 294.0$\pm$0.2 \\
H$_2$O + Stellar activity (spots+faculae) + clouds + Na & $3556^{+653}_{-574}$ & $1.17^{+0.02}_{-0.09}$ & $-6.2^{+2.3}_{-0.3}$ & 295.2$\pm$0.3 \\
H$_2$O + Stellar activity (spots only) + clouds + Na & $4219^{+666}_{-623}$ & $1.07^{+0.05}_{-0.04}$ & $-3.8^{+0.9}_{-0.9}$ & 293.6$\pm$0.2 \\
\hline
\multicolumn{5}{c} {Case 2: STIS G430L visit 2 + STIS G750L + WFC3} \\
\hline
Model setup & $T$ [K] & $R_\text{p}$ [R$_\text{J}$] & $\log_{10}$(H$_2$O) & ln(E) \\
\hline
H$_2$O (base model) & $1565^{+148}_{-172}$ & $1.274^{+0.004}_{-0.003}$ & $-6.5^{+0.2}_{-0.3}$ & 272.9$\pm$0.2 \\
H$_2$O + Na & $1317^{+189}_{-186}$ & $1.276^{+0.004}_{-0.004}$ & $-5.8^{+0.3}_{-0.3}$ & 289.3$\pm$0.2 \\
H$_2$O + clouds & $1637^{+286}_{-196}$ & $1.272^{+0.005}_{-0.008}$ & $-6.5^{+0.2}_{-0.3}$ & 272.1$\pm$0.2 \\
H$_2$O + clouds + Na & $1490^{+448}_{-273}$ & $1.271^{+0.007}_{-0.017}$ & $-5.8^{+0.3}_{-0.2}$ & 289.2$\pm$0.2 \\ \cline{1-1}
H$_2$O + Stellar activity (spots+faculae) & $2832^{+384}_{-377}$ & $1.20^{+0.02}_{-0.01}$ & ${-6.4}^{+0.2}_{-0.2}$ & 292.8$\pm$0.3 \\
H$_2$O + Stellar activity (spots only) & $1666^{+185}_{-189}$ & $1.252^{+0.008}_{-0.009}$ & $-6.1^{+0.2}_{-0.2}$ & 282.6$\pm$0.2 \\
H$_2$O + Stellar activity (spots+faculae) + Na & $2758^{+273}_{-448}$ & $1.20^{+0.02}_{-0.01}$ & $-6.4^{+0.2}_{-0.2}$ & 292.4$\pm$0.3 \\
H$_2$O + Stellar activity (spots only) + Na & $1365^{+193}_{-194}$ & $1.270^{+0.006}_{-0.007}$ & $-5.8^{+0.3}_{-0.3}$ & 288.1$\pm$0.2 \\
H$_2$O + Stellar activity (spots+faculae) + clouds & $4017^{+660}_{-528}$ & $1.10^{+0.06}_{-0.07}$ & $-5^{+1}_{-2}$ & 293.0$\pm$0.3 \\
H$_2$O + Stellar activity (spots only) + clouds & $4290^{+696}_{-636}$ & $1.05^{+0.05}_{-0.05}$ & $-3.3^{+0.7}_{-0.8}$ & 291.3$\pm$0.3 \\
H$_2$O + Stellar activity (spots+faculae) + clouds + Na & $4091^{+658}_{-583}$ & $1.09^{+0.07}_{-0.06}$ & $-4^{+1}_{-1}$ & 293.9$\pm$0.2 \\
H$_2$O + Stellar activity (spots only) + clouds + Na & $4013^{+809}_{-592}$ & $1.08^{+0.04}_{-0.05}$ & $-3.9^{+1.2}_{-0.7}$ & 291.8$\pm$0.3 \\
\hline \hline   
\end{tabular}
\label{tab:HAT-P-1_retrieval_results}
\end{table}

\begin{table}[htp]
\centering
\caption{Results of the retrieval models tested on HAT-P-12\,b, split by Case number.} 
\begin{tabular}{|c|c|c|c|c|}
\multicolumn{5}{c} {HAT-P-12\,b} \\ \hline \hline
\multicolumn{5}{c} {Case 1: STIS G430L visit 1 + STIS G750L + WFC3} \\ \hline
Model setup & $T$ [K] & $R_\text{p}$ [R$_\text{J}$] & $\log_{10}$(H$_2$O) & ln(E) \\
\hline
H$_2$O (base model) & $306^{+166}_{-123}$ & $0.924^{+0.004}_{-0.005}$ & $-4.8^{+1.1}_{-0.7}$ & 278.7$\pm$0.2 \\
H$_2$O + clouds & $361^{+243}_{-157}$ & $0.922^{+0.005}_{-0.008}$ & $-4.9^{+1.1}_{-0.9}$ & 278.9$\pm$0.2 \\ \cline{1-1}
H$_2$O + Stellar activity & $262^{+188}_{-107}$ & $0.924^{+0.003}_{-0.005}$ & $-4.6^{+1.3}_{-0.8}$ & 275.1$\pm$0.2 \\
H$_2$O + Stellar activity + clouds & $645^{+783}_{-307}$ & $0.90^{+0.02}_{-0.06}$ & $-5^{+2}_{-3}$ & 275.5$\pm$0.2 \\
\hline
\multicolumn{5}{c} {Case 2: STIS G430L visit 2 + STIS G750L + WFC3} \\
\hline
Model setup & $T$ [K] & $R_\text{p}$ [R$_\text{J}$] & $\log_{10}$(H$_2$O) & ln(E) \\
\hline
H$_2$O (base model) & $151^{+46}_{-31}$ & $0.928^{+0.001}_{-0.002}$ & $-3^{+1}_{-1}$ & 274.8$\pm$0.2 \\
H$_2$O + clouds & $390^{+552}_{-222}$ & $0.91^{+0.02}_{-0.05}$ & $-3^{+1}_{-2}$ & 276.1$\pm$0.2 \\ \cline{1-1}
H$_2$O + Stellar activity & $211^{+100}_{-74}$ & $0.926^{+0.003}_{-0.004}$ & $-4.1^{+1.5}_{-0.9}$ & 274.6$\pm$0.2 \\
H$_2$O + Stellar activity + clouds & $353^{+1148}_{-174}$ & $0.91^{+0.02}_{-0.08}$ & $-4^{+2}_{-3}$ & 275$\pm$0.2 \\
\hline \hline   
\end{tabular}
\label{tab:HAT-P-12_retrieval_results}
\end{table}

\begin{table}[htp]
\centering
\caption{Results of the retrieval models tested on HAT-P-18\,b.} 
\begin{tabular}{|c|c|c|c|c|}
\multicolumn{5}{c} {HAT-P-18\,b} \\ \hline \hline
Model setup & $T$ [K] & $R_\text{p}$ [R$_\text{J}$] & $\log_{10}$(H$_2$O) & ln(E) \\
\hline
H$_2$O (base model) & $432^{+172}_{-199}$ & $0.940^{+0.006}_{-0.005}$ & $-5.2^{+1.1}_{-0.6}$ & 264.1$\pm$0.2 \\
H$_2$O + clouds & $540^{+182}_{-191}$ & $0.936^{+0.007}_{-0.006}$ & $-5.5^{+0.7}_{-0.5}$ & 264.4$\pm$0.2 \\ 
H$_2$O + clouds + Na & $383^{+152}_{-88}$ & $0.942^{+0.004}_{-0.009}$ & $-4.0^{+0.7}_{-0.7}$ & 274.9$\pm$0.2 \\ \cline{1-1}
H$_2$O + Stellar activity & $298^{+428}_{-175}$ & $0.90^{+0.01}_{-0.01}$ & $-6^{+1}_{-2}$ & 274.8$\pm$0.2 \\
H$_2$O + Stellar activity + clouds & $1757^{+2142}_{-941}$ & $0.80^{+0.07}_{-0.16}$ & $-8^{+2}_{-2}$ & 278.7$\pm$0.2 \\
H$_2$O + Stellar activity + clouds + Na & $1357^{+1304}_{-990}$ & $0.83^{+0.06}_{-0.12}$ & $-7^{+3}_{-3}$ & 278.8$\pm$0.2 \\
\hline \hline   
\end{tabular}
\label{tab:HAT-P-18_retrieval_results}
\end{table}

\begin{table}[htp]
\centering
\caption{Results of the retrieval models tested on HAT-P-26\,b, split by Case number.} 
\begin{tabular}{|c|c|c|c|c|}
\multicolumn{5}{c} {HAT-P-26\,b} \\ \hline \hline
\multicolumn{5}{c} {Case 1: STIS G430L visit 1 + STIS G750L + WFC3} \\ \hline
Model setup & $T$ [K] & $R_\text{p}$ [R$_\text{J}$] & $\log_{10}$(H$_2$O) & ln(E) \\
\hline
H$_2$O (base model) (sol 0) & $480^{+61}_{-48}$ & $0.522^{+0.001}_{-0.002}$ & $-3.1^{+0.4}_{-0.5}$ & 401.1$\pm$0.2 \\
H$_2$O (base model) (sol 1) & $474^{+33}_{-40}$ & $0.522^{+0.002}_{-0.001}$ & $-1.15^{+0.09}_{-0.16}$ & 400.3$\pm$0.2 \\
H$_2$O + clouds & $793^{+190}_{-199}$ & $0.49^{+0.02}_{-0.02}$ & $-3.0^{+1.3}_{-0.9}$ & 403.4$\pm$0.2 \\
H$_2$O + clouds + K + Na & $802^{+184}_{-152}$ & $0.49^{+0.01}_{-0.02}$ & $-2.4^{+0.8}_{-1.1}$ & 402.7$\pm$0.2 \\ \cline{1-1}
H$_2$O + Stellar activity & $453^{+81}_{-66}$ & $0.525^{+0.004}_{-0.005}$ & $-3.0^{+0.9}_{-0.6}$ & 401.2$\pm$0.2 \\
H$_2$O + Stellar activity + clouds & $881^{+258}_{-244}$ & $0.48^{+0.03}_{-0.03}$ & $-3.0^{+1.1}_{-0.9}$ & 402.9$\pm$0.2 \\
H$_2$O + Stellar activity + clouds + K + Na & $924^{+260}_{-271}$ & $0.47^{+0.03}_{-0.03}$ & $-2.5^{+0.9}_{-1.1}$ & 401.5$\pm$0.2 \\
\hline
\multicolumn{5}{c} {Case 2: STIS G430L visit 2 + STIS G750L + WFC3} \\
\hline
Model setup & $T$ [K] & $R_\text{p}$ [R$_\text{J}$] & $\log_{10}$(H$_2$O) & ln(E) \\
\hline
H$_2$O (base model) & $575^{+104}_{-97}$ & 0.520$\pm$0.003 & $-3.8^{+0.6}_{-0.5}$ & 398.0$\pm$0.2 \\
H$_2$O + clouds & $748^{+198}_{-168}$ & $0.511^{+0.008}_{-0.011}$ & $-4.2^{+0.7}_{-0.5}$ & 398.3$\pm$0.2 \\
H$_2$O + clouds + K + Na & $748^{+224}_{-182}$ & $0.510^{+0.009}_{-0.013}$ & $-4.0^{+0.8}_{-0.6}$ & 396.8$\pm$0.2 \\ \cline{1-1}
H$_2$O + Stellar activity & $572^{+112}_{-101}$ & $0.520^{+0.005}_{-0.005}$ & $-3.8^{+0.6}_{-0.5}$ & 395.9$\pm$0.2 \\
H$_2$O + Stellar activity + clouds & $1049^{+292}_{-298}$ & $0.49^{+0.02}_{-0.03}$ & $-4.2^{+0.8}_{-0.6}$ & 398.6$\pm$0.2 \\
H$_2$O + Stellar activity + clouds + K + Na & $1107^{+258}_{-240}$ & $0.46^{+0.03}_{-0.03}$ & $-3^{+1}_{-1}$ & 397.5$\pm$0.2 \\
\hline \hline   
\end{tabular}
\label{tab:HAT-P-26_retrieval_results}
\end{table}

\begin{table}[htp]
\centering
\caption{Results of the retrieval models tested on HD 209458\,b, split by Case number.} 
\begin{tabular}{|c|c|c|c|c|}
\multicolumn{5}{c} {HD 209458\,b} \\ \hline \hline
\multicolumn{5}{c} {Case 1: STIS G430L visit 1 + STIS G750L visit 1 + WFC3} \\ \hline
Model setup & $T$ [K] & $R_\text{p}$ [R$_\text{J}$] & $\log_{10}$(H$_2$O) & ln(E) \\
\hline
H$_2$O (base model) & $754^{+87}_{-115}$ & $1.343^{+0.002}_{-0.002}$ & $-5.4^{+0.3}_{-0.2}$ & 314.1$\pm$0.2 \\
H$_2$O + clouds & $864^{+217}_{-126}$ & $1.340^{+0.003}_{-0.006}$ & $-5.6^{+0.3}_{-0.3}$ & 315.7$\pm$0.2 \\\cline{1-1}
H$_2$O + Stellar activity & $748^{+84}_{-117}$ & $1.343^{+0.002}_{-0.002}$ & $-5.4^{+0.3}_{-0.2}$ & 314.5$\pm$0.2 \\
H$_2$O + Stellar activity + clouds & $882^{+214}_{-118}$ & $1.339^{+0.003}_{-0.006}$ & ${-5.6}^{+0.3}_{-0.3}$ & 316.1$\pm$0.3 \\
\hline
\multicolumn{5}{c} {Case 2: STIS G430L visit 1 + STIS G750L visit 2 + WFC3} \\
\hline
Model setup & $T$ [K] & $R_\text{p}$ [R$_\text{J}$] & $\log_{10}$(H$_2$O) & ln(E) \\
\hline
H$_2$O (base model) & $898^{+67}_{-71}$ & $1.338^{+0.002}_{-0.001}$ & ${-5.3}^{+0.2}_{-0.2}$ & 318.3$\pm$0.2 \\
H$_2$O + clouds (sol0) & $1166^{+349}_{-252}$ & $1.330^{+0.007}_{-0.009}$ & $-5.5^{+0.3}_{-0.3}$ & 319.7$\pm$0.2 \\
H$_2$O + clouds (sol1) & $2167^{+223}_{-213}$ & $1.300^{+0.005}_{-0.005}$ & $-5.84^{+0.07}_{-0.09}$ & 311.0$\pm$0.3 \\ \cline{1-1}
H$_2$O + Stellar activity & $900^{+66}_{-61}$ & $1.338^{+0.001}_{-0.001}$ & $-5.3^{+0.2}_{-0.2}$ & 319.4$\pm$0.2 \\
H$_2$O + Stellar activity + clouds & $1211^{+291}_{-240}$ & $1.329^{+0.007}_{-0.008}$ & $-5.6^{+0.3}_{-0.2}$ & 319.4$\pm$0.2 \\
\hline
\multicolumn{5}{c} {Case 3: STIS G430L visit 2 + STIS G750L visit 1 + WFC3} \\ \hline
Model setup & $T$ [K] & $R_\text{p}$ [R$_\text{J}$] & $\log_{10}$(H$_2$O) & ln(E) \\
\hline
H$_2$O (base model) & $275^{+71}_{-53}$ & $1.352^{+0.001}_{-0.002}$ & $-3.7^{+0.6}_{-0.5}$ & 291.8$\pm$0.2 \\
H$_2$O + clouds & $286^{+81}_{-63}$ & $1.351^{+0.002}_{-0.002}$ & $-3.7^{+0.6}_{-0.5}$ & 291.2$\pm$0.2 \\
H$_2$O + Na & $221^{+24}_{-29}$ & $1.3523^{+0.0009}_{-0.0010}$ & $-1.8^{+0.3}_{-0.5}$ & 303.3$\pm$0.3 \\
H$_2$O + clouds + Na & $684^{+96}_{-238}$ & $1.341^{+0.005}_{-0.003}$ & $-1.9^{+0.4}_{-0.3}$ & 308.0$\pm$0.3 \\ \cline{1-1}
H$_2$O + Stellar activity & $273^{+70}_{-58}$ & $1.352^{+0.001}_{-0.002}$ & $-3.6^{+0.6}_{-0.5}$ & 291.8$\pm$0.2 \\
H$_2$O + Stellar activity + clouds & $290^{+79}_{-63}$ & $1.351^{+0.002}_{-0.002}$ & $-3.7^{+0.5}_{-0.5}$ & 291.2$\pm$0.2 \\
H$_2$O + Stellar activity + Na & $263^{+57}_{-49}$ & $1.352^{+0.001}_{-0.001}$ & $-1.5^{+0.3}_{-0.5}$ & 306.0$\pm$0.2 \\
H$_2$O + Stellar activity + clouds + Na & $333^{+100}_{-71}$ & $1.349^{+0.003}_{-0.003}$ & $-1.5^{+0.3}_{-0.4}$ & 306.8$\pm$0.3 \\
\hline
\multicolumn{5}{c} {Case 4: STIS G430L visit 2 + STIS G750L visit 2 + WFC3} \\ \hline
Model setup & $T$ [K] & $R_\text{p}$ [R$_\text{J}$] & $\log_{10}$(H$_2$O) & ln(E) \\
\hline
H$_2$O (base model) & $374^{+89}_{-73}$ & $1.348^{+0.002}_{-0.002}$ & $-3.6^{+0.6}_{-0.5}$ & 294.4$\pm$0.2 \\
H$_2$O + clouds & $463^{+280}_{-117}$ & $1.345^{+0.003}_{-0.009}$ & $-3.8^{+0.6}_{-0.6}$ & 296$\pm$0.2 \\
H$_2$O + Na & $462^{+31}_{-44}$ & $1.346^{+0.001}_{-0.002}$ & $-1.07^{+0.05}_{-0.08}$ & 311$\pm$0.3 \\
H$_2$O + clouds + Na & $949^{+244}_{-206}$ & $1.31^{+0.01}_{-0.01}$ & $-3.1^{+0.6}_{-0.5}$ & 324.9$\pm$0.3 \\ \cline{1-1}
H$_2$O + Stellar activity & $373^{+86}_{-63}$ & $1.348^{+0.002}_{-0.002}$ & $-3.6^{+0.5}_{-0.5}$ & 295.2$\pm$0.2 \\
H$_2$O + Stellar activity + clouds & $497^{+284}_{-123}$ & $1.344^{+0.003}_{-0.012}$ & $-3.9^{+0.6}_{-0.6}$ & 296.5$\pm$0.2 \\
H$_2$O + Stellar activity + Na & $441^{+29}_{-68}$ & $1.346^{+0.002}_{-0.001}$ & $-1.10^{+0.06}_{-0.14}$ & 309.9$\pm$0.3 \\
H$_2$O + Stellar activity + clouds + Na & $953^{+238}_{-194}$ & $1.31^{+0.01}_{-0.01}$ & $-3.1^{+0.5}_{-0.5}$ & 325.6$\pm$0.2 \\
\hline
\hline \hline   
\end{tabular}
\label{tab:HD209458_retrieval_results}
\end{table}

\begin{table}[htp]
\centering
\caption{Results of the retrieval models tested on WASP-17\,b, split by Case number.} 
\begin{tabular}{|c|c|c|c|c|}
\multicolumn{5}{c} {WASP-17\,b} \\ \hline \hline
\multicolumn{5}{c} {Case 1: STIS G430L visit 1 + STIS G750L + WFC3} \\ \hline
Model setup & $T$ [K] & $R_\text{p}$ [R$_\text{J}$] & $\log_{10}$(H$_2$O) & ln(E) \\
\hline
H$_2$O (base model) & $1094^{+85}_{-84}$ & $1.823^{+0.004}_{-0.004}$ & $-4.9^{+0.2}_{-0.2}$ & 367.5$\pm$0.2 \\
H$_2$O + clouds (sol 0) & $1734^{+221}_{-214}$ & $1.75^{+0.02}_{-0.02}$ & $-5.0^{+0.3}_{-0.2}$ & 375.7$\pm$0.2 \\ 
H$_2$O + clouds (sol 1) & $1546^{+227}_{-173}$ & $1.66^{+0.03}_{-0.04}$ & $-1.76^{+0.07}_{-0.11}$ & 371.9$\pm$0.2 \\
H$_2$O + clouds + K (sol 0) & $1401^{+116}_{-118}$ & $1.69^{+0.02}_{-0.03}$ & $-1.8^{+0.1}_{-0.3}$ & 371.6$\pm$0.2 \\
H$_2$O + clouds + K (sol 1) & $1720^{+229}_{-234}$ & $1.75^{+0.02}_{-0.02}$ & $-5.0^{+0.3}_{-0.2}$ & 375.1$\pm$0.2 \\ \cline{1-1}
H$_2$O + Stellar activity & $929^{+102}_{-101}$ & $1.855^{+0.011}_{-0.010}$ & $-4.8^{+0.3}_{-0.3}$ & 374.2$\pm$0.2 \\
H$_2$O + Stellar activity + clouds & $1510^{+395}_{-305}$ & $1.79^{+0.03}_{-0.04}$ & $-5.1^{+0.3}_{-0.3}$ & 377.3$\pm$0.2 \\
H$_2$O + Stellar activity + clouds + K & $1508^{+309}_{-303}$ & $1.79^{+0.03}_{-0.04}$ & $-5.0^{+0.4}_{-0.3}$ & 376.9$\pm$0.2 \\
\hline
\multicolumn{5}{c} {Case 2: STIS G430L visit 2 + STIS G750L + WFC3} \\
\hline
Model setup & $T$ [K] & $R_\text{p}$ [R$_\text{J}$] & $\log_{10}$(H$_2$O) & ln(E) \\
\hline
H$_2$O (base model) & $952^{+46}_{-38}$ & $1.805^{+0.005}_{-0.006}$ & $-1.01^{+0.01}_{-0.02}$ & 380.4$\pm$0.3 \\
H$_2$O + clouds & $1131^{+55}_{-96}$ & $1.78^{+0.01}_{-0.01}$ & $-1.05^{+0.03}_{-0.09}$ & 383.1$\pm$0.3 \\
H$_2$O + clouds + K & $1098^{+109}_{-85}$ & $1.78^{+0.01}_{-0.02}$ & $-1.03^{+0.02}_{-0.05}$ & 384.2$\pm$0.3 \\ \cline{1-1}
H$_2$O + Stellar activity & $469^{+115}_{-60}$ & $1.885^{+0.007}_{-0.011}$ & $-2^{+1}_{-1}$ & 386.7$\pm$0.2 \\
H$_2$O + Stellar activity + clouds & $975^{+163}_{-141}$ & $1.79^{+0.03}_{-0.03}$ & $-1.5^{+0.3}_{-0.7}$ & 389.7$\pm$0.3 \\
H$_2$O + Stellar activity + clouds + K & $1026^{+142}_{-133}$ & $1.79^{+0.02}_{-0.03}$ & $-1.2^{+0.1}_{-0.4}$ & 391.4$\pm$0.3 \\
\hline \hline   
\end{tabular}
\label{tab:WASP-17_retrieval_results}
\end{table}

\begin{table}[htp]
\centering
\caption{Results of the retrieval models tested on WASP-29\,b, split by Case number.} 
\begin{tabular}{|c|c|c|c|c|}
\multicolumn{5}{c} {WASP-29\,b} \\ \hline \hline
\multicolumn{5}{c} {Case 1: STIS G430L visit 1 + STIS G750L + WFC3} \\ \hline
Model setup & $T$ [K] & $R_\text{p}$ [R$_\text{J}$] & $\log_{10}$(H$_2$O) & ln(E) \\
\hline
H$_2$O (base model) & $251^{+160}_{-95}$ & $0.765^{+0.001}_{-0.003}$ & $-8^{+2}_{-3}$ & 306.9$\pm$0.2 \\
H$_2$O + clouds & $465^{+2216}_{-255}$ & $0.760^{+0.006}_{-0.103}$ & $-9^{+3}_{-2}$ & 307.1$\pm$0.2 \\ 
H$_2$O + clouds + Na (sol 0) & $234^{+149}_{-80}$ & $0.765^{+0.001}_{-0.003}$ & $-8^{+2}_{-2}$ & 307.5$\pm$0.2 \\
H$_2$O + clouds + Na (sol 1) & $4136^{+676}_{-800}$ & $0.63^{+0.02}_{-0.01}$ & $-8^{+1}_{-2}$ & 305.3$\pm$0.2 \\ \cline{1-1}
H$_2$O + Stellar activity & $290^{+215}_{-122}$ & $0.762^{+0.003}_{-0.004}$ & $-8^{+2}_{-2}$ & 305.2$\pm$0.2 \\
H$_2$O + Stellar activity + clouds & $1919^{+2239}_{-1306}$ & $0.68^{+0.06}_{-0.10}$ & $-8^{+2}_{-2}$ & 306.4$\pm$0.2 \\
H$_2$O + Stellar activity + clouds + Na & $2124^{+1692}_{-1397}$ & $0.67^{+0.06}_{-0.08}$ & $-8^{+3}_{-3}$ & 306.5$\pm$0.2 \\
\hline
\multicolumn{5}{c} {Case 2: STIS G430L visit 2 + STIS G750L + WFC3} \\
\hline
Model setup & $T$ [K] & $R_\text{p}$ [R$_\text{J}$] & $\log_{10}$(H$_2$O) & ln(E) \\
\hline
H$_2$O (base model) & $229^{+129}_{-83}$ & $0.766^{+0.001}_{-0.002}$ & $-8^{+2}_{-3}$ & 300.5$\pm$0.2 \\
H$_2$O + clouds & $517^{+2815}_{-318}$ & $0.75^{+0.01}_{-0.10}$ & $-8^{+3}_{-2}$ & 301.0$\pm$0.2 \\ 
H$_2$O + clouds + Na (sol 0) & $200^{+87}_{-62}$ & $0.766^{+0.001}_{-0.002}$ & $-8^{+2}_{-3}$ & 300.7$\pm$0.2 \\
H$_2$O + clouds + Na (sol 1) & $4767^{+405}_{-642}$ & $0.59^{+0.03}_{-0.02}$ & $-9^{+3}_{-1}$ & 299.9$\pm$0.2 \\ \cline{1-1}
H$_2$O + Stellar activity & $232^{+165}_{-92}$ & $0.764^{+0.002}_{-0.003}$ & $-8^{+2}_{-2}$ & 297.5$\pm$0.2 \\
H$_2$O + Stellar activity + clouds & $1505^{+2448}_{-845}$ & $0.69^{+0.05}_{-0.08}$ & $-8^{+3}_{-2}$ & 298.8$\pm$0.2 \\
H$_2$O + Stellar activity + clouds + Na & $2116^{+1909}_{-1237}$ & $0.67^{+0.06}_{-0.08}$ & $-8^{+2}_{-2}$ & 298.7$\pm$0.2 \\
\hline \hline   
\end{tabular}
\label{tab:WASP-29_retrieval_results}
\end{table}

\begin{table}[htp]
\centering
\caption{Results of the retrieval models tested on WASP-31\,b, split by Case number.} 
\begin{tabular}{|c|c|c|c|c|}
\multicolumn{5}{c} {WASP-31\,b} \\ \hline \hline
\multicolumn{5}{c} {Case 1: STIS G430L visit 1 + STIS G750L + WFC3} \\ \hline
Model setup & $T$ [K] & $R_\text{p}$ [R$_\text{J}$] & $\log_{10}$(H$_2$O) & ln(E) \\
\hline
H$_2$O (base model) & $380^{+252}_{-140}$ & $1.514^{+0.008}_{-0.011}$ & $-3^{+1}_{-1}$ & 273.6$\pm$0.2 \\
H$_2$O + clouds & $1254^{+953}_{-642}$ & $1.45^{+0.05}_{-0.09}$ & $-4^{+1}_{-1}$ & 275.0$\pm$0.2 \\ \cline{1-1}
H$_2$O + Stellar activity & $566^{+282}_{-266}$ & $1.510^{+0.010}_{-0.011}$ & $-4.4^{+1.9}_{-0.8}$ & 271.8$\pm$0.2 \\
H$_2$O + Stellar activity + clouds & $1614^{+862}_{-772}$ & $1.43^{+0.06}_{-0.10}$ & $-5^{+1}_{-1}$ & 272.7$\pm$0.2 \\
\hline
\multicolumn{5}{c} {Case 2: STIS G430L visit 2 + STIS G750L + WFC3} \\
\hline
Model setup & $T$ [K] & $R_\text{p}$ [R$_\text{J}$] & $\log_{10}$(H$_2$O) & ln(E) \\
\hline
H$_2$O (base model) (sol 0) & $634^{+164}_{-168}$ & $1.506^{+0.007}_{-0.007}$ & $-4.5^{+0.7}_{-0.5}$ & 270.5$\pm$0.2 \\
H$_2$O (base model) (sol 1) & $1093^{+119}_{-85}$ & $1.491^{+0.004}_{-0.005}$ & ${-5.88}^{+0.1}_{-0.1}$ & 268.1$\pm$0.2 \\
H$_2$O + clouds (sol 0) & $1738^{+788}_{-643}$ & $1.45^{+0.03}_{-0.04}$ & $-5.8^{+0.5}_{-0.3}$ & 272.9$\pm$0.2 \\
H$_2$O + clouds (sol 1) & $5075^{+202}_{-219}$ & $1.303^{+0.009}_{-0.013}$ & $-6.4^{+0.2}_{-1.7}$ & 269.1$\pm$0.2 \\ \cline{1-1}
H$_2$O + Stellar activity & $946^{+218}_{-290}$ & $1.430^{+0.012}_{-0.009}$ & $-5.3^{+0.7}_{-0.5}$ & 
269.6$\pm$0.2 \\
H$_2$O + Stellar activity + clouds & $2075^{+1072}_{-724}$ & $1.43^{+0.04}_{-0.06}$ & $-5.7^{+0.5}_{-0.4}$ & 271.6$\pm$0.2 \\
\hline \hline   
\end{tabular}
\label{tab:WASP-31_retrieval_results}
\end{table}

\begin{table}[htp]
\centering
\caption{Results of the retrieval models tested on WASP-52\,b, split by Case number.} 
\begin{tabular}{|c|c|c|c|c|}
\multicolumn{5}{c} {WASP-52\,b} \\ \hline \hline
\multicolumn{5}{c} {Case 1: STIS G430L visit 1 + STIS G750L + WFC3} \\ \hline
Model setup & $T$ [K] & $R_\text{p}$ [R$_\text{J}$] & $\log_{10}$(H$_2$O) & ln(E) \\
\hline
H$_2$O (base model) & $819^{+43}_{-70}$ & $1.244^{+0.002}_{-0.002}$ & ${-5.4}^{+0.2}_{-0.2}$ & 239.4$\pm$0.2 \\
H$_2$O + clouds (sol 0) & $4990^{+114}_{-151}$ & $1.127^{+0.005}_{-0.007}$ & $-10.7^{+0.8}_{-0.8}$ & 227.5$\pm$0.2 \\
H$_2$O + clouds (sol 1) & $4862^{+292}_{-229}$ & $1.132^{+0.004}_{-0.010}$ & $-7.5^{+0.3}_{-0.3}$ & 227.4$\pm$0.2 \\ \cline{1-1}
H$_2$O + Stellar activity & $1228^{+170}_{-168}$ & $1.206^{+0.007}_{-0.007}$ & $-5.7^{+0.2}_{-0.2}$ & 266.2$\pm$0.2 \\
H$_2$O + Stellar activity + clouds & $1207^{+213}_{-156}$ & $1.207^{+0.007}_{-0.009}$ & $-5.7^{+0.3}_{-0.2}$ & 265.6$\pm$0.3 \\
\hline
\multicolumn{5}{c} {Case 2: STIS G430L visit 2 + STIS G750L + WFC3} \\
\hline
Model setup & $T$ [K] & $R_\text{p}$ [R$_\text{J}$] & $\log_{10}$(H$_2$O) & ln(E) \\
\hline
H$_2$O (base model) & $791^{+51}_{-93}$ & $1.244^{+0.002}_{-0.002}$ & $-5.3^{+0.3}_{-0.2}$ & 242.6$\pm$0.2 \\
H$_2$O + clouds & $793^{+51}_{-92}$ & $1.244^{+0.002}_{-0.002}$ & $-5.3^{+0.3}_{-0.2}$ & 242.0$\pm$0.2 \\ \cline{1-1}
H$_2$O + Stellar activity & $1217^{+167}_{-172}$ & $1.202^{+0.007}_{-0.007}$ & $-5.7^{+0.3}_{-0.2}$ & 263.8$\pm$0.3 \\
H$_2$O + Stellar activity + clouds & $1226^{+205}_{-189}$ & $1.201^{+0.008}_{-0.008}$ & $-5.7^{+0.3}_{-0.3}$ & 263.2$\pm$0.3 \\
\hline \hline   
\end{tabular}
\label{tab:WASP-52_retrieval_results}
\end{table}

\begin{table}[htp]
\centering
\caption{Results of the retrieval models tested on WASP-69\,b, split by Case number.} 
\begin{tabular}{|c|c|c|c|c|}
\multicolumn{5}{c} {WASP-69\,b} \\ \hline \hline
\multicolumn{5}{c} {Case 1: STIS G430L visit 1 + STIS G750L + WFC3} \\ \hline
Model setup & $T$ [K] & $R_\text{p}$ [R$_\text{J}$] & $\log_{10}$(H$_2$O) & ln(E) \\
\hline
H$_2$O (base model) &$253^{+72}_{-55}$ & $1.009^{+0.002}_{-0.002}$ & $-4.3^{+0.5}_{-0.5}$ & 311.7$\pm$0.2 \\
H$_2$O + clouds (sol 0) & $400^{+156}_{-139}$ & $1.004^{+0.005}_{-0.006}$ & $-4.9^{+0.6}_{-0.5}$ & 312.6$\pm$0.2 \\
H$_2$O + clouds (sol 1) & $750^{+184}_{-121}$ & $0.95^{+0.01}_{-0.02}$ & $-2.8^{+0.3}_{-0.4}$ & 298.6$\pm$0.2 \\ \cline{1-1}
H$_2$O + Stellar activity & $152^{+65}_{-29}$ & $1.009^{+0.001}_{-0.001}$ & $-3.1^{+1.0}_{-0.9}$ & 309.7$\pm$0.3 \\
H$_2$O + Stellar activity + clouds & $597^{+257}_{-189}$ & $0.98^{+0.01}_{-0.02}$ & $-4^{+1}_{-1}$ & 314.1$\pm$0.3 \\
\hline
\multicolumn{5}{c} {Case 2: STIS G430L visit 2 + STIS G750L + WFC3} \\
\hline
Model setup & $T$ [K] & $R_\text{p}$ [R$_\text{J}$] & $\log_{10}$(H$_2$O) & ln(E) \\
\hline
H$_2$O (base model) & $767^{+51}_{-64}$ & $0.996^{+0.002}_{-0.001}$ & $-6.0^{+0.1}_{-0.1}$ & 286.9$\pm$0.2 \\
H$_2$O + clouds (sol 0) & $811^{+31}_{-62}$ & $0.994^{+0.002}_{-0.001}$ & $-6.1^{+0.2}_{-0.2}$ & 289.2$\pm$0.3 \\ 
H$_2$O + clouds (sol 1) & $4909^{+294}_{-444}$ & $0.850^{+0.013}_{-0.009}$ & $-9^{+2}_{-2}$ & 288.6$\pm$0.2 \\
\cline{1-1}
H$_2$O + Stellar activity & $682^{+83}_{-79}$ & $0.995^{+0.002}_{-0.002}$ & $-5.7^{+0.2}_{-0.2}$ & 289.7$\pm$0.3 \\
H$_2$O + Stellar activity + clouds & $623^{+329}_{-215}$ & $0.97^{+0.02}_{-0.03}$ & $-3^{+1}_{-1}$ & 313.8$\pm$0.3 \\
\hline \hline   
\end{tabular}
\label{tab:WASP-69_retrieval_results}
\end{table}

\begin{table}[htp]
\centering
\caption{Results of the retrieval models tested on WASP-74\,b, split by Case number.} 
\begin{tabular}{|c|c|c|c|c|}
\multicolumn{5}{c} {WASP-74\,b} \\ \hline \hline
\multicolumn{5}{c} {Case 1: STIS G430L visit 1 + STIS G750L + WFC3} \\ \hline
Model setup & $T$ [K] & $R_\text{p}$ [R$_\text{J}$] & $\log_{10}$(H$_2$O) & ln(E) \\
\hline
H$_2$O (base model) & $979^{+368}_{-231}$ & $1.512^{+0.005}_{-0.007}$ & $-6.2^{+0.6}_{-1.2}$ & 304.5$\pm$0.2 \\
H$_2$O + clouds (sol 0) & $4840^{+451}_{-838}$ & $1.43^{+0.02}_{-0.01}$ & $-10^{+2}_{-2}$ & 305.9$\pm$0.2 \\ 
H$_2$O + clouds (sol 1) & $951^{+421}_{-161}$ & $1.512^{+0.004}_{-0.007}$ & $-6.1^{+0.5}_{-0.4}$ & 302.9$\pm$0.2 \\
H$_2$O + clouds + K (sol 0) & $4916^{+343}_{-539}$ & $1.433^{+0.012}_{-0.009}$ & $-10^{+2}_{-2}$ & 306.5$\pm$0.2 \\ 
H$_2$O + clouds + K (sol 1) & $1712^{+139}_{-322}$ & $1.495^{+0.008}_{-0.003}$ & $-6.4^{+0.4}_{-0.4}$ & 303.3$\pm$0.2 \\ \cline{1-1}
H$_2$O + Stellar activity & $791^{+328}_{-251}$ & $1.500^{+0.006}_{-0.008}$ & $-5.2^{+0.7}_{-0.6}$ & 311.8$\pm$0.2 \\
H$_2$O + Stellar activity + clouds & $958^{+691}_{-390}$ & $1.49^{+0.01}_{-0.04}$ & $-5^{+2}_{-1}$ & 312.4$\pm$0.2 \\
H$_2$O + Stellar activity + clouds + K & $1058^{+675}_{-411}$ & $1.49^{+0.02}_{-0.03}$ & $-4.9^{+1.5}_{-0.9}$ & 312.6$\pm$0.2 \\
\hline
\multicolumn{5}{c} {Case 2: STIS G430L visit 2 + STIS G750L + WFC3} \\
\hline
Model setup & $T$ [K] & $R_\text{p}$ [R$_\text{J}$] & $\log_{10}$(H$_2$O) & ln(E) \\
\hline
H$_2$O (base model) & $724^{+198}_{-207}$ & $1.515^{+0.004}_{-0.004}$ & $-5.4^{+0.7}_{-0.7}$ & 302.9$\pm$0.2 \\
H$_2$O + clouds (sol 0) & $4811^{+446}_{-692}$ & $1.43^{+0.01}_{-0.01}$ & $-10^{+2}_{-2}$ & 302.6$\pm$0.2 \\
H$_2$O + clouds (sol 1) & $769^{+214}_{-201}$ & $1.514^{+0.004}_{-0.004}$ & $-5.4^{+0.7}_{-0.5}$ & 302.1$\pm$0.2 \\
H$_2$O + clouds + K (sol 0) & $904^{+836}_{-215}$ & $1.510^{+0.004}_{-0.020}$ & $-5.5^{+0.6}_{-0.6}$ & 304.0$\pm$0.2 \\ 
H$_2$O + clouds + K (sol 1) & $4846^{+448}_{-386}$ & $1.425^{+0.008}_{-0.012}$ & $-9^{+2}_{-2}$ & 302.6$\pm$0.2 \\ \cline{1-1}
H$_2$O + Stellar activity & $554^{+235}_{-199}$ & $1.508^{+0.005}_{-0.005}$ & $-4.4^{+1.1}_{-0.8}$ & 303.5$\pm$0.2 \\
H$_2$O + Stellar activity + clouds & $925^{+703}_{-404}$ & $1.49^{+0.02}_{-0.04}$ & $-4^{+2}_{-2}$ & 304.1$\pm$0.2 \\
H$_2$O + Stellar activity + clouds + K & $832^{+718}_{-345}$ & $1.50^{+0.01}_{-0.03}$ & $-4^{+2}_{-1}$ & 305.2$\pm$0.2 \\
\hline \hline   
\end{tabular}
\label{tab:WASP-74_retrieval_results}
\end{table}

\begin{table}[htp]
\centering
\caption{Results of the retrieval models tested on WASP-80\,b, split by Case number.} 
\begin{tabular}{|c|c|c|c|c|}
\multicolumn{5}{c} {WASP-80\,b} \\ \hline \hline
\multicolumn{5}{c} {Case 1: STIS G430L visit 1 + STIS G750L + WFC3} \\ \hline
Model setup & $T$ [K] & $R_\text{p}$ [R$_\text{J}$] & $\log_{10}$(H$_2$O) & ln(E) \\
\hline
H$_2$O (base model) & $336^{+240}_{-168}$ & $0.972^{+0.002}_{-0.003}$ & $-4.8^{+1.0}_{-0.7}$ & 289.6$\pm$0.2 \\
H$_2$O + clouds & $537^{+294}_{-254}$ & $0.967^{+0.005}_{-0.013}$ & $-5^{+2}_{-1}$ & 291.0$\pm$0.2 \\ \cline{1-1}
H$_2$O + Stellar activity & $413^{+178}_{-197}$ & $0.971^{+0.002}_{-0.002}$ & $-5.2^{+0.9}_{-0.5}$ & 288.7$\pm$0.3 \\
H$_2$O + Stellar activity + clouds & $863^{+451}_{-275}$ & $0.95^{+0.01}_{-0.02}$ & $-4^{+1}_{-1}$ & 289.7$\pm$0.2 \\
\hline
\multicolumn{5}{c} {Case 2: STIS G430L visit 2 + STIS G750L + WFC3} \\
\hline
Model setup & $T$ [K] & $R_\text{p}$ [R$_\text{J}$] & $\log_{10}$(H$_2$O) & ln(E) \\
\hline
H$_2$O (base model) & $765^{+65}_{-114}$ & $0.9674^{+0.0012}_{-0.0007}$ & $-5.9^{+0.3}_{-0.2}$ & 288.1$\pm$0.2 \\
H$_2$O + clouds (sol 0) & $788^{+63}_{-100}$ & $0.967^{+0.001}_{-0.001}$ & $-5.9^{+0.2}_{-0.3}$ & 288.6$\pm$0.2 \\
H$_2$O + clouds (sol 1) & $5042^{+303}_{-610}$ & $0.904^{+0.007}_{-0.005}$ & $-10^{+1}_{-2}$ & 285.6$\pm$0.2 \\ \cline{1-1}
H$_2$O + Stellar activity & $759^{+69}_{-108}$ & $0.967^{+0.001}_{-0.001}$ & $-5.8^{+0.2}_{-0.2}$ & 284.9$\pm$0.3 \\
H$_2$O + Stellar activity + clouds & $970^{+357}_{-237}$ & $0.961^{+0.004}_{-0.007}$ & $-6.0^{+0.5}_{-0.3}$ & 286.3$\pm$0.3 \\
\hline \hline   
\end{tabular}
\label{tab:WASP-80_retrieval_results}
\end{table}

\begin{table}[htp]
\centering
\caption{Results of the retrieval models tested on WASP-101\,b, split by Case number.} 
\begin{tabular}{|c|c|c|c|c|}
\multicolumn{5}{c} {WASP-101\,b} \\ \hline \hline
\multicolumn{5}{c} {Case 1: STIS G430L visit 1 + STIS G750L + WFC3} \\ \hline
Model setup & $T$ [K] & $R_\text{p}$ [R$_\text{J}$] & $\log_{10}$(H$_2$O) & ln(E) \\
\hline
H$_2$O (base model) & $784^{+56}_{-83}$ & $1.346^{+0.003}_{-0.002}$ & $-9^{+2}_{-2}$ & 286.2$\pm$0.2 \\
H$_2$O + clouds (sol 0) & $5172^{+222}_{-282}$ & $1.200^{+0.008}_{-0.006}$ & $-10^{+1}_{-1}$ & 293.1$\pm$0.2 \\
H$_2$O + clouds (sol 1) & $4517^{+111}_{-190}$ & $1.224^{+0.005}_{-0.003}$ & $-10^{+1}_{-1}$ & 291.1$\pm$0.2 \\
H$_2$O + clouds + K (sol 0) & $5165^{+230}_{-271}$ & $1.200^{+0.007}_{-0.007}$ & $-10^{+2}_{-1}$ & 292.9$\pm$0.2 \\
H$_2$O + clouds + K (sol 1) & $4523^{+180}_{-187}$ & $1.223^{+0.005}_{-0.005}$ & $-10.2^{+0.9}_{-1.2}$ & 290.8$\pm$0.2 \\ \cline{1-1}
H$_2$O + Stellar activity & $185^{+127}_{-56}$ & $1.345^{+0.003}_{-0.004}$ & $-3^{+1}_{-2}$ & 295.9$\pm$0.2 \\
H$_2$O + Stellar activity + clouds & $1250^{+1485}_{-609}$ & $1.29^{+0.03}_{-0.09}$ & $-7^{+2}_{-3}$ & 298.6$\pm$0.2 \\
H$_2$O + Stellar activity + clouds + K & $1445^{+1067}_{-861}$ & $1.28^{+0.04}_{-0.09}$ & $-7^{+2}_{-3}$ & 298.3$\pm$0.2 \\
\hline
\multicolumn{5}{c} {Case 2: STIS G430L visit 2 + STIS G750L + WFC3} \\
\hline
Model setup & $T$ [K] & $R_\text{p}$ [R$_\text{J}$] & $\log_{10}$(H$_2$O) & ln(E) \\
\hline
H$_2$O (base model) & $424^{+163}_{-174}$ & $1.353^{+0.005}_{-0.004}$ & $-6^{+1}_{-3}$ & 303.0$\pm$0.2 \\
H$_2$O + clouds (sol 0) & $4010^{+1091}_{-888}$ & $1.22^{+0.03}_{-0.04}$ & $-10^{+1}_{-1}$ & 304.2$\pm$0.2 \\
H$_2$O + clouds (sol 1) & $577^{+150}_{-69}$ & $1.348^{+0.002}_{-0.007}$ & $-6.3^{+0.4}_{-0.5}$ & 302.1$\pm$0.2 \\ 
H$_2$O + clouds + K & $1572^{+1674}_{-1096}$ & $1.31^{+0.04}_{-0.06}$ & $-9^{+2}_{-2}$ & 303.9$\pm$0.2 \\ \cline{1-1}
H$_2$O + Stellar activity & $175^{+91}_{-48}$ & $1.349^{+0.003}_{-0.005}$ & $-4^{+2}_{-2}$ & 305.7$\pm$0.2 \\
H$_2$O + Stellar activity + clouds & $2080^{+1758}_{-1442}$ & $1.23^{+0.09}_{-0.13}$ & $-8^{+3}_{-3}$ & 308.2$\pm$0.2 \\
H$_2$O + Stellar activity + clouds + K & $1137^{+1192}_{-655}$ & $1.27^{+0.05}_{-0.09}$ & $-6^{+3}_{-3}$ & 308.0$\pm$0.2 \\
\hline \hline   
\end{tabular}
\label{tab:WASP-101_retrieval_results}
\end{table}

\begin{table}[htp]
\centering
\caption{Results of the retrieval models tested on WASP-121\,b, split by Case number.} 
\begin{tabular}{|c|c|c|c|c|}
\multicolumn{5}{c} {WASP-121\,b} \\ \hline \hline
\multicolumn{5}{c} {Case 1: STIS G430L visit 1 + STIS G750L + WFC3} \\ \hline
Model setup & $T$ [K] & $R_\text{p}$ [R$_\text{J}$] & $\log_{10}$(H$_2$O) & ln(E) \\
\hline
H$_2$O (base model) & $316^{+148}_{-126}$ & $1.708^{+0.003}_{-0.004}$ & $-1.2^{+0.1}_{-0.6}$ & 293.6$\pm$0.2 \\
H$_2$O + clouds (sol 0) & $4848^{+190}_{-408}$ & $1.48^{+0.02}_{-0.03}$ & $-3.1^{+0.3}_{-0.6}$ & 303.6$\pm$0.2 \\
H$_2$O + clouds (sol 1) & $2495^{+314}_{-282}$ & $1.59^{+0.03}_{-0.02}$ & $-3.5^{+0.8}_{-0.9}$ & 307.5$\pm$0.2 \\
\cline{1-1}
H$_2$O + Stellar activity (sol 0) & $1091^{+151}_{-134}$ & $1.700^{+0.004}_{-0.005}$ & $-5.9^{+0.3}_{-0.2}$ & 297.2$\pm$0.2 \\
H$_2$O + Stellar activity (sol 1) & $1986^{+130}_{-146}$ & $1.659^{+0.004}_{-0.005}$ & $-6.5^{+0.2}_{-0.2}$ & 306.3$\pm$0.3 \\
H$_2$O + Stellar activity + clouds & $3843^{+982}_{-845}$ & $1.52^{+0.06}_{-0.05}$ & $-3.2^{+0.6}_{-0.8}$ & 307.6$\pm$0.3 \\
\hline
\multicolumn{5}{c} {Case 2: STIS G430L visit 2 + STIS G750L + WFC3} \\
\hline
Model setup & $T$ [K] & $R_\text{p}$ [R$_\text{J}$] & $\log_{10}$(H$_2$O) & ln(E) \\
\hline
H$_2$O (base model) (sol 0) & $214^{+38}_{-57}$ & $1.7103^{+0.0013}_{-0.0008}$ & $-1.8^{+0.6}_{-1.0}$ & 302.8$\pm$0.2 \\
H$_2$O (base model) (sol 1) & $1072^{+122}_{-138}$ & $1.695^{+0.003}_{-0.002}$ & $-5.7^{+0.3}_{-0.2}$ & 306.5$\pm$0.2 \\
H$_2$O + clouds & $2429^{+418}_{-357}$ & $1.62^{+0.03}_{-0.04}$ & $-4^{+1}_{-1}$ & 314.2$\pm$0.2 \\ \cline{1-1}
H$_2$O + Stellar activity (sol 0) & $2273^{+230}_{-306}$ & $1.648^{+0.012}_{-0.008}$ & $-6.5^{+0.2}_{-0.2}$ & 313.0$\pm$0.3 \\
H$_2$O + Stellar activity (sol 1) & $1376^{+195}_{-187}$ & $1.689^{+0.005}_{-0.005}$ & $-6.1^{+0.2}_{-0.2}$ & 304.9$\pm$0.3 \\
H$_2$O + Stellar activity + clouds & $3097^{+619}_{-472}$ & $1.57^{+0.03}_{-0.03}$ & $-3.8^{+0.7}_{-0.6}$ & 314.2$\pm$0.3 \\
\hline \hline   
\end{tabular}
\label{tab:WASP-121_retrieval_results}
\end{table}

\begin{table}[htp]
\centering
\caption{Results of the retrieval models tested on WASP-127\,b.} 
\begin{tabular}{|c|c|c|c|c|}
\multicolumn{5}{c} {WASP-127\,b} \\ \hline \hline
Model setup & $T$ [K] & $R_\text{p}$ [R$_\text{J}$] & $\log_{10}$(H$_2$O) & ln(E) \\
\hline
H$_2$O (base model) (sol 0) & $531^{+15}_{-13}$ & $1.302^{+0.001}_{-0.002}$ & ${-3.6}^{+0.1}_{-0.1}$ & 255.7$\pm$0.2 \\
H$_2$O (base model) (sol 1) & ${474}^{+8}_{-8}$ & $1.3067^{+0.0009}_{-0.0007}$ & $-3.34^{+0.04}_{-0.07}$ & 253.4$\pm$0.2 \\
H$_2$O + clouds & $1071^{+137}_{-148}$ & $1.24^{+0.02}_{-0.02}$ & $-4.8^{+0.2}_{-0.2}$ & 268.6$\pm$0.2
\\
H$_2$O + clouds + K + Na & $1101^{+125}_{-137}$ & $1.23^{+0.02}_{-0.01}$ & $-4.8^{+0.2}_{-0.2}$ & 273.7$\pm$0.3 \\ \cline{1-1}
H$_2$O + Stellar activity & $630^{+91}_{-76}$ & $1.297^{+0.007}_{-0.012}$ & $-4.3^{+0.4}_{-0.4}$ & 255.2$\pm$0.3 \\
H$_2$O + Stellar activity + clouds & $1527^{+144}_{-134}$ & $1.15^{+0.02}_{-0.04}$ & $-4.9^{+0.7}_{-0.3}$ & 280.9$\pm$0.3 \\
H$_2$O + Stellar activity + clouds + Na + K & $1521^{+140}_{-143}$ & $1.14^{+0.03}_{-0.05}$ & $-4.7^{+1.1}_{-0.4}$ & 281.5$\pm$0.3 \\
\hline \hline   
\end{tabular}
\label{tab:WASP-127_retrieval_results}
\end{table}

\clearpage

\section{FURTHER INFORMATION ON INDIVIDUAL PLANETS}
\label{appendix:individual_planets}

\subsection{GJ 436 b}
\label{appendix:GJ 436b}
\noindent \textbf{Discovery, Observations and Previous Findings:} Discovered through a radial velocity survey \citep{2004ApJ...617..580B} and then confirmed to orbit a main-sequence M dwarf \citep{2007A&A...472L..13G} in a near-grazing transit \citep{2008ApJ...677L..59R}. Initial spectroscopic observations performed with NICMOS returned a flat transmission spectrum with no signal in the water band at 1.4$\mu$m \citep{2009MNRAS.393L...6P}. Similarly, \textit{Hubble Space Telescope} observations confirmed the featureless nature of the spectrum and ruled out a cloud-free hydrogen-dominated atmosphere \citep{2014Natur.505...66K}. Additional optical data from HST/STIS did not find any alkali signatures nor scattering hazes, further confirming a featureless spectrum at shorter wavelengths \citep{2018AJ....155...66L}. The host M dwarf is leading the exospheric escape of the planetary companion atmosphere \citep{2011A&A...529A..80E}: ultraviolet observations conducted with STIS/FUV-MAMA and Chandra, showed how the UV transit starts about two hours before, and ends more than three hours after the approximately one hour optical transit. The escaping exosphere produces a transit depth of 56.3$\pm$3.5\%  deeper and longer than the 0.69\% optical transit depth \citep{2015Natur.522..459E}. The large exospheric hydrogen tail keeps occulting the star for 10-25 hours after primary transit \citep{2017A&A...605L...7L}, driving a mass-loss rate in the range of about 10$^{8}$-10$^{9}$ grams per second \citep{2015Natur.522..459E}.
\newline
\textbf{Chemistry Results:} There is no evidence of alkali features, likely attributable to a dense high-altitude cloud layer enveloping the planet's atmosphere. This opaque layer also attenuates the water feature at 1.4 $\mu$m, which is retrieved in the range between -8 and -7 in volume mixing ratio.
\newline
\textbf{Stellar Activity:} Reported S-indexes ($S_\text{HK}$) and $\log(R'_\text{HK})$ values: a value of $S_\text{HK}$ = 0.620 and a corresponding $\log(R'_\text{HK})$ = -5.298 were reported in \citet{Knutson2010}, \citet{Sreejith2020} report a $\log(R'_\text{HK})$ = -5.090$\pm$0.001 and most recently, \citet{Claudi2024} report an S-index of $S_\text{HK}$ = 0.7$\pm$0.3 and a corresponding $\log(R'_\text{HK})$ = -5.33$\pm$0.03. Considering the Ca H\&K lines in isolation, all of the reported values in the literature are consistent and indicative of a lower level of stellar activity for GJ 436. Multiple observations spanning almost a full proposed stellar activity cycle (7.8$\pm$0.1 yrs), revealed large variations 38\%$\pm$3\% in the summed flux of the major FUV emission lines, and a multitude of brief flares \citep{2023AJ....165..146L}. Spitzer Space Telescope showed variations in transit depth in the same bandpass across different visits, an indication for occultation of stellar active regions \citep{2011ApJ...735...27K}. Although our Bayes factor results suggest a low preference for an active star model, the \textit{SAD} favours a substantial degree of stellar contamination. However, caution in the interpretation of this metric is advised due to the considerable uncertainties associated with the optical data. 

\subsection{GJ 3470 b}
\label{appendix:GJ 3470b}
\noindent \textbf{Discovery, Observations and Previous Findings:} The hot Uranus GJ 3470\,b was initially identified through a radial velocity survey conducted with HARPS and subsequently validated as a transiting planet through observations from the TRAPPIST campaign \citep{2012A&A...546A..27B}. This exoplanet, situated approximately 25.2 parsecs away from Earth, orbits a M1.5 dwarf star in a polar trajectory \citep{lepine2005catalog}. Recent investigations \citep{2022ApJ...931L..15S} have unveiled evidence of a long-term acceleration within the system, hinting at the potential presence of an outer companion in the system. Initial atmospheric studies, mostly from ground-based spectrographs, hinted at optical slopes due to molecular hydrogen or small-sized ($\leq$0.1 $\mu$m) hazes \citep{2013ApJ...770...95F} and flat infrared spectra explained by high-altitude clouds and haze \citep{2013A&A...559A..33C}, both scenarios later confirmed by \cite{2015ApJ...814..102D}. More recently, by combining STIS, WFC3 and Spitzer transmission observations, \cite{2019NatAs...3..813B} reported a robust water detection, albeit with a reduced amplitude. Additionally, they find a sharp drop in the cloud opacity at 2-3 $\mu$m characteristic of finite-sized Mie scattering aerosol particles. Transit observations in the far-ultraviolet (FUV) using STIS/G140M and covering the Lyman-$\alpha$ line revealed an expansive exosphere of neutral hydrogen surrounding GJ 3470\,b. Its blue-shifted signature suggests that neutral hydrogen atoms are propelled away from the star by the intense radiation pressure and rapidly undergo photoionization. This process leads to a significant mass loss \citep{2018A&A...620A.147B}, a process further supported by the identification of He{\sc i} 10830 $\AA$ absorption in the GJ 3470\,b's exosphere \citep{2020ApJ...894...97N}. 
\newline
\textbf{Chemistry Results:} There is no evidence of alkali features, likely attributable to a dense high-altitude cloud layer enveloping the planet's atmosphere. This opaque layer also attenuates the water feature at 1.4 $\mu$m, which is retrieved in the range between -6.3 and -4.4 in volume mixing ratio. 
\newline
\textbf{Stellar Activity:} Reported S-indexes ($S_\text{HK}$) and $\log(R'_\text{HK})$ values: \citet{Sreejith2020} report a $\log(R'_\text{HK})$ = -4.86. \citet{Claudi2024} report an S-index of $S_\text{HK}$ = 1.8$\pm$0.3 and a corresponding $\log(R'_\text{HK})$ = -5.06$\pm$0.08. The inconsistency between these reported measurements could be due to differing activity levels at each epoch, although broadly speaking both of the above literature values are indicative of a low-to-moderate activity regime for GJ 3470. STIS observations as analysed by \cite{2019NatAs...3..813B} do not display any signs of stellar activity, while our interpretation indicate a low level of activity across the stellar metrics we defined, with substantial spectral temporal variations. The monitoring of flaring activity with the Cosmic Origins Spectrograph (COS) provided the detection of three flares \citep{2021A&A...650A..73B}, on top of confirming the absorption signature originating from GJ 3470\,b's exosphere in the stellar Lyman-$\alpha$ line.

\subsection{HAT-P-1 b}
\label{appendix:hat1b}
\noindent \textbf{Discovery, Observations and Previous Findings:} HAT-P-1\,b belongs to the planetary system ADS 16402, consisting of a pair of G0 main-sequence stars with an age of about 3 Gyr. The hot-Jupiter transits one of these twin stars and has a mass of approximately 0.53 M$_\text{J}$ and a radius of $\sim$1.36 R$_\text{J}$ \citep{2007ApJ...656..552B}. The study by \cite{2013MNRAS.435.3481W} presents the first atmospheric characterisation of HAT-P-1\,b using WFC3/G141 spatial scanning observations, revealing a significant 1.4 $\mu$m water absorption above the 5$\sigma$ level. By adding STIS data sets to the pre-existing WFC3 data set, \cite{2014MNRAS.437...46N} detect a strong absorption signature short-ward of 0.55 $\mu$m, including atmospheric sodium absorption. However, no evidence for potassium absorption is found, until \cite{2015MNRAS.450..192W} detects a K signature in the atmosphere of HAT-P-1\,b using optical transit narrow-band photometry. HAT-P-1\,b seems to have a partially clear atmosphere \citep{2015ApJ...811...55M} with a $\sim$20\% cloud coverage dominated by enhanced haze \citep{2022AJ....164..173C}.
\newline
\textbf{Chemistry Results:} We find evidence for sodium absorption and a clear water feature. The atmosphere is partially free from opaque clouds as this thick layer seems to affect the near-infrared wavelengths more than the optical. In the visible wavelengths, Mie scattering features from clouds or hazes are needed to better fit this data, confirming their presence.
\newline
\textbf{Stellar Activity:} Reported S-indexes ($S_\text{HK}$) and $\log(R'_\text{HK})$ values: In the discovery paper, \citet{bakos2007hat} report a value of $\log(R'_\text{HK})$ = -4.72$\pm$0.05. \citet{Knutson2010} calculate an S-index of $S_\text{HK}$ = 0.158 and a corresponding $\log(R'_\text{HK})$ = -4.984. \citet{Danielski2022} find an S-index of $S_\text{HK}$ = 0.15 $\pm$0.00 and a corresponding $\log(R'_\text{HK})$ = -5.11$\pm$0.07, and \citet{Claudi2024} report an S-index of $S_\text{HK}$ = 0.15$\pm$0.01 and a corresponding $\log(R'_\text{HK})$ = -5.0$\pm$0.1, or $\log(R'_\text{HK})$ = -4.9$\pm$0.1 when the correction detailed in \citet{Mittag2013} is applied. The literature values are consistent and all imply a low-to-moderate level of stellar activity for HAT-P-1. Our Bayes factor results return two of the highest values across the whole sample, indicating a high level of stellar activity. However, the \textit{SAD} and \textit{SAT} indicate a low preference for the stellar contamination model and show only minor temporal deviations among the STIS G430L observations which were taken four months apart.

\subsection{HAT-P-12 b}
\label{appendix:hat12b}
\noindent \textbf{Discovery, Observations and Previous Findings:} Initial studies conducted on HAT-P-12\,b have characterised it as a H/He-dominated planet with a core mass around 10 $M_{\mathrm{\oplus}}$, making it the least massive gas giant of its kind known at the time, surpassing the record previously held by Saturn \citep{2009ApJ...706..785H}. Contrary to that, \cite{2013ApJ...778..183L} found that HAT-P-12\,b's transmission spectrum lacked the anticipated water absorption feature for a hydrogen-dominated atmosphere. Instead, a model with high-altitude clouds provided the best fit, with the authors confidently ruling out a cloud-free, hydrogen-dominated atmosphere. Further investigations by \cite{2015A&A...583A.138M} and \cite{2017MNRAS.472.3871T}, supported the notion of a flat spectrum across optical wavelengths, suggesting the presence of clouds in the planet's atmosphere. Subsequently, \cite{2018A&A...620A.142A} reexamined previous data sets using consistent methodologies, revealing a low-amplitude spectral slope and a tentative detection of potassium absorption. \cite{2020AJ....159..234W} employed a combination of STIS, WFC3, and Spitzer observations, detecting a muted water vapour absorption feature at 1.4 $\mu$m, attenuated by clouds, and a Rayleigh scattering slope in the optical indicative of small particles. Additionally, \cite{2020A&A...642A..98Y} reported a relatively flat spectrum without significant sodium or potassium absorption features, confirming the presence of high-altitude clouds. \cite{2021A&A...656A.114J} retrieved no alkali absorption signatures but reported tentative molecular absorption features of \ce{H2O}, \ce{CH4} and \ce{NH3}. 
\newline
\textbf{Chemistry Results:} Best fit model includes an opaque layer of clouds that completely covers the alkali signatures and attenuates the water feature in the near-infrared.
\newline 
\textbf{Stellar Activity:} Reported S-indexes ($S_\text{HK}$) and $\log(R'_\text{HK})$ values: \citet{Knutson2010} report an S-index of $S_\text{HK}$ = 0.253 and a corresponding $\log(R'_\text{HK})$ = -5.104. Observations from two different epochs were analysed by \citet{Mancini2018} yielding values of $S_\text{HK}$ = 0.36$\pm$0.05 and corresponding $\log(R'_\text{HK})$ = -4.88$\pm$0.07 (14.03.2015) and $S_\text{HK}$ = 0.38$\pm$0.08 and corresponding $\log(R'_\text{HK})$ = -4.87$\pm$0.09 (24.04.2015) respectively. \citet{Claudi2024} find an S-index of $S_\text{HK}$ = 0.36$\pm$0.06 and a corresponding $\log(R'_\text{HK})$ = -4.91$\pm$0.08, or $\log(R'_\text{HK})$ = -4.54$\pm$0.09 when the correction detailed in \citet{Mittag2013} is applied. Some variation is observed in the literature values but broadly they are suggestive of low-to-moderate activity for HAT-P-12. \cite{2021A&A...656A.114J} report that the discrepant chemical results among different transmission studies can be attributed to varying levels of unocculted stellar spots and faculae. Using the GTC OSIRIS observatory centred in the optical (500-900 nm), they infer a photosphere covered by 33\% of spots and 9\% of faculae. We find an overall quiet star across all our metrics, even though the \textit{SAT} reported is moderately high within our the sample at 6\% $\pm$ 2\%, indicating a relatively large temporal variability across 20 days during which the system was observed with STIS G430L.

\subsection{HAT-P-18 b}
\noindent \textbf{Discovery, Observations and Previous Findings:} HAT-P-18\,b, part of a group of low-density Saturn-mass planets, orbits a metal-rich K2 dwarf star \citep{2011ApJ...726...52H}. The planet follows a retrograde orbit, with a sky-projected angle ($\lambda$) between the stellar spin axis and the planet orbital axis measured at 132$\pm$15 degrees \citep{2014A&A...564L..13E}. Ground-based transmission spectroscopy of HAT-P-18\,b using the ACAM instrument on the William Herschel Telescope reveals a blueward slope in the optical transmission spectrum consistent with Rayleigh scattering at the planet's equilibrium temperature (852 K). No enhanced sodium absorption is detected, suggesting a high-altitude haze masking the feature and contributing to the Rayleigh slope \citep{2017MNRAS.468.3907K}. Furthermore, studies using the metastable helium line at 1083 nm indicate an atmospheric mass loss rate for HAT-P-18\,b between ${8}^{+3}_{-2}$$\times$10$^{-5}$ M$_\text{J}$ Gyr${-1}$ and ${2.6}^{+0.5}_{-0.6}$$\times$10$^{-3}$ M$_\text{J}$ Gyr${-1}$  \citep{2021ApJ...909L..10P}. Recent JWST/NIRISS observations spanning from 0.6 to 2.8 $\mu$m reveal a hazy atmosphere with clear water and escaping helium tail features, along with a distinctive signature of a star spot crossing event around mid-transit \citep{2022ApJ...940L..35F}. Further investigations by \cite{2024MNRAS.528.3354F} revealed carbon dioxide absorption and opaque clouds at a high significance levels, alongside hints of sodium.
\newline
\textbf{Chemistry Results:} We find evidence for clouds and sodium.
\newline
\textbf{Stellar Activity:} Reported S-indexes ($S_\text{HK}$) and $\log(R'_\text{HK})$ values: \citet{Claudi2024} report an S-index of $S_\text{HK}$ = 0.33$\pm$0.08 and corresponding $\log(R'_\text{HK})$ = -4.9$\pm$0.1, or $\log(R'_\text{HK})$ = -4.5$\pm$0.2 when the correction detailed in \citet{Mittag2013} is applied. These values are consistent with a moderate-to-high level of activity for HAT-P-12. Recent JWST observations with NIRISS are also suggestive of the presence of stellar contamination \citep{2024MNRAS.528.3354F}. Both the Bayes factor and the \textit{SAD} metric agree to a moderate-to-high level of stellar contamination.

\subsection{HAT-P-26 b}
\noindent \textbf{Discovery, Observations and Previous Findings:} The exoplanet HAT-P-26\,b was first discovered in 2011 as it transited its K1 dwarf host star \citep{2011ApJ...728..138H}. The system exhibits transit timing variations, as noted by \cite{2022A&A...664A.162M} and \cite{2023AJ....166..223A}, suggesting the potential influence of a third celestial body. In an effort to understand the atmospheric composition of the Neptune-mass planet HAT-P-26\,b, \cite{2016ApJ...817..141S} utilised data from LDSS-3C and the Spitzer Space Telescope. Their findings suggest the presence of water vapour, an absence of potassium, and either a high-metallicity, cloud-free atmosphere or a solar-metallicity atmosphere with a cloud deck at approximately 10 mbar. Building on this, \cite{2017Sci...356..628W} employed STIS/G750L, WFC3/G102 and G141 together with Spitzer observations and detected prominent \ce{H2O} absorption bands in the transmission spectrum. The measured water abundance served as a proxy for metallicity, indicating HAT-P-26\,b’s atmospheric heavy element content, which suggests a primordial atmosphere acquired late in its disk lifetime. Further insights into the atmospheric composition reported water, O/H at 20 times solar, and C/O less than 0.33 \citep{2019MNRAS.486.1292M}. Additionally, the authors identified potential metal hydrides (TiH, CrH, or ScH) features at significant confidence levels, suggesting the need for strong disequilibrium processes or external replenishment to maintain gas-phase metal hydrides at the planet equilibrium temperature. 
\newline
\textbf{Chemistry Results:} Despite the atmosphere of the planet being covered in a grey cloud layer that masks potential alkali features, the planet is rich in water vapour. We run an additional retrievals on both spectral cases by including TiH, ScH and CrH to match the setup used by \cite{2019MNRAS.486.1292M} for comparison. We retrieve only upper bounds for TiH and CrH and a potential tentative signature of ScH, but no strong evidence for any of these metal hydrides.
\newline
\textbf{Stellar Activity:} Reported S-indexes ($S_\text{HK}$) and $\log(R'_\text{HK})$ values: In the discovery paper, \citet{2011ApJ...728..138H} report an S-index of $S_\text{HK}$ = 0.182 and a corresponding $\log(R'_\text{HK})$ = -4.992. \citet{Claudi2024} determine an S-index of $S_\text{HK}$ = 0.17$\pm$0.08 and a corresponding $\log(R'_\text{HK})$ = -5.0$\pm$0.2, or $\log(R'_\text{HK})$ = -4.9$\pm$0.3 when the correction detailed in \citet{Mittag2013} is applied. Literature values are consistent with a lower level of stellar activity for HAT-P-26. The photometric follow-up of the host star did not reveal evidence of spot modulation \citep{2019A&A...628A.116V}. Despite our data revealing the strongest temporal variations among G430L visits across the entire sample, the star appears to be quiet according to the rest of the stellar activity metrics we defined.

\subsection{HD 209458 b}
\label{appendix:hd209458b}
\noindent \textbf{Discovery, Observations and Previous Findings:} HD 209458\,b, one of the earliest exoplanets discovered overall and the first discovered using the transit technique \citep{2000ApJ...529L..45C}, remains one of the most extensively studied exoplanets. This hot-Jupiter boasts an equilibrium temperature nearing 1500 K, a mass approximately 0.7 times that of Jupiter, and a radius equivalent to 1.36 R$_\text{J}$ \citep{2008ApJ...677.1324T}. Initial high-resolution atmospheric flow simulations of HD 209458\,b revealed notable spatio-temporal variability, challenging the conventional understanding of a simple permanent day/night dichotomy. The planet's global circulation features polar vortexes around each pole and a banded structure with roughly three broad east-west jets. The significant temperature differences induced by strong jets, up to $\sim$1000 K, suggest potential observable atmospheric variability in spectral and photometric signatures \citep{2003ApJ...587L.117C}. Transit observations using STIS G140L detected absorption in H{\sc i}, O{\sc i}, and C{\sc ii}, which indicate the presence of oxygen and carbon in HD 209458\,b's extended upper atmosphere \citep{2004ApJ...604L..69V}. Subsequent studies with STIS data identified Rayleigh scattering \citep{2008A&A...485..865L} and Na absorption \citep{2008ApJ...686..658S}, further corroborated at high spectral resolution by \cite{2008A&A...487..357S}. The first water detection in the atmosphere of HD 209458\,b was reported using WFC3 spatial scanning data \citep{2013ApJ...774...95D}. Confirming potential detections of \ce{CO} and \ce{H2O}, a study utilising high-resolution spectroscopy from VLT CRIRES reported additional \ce{HCN} signatures \citep{2018ApJ...863L..11H}.
\newline
\textbf{Chemistry:} While we detect evidence of sodium in scenarios 3 and 4, it is likely attributable to the fortunate alignment of STIS G430L visit 2 with STIS G750L visit 1 and visit 2, resulting in a significant feature centred around 0.6 $\mu$m. However, we do not consider this to be a genuine feature. A dense layer of clouds is retrieved independently of the spectral case considered.
\newline
\textbf{Stellar Activity:} Reported S-indexes ($S_\text{HK}$) and $\log(R'_\text{HK})$ values: \citet{Knutson2010} report an S-index of $S_\text{HK}$ = 0.160 and corresponding $\log(R'_\text{HK})$ = -4.970. \citet{Sreejith2020} report a value of $\log(R'_\text{HK})$ = -4.92$\pm$0.03. \citet{Claudi2024} report an S-index of $S_\text{HK}$ = 0.156$\pm$0.003 and a corresponding $\log(R'_\text{HK})$ = -4.97$\pm$0.02, or $\log(R'_\text{HK})$ = -4.86$\pm$0.03 when the correction detailed in \citet{Mittag2013} is applied. Literature values are consistent with each other and indicative of a low level of stellar activity for HD 209458. All of our activity indicators prefer a quiet star over an active one.

\subsection{WASP-17 b}
\label{appendix:wasp17b}
\noindent \textbf{Discovery, Observations and Previous Findings:} Among the exoplanet population of inflated hot-Jupiters, WASP-17\,b \citep{2010ApJ...709..159A} is one of
the least dense planets discovered so far, showcasing a retrograde orbit \citep{2010ApJ...722L.224B}. Its large scale height allowed for a variety of transmission studies to be extremely successful, both from space and from ground-based facilities. Sodium was detected by \cite{2011MNRAS.412.2376W, 2012MNRAS.426.2483Z, 2018A&A...618A..98K}, potassium by \cite{2016A&A...596A..47S}, water was confirmed by \citep{sing2016continuum, 2022MNRAS.512.4185A} and \cite{sabatransmission} who also tentatively detected AlO and TiH in a cloud-free atmosphere. More recently, a single transit obtained with the MIRI LRS instrument on \textit{JWST} was able to identify absorption from \ce{SiO2} (quartz) clouds at 8.6 $\mu$m. These \ce{SiO2} clouds consist of small particles measuring approximately 0.01 $\mu$m, and they extend to elevated altitudes within the planetary atmosphere \citep{2023ApJ...956L..29G}.
\newline
\textbf{Chemistry Results:} We find reasonable evidence for clouds in both scenarios but the retrieved values are not consistent between visits ($\log_{10}$($P_{\text{clouds}}$) = 3.8 Pa (Case 1) and $\log_{10}$($P_{\text{clouds}}$) = 1.4 Pa (Case 2)). No detection of K in either Case.
\newline
\textbf{Stellar Activity:} Reported S-indexes ($S_\text{HK}$) and $\log(R'_\text{HK})$ values: \citet{Knutson2010} report an S-index of $S_\text{HK}$ = 0.121 and a corresponding $\log(R'_\text{HK})$ = -5.331. \citet{Claudi2024} derive a $\log(R'_\text{HK})$ = -5.0183 from the S-index obtained by \citet{Knutson2010}. Both of these literature values are indicative of a lower activity level for WASP-17. In contrast to this, our results hint at a moderate to strong level of activity for these data sets with a high temporal variation of the optical data.

\subsection{WASP-29 b}
\noindent \textbf{Discovery, Observations and Previous Findings:} WASP-29\,b, a Saturn-sized planet, was discovered in orbit around a K4 dwarf star with a V magnitude of 11.3, completing a transit every 3.9 days \citep{2010ApJ...723L..60H}. Observations using the Gemini Telescope investigated the transmission spectrum of WASP-29\,b in the wavelength range of 515 to 720 nm and reported the absence of spectral features including the non-detection of a pressure-broadened Na feature \citep{2013MNRAS.428.3680G}. These results indicated that WASP-29\,b's atmosphere is likely devoid of sodium or that clouds and hazes play a significant role, possibly masking the broad wings of the Na feature. The evidence suggests that Na may not be present in the atmosphere, especially considering WASP-29\,b's equilibrium temperature of 970 K, at which Na can form various compounds. Moreover, recent findings based on data from the Hubble and the Spitzer Space Telescopes reveal compelling evidence of aerosol opacity along the terminator of WASP-29\,b \citep{2022AJ....164...30W}. The flat spectrum across the visible and near-infrared regions suggests the existence of condensate clouds at low pressures. 
\newline
\textbf{Chemistry Results:} Flat spectrum from optical to infrared wavelengths indicative of opaque clouds at a pressure of tens of mbar.
\newline
\textbf{Stellar Activity:} Reported S-indexes ($S_\text{HK}$) and $\log(R'_\text{HK})$ values: No S-index or $\log(R'_\text{HK})$ values found in the literature. Observations of WASP-29 with HST-COS reveal a low X-ray emission flux for a star of its spectral type (K4V). This combined with a lack of observed flares and little variability in the FUV could imply that WASP-29 is old and inactive \citep{2021A&A...649A..40D}. Despite not detecting strong evidence for brightness variations on the host star, it remains challenging to confidently exclude the possibility of lower-level stellar variability. We find no evidence for stellar contamination, albeit moderately high temporal variations across G430L visits taken a week apart.

\subsection{WASP-31 b}
\noindent \textbf{Discovery, Observations and Previous Findings:} \citep{anderson2011wasp} reported the discovery of a low-density planet transiting a metal-poor, late-F-type dwarf star, WASP-31, which is part of a visual double star located approximately 35'' away. The companion star is a mid-to-late K-type star, with proper motions indicating a common proper motion pair. \citep{2015MNRAS.446.2428S} conducted a spectral analysis of the planet atmosphere covering 0.3-1.7 $\mu$m at a resolution R $\sim$ 70, combined with Spitzer photometry for a full optical to IR coverage. The spectrum revealed a predominant cloud deck with a flat transmission spectrum, notably apparent at wavelengths $>$ 0.52 $\mu$m. This cloud deck, situated at high altitudes and low pressures, covers the expected optical Na line and near-IR \ce{H2O} features. Although Na{\sc i} absorption was not definitively identified, a robust potassium feature was detected at the 4.2$\sigma$ confidence level. The absence of Na and the strong presence of K suggest a sub-solar Na/K abundance ratio. A distinct Rayleigh scattering signature was identified at short wavelengths. The spectrum is explained by two aerosol size populations, with a smaller sub-micron grain population reaching high altitudes producing a blue Rayleigh scattering signature, overlaying a larger, lower-lying population responsible for the flat cloud deck at longer wavelengths. \citep{2017MNRAS.467.4591G} confirmed the previously inferred cloud deck using FORS2 observations, aligning with earlier HST/STIS measurements. The Rayleigh scattering signature at short wavelengths and the cloud deck at longer wavelengths were successfully re-detected, but the large potassium feature previously detected using STIS was seemingly ruled out by FORS2 observations. In an effort to resolve discrepancies, \citep{2019MNRAS.482..606G} presented high-resolution observations (R $>$ 80,000) of WASP-31\,b obtained with the UVES spectrograph. These observations did not detect K absorption at the level previously reported with HST, consistent with VLT observations. More recently, \citep{2020AJ....160..230M} presented a new optical (400-950 nm) transmission spectrum of WASP-31\,b observed with IMACS on the Magellan Baade Telescope. The best-fit model suggested a scattering slope consistent with a Rayleigh slope, high-altitude clouds and muted \ce{H2O} features. Finally, \citep{2021A&A...646A..17B} modelled existing STIS, WFC3, and Spitzer data sets, reporting evidence for spectroscopic signatures of chromium hydride (CrH), \ce{H2O}, and K in WASP-31\,b.
\newline
\textbf{Chemistry Results:} As we see no hints for alkali features, we do not include these in our retrievals. If present at all, their apparent absence may be due to the presence of a thick layer of clouds that weakens the amplitude of the water feature in the near-infrared.
\newline
\textbf{Stellar Activity:} Reported S-indexes ($S_\text{HK}$) and $\log(R'_\text{HK})$ values: \citet{sing2016continuum} state a value of $\log(R'_\text{HK})$ = -5.225. \citet{Danielski2022} find an S-index $S_\text{HK}$ = 0.15$\pm$0.00 but they note that WASP-31 is outside of the magnitude range required for their self-consistent conversion to $\log(R'_\text{HK})$ and therefore do not calculate this. \citet{Claudi2024} find an S-index of $S_\text{HK}$ = 0.15$\pm$0.02 and a corresponding $\log(R'_\text{HK})$ = -5.0$\pm$0.1, or $\log(R'_\text{HK})$ = -4.9$\pm$0.2 when the correction detailed in \citet{Mittag2013} is applied. All of the reported values in the literature are consistent with a low level of activity for WASP-31. Bayes factor and \textit{SAD} indicate absent-to-low levels of contamination in addition to a small \textit{SAT} value.

\subsection{WASP-52 b}
\noindent \textbf{Discovery, Observations and Previous Findings:} The hot-Jupiter WASP-52\,b represents another example of inflated planet with low density. Its orbit aligns with the rotational direction of the host star, albeit with a slight misalignment relative to the stellar equator  \citep{hebrard2013wasp}. Utilising observations from the Herschel Telescope \cite{2016MNRAS.463.2922K} determined that Rayleigh scattering is not the primary source of opacity in the planetary atmosphere. Instead, the transmission spectrum suggests grey cloud formations. The transit data obtained from additional ground-based observations, reported a notable sodium absorption signature \citep{2017A&A...600L..11C} among an otherwise featureless spectrum, further ruling out a cloud-free atmosphere \citep{2017MNRAS.470..742L}. A complete optical to NIR spectrum obtained with space-based telescopes displays again a featureless spectra with evidence for sodium absorption \citep{2018AJ....156..298A}. At high spectral resolutions, excess absorption is detected at the Na doublet, the H$\alpha$ line, and the K D1 line \citep{2020A&A...635A.171C}, as well as at the neutral helium line \citep{2022AJ....164...24K}, estimating an atmospheric loss rate equal to 0.5\% of the planet mass per Gyr. 
\newline
\textbf{Chemistry Results:} No evidence of Na and K. On both spectral scenarios, the cloud pressure tends towards the upper bound ($\log_{10}$($P_{\text{clouds}}$) $>$ 5), so if present at all clouds would be at high pressures/low altitudes.
\newline
\textbf{Stellar Activity:} Reported S-indexes ($S_\text{HK}$) and $\log(R'_\text{HK})$ values: In the discovery paper, \citet{hebrard2013wasp} report a value of $\log(R'_\text{HK})$ = -4.4$\pm$0.2. \citet{Claudi2024} convert the S-index of $S_\text{HK}$=0.585 from \citet{hebrard2013wasp} to a corresponding $\log(R'_\text{HK})$ = -4.4180, or $\log(R'_\text{HK})$ = -4.1318 when the correction detailed in \citet{Mittag2013} is applied. The literature values are all consistent with a moderate-to-high level of activity for WASP-52. The occasional contamination from unocculted spots during transit is noted in various studies (e.g. \cite{2018AJ....156..122M}), with an estimated star spots temperature $<$ 3000 K and a 5\% fractional coverage consistent with expectations for the host star's spectral type \citep{2020MNRAS.491.5361B}. More recently, \cite {2023A&A...674A.174C} reported the absence of star spot occultation in WASP-52. This lack of detection is attributed to the star's relatively cool temperature, elevated magnetic activity reducing center-to-limb variations (CLV), and a projected obliquity near 0 degrees, thereby minimising the probability of the transit chord crossing multiple stellar latitudes. WASP-52\,b is the planet with the highest level of stellar contamination according to the Bayes factor; both spectral scenarios score a large \textit{SAD} compared to minimal temporal variations observed over a one-month period.

\subsection{WASP-69 b}
\noindent \textbf{Discovery, Observations and Previous Findings:} WASP-69\,b, an inflated Saturn-mass exoplanet, orbits a mid-K dwarf in a 3.868-day period, as reported by \cite{2014MNRAS.445.1114A}. The first atmospheric characterisation was performed at high-resolution using HARPS-N, leading to sodium detection \citep{2017A&A...608A.135C}. Subsequent high-dispersion studies identified Na{\sc i} D2 and D1 lines in the atmosphere of WASP-69\,b, suggesting strong Rayleigh scattering, solar to super-solar water abundance, and a highly muted sodium feature \citep{2021A&A...656A.142K}. More recently, \cite{2022A&A...665A.104G} detected \ce{CH4}, \ce{NH3}, \ce{CO}, \ce{C2H2}, and \ce{H2O} in the atmosphere of WASP-69\,b, marking the first simultaneous detection of five molecules in the atmosphere of a warm giant planet. Additionally, \cite{2018Sci...362.1388N} and \cite{2020AJ....159..278V} observed excess absorption in the helium triplet at 1083 nm during and post transit, interpreting it as atmospheric escape in a comet-like form. Reportedly, WASP-69\,b loses its atmosphere at a rate of (0.9$\pm$0.5)$\times$10$^{11}$ gs$^{-1}$ \citep{2023A&A...673A.140L}. From space, WFC3 observations highlighted a muted water feature at 1.4 $\mu$m in the planet's spectrum, with an overlying slope indicating a larger planet radius toward bluer wavelengths \citep{tsiaras2018population}. The inclusion of STIS data revealed direct evidence of aerosols at $\sim$1 microbar pressures in the atmosphere of WASP-69\,b \citep{2021AJ....162...91E}. An optical ground-based spectrum also revealed an increase in transit depth toward bluer wavelengths, consistent with space-based measurements obtained by the \textit{Hubble Space Telescope} \citep{2020A&A...641A.158M}. The origin of the detected slope is related to either Rayleigh scattering in the planet's atmosphere or a signal induced by stellar activity. Observations from the Southern Astrophysical Research Telescope (SOAR) by \cite{2023MNRAS.521.5860O} confirmed a Rayleigh scattering in WASP-69\,b's atmosphere, and the retrieval analysis tentatively identified a \ce{TiO} absorption feature in the transmission spectrum.
\newline
\textbf{Chemistry Results:} No evidence for alkali features; muted water feature and high altitude gray clouds at 63 mbar.
\newline
\textbf{Stellar Activity:} Reported S-indexes ($S_\text{HK}$) and $\log(R'_\text{HK})$ values: \citet{Sreejith2020} report a $\log(R'_\text{HK})$ = -4.54$\pm$0.04. \citet{Claudi2024} report an S-index of $S_\text{HK}$ = 0.67$\pm$0.01 and a corresponding $\log(R'_\text{HK})$ = -4.522$\pm$0.008, or $\log(R'_\text{HK})$ = -4.163$\pm$0.009 when the correction detailed in \citet{Mittag2013} is applied. These reported values are consistent with a moderate-to-high level of activity for WASP-69. The Bayes factor prefers the base model over the stellar model in both scenarios, however the inclusion of clouds strongly drives the preference towards the stellar activity model. Overall we find multiple indications for a quiet star.

\subsection{WASP-74 b}
\noindent \textbf{Discovery, Observations and Previous Findings:} WASP-74\,b, with a mass of 0.95 M$_\text{J}$ and a moderately inflated radius of 1.5 R$_\text{J}$, orbits a slightly evolved F9 star in a 2-day period \citep{2015AJ....150...18H}. By combining optical data with near-infrared observations from the \textit{Hubble Space Telescope}, \cite{2019MNRAS.485.5168M} found the likely presence of optical absorbers, such as TiO and VO gases, in the atmosphere of the planet. However, later studies were discrepant in revealing a steeper slope than expected from Rayleigh scattering alone \citep{2020A&A...642A..50L} and no indications of TiO/VO or super-Rayleigh scattering \citep{2021AJ....162..271F}. The transmission spectrum indicates an atmosphere depleted in water \citep{tsiaras2018population}, consistent with a primarily featureless optical slope, favouring hazes in the upper atmospheric layers \citep{2023MNRAS.521.2163S}.
\newline
\textbf{Chemistry Results:} Weak evidence for potassium but none for sodium. Small amplitude water feature covered by thick clouds.
\newline
\textbf{Stellar Activity:} Reported S-indexes ($S_\text{HK}$) and $\log(R'_\text{HK})$ values: No S-index or $\log(R'_\text{HK})$ values found in the literature. Ground-based photometric monitoring from 2018-2021 showed very little variability suggesting that WASP-74 is magnetically inactive \citep{2021AJ....162..271F}. According to the Bayes factor and the \textit{SAD}, Case 1 and Case 2 report a strong and an inconclusive preference for stellar activity respectively, with moderate temporal variations in the order of 4.2\%. We cannot definitively ascertain whether the star is active or inactive.

\subsection{WASP-80 b}
\noindent \textbf{Discovery, Observations and Previous Findings:} The gas giant WASP-80\,b is similar in brightness and colour to a T4-dwarf but possesses a cooler temperature ($\sim$800 K) \citep{2015MNRAS.450.2279T}. Almost all atmospheric studies conducted on this planets agreed on the planet possessing a thick cloud layer \citep{2014ApJ...790..108F}, an atmosphere depleted of molecular species in the infrared \citep{2017MNRAS.468.3123S} and no evidence for alkali signatures \citep{2018A&A...609A..33P, 2018MNRAS.474..876K}. The discernible but muted water absorption feature at 1.4 $\mu$m turns into a steep optical spectral slope, possibly caused by fine-particle aerosols or contamination from unocculted spots on the host star \citep{2022AJ....164...30W}. The cross-correlation technique applied to GIANO-B near-infrared observations indicate the presence of \ce{H2O}, \ce{CH4}, \ce{NH3}, and \ce{HCN} with high significance, a tentative detection of \ce{CO2}, and inconclusive results for \ce{C2H2} and CO \citep{2022AJ....164..101C}. No helium absorption was detected \citep{2022A&A...658A.136F}, likely due to a solar He/H abundance ratio combined with a strong stellar wind or by a subsolar He/H abundance ratio, or a combination of both \citep{2023A&A...673A..37F}. 
\newline
\textbf{Chemistry Results:} The atmosphere exhibits no signs of alkali features, with a water absorption signal obscured by a dense layer of opaque clouds. The optical slope appears steeper in Case 2 compared to Case 1, potentially indicating the presence of optical hazes, though we have not conducted tests to confirm this.
\newline
\textbf{Stellar Activity:} Reported S-indexes ($S_\text{HK}$) and $\log(R'_\text{HK})$ values: \citet{Claudi2024} report an S-index of $S_\text{HK}$ = 1.77$\pm$0.09 and a corresponding $\log(R'_\text{HK})$ = -4.82$\pm$0.02, or $\log(R'_\text{HK})$ = -4.171$\pm$0.002 when the correction detailed in \citet{Mittag2013} is applied. There is a large difference between the derived $\log(R'_\text{HK})$ values depending on whether or not the correction is used making it harder to categorise the expected activity level of WASP-80. Taking the $\log(R'_\text{HK})$ value derived solely with the \citet{Noyes1984} method indicates a low-to-moderate level of activity, whereas, following the \citet{Mittag2013} correction, the implied activity level is much higher. WASP-80 is a potentially active K7V dwarf \citep{triaud2013} with a strong chromospheric activity (log R$_{\text{'HK}}$=-4.495). Despite this, the non-detection of stellar spot crossings implies that, if present, spots must be small and symmetrically distributed on the stellar surface \citep{2014A&A...562A.126M}. Our findings are in agreement with a low level of activity for WASP-80, confirmed by a negative Bayes factor, a maximum \textit{SAD} of 1.2$\pm$0.5 and a low \textit{SAT} in the range of 3.4\%.

\subsection{WASP-101 b}
\noindent \textbf{Discovery, Observations and Previous Findings:} The bloated low-mass planet WASP-101\,b orbits a main sequence F6 star \citep{2014MNRAS.440.1982H}. \textit{Hubble Space Telescope} (HST) Wide Field Camera 3 observations revealed that the near-infrared transmission spectrum of WASP-101\,b lacks significant H2O absorption features \citep{2017ApJ...835L..12W}. The addition of HST/STIS data confirms a flat and featureless transmission spectrum for WASP-101\,b, decisively ruling out a clear atmosphere. The spectrum aligns with high-altitude silicate clouds at pressures lower than 100 $\mu$bar. This cloud layer obstructs views into deeper atmospheric layers, marking WASP-101\,b as the cloudiest observed gas giant to date \citep{2023MNRAS.522..582R}.
\newline
\textbf{Chemistry Results:} A thick layer of clouds heavily mutes K and water features. No obvious evidence of Na.
\newline
\textbf{Stellar Activity:} Reported S-indexes ($S_\text{HK}$) and $\log(R'_\text{HK})$ values: No S-index or $\log(R'_\text{HK})$ values found in the literature. Photometric observations conducted over five observing seasons showed minimal variability suggestive of a very low/negligible level of activity for WASP-101 \citep{Rathcke2023}. There is a moderate to strong preference for an active star according to the Bayes factor. However, the \textit{SAD} indicates minimal preference for the stellar contamination model as the model that minimises the distance to the data. We find a low spectral variability across observations taken four months apart.

\subsection{WASP-121 b}
\noindent \textbf{Discovery, Observations and Previous Findings:} WASP-121\,b, a short-period transiting ultra-hot-Jupiter, was found to possess a significant orbital misalignment \citep{delrez2016w121} consistent to a nearly polar orbit \citep{2020A&A...635A.205B}. Using WFC3 data and archival ground based optical data, \cite{2016ApJ...822L...4E} identified a strong water absorption feature and the presence of titanium and vanadium oxide, challenging the notion that high-altitude haze alone could explain the observed optical slope. Furthermore, by adding STIS CCD data, \cite{2018AJ....156..283E} ruled out a gray cloud deck and proposed the possibility of VO spectral bands. Various observations using different instruments have contributed to our understanding of WASP-121\,b's atmosphere. \cite{2019AJ....158...91S} utilised the HST STIS NUV-MAMA detector to identify Mg{\sc ii} and Fe{\sc ii} in the planetary exosphere, suggesting hydrodynamic escape or magnetic confinement of ionised species at high altitudes. Multiple studies have reported the detection of various metals such as iron \citep{2020MNRAS.493.2215G, 2020ApJ...897L...5B}, chromium \citep{2020MNRAS.494..363C} and vanadium oxide in the planet's atmosphere, further supporting the presence of a stratospheric layer \citep{2020A&A...637A..36B}. Additional observations at high resolution detected significant absorption in the transmission spectrum, including Na, H, K, Li, Ca, Mg, Fe, Cr, V, Ni, VO, Sc \citep{2020A&A...635A.205B, 2020MNRAS.494..363C, 2020ApJ...897L...5B, 2020A&A...641A.123H, 2021A&A...645A..24B, 2021MNRAS.506.3853M} and Ba, the heaviest species detected in any exoplanetary atmosphere to date \citep{2022A&A...666L..10A}. Non-detections of Ti and TiO support the hypothesis that titanium is depleted via a cold-trap mechanism, as proposed by \cite{2020A&A...641A.123H}.
\newline
\textbf{Chemistry Results:} We retrieve a strong water feature and a layer of high altitude clouds. Our Case 1 spectra displays an offset of G430L with respect to the rest of the spectrum, so we are less confident about these results. No alkali features can be observed. 
\newline
\textbf{Stellar Activity:} Reported S-indexes ($S_\text{HK}$) and $\log(R'_\text{HK})$ values: Observations of two transits a few months apart yielded $\log(R'_\text{HK})$ values of -4.81$\pm$0.01 (30.11.2018) and -4.87$\pm$0.01 (06.01.2019) respectively \citep{Borsa2021}. \citet{Danielski2022} report an S-index of $S_\text{HK}$ = 0.18$\pm$0.00 and a corresponding $\log(R'_\text{HK})$ = -5.05$\pm$0.07. These reported values are all indicative of a low-to-moderate level of activity for WASP-121. Moderate to strong levels of activity as indicated by the Bayes factor, low levels of activity and temporal variations according to the \textit{SAD} and \textit{SAT} respectively.

\subsection{WASP-127 b} 
\noindent \textbf{Discovery, Observations and Previous Findings:} Discovered through the SuperWASP survey \citep{2006MNRAS.372.1117C}, the heavily inflated super-Neptune WASP-127\,b, situated at the edge of the Neptune desert, orbits a slowly-rotating bright G5 star on a retrograde misaligned orbit \citep{2020A&A...644A.155A}. With a mass of 0.18$\pm$0.02 M$_{\text{J}}$ and a radius of 1.37$\pm$0.04 R$_{\text{J}}$ it is one of the least massive planets found by the WASP project \citep{2017A&A...599A...3L}. The first transmission spectrum for WASP-127\,b presented a strong Rayleigh slope at blue wavelengths within an overall cloud-free atmosphere, while the redder optical wavelengths were best explained with TiO and VO absorption features \citep{2017A&A...602L..15P}. Although debated by some authors \citep{2020A&A...643A..45S}, the presence of pressure-broadened spectral profiles of Na has been reported by various studies \citep{2018A&A...616A.145C, 2019AJ....158..120Z, 2020A&A...644A.155A}. Initial WFC3 observations indicated a strong water absorption and a cloudy atmosphere for WASP-127\,b \citep{skaf2020ares}. Later investigations on a complete optical to infrared spectrum obtained with the Hubble and Spitzer Space Telescopes revealed absorption features from Na, \ce{H2O} (also detected with the SPIRou spectrograph \citep{2023MNRAS.522.5062B}), and \ce{CO2}, along with wavelength-dependent scattering from small-particle condensates \citep{2021MNRAS.500.4042S}. Regarding helium, \cite{2020A&A...640A..29D} did not find a significant in-transit absorption signal around WASP-127\,b, attributing it to unfavourable photoionization conditions. The lack of a detectable He{\sc i} feature is supported by the low coronal and chromospheric activity of the host star and the old age of the system, leading to a relatively mild high-energy environment. 
\newline
\textbf{Chemistry Results:} We retrieve a layer of wavelength-independent clouds and a clear water vapour signature. There is weak evidence for Na, while the K feature is muted, if present at all.
\newline
\textbf{Stellar Activity:} Reported S-indexes ($S_\text{HK}$) and $\log(R'_\text{HK})$ values: \citet{Danielski2022} find an S-index of $S_\text{HK}$ = 0.14$\pm$0.00 and a corresponding $\log(R'_\text{HK})$ = -5.12$\pm$0.09. \citet{Claudi2024} report an S-index of $S_\text{HK}$ = 0.156$\pm$0.004 and a corresponding $\log(R'_\text{HK})$ = -5.02$\pm$0.03 or $\log(R'_\text{HK})$ = -4.79$\pm$0.03 when the correction detailed in \citet{Mittag2013} is applied. These literature values are both indicative of a lower level of activity for WASP-127. Our Bayes factor indicates an inconclusive preference for stellar contamination, coupled with the distance metric which is the lowest among the whole sample. However, a stronger preference for the active star model (Bayes factor = 7.75) is obtained when considering gray clouds and alkali metals in the retrieval setup, suggesting the potential presence of stellar activity.

\clearpage
\bibliography{sample631}{}

\begin{thebibliography}{}
\expandafter\ifx\csname natexlab\endcsname\relax\def\natexlab#1{#1}\fi
\providecommand{\url}[1]{\href{#1}{#1}}
\providecommand{\dodoi}[1]{doi:~\href{http://doi.org/#1}{\nolinkurl{#1}}}
\providecommand{\doeprint}[1]{\href{http://ascl.net/#1}{\nolinkurl{http://ascl.net/#1}}}
\providecommand{\doarXiv}[1]{\href{https://arxiv.org/abs/#1}{\nolinkurl{https://arxiv.org/abs/#1}}}

\bibitem[{{A-thano} {et~al.}(2023){A-thano}, {Awiphan}, {Jiang}, {Kerins}, {Priyadarshi}, {McDonald}, {Joshi}, {Chulikorn}, {Hayes}, {Charles}, {Huang}, {Rattanamala}, {Yeh}, \& {Dhillon}}]{2023AJ....166..223A}
{A-thano}, N., {Awiphan}, S., {Jiang}, I.-G., {et~al.} 2023, \aj, 166, 223, \dodoi{10.3847/1538-3881/acfeea}

\bibitem[{Abel {et~al.}(2011)Abel, Frommhold, Li, \& Hunt}]{PMID:21207941}
Abel, M., Frommhold, L., Li, X., \& Hunt, K. L.~C. 2011, The journal of physical chemistry. A, 115, 6805—6812, \dodoi{10.1021/jp109441f}

\bibitem[{Abel {et~al.}(2012)Abel, Frommhold, Li, \& Hunt}]{PMID:22299883}
---. 2012, The Journal of chemical physics, 136, 044319, \dodoi{10.1063/1.3676405}

\bibitem[{{Ahrer} {et~al.}(2023){Ahrer}, {Stevenson}, {Mansfield}, {Moran}, {Brande}, {Morello}, {Murray}, {Nikolov}, {Petit dit de la Roche}, {Schlawin}, {Wheatley}, {Zieba}, {Batalha}, {Damiano}, {Goyal}, {Lendl}, {Lothringer}, {Mukherjee}, {Ohno}, {Batalha}, {Battley}, {Bean}, {Beatty}, {Benneke}, {Berta-Thompson}, {Carter}, {Cubillos}, {Daylan}, {Espinoza}, {Gao}, {Gibson}, {Gill}, {Harrington}, {Hu}, {Kreidberg}, {Lewis}, {Line}, {L{\'o}pez-Morales}, {Parmentier}, {Powell}, {Sing}, {Tsai}, {Wakeford}, {Welbanks}, {Alam}, {Alderson}, {Allen}, {Anderson}, {Barstow}, {Bayliss}, {Bell}, {Blecic}, {Bryant}, {Burleigh}, {Carone}, {Casewell}, {Changeat}, {Chubb}, {Crossfield}, {Crouzet}, {Decin}, {D{\'e}sert}, {Feinstein}, {Flagg}, {Fortney}, {Gizis}, {Heng}, {Iro}, {Kempton}, {Kendrew}, {Kirk}, {Knutson}, {Komacek}, {Lagage}, {Leconte}, {Lustig-Yaeger}, {MacDonald}, {Mancini}, {May}, {Mayne}, {Miguel}, {Mikal-Evans}, {Molaverdikhani}, {Palle}, {Piaulet}, {Rackham}, {Redfield}, {Rogers}, {Roy}, {Rustamkulov},
  {Shkolnik}, {Sotzen}, {Taylor}, {Tremblin}, {Tucker}, {Turner}, {de Val-Borro}, {Venot}, \& {Zhang}}]{2023Natur.614..653A}
{Ahrer}, E.-M., {Stevenson}, K.~B., {Mansfield}, M., {et~al.} 2023, \nat, 614, 653, \dodoi{10.1038/s41586-022-05590-4}

\bibitem[{{Al-Refaie} {et~al.}(2022){Al-Refaie}, {Changeat}, {Venot}, {Waldmann}, \& {Tinetti}}]{2022ApJ...932..123A}
{Al-Refaie}, A.~F., {Changeat}, Q., {Venot}, O., {Waldmann}, I.~P., \& {Tinetti}, G. 2022, \apj, 932, 123, \dodoi{10.3847/1538-4357/ac6dcd}

\bibitem[{Al-Refaie {et~al.}(2021)Al-Refaie, Changeat, Waldmann, \& Tinetti}]{Al_Refaie_2021}
Al-Refaie, A.~F., Changeat, Q., Waldmann, I.~P., \& Tinetti, G. 2021, The Astrophysical Journal, 917, 37, \dodoi{10.3847/1538-4357/ac0252}

\bibitem[{{Alam} {et~al.}(2018){Alam}, {Nikolov}, {L{\'o}pez-Morales}, {Sing}, {Goyal}, {Henry}, {Sanz-Forcada}, {Williamson}, {Evans}, {Wakeford}, {Bruno}, {Ballester}, {Stevenson}, {Lewis}, {Barstow}, {Bourrier}, {Buchhave}, {Ehrenreich}, \& {Garc{\'\i}a Mu{\~n}oz}}]{2018AJ....156..298A}
{Alam}, M.~K., {Nikolov}, N., {L{\'o}pez-Morales}, M., {et~al.} 2018, \aj, 156, 298, \dodoi{10.3847/1538-3881/aaee89}

\bibitem[{{Alderson} {et~al.}(2022){Alderson}, {Wakeford}, {MacDonald}, {Lewis}, {May}, {Grant}, {Sing}, {Stevenson}, {Fowler}, {Goyal}, {Batalha}, \& {Kataria}}]{2022MNRAS.512.4185A}
{Alderson}, L., {Wakeford}, H.~R., {MacDonald}, R.~J., {et~al.} 2022, \mnras, 512, 4185, \dodoi{10.1093/mnras/stac661}

\bibitem[{{Alderson} {et~al.}(2023){Alderson}, {Wakeford}, {Alam}, {Batalha}, {Lothringer}, {Adams Redai}, {Barat}, {Brande}, {Damiano}, {Daylan}, {Espinoza}, {Flagg}, {Goyal}, {Grant}, {Hu}, {Inglis}, {Lee}, {Mikal-Evans}, {Ramos-Rosado}, {Roy}, {Wallack}, {Batalha}, {Bean}, {Benneke}, {Berta-Thompson}, {Carter}, {Changeat}, {Col{\'o}n}, {Crossfield}, {D{\'e}sert}, {Foreman-Mackey}, {Gibson}, {Kreidberg}, {Line}, {L{\'o}pez-Morales}, {Molaverdikhani}, {Moran}, {Morello}, {Moses}, {Mukherjee}, {Schlawin}, {Sing}, {Stevenson}, {Taylor}, {Aggarwal}, {Ahrer}, {Allen}, {Barstow}, {Bell}, {Blecic}, {Casewell}, {Chubb}, {Crouzet}, {Cubillos}, {Decin}, {Feinstein}, {Fortney}, {Harrington}, {Heng}, {Iro}, {Kempton}, {Kirk}, {Knutson}, {Krick}, {Leconte}, {Lendl}, {MacDonald}, {Mancini}, {Mansfield}, {May}, {Mayne}, {Miguel}, {Nikolov}, {Ohno}, {Palle}, {Parmentier}, {Petit dit de la Roche}, {Piaulet}, {Powell}, {Rackham}, {Redfield}, {Rogers}, {Rustamkulov}, {Tan}, {Tremblin}, {Tsai}, {Turner}, {de Val-Borro},
  {Venot}, {Welbanks}, {Wheatley}, \& {Zhang}}]{2023Natur.614..664A}
{Alderson}, L., {Wakeford}, H.~R., {Alam}, M.~K., {et~al.} 2023, \nat, 614, 664, \dodoi{10.1038/s41586-022-05591-3}

\bibitem[{{Alexoudi} {et~al.}(2018){Alexoudi}, {Mallonn}, {von Essen}, {Turner}, {Keles}, {Southworth}, {Mancini}, {Ciceri}, {Granzer}, {Denker}, {Dineva}, \& {Strassmeier}}]{2018A&A...620A.142A}
{Alexoudi}, X., {Mallonn}, M., {von Essen}, C., {et~al.} 2018, \aap, 620, A142, \dodoi{10.1051/0004-6361/201833691}

\bibitem[{{Allard} {et~al.}(2012){Allard}, {Homeier}, \& {Freytag}}]{2012RSPTA.370.2765A}
{Allard}, F., {Homeier}, D., \& {Freytag}, B. 2012, Philosophical Transactions of the Royal Society of London Series A, 370, 2765, \dodoi{10.1098/rsta.2011.0269}

\bibitem[{Allard {et~al.}(2019)Allard, Spiegelman, Leininger, \& Molliere}]{allard2019new}
Allard, N., Spiegelman, F., Leininger, T., \& Molliere, P. 2019, Astronomy \& Astrophysics, 628, A120

\bibitem[{{Allart} {et~al.}(2018){Allart}, {Bourrier}, {Lovis}, {Ehrenreich}, {Spake}, {Wyttenbach}, {Pino}, {Pepe}, {Sing}, \& {Lecavelier des Etangs}}]{2018Sci...362.1384A}
{Allart}, R., {Bourrier}, V., {Lovis}, C., {et~al.} 2018, Science, 362, 1384, \dodoi{10.1126/science.aat5879}

\bibitem[{{Allart} {et~al.}(2020){Allart}, {Pino}, {Lovis}, {Sousa}, {Casasayas-Barris}, {Zapatero Osorio}, {Cretignier}, {Palle}, {Pepe}, {Cristiani}, {Rebolo}, {Santos}, {Borsa}, {Bourrier}, {Demangeon}, {Ehrenreich}, {Lavie}, {Lendl}, {Lillo-Box}, {Micela}, {Oshagh}, {Sozzetti}, {Tabernero}, {Adibekyan}, {Allende Prieto}, {Alibert}, {Amate}, {Benz}, {Bouchy}, {Cabral}, {Dekker}, {D'Odorico}, {Di Marcantonio}, {Dumusque}, {Figueira}, {Genova Santos}, {Gonz{\'a}lez Hern{\'a}ndez}, {Lo Curto}, {Manescau}, {Martins}, {M{\'e}gevand}, {Mehner}, {Molaro}, {Nunes}, {Poretti}, {Riva}, {Su{\'a}rez Mascare{\~n}o}, {Udry}, \& {Zerbi}}]{2020A&A...644A.155A}
{Allart}, R., {Pino}, L., {Lovis}, C., {et~al.} 2020, \aap, 644, A155, \dodoi{10.1051/0004-6361/202039234}

\bibitem[{{Ambikasaran} {et~al.}(2015){Ambikasaran}, {Foreman-Mackey}, {Greengard}, {Hogg}, \& {O'Neil}}]{2015ITPAM..38..252A}
{Ambikasaran}, S., {Foreman-Mackey}, D., {Greengard}, L., {Hogg}, D.~W., \& {O'Neil}, M. 2015, IEEE Transactions on Pattern Analysis and Machine Intelligence, 38, 252, \dodoi{10.1109/TPAMI.2015.2448083}

\bibitem[{Anderson {et~al.}(2011)Anderson, Cameron, Hellier, Lendl, Lister, Maxted, Queloz, Smalley, Smith, Triaud, {et~al.}}]{anderson2011wasp}
Anderson, D., Cameron, A.~C., Hellier, C., {et~al.} 2011, Astronomy \& Astrophysics, 531, A60

\bibitem[{{Anderson} {et~al.}(2010){Anderson}, {Hellier}, {Gillon}, {Triaud}, {Smalley}, {Hebb}, {Collier Cameron}, {Maxted}, {Queloz}, {West}, {Bentley}, {Enoch}, {Horne}, {Lister}, {Mayor}, {Parley}, {Pepe}, {Pollacco}, {S{\'e}gransan}, {Udry}, \& {Wilson}}]{2010ApJ...709..159A}
{Anderson}, D.~R., {Hellier}, C., {Gillon}, M., {et~al.} 2010, \apj, 709, 159, \dodoi{10.1088/0004-637X/709/1/159}

\bibitem[{{Anderson} {et~al.}(2014){Anderson}, {Collier Cameron}, {Delrez}, {Doyle}, {Faedi}, {Fumel}, {Gillon}, {G{\'o}mez Maqueo Chew}, {Hellier}, {Jehin}, {Lendl}, {Maxted}, {Pepe}, {Pollacco}, {Queloz}, {S{\'e}gransan}, {Skillen}, {Smalley}, {Smith}, {Southworth}, {Triaud}, {Turner}, {Udry}, \& {West}}]{2014MNRAS.445.1114A}
{Anderson}, D.~R., {Collier Cameron}, A., {Delrez}, L., {et~al.} 2014, \mnras, 445, 1114, \dodoi{10.1093/mnras/stu1737}

\bibitem[{Anisman {et~al.}(2020)Anisman, Edwards, Changeat, Venot, Al-Refaie, Tsiaras, \& Tinetti}]{anisman2020wasp}
Anisman, L.~O., Edwards, B., Changeat, Q., {et~al.} 2020, The Astronomical Journal, 160, 233

\bibitem[{Armstrong {et~al.}(2016)Armstrong, de~Mooij, Barstow, Osborn, Blake, \& Saniee}]{armstrong2016variability}
Armstrong, D.~J., de~Mooij, E., Barstow, J., {et~al.} 2016, Nature Astronomy, 1, 1

\bibitem[{{Astropy Collaboration} {et~al.}(2013){Astropy Collaboration}, {Robitaille}, {Tollerud}, {Greenfield}, {Droettboom}, {Bray}, {Aldcroft}, {Davis}, {Ginsburg}, {Price-Whelan}, {Kerzendorf}, {Conley}, {Crighton}, {Barbary}, {Muna}, {Ferguson}, {Grollier}, {Parikh}, {Nair}, {Unther}, {Deil}, {Woillez}, {Conseil}, {Kramer}, {Turner}, {Singer}, {Fox}, {Weaver}, {Zabalza}, {Edwards}, {Azalee Bostroem}, {Burke}, {Casey}, {Crawford}, {Dencheva}, {Ely}, {Jenness}, {Labrie}, {Lim}, {Pierfederici}, {Pontzen}, {Ptak}, {Refsdal}, {Servillat}, \& {Streicher}}]{2013A&A...558A..33A}
{Astropy Collaboration}, {Robitaille}, T.~P., {Tollerud}, E.~J., {et~al.} 2013, \aap, 558, A33, \dodoi{10.1051/0004-6361/201322068}

\bibitem[{{Astropy Collaboration} {et~al.}(2018){Astropy Collaboration}, {Price-Whelan}, {Sip{\H{o}}cz}, {G{\"u}nther}, {Lim}, {Crawford}, {Conseil}, {Shupe}, {Craig}, {Dencheva}, {Ginsburg}, {VanderPlas}, {Bradley}, {P{\'e}rez-Su{\'a}rez}, {de Val-Borro}, {Aldcroft}, {Cruz}, {Robitaille}, {Tollerud}, {Ardelean}, {Babej}, {Bach}, {Bachetti}, {Bakanov}, {Bamford}, {Barentsen}, {Barmby}, {Baumbach}, {Berry}, {Biscani}, {Boquien}, {Bostroem}, {Bouma}, {Brammer}, {Bray}, {Breytenbach}, {Buddelmeijer}, {Burke}, {Calderone}, {Cano Rodr{\'\i}guez}, {Cara}, {Cardoso}, {Cheedella}, {Copin}, {Corrales}, {Crichton}, {D'Avella}, {Deil}, {Depagne}, {Dietrich}, {Donath}, {Droettboom}, {Earl}, {Erben}, {Fabbro}, {Ferreira}, {Finethy}, {Fox}, {Garrison}, {Gibbons}, {Goldstein}, {Gommers}, {Greco}, {Greenfield}, {Groener}, {Grollier}, {Hagen}, {Hirst}, {Homeier}, {Horton}, {Hosseinzadeh}, {Hu}, {Hunkeler}, {Ivezi{\'c}}, {Jain}, {Jenness}, {Kanarek}, {Kendrew}, {Kern}, {Kerzendorf}, {Khvalko}, {King}, {Kirkby}, {Kulkarni},
  {Kumar}, {Lee}, {Lenz}, {Littlefair}, {Ma}, {Macleod}, {Mastropietro}, {McCully}, {Montagnac}, {Morris}, {Mueller}, {Mumford}, {Muna}, {Murphy}, {Nelson}, {Nguyen}, {Ninan}, {N{\"o}the}, {Ogaz}, {Oh}, {Parejko}, {Parley}, {Pascual}, {Patil}, {Patil}, {Plunkett}, {Prochaska}, {Rastogi}, {Reddy Janga}, {Sabater}, {Sakurikar}, {Seifert}, {Sherbert}, {Sherwood-Taylor}, {Shih}, {Sick}, {Silbiger}, {Singanamalla}, {Singer}, {Sladen}, {Sooley}, {Sornarajah}, {Streicher}, {Teuben}, {Thomas}, {Tremblay}, {Turner}, {Terr{\'o}n}, {van Kerkwijk}, {de la Vega}, {Watkins}, {Weaver}, {Whitmore}, {Woillez}, {Zabalza}, \& {Astropy Contributors}}]{2018AJ....156..123A}
{Astropy Collaboration}, {Price-Whelan}, A.~M., {Sip{\H{o}}cz}, B.~M., {et~al.} 2018, \aj, 156, 123, \dodoi{10.3847/1538-3881/aabc4f}

\bibitem[{{Azevedo Silva} {et~al.}(2022){Azevedo Silva}, {Demangeon}, {Santos}, {Allart}, {Borsa}, {Cristo}, {Esparza-Borges}, {Seidel}, {Palle}, {Sousa}, {Tabernero}, {Zapatero Osorio}, {Cristiani}, {Pepe}, {Rebolo}, {Adibekyan}, {Alibert}, {Barros}, {Bouchy}, {Bourrier}, {Lo Curto}, {Di Marcantonio}, {D'Odorico}, {Ehrenreich}, {Figueira}, {Gonz{\'a}lez Hern{\'a}ndez}, {Lovis}, {Martins}, {Mehner}, {Micela}, {Molaro}, {Mounzer}, {Nunes}, {Sozzetti}, {Su{\'a}rez Mascare{\~n}o}, \& {Udry}}]{2022A&A...666L..10A}
{Azevedo Silva}, T., {Demangeon}, O.~D.~S., {Santos}, N.~C., {et~al.} 2022, \aap, 666, L10, \dodoi{10.1051/0004-6361/202244489}

\bibitem[{{Bakos} {et~al.}(2007{\natexlab{a}}){Bakos}, {Noyes}, {Kov{\'a}cs}, {Latham}, {Sasselov}, {Torres}, {Fischer}, {Stefanik}, {Sato}, {Johnson}, {P{\'a}l}, {Marcy}, {Butler}, {Esquerdo}, {Stanek}, {L{\'a}z{\'a}r}, {Papp}, {S{\'a}ri}, \& {Sip{\H{o}}cz}}]{bakos2007hat}
{Bakos}, G.~{\'A}., {Noyes}, R.~W., {Kov{\'a}cs}, G., {et~al.} 2007{\natexlab{a}}, \apj, 656, 552, \dodoi{10.1086/509874}

\bibitem[{{Bakos} {et~al.}(2007{\natexlab{b}}){Bakos}, {Noyes}, {Kov{\'a}cs}, {Latham}, {Sasselov}, {Torres}, {Fischer}, {Stefanik}, {Sato}, {Johnson}, {P{\'a}l}, {Marcy}, {Butler}, {Esquerdo}, {Stanek}, {L{\'a}z{\'a}r}, {Papp}, {S{\'a}ri}, \& {Sip{\H{o}}cz}}]{2007ApJ...656..552B}
---. 2007{\natexlab{b}}, \apj, 656, 552, \dodoi{10.1086/509874}

\bibitem[{{Bakos} {et~al.}(2010){Bakos}, {Torres}, {P{\'a}l}, {Hartman}, {Kov{\'a}cs}, {Noyes}, {Latham}, {Sasselov}, {Sip{\H{o}}cz}, {Esquerdo}, {Fischer}, {Johnson}, {Marcy}, {Butler}, {Isaacson}, {Howard}, {Vogt}, {Kov{\'a}cs}, {Fernandez}, {Mo{\'o}r}, {Stefanik}, {L{\'a}z{\'a}r}, {Papp}, \& {S{\'a}ri}}]{bakos2010hat}
{Bakos}, G.~{\'A}., {Torres}, G., {P{\'a}l}, A., {et~al.} 2010, \apj, 710, 1724, \dodoi{10.1088/0004-637X/710/2/1724}

\bibitem[{{Ballerini} {et~al.}(2012){Ballerini}, {Micela}, {Lanza}, \& {Pagano}}]{2012A&A...539A.140B}
{Ballerini}, P., {Micela}, G., {Lanza}, A.~F., \& {Pagano}, I. 2012, \aap, 539, A140, \dodoi{10.1051/0004-6361/201117102}

\bibitem[{{Balona}(2015)}]{2015MNRAS.447.2714B}
{Balona}, L.~A. 2015, \mnras, 447, 2714, \dodoi{10.1093/mnras/stu2651}

\bibitem[{{Balona}(2022)}]{2022arXiv221210776B}
---. 2022, arXiv e-prints, arXiv:2212.10776, \dodoi{10.48550/arXiv.2212.10776}

\bibitem[{{Barclay} {et~al.}(2021){Barclay}, {Kostov}, {Col{\'o}n}, {Quintana}, {Schlieder}, {Louie}, {Gilbert}, \& {Mullally}}]{2021AJ....162..300B}
{Barclay}, T., {Kostov}, V.~B., {Col{\'o}n}, K.~D., {et~al.} 2021, \aj, 162, 300, \dodoi{10.3847/1538-3881/ac2824}

\bibitem[{{Barstow} {et~al.}(2022){Barstow}, {Changeat}, {Chubb}, {Cubillos}, {Edwards}, {MacDonald}, {Min}, \& {Waldmann}}]{2022ExA....53..447B}
{Barstow}, J.~K., {Changeat}, Q., {Chubb}, K.~L., {et~al.} 2022, Experimental Astronomy, 53, 447, \dodoi{10.1007/s10686-021-09821-w}

\bibitem[{{Barstow} {et~al.}(2020){Barstow}, {Changeat}, {Garland}, {Line}, {Rocchetto}, \& {Waldmann}}]{2020MNRAS.493.4884B}
{Barstow}, J.~K., {Changeat}, Q., {Garland}, R., {et~al.} 2020, \mnras, 493, 4884, \dodoi{10.1093/mnras/staa548}

\bibitem[{{Basri} {et~al.}(2013){Basri}, {Walkowicz}, \& {Reiners}}]{2013ApJ...769...37B}
{Basri}, G., {Walkowicz}, L.~M., \& {Reiners}, A. 2013, \apj, 769, 37, \dodoi{10.1088/0004-637X/769/1/37}

\bibitem[{{Bastien} {et~al.}(2013){Bastien}, {Stassun}, {Basri}, \& {Pepper}}]{2013Natur.500..427B}
{Bastien}, F.~A., {Stassun}, K.~G., {Basri}, G., \& {Pepper}, J. 2013, \nat, 500, 427, \dodoi{10.1038/nature12419}

\bibitem[{{Bayliss} {et~al.}(2010){Bayliss}, {Winn}, {Mardling}, \& {Sackett}}]{2010ApJ...722L.224B}
{Bayliss}, D. D.~R., {Winn}, J.~N., {Mardling}, R.~A., \& {Sackett}, P.~D. 2010, \apjl, 722, L224, \dodoi{10.1088/2041-8205/722/2/L224}

\bibitem[{{Ben-Yami} {et~al.}(2020){Ben-Yami}, {Madhusudhan}, {Cabot}, {Constantinou}, {Piette}, {Gandhi}, \& {Welbanks}}]{2020ApJ...897L...5B}
{Ben-Yami}, M., {Madhusudhan}, N., {Cabot}, S. H.~C., {et~al.} 2020, \apjl, 897, L5, \dodoi{10.3847/2041-8213/ab94aa}

\bibitem[{{Benneke} {et~al.}(2019){Benneke}, {Knutson}, {Lothringer}, {Crossfield}, {Moses}, {Morley}, {Kreidberg}, {Fulton}, {Dragomir}, {Howard}, {Wong}, {D{\'e}sert}, {McCullough}, {Kempton}, {Fortney}, {Gilliland}, {Deming}, \& {Kammer}}]{2019NatAs...3..813B}
{Benneke}, B., {Knutson}, H.~A., {Lothringer}, J., {et~al.} 2019, Nature Astronomy, 3, 813, \dodoi{10.1038/s41550-019-0800-5}

\bibitem[{{Berdyugina}(2005)}]{berdyugina2005}
{Berdyugina}, S.~V. 2005, Living Reviews in Solar Physics, 2, 8, \dodoi{10.12942/lrsp-2005-8}

\bibitem[{{Biddle} {et~al.}(2014){Biddle}, {Pearson}, {Crossfield}, {Fulton}, {Ciceri}, {Eastman}, {Barman}, {Mann}, {Henry}, {Howard}, {Williamson}, {Sinukoff}, {Dragomir}, {Vican}, {Mancini}, {Southworth}, {Greenberg}, {Turner}, {Thompson}, {Taylor}, {Levine}, \& {Webber}}]{2014MNRAS.443.1810B}
{Biddle}, L.~I., {Pearson}, K.~A., {Crossfield}, I. J.~M., {et~al.} 2014, \mnras, 443, 1810, \dodoi{10.1093/mnras/stu1199}

\bibitem[{{Bonfils} {et~al.}(2012){Bonfils}, {Gillon}, {Udry}, {Armstrong}, {Bouchy}, {Delfosse}, {Forveille}, {Fumel}, {Jehin}, {Lendl}, {Lovis}, {Mayor}, {McCormac}, {Neves}, {Pepe}, {Perrier}, {Pollaco}, {Queloz}, \& {Santos}}]{2012A&A...546A..27B}
{Bonfils}, X., {Gillon}, M., {Udry}, S., {et~al.} 2012, \aap, 546, A27, \dodoi{10.1051/0004-6361/201219623}

\bibitem[{{Bonomo} {et~al.}(2017){Bonomo}, {Desidera}, {Benatti}, {Borsa}, {Crespi}, {Damasso}, {Lanza}, {Sozzetti}, {Lodato}, {Marzari}, {Boccato}, {Claudi}, {Cosentino}, {Covino}, {Gratton}, {Maggio}, {Micela}, {Molinari}, {Pagano}, {Piotto}, {Poretti}, {Smareglia}, {Affer}, {Biazzo}, {Bignamini}, {Esposito}, {Giacobbe}, {H{\'e}brard}, {Malavolta}, {Maldonado}, {Mancini}, {Martinez Fiorenzano}, {Masiero}, {Nascimbeni}, {Pedani}, {Rainer}, \& {Scandariato}}]{2017A&A...602A.107B}
{Bonomo}, A.~S., {Desidera}, S., {Benatti}, S., {et~al.} 2017, \aap, 602, A107, \dodoi{10.1051/0004-6361/201629882}

\bibitem[{{Boro Saikia} {et~al.}(2018){Boro Saikia}, {Marvin}, {Jeffers}, {Reiners}, {Cameron}, {Marsden}, {Petit}, {Warnecke}, \& {Yadav}}]{2018A&A...616A.108B}
{Boro Saikia}, S., {Marvin}, C.~J., {Jeffers}, S.~V., {et~al.} 2018, \aap, 616, A108, \dodoi{10.1051/0004-6361/201629518}

\bibitem[{{Borsa} {et~al.}(2021{\natexlab{a}}){Borsa}, {Allart}, {Casasayas-Barris}, {Tabernero}, {Zapatero Osorio}, {Cristiani}, {Pepe}, {Rebolo}, {Santos}, {Adibekyan}, {Bourrier}, {Demangeon}, {Ehrenreich}, {Pall{\'e}}, {Sousa}, {Lillo-Box}, {Lovis}, {Micela}, {Oshagh}, {Poretti}, {Sozzetti}, {Allende Prieto}, {Alibert}, {Amate}, {Benz}, {Bouchy}, {Cabral}, {Dekker}, {D'Odorico}, {Di Marcantonio}, {Figueira}, {Genova Santos}, {Gonz{\'a}lez Hern{\'a}ndez}, {Lo Curto}, {Manescau}, {Martins}, {M{\'e}gevand}, {Mehner}, {Molaro}, {Nunes}, {Riva}, {Su{\'a}rez Mascare{\~n}o}, {Udry}, \& {Zerbi}}]{2021A&A...645A..24B}
{Borsa}, F., {Allart}, R., {Casasayas-Barris}, N., {et~al.} 2021{\natexlab{a}}, \aap, 645, A24, \dodoi{10.1051/0004-6361/202039344}

\bibitem[{{Borsa} {et~al.}(2021{\natexlab{b}}){Borsa}, {Allart}, {Casasayas-Barris}, {Tabernero}, {Zapatero Osorio}, {Cristiani}, {Pepe}, {Rebolo}, {Santos}, {Adibekyan}, {Bourrier}, {Demangeon}, {Ehrenreich}, {Pall{\'e}}, {Sousa}, {Lillo-Box}, {Lovis}, {Micela}, {Oshagh}, {Poretti}, {Sozzetti}, {Allende Prieto}, {Alibert}, {Amate}, {Benz}, {Bouchy}, {Cabral}, {Dekker}, {D'Odorico}, {Di Marcantonio}, {Figueira}, {Genova Santos}, {Gonz{\'a}lez Hern{\'a}ndez}, {Lo Curto}, {Manescau}, {Martins}, {M{\'e}gevand}, {Mehner}, {Molaro}, {Nunes}, {Riva}, {Su{\'a}rez Mascare{\~n}o}, {Udry}, \& {Zerbi}}]{Borsa2021}
---. 2021{\natexlab{b}}, \aap, 645, A24, \dodoi{10.1051/0004-6361/202039344}

\bibitem[{Bostroem \& Proffitt(2011)}]{bostroem2011stis}
Bostroem, K., \& Proffitt, C. 2011, STIS Data Handbook

\bibitem[{{Boucher} {et~al.}(2023){Boucher}, {Lafreni{\'e}re}, {Pelletier}, {Darveau-Bernier}, {Radica}, {Allart}, {Artigau}, {Cook}, {Debras}, {Doyon}, {Gaidos}, {Benneke}, {Cadieux}, {Carmona}, {Cloutier}, {Cort{\'e}s-Zuleta}, {Cowan}, {Delfosse}, {Donati}, {Fouqu{\'e}}, {Forveille}, {Grankin}, {H{\'e}brard}, {Martins}, {Martioli}, {Masson}, \& {Vinatier}}]{2023MNRAS.522.5062B}
{Boucher}, A., {Lafreni{\'e}re}, D., {Pelletier}, S., {et~al.} 2023, \mnras, 522, 5062, \dodoi{10.1093/mnras/stad1247}

\bibitem[{{Bourrier} {et~al.}(2018){Bourrier}, {Lecavelier des Etangs}, {Ehrenreich}, {Sanz-Forcada}, {Allart}, {Ballester}, {Buchhave}, {Cohen}, {Deming}, {Evans}, {Garc{\'\i}a Mu{\~n}oz}, {Henry}, {Kataria}, {Lavvas}, {Lewis}, {L{\'o}pez-Morales}, {Marley}, {Sing}, \& {Wakeford}}]{2018A&A...620A.147B}
{Bourrier}, V., {Lecavelier des Etangs}, A., {Ehrenreich}, D., {et~al.} 2018, \aap, 620, A147, \dodoi{10.1051/0004-6361/201833675}

\bibitem[{{Bourrier} {et~al.}(2020{\natexlab{a}}){Bourrier}, {Ehrenreich}, {Lendl}, {Cretignier}, {Allart}, {Dumusque}, {Cegla}, {Su{\'a}rez-Mascare{\~n}o}, {Wyttenbach}, {Hoeijmakers}, {Melo}, {Kuntzer}, {Astudillo-Defru}, {Giles}, {Heng}, {Kitzmann}, {Lavie}, {Lovis}, {Murgas}, {Nascimbeni}, {Pepe}, {Pino}, {Segransan}, \& {Udry}}]{2020A&A...635A.205B}
{Bourrier}, V., {Ehrenreich}, D., {Lendl}, M., {et~al.} 2020{\natexlab{a}}, \aap, 635, A205, \dodoi{10.1051/0004-6361/201936640}

\bibitem[{{Bourrier} {et~al.}(2020{\natexlab{b}}){Bourrier}, {Kitzmann}, {Kuntzer}, {Nascimbeni}, {Lendl}, {Lavie}, {Hoeijmakers}, {Pino}, {Ehrenreich}, {Heng}, {Allart}, {Cegla}, {Dumusque}, {Melo}, {Astudillo-Defru}, {Caldwell}, {Cretignier}, {Giles}, {Henze}, {Jenkins}, {Lovis}, {Murgas}, {Pepe}, {Ricker}, {Rose}, {Seager}, {Segransan}, {Su{\'a}rez-Mascare{\~n}o}, {Udry}, {Vanderspek}, \& {Wyttenbach}}]{2020A&A...637A..36B}
{Bourrier}, V., {Kitzmann}, D., {Kuntzer}, T., {et~al.} 2020{\natexlab{b}}, \aap, 637, A36, \dodoi{10.1051/0004-6361/201936647}

\bibitem[{{Bourrier} {et~al.}(2021){Bourrier}, {dos Santos}, {Sanz-Forcada}, {Garc{\'\i}a Mu{\~n}oz}, {Henry}, {Lavvas}, {Lecavelier}, {L{\'o}pez-Morales}, {Mikal-Evans}, {Sing}, {Wakeford}, \& {Ehrenreich}}]{2021A&A...650A..73B}
{Bourrier}, V., {dos Santos}, L.~A., {Sanz-Forcada}, J., {et~al.} 2021, \aap, 650, A73, \dodoi{10.1051/0004-6361/202140487}

\bibitem[{{Braam} {et~al.}(2021){Braam}, {van der Tak}, {Chubb}, \& {Min}}]{2021A&A...646A..17B}
{Braam}, M., {van der Tak}, F. F.~S., {Chubb}, K.~L., \& {Min}, M. 2021, \aap, 646, A17, \dodoi{10.1051/0004-6361/202039509}

\bibitem[{{Brande} {et~al.}(2024){Brande}, {Crossfield}, {Kreidberg}, {Morley}, {Barman}, {Benneke}, {Christiansen}, {Dragomir}, {Fortney}, {Greene}, {Hardegree-Ullman}, {Howard}, {Knutson}, {Lothringer}, \& {Mikal-Evans}}]{2024ApJ...961L..23B}
{Brande}, J., {Crossfield}, I. J.~M., {Kreidberg}, L., {et~al.} 2024, \apjl, 961, L23, \dodoi{10.3847/2041-8213/ad1b5c}

\bibitem[{{Brown} {et~al.}(2001){Brown}, {Charbonneau}, {Gilliland}, {Noyes}, \& {Burrows}}]{2001ApJ...552..699B}
{Brown}, T.~M., {Charbonneau}, D., {Gilliland}, R.~L., {Noyes}, R.~W., \& {Burrows}, A. 2001, \apj, 552, 699, \dodoi{10.1086/320580}

\bibitem[{{Bruno} {et~al.}(2020){Bruno}, {Lewis}, {Alam}, {L{\'o}pez-Morales}, {Barstow}, {Wakeford}, {Sing}, {Henry}, {Ballester}, {Bourrier}, {Buchhave}, {Cohen}, {Mikal-Evans}, {Garc{\'\i}a Mu{\~n}oz}, {Lavvas}, \& {Sanz-Forcada}}]{2020MNRAS.491.5361B}
{Bruno}, G., {Lewis}, N.~K., {Alam}, M.~K., {et~al.} 2020, \mnras, 491, 5361, \dodoi{10.1093/mnras/stz3194}

\bibitem[{{Buchner}(2016)}]{2016S&C....26..383B}
{Buchner}, J. 2016, Statistics and Computing, 26, 383, \dodoi{10.1007/s11222-014-9512-y}

\bibitem[{{Buchner}(2019)}]{2019PASP..131j8005B}
---. 2019, \pasp, 131, 108005, \dodoi{10.1088/1538-3873/aae7fc}

\bibitem[{{Buchner}(2021)}]{2021JOSS....6.3001B}
---. 2021, The Journal of Open Source Software, 6, 3001, \dodoi{10.21105/joss.03001}

\bibitem[{Buchner {et~al.}(2014)Buchner, Georgakakis, Nandra, Hsu, Rangel, Brightman, Merloni, Salvato, Donley, \& Kocevski}]{buchner2014x}
Buchner, J., Georgakakis, A., Nandra, K., {et~al.} 2014, Astronomy \& Astrophysics, 564, A125

\bibitem[{{Buchner} {et~al.}(2014){Buchner}, {Georgakakis}, {Nandra}, {Hsu}, {Rangel}, {Brightman}, {Merloni}, {Salvato}, {Donley}, \& {Kocevski}}]{2014A&A...564A.125B}
{Buchner}, J., {Georgakakis}, A., {Nandra}, K., {et~al.} 2014, \aap, 564, A125, \dodoi{10.1051/0004-6361/201322971}

\bibitem[{{Buldyreva} {et~al.}(2022){Buldyreva}, {Yurchenko}, \& {Tennyson}}]{2022RASTI...1...43B}
{Buldyreva}, J., {Yurchenko}, S.~N., \& {Tennyson}, J. 2022, RAS Techniques and Instruments, 1, 43, \dodoi{10.1093/rasti/rzac004}

\bibitem[{{Butler} {et~al.}(2004){Butler}, {Vogt}, {Marcy}, {Fischer}, {Wright}, {Henry}, {Laughlin}, \& {Lissauer}}]{2004ApJ...617..580B}
{Butler}, R.~P., {Vogt}, S.~S., {Marcy}, G.~W., {et~al.} 2004, \apj, 617, 580, \dodoi{10.1086/425173}

\bibitem[{{Cabot} {et~al.}(2020){Cabot}, {Madhusudhan}, {Welbanks}, {Piette}, \& {Gandhi}}]{2020MNRAS.494..363C}
{Cabot}, S. H.~C., {Madhusudhan}, N., {Welbanks}, L., {Piette}, A., \& {Gandhi}, S. 2020, \mnras, 494, 363, \dodoi{10.1093/mnras/staa748}

\bibitem[{{Carleo} {et~al.}(2022){Carleo}, {Giacobbe}, {Guilluy}, {Cubillos}, {Bonomo}, {Sozzetti}, {Brogi}, {Gandhi}, {Fossati}, {Turrini}, {Biazzo}, {Borsa}, {Lanza}, {Malavolta}, {Maggio}, {Mancini}, {Micela}, {Pino}, {Poretti}, {Rainer}, {Scandariato}, {Schisano}, {Andreuzzi}, {Bignamini}, {Cosentino}, {Fiorenzano}, {Harutyunyan}, {Molinari}, {Pedani}, {Redfield}, \& {Stoev}}]{2022AJ....164..101C}
{Carleo}, I., {Giacobbe}, P., {Guilluy}, G., {et~al.} 2022, \aj, 164, 101, \dodoi{10.3847/1538-3881/ac80bf}

\bibitem[{{Carter} {et~al.}(2020){Carter}, {Nikolov}, {Sing}, {Alam}, {Goyal}, {Mikal-Evans}, {Wakeford}, {Henry}, {Morrell}, {L{\'o}pez-Morales}, {Smalley}, {Lavvas}, {Barstow}, {Garc{\'\i}a Mu{\~n}oz}, {Gibson}, \& {Wilson}}]{2020MNRAS.494.5449C}
{Carter}, A.~L., {Nikolov}, N., {Sing}, D.~K., {et~al.} 2020, \mnras, 494, 5449, \dodoi{10.1093/mnras/staa1078}

\bibitem[{{Casasayas-Barris} {et~al.}(2017){Casasayas-Barris}, {Palle}, {Nowak}, {Yan}, {Nortmann}, \& {Murgas}}]{2017A&A...608A.135C}
{Casasayas-Barris}, N., {Palle}, E., {Nowak}, G., {et~al.} 2017, \aap, 608, A135, \dodoi{10.1051/0004-6361/201731956}

\bibitem[{{Cassan} {et~al.}(2012){Cassan}, {Kubas}, {Beaulieu}, {Dominik}, {Horne}, {Greenhill}, {Wambsganss}, {Menzies}, {Williams}, {J{\o}rgensen}, {Udalski}, {Bennett}, {Albrow}, {Batista}, {Brillant}, {Caldwell}, {Cole}, {Coutures}, {Cook}, {Dieters}, {Dominis Prester}, {Donatowicz}, {Fouqu{\'e}}, {Hill}, {Kains}, {Kane}, {Marquette}, {Martin}, {Pollard}, {Sahu}, {Vinter}, {Warren}, {Watson}, {Zub}, {Sumi}, {Szyma{\'n}ski}, {Kubiak}, {Poleski}, {Soszynski}, {Ulaczyk}, {Pietrzy{\'n}ski}, \& {Wyrzykowski}}]{2012Natur.481..167C}
{Cassan}, A., {Kubas}, D., {Beaulieu}, J.~P., {et~al.} 2012, \nat, 481, 167, \dodoi{10.1038/nature10684}

\bibitem[{{Cegla} {et~al.}(2023){Cegla}, {Roguet-Kern}, {Lendl}, {Akinsanmi}, {McCormac}, {Oshagh}, {Wheatley}, {Chen}, {Allart}, {Mortier}, {Bourrier}, {Buchschacher}, {Lovis}, {Sosnowska}, {Sulis}, {Turner}, {Casasayas-Barris}, {Palle}, {Yan}, {Burleigh}, {Casewell}, {Goad}, {Hawthorn}, \& {Wyttenbach}}]{2023A&A...674A.174C}
{Cegla}, H.~M., {Roguet-Kern}, N., {Lendl}, M., {et~al.} 2023, \aap, 674, A174, \dodoi{10.1051/0004-6361/202245523}

\bibitem[{{Chachan} {et~al.}(2019){Chachan}, {Knutson}, {Gao}, {Kataria}, {Wong}, {Henry}, {Benneke}, {Zhang}, {Barstow}, {Bean}, {Mikal-Evans}, {Lewis}, {Mansfield}, {L{\'o}pez-Morales}, {Nikolov}, {Sing}, \& {Wakeford}}]{2019AJ....158..244C}
{Chachan}, Y., {Knutson}, H.~A., {Gao}, P., {et~al.} 2019, \aj, 158, 244, \dodoi{10.3847/1538-3881/ab4e9a}

\bibitem[{{Changeat} {et~al.}(2022){Changeat}, {Edwards}, {Al-Refaie}, {Tsiaras}, {Skinner}, {Cho}, {Yip}, {Anisman}, {Ikoma}, {Bieger}, {Venot}, {Shibata}, {Waldmann}, \& {Tinetti}}]{2022ApJS..260....3C}
{Changeat}, Q., {Edwards}, B., {Al-Refaie}, A.~F., {et~al.} 2022, \apjs, 260, 3, \dodoi{10.3847/1538-4365/ac5cc2}

\bibitem[{{Changeat} {et~al.}(2024){Changeat}, {Skinner}, {Cho}, {N{\"a}ttil{\"a}}, {Waldmann}, {Al-Refaie}, {Dyrek}, {Edwards}, {Mikal-Evans}, {Joshua}, {Morello}, {Skaf}, {Tsiaras}, {Venot}, \& {Yip}}]{2024ApJS..270...34C}
{Changeat}, Q., {Skinner}, J.~W., {Cho}, J.~Y.~K., {et~al.} 2024, \apjs, 270, 34, \dodoi{10.3847/1538-4365/ad1191}

\bibitem[{{Charbonneau} {et~al.}(2000){Charbonneau}, {Brown}, {Latham}, \& {Mayor}}]{2000ApJ...529L..45C}
{Charbonneau}, D., {Brown}, T.~M., {Latham}, D.~W., \& {Mayor}, M. 2000, \apjl, 529, L45, \dodoi{10.1086/312457}

\bibitem[{Charnay {et~al.}(2021)Charnay, Blain, B{\'e}zard, Leconte, Turbet, \& Falco}]{charnay2021formation}
Charnay, B., Blain, D., B{\'e}zard, B., {et~al.} 2021, Astronomy \& Astrophysics, 646, A171

\bibitem[{{Chen} {et~al.}(2020){Chen}, {Casasayas-Barris}, {Pall{\'e}}, {Yan}, {Stangret}, {Cegla}, {Allart}, \& {Lovis}}]{2020A&A...635A.171C}
{Chen}, G., {Casasayas-Barris}, N., {Pall{\'e}}, E., {et~al.} 2020, \aap, 635, A171, \dodoi{10.1051/0004-6361/201936986}

\bibitem[{{Chen} {et~al.}(2017){Chen}, {Pall{\'e}}, {Nortmann}, {Murgas}, {Parviainen}, \& {Nowak}}]{2017A&A...600L..11C}
{Chen}, G., {Pall{\'e}}, E., {Nortmann}, L., {et~al.} 2017, \aap, 600, L11, \dodoi{10.1051/0004-6361/201730736}

\bibitem[{{Chen} {et~al.}(2022){Chen}, {Wang}, {van Boekel}, \& {Pall{\'e}}}]{2022AJ....164..173C}
{Chen}, G., {Wang}, H., {van Boekel}, R., \& {Pall{\'e}}, E. 2022, \aj, 164, 173, \dodoi{10.3847/1538-3881/ac8df6}

\bibitem[{{Chen} {et~al.}(2018){Chen}, {Pall{\'e}}, {Welbanks}, {Prieto-Arranz}, {Madhusudhan}, {Gandhi}, {Casasayas-Barris}, {Murgas}, {Nortmann}, {Crouzet}, {Parviainen}, \& {Gandolfi}}]{2018A&A...616A.145C}
{Chen}, G., {Pall{\'e}}, E., {Welbanks}, L., {et~al.} 2018, \aap, 616, A145, \dodoi{10.1051/0004-6361/201833033}

\bibitem[{{Cho} {et~al.}(2003){Cho}, {Menou}, {Hansen}, \& {Seager}}]{2003ApJ...587L.117C}
{Cho}, J. Y.~K., {Menou}, K., {Hansen}, B. M.~S., \& {Seager}, S. 2003, \apjl, 587, L117, \dodoi{10.1086/375016}

\bibitem[{{Christian} {et~al.}(2006){Christian}, {Pollacco}, {Skillen}, {Street}, {Keenan}, {Clarkson}, {Collier Cameron}, {Kane}, {Lister}, {West}, {Enoch}, {Evans}, {Fitzsimmons}, {Haswell}, {Hellier}, {Hodgkin}, {Horne}, {Irwin}, {Norton}, {Osborne}, {Ryans}, {Wheatley}, \& {Wilson}}]{2006MNRAS.372.1117C}
{Christian}, D.~J., {Pollacco}, D.~L., {Skillen}, I., {et~al.} 2006, \mnras, 372, 1117, \dodoi{10.1111/j.1365-2966.2006.10913.x}

\bibitem[{Chubb {et~al.}(2021)Chubb, Rocchetto, Yurchenko, Min, Waldmann, Barstow, Molli{\`e}re, Al-Refaie, Phillips, \& Tennyson}]{chubb2021exomolop}
Chubb, K.~L., Rocchetto, M., Yurchenko, S.~N., {et~al.} 2021, Astronomy \& Astrophysics, 646, A21

\bibitem[{{Claret}(2000)}]{Claret2000}
{Claret}, A. 2000, \aap, 363, 1081

\bibitem[{{Claudi} {et~al.}(2024){Claudi}, {Bruno}, {Fossati}, {Lanza}, {Maggio}, {Micela}, {Maldonado}, {Benatti}, {Biazzo}, {Bignamini}, {Cabona}, {Carleo}, {Danielski}, {Desidera}, {Malavolta}, {Mancini}, {Montalto}, {Nardiello}, {Rainer}, {Scandariato}, {Sozzetti}, {Cosentino}, {Covino}, {Di Fabrizio}, {Ghedina}, {Lorenzi}, {Molinari}, {Molinaro}, {Pagano}, {Piotto}, \& {Poretti}}]{Claudi2024}
{Claudi}, R., {Bruno}, G., {Fossati}, L., {et~al.} 2024, \aap, 682, A136, \dodoi{10.1051/0004-6361/202347079}

\bibitem[{{Cohen} {et~al.}(2023){Cohen}, {Bollasina}, {Sergeev}, {Palmer}, \& {Mayne}}]{2023PSJ.....4...68C}
{Cohen}, M., {Bollasina}, M.~A., {Sergeev}, D.~E., {Palmer}, P.~I., \& {Mayne}, N.~J. 2023, \psj, 4, 68, \dodoi{10.3847/PSJ/acc9c4}

\bibitem[{Collette(2013)}]{collette2013python}
Collette, A. 2013, Python and HDF5: unlocking scientific data (" O'Reilly Media, Inc.")

\bibitem[{{Cowan} \& {Agol}(2011)}]{2011ApJ...729...54C}
{Cowan}, N.~B., \& {Agol}, E. 2011, \apj, 729, 54, \dodoi{10.1088/0004-637X/729/1/54}

\bibitem[{{Crossfield} {et~al.}(2013){Crossfield}, {Barman}, {Hansen}, \& {Howard}}]{2013A&A...559A..33C}
{Crossfield}, I. J.~M., {Barman}, T., {Hansen}, B. M.~S., \& {Howard}, A.~W. 2013, \aap, 559, A33, \dodoi{10.1051/0004-6361/201322278}

\bibitem[{{Danielski} {et~al.}(2014){Danielski}, {Deroo}, {Waldmann}, {Hollis}, {Tinetti}, \& {Swain}}]{2014ApJ...785...35D}
{Danielski}, C., {Deroo}, P., {Waldmann}, I.~P., {et~al.} 2014, \apj, 785, 35, \dodoi{10.1088/0004-637X/785/1/35}

\bibitem[{{Danielski} {et~al.}(2022){Danielski}, {Brucalassi}, {Benatti}, {Campante}, {Delgado-Mena}, {Rainer}, {Sacco}, {Adibekyan}, {Biazzo}, {Bossini}, {Bruno}, {Casali}, {Kabath}, {Magrini}, {Micela}, {Morello}, {Palladino}, {Sanna}, {Sarkar}, {Sousa}, {Tsantaki}, {Turrini}, \& {Van der Swaelmen}}]{Danielski2022}
{Danielski}, C., {Brucalassi}, A., {Benatti}, S., {et~al.} 2022, Experimental Astronomy, 53, 473, \dodoi{10.1007/s10686-021-09765-1}

\bibitem[{{Delrez} {et~al.}(2016){Delrez}, {Santerne}, {Almenara}, {Anderson}, {Collier-Cameron}, {D{\'\i}az}, {Gillon}, {Hellier}, {Jehin}, {Lendl}, {Maxted}, {Neveu-VanMalle}, {Pepe}, {Pollacco}, {Queloz}, {S{\'e}gransan}, {Smalley}, {Smith}, {Triaud}, {Udry}, {Van Grootel}, \& {West}}]{delrez2016w121}
{Delrez}, L., {Santerne}, A., {Almenara}, J.~M., {et~al.} 2016, \mnras, 458, 4025, \dodoi{10.1093/mnras/stw522}

\bibitem[{{Deming} {et~al.}(2011){Deming}, {Sada}, {Jackson}, {Peterson}, {Agol}, {Knutson}, {Jennings}, {Haase}, \& {Bays}}]{2011ApJ...740...33D}
{Deming}, D., {Sada}, P.~V., {Jackson}, B., {et~al.} 2011, \apj, 740, 33, \dodoi{10.1088/0004-637X/740/1/33}

\bibitem[{{Deming} {et~al.}(2013){Deming}, {Wilkins}, {McCullough}, {Burrows}, {Fortney}, {Agol}, {Dobbs-Dixon}, {Madhusudhan}, {Crouzet}, {Desert}, {Gilliland}, {Haynes}, {Knutson}, {Line}, {Magic}, {Mandell}, {Ranjan}, {Charbonneau}, {Clampin}, {Seager}, \& {Showman}}]{2013ApJ...774...95D}
{Deming}, D., {Wilkins}, A., {McCullough}, P., {et~al.} 2013, \apj, 774, 95, \dodoi{10.1088/0004-637X/774/2/95}

\bibitem[{{dos Santos} {et~al.}(2020){dos Santos}, {Ehrenreich}, {Bourrier}, {Allart}, {King}, {Lendl}, {Lovis}, {Margheim}, {Mel{\'e}ndez}, {Seidel}, \& {Sousa}}]{2020A&A...640A..29D}
{dos Santos}, L.~A., {Ehrenreich}, D., {Bourrier}, V., {et~al.} 2020, \aap, 640, A29, \dodoi{10.1051/0004-6361/202038802}

\bibitem[{{dos Santos} {et~al.}(2021){dos Santos}, {Bourrier}, {Ehrenreich}, {Sanz-Forcada}, {L{\'o}pez-Morales}, {Sing}, {Garc{\'\i}a Mu{\~n}oz}, {Henry}, {Lavvas}, {Lecavelier des Etangs}, {Mikal-Evans}, {Vidal-Madjar}, \& {Wakeford}}]{2021A&A...649A..40D}
{dos Santos}, L.~A., {Bourrier}, V., {Ehrenreich}, D., {et~al.} 2021, \aap, 649, A40, \dodoi{10.1051/0004-6361/202140491}

\bibitem[{{Dragomir} {et~al.}(2015){Dragomir}, {Benneke}, {Pearson}, {Crossfield}, {Eastman}, {Barman}, \& {Biddle}}]{2015ApJ...814..102D}
{Dragomir}, D., {Benneke}, B., {Pearson}, K.~A., {et~al.} 2015, \apj, 814, 102, \dodoi{10.1088/0004-637X/814/2/102}

\bibitem[{{Edwards} {et~al.}(2019){Edwards}, {Rice}, {Zingales}, {Tessenyi}, {Waldmann}, {Tinetti}, {Pascale}, {Savini}, \& {Sarkar}}]{2019ExA....47...29E}
{Edwards}, B., {Rice}, M., {Zingales}, T., {et~al.} 2019, Experimental Astronomy, 47, 29, \dodoi{10.1007/s10686-018-9611-4}

\bibitem[{{Edwards} {et~al.}(2021){Edwards}, {Changeat}, {Mori}, {Anisman}, {Morvan}, {Yip}, {Tsiaras}, {Al-Refaie}, {Waldmann}, \& {Tinetti}}]{2021AJ....161...44E}
{Edwards}, B., {Changeat}, Q., {Mori}, M., {et~al.} 2021, \aj, 161, 44, \dodoi{10.3847/1538-3881/abc6a5}

\bibitem[{Edwards {et~al.}(2022)Edwards, Changeat, Tsiaras, Yip, Al-Refaie, Anisman, Bieger, Gressier, Shibata, Skaf, {et~al.}}]{edwards2022exploring}
Edwards, B., Changeat, Q., Tsiaras, A., {et~al.} 2022, arXiv preprint arXiv:2211.00649

\bibitem[{{Ehrenreich} {et~al.}(2011){Ehrenreich}, {Lecavelier Des Etangs}, \& {Delfosse}}]{2011A&A...529A..80E}
{Ehrenreich}, D., {Lecavelier Des Etangs}, A., \& {Delfosse}, X. 2011, \aap, 529, A80, \dodoi{10.1051/0004-6361/201016162}

\bibitem[{{Ehrenreich} {et~al.}(2015){Ehrenreich}, {Bourrier}, {Wheatley}, {Lecavelier des Etangs}, {H{\'e}brard}, {Udry}, {Bonfils}, {Delfosse}, {D{\'e}sert}, {Sing}, \& {Vidal-Madjar}}]{2015Natur.522..459E}
{Ehrenreich}, D., {Bourrier}, V., {Wheatley}, P.~J., {et~al.} 2015, \nat, 522, 459, \dodoi{10.1038/nature14501}

\bibitem[{{Espinoza} {et~al.}(2019){Espinoza}, {Kossakowski}, \& {Brahm}}]{2019MNRAS.490.2262E}
{Espinoza}, N., {Kossakowski}, D., \& {Brahm}, R. 2019, \mnras, 490, 2262, \dodoi{10.1093/mnras/stz2688}

\bibitem[{Esposito {et~al.}(2014)Esposito, Covino, Mancini, Harutyunyan, Southworth, Biazzo, Gandolfi, Lanza, Barbieri, Bonomo, {et~al.}}]{esposito2014gaps}
Esposito, M., Covino, E., Mancini, L., {et~al.} 2014, Astronomy \& Astrophysics, 564, L13

\bibitem[{{Esposito} {et~al.}(2014){Esposito}, {Covino}, {Mancini}, {Harutyunyan}, {Southworth}, {Biazzo}, {Gandolfi}, {Lanza}, {Barbieri}, {Bonomo}, {Borsa}, {Claudi}, {Cosentino}, {Desidera}, {Gratton}, {Pagano}, {Sozzetti}, {Boccato}, {Maggio}, {Micela}, {Molinari}, {Nascimbeni}, {Piotto}, {Poretti}, \& {Smareglia}}]{2014A&A...564L..13E}
{Esposito}, M., {Covino}, E., {Mancini}, L., {et~al.} 2014, \aap, 564, L13, \dodoi{10.1051/0004-6361/201423735}

\bibitem[{{Estrela} {et~al.}(2021){Estrela}, {Swain}, {Roudier}, {West}, {Sedaghati}, \& {Valio}}]{2021AJ....162...91E}
{Estrela}, R., {Swain}, M.~R., {Roudier}, G.~M., {et~al.} 2021, \aj, 162, 91, \dodoi{10.3847/1538-3881/ac0c7c}

\bibitem[{{Evans} {et~al.}(2016){Evans}, {Sing}, {Wakeford}, {Nikolov}, {Ballester}, {Drummond}, {Kataria}, {Gibson}, {Amundsen}, \& {Spake}}]{2016ApJ...822L...4E}
{Evans}, T.~M., {Sing}, D.~K., {Wakeford}, H.~R., {et~al.} 2016, \apjl, 822, L4, \dodoi{10.3847/2041-8205/822/1/L4}

\bibitem[{{Evans} {et~al.}(2018){Evans}, {Sing}, {Goyal}, {Nikolov}, {Marley}, {Zahnle}, {Henry}, {Barstow}, {Alam}, {Sanz-Forcada}, {Kataria}, {Lewis}, {Lavvas}, {Ballester}, {Ben-Jaffel}, {Blumenthal}, {Bourrier}, {Drummond}, {Garc{\'\i}a Mu{\~n}oz}, {L{\'o}pez-Morales}, {Tremblin}, {Ehrenreich}, {Wakeford}, {Buchhave}, {Lecavelier des Etangs}, {H{\'e}brard}, \& {Williamson}}]{2018AJ....156..283E}
{Evans}, T.~M., {Sing}, D.~K., {Goyal}, J.~M., {et~al.} 2018, \aj, 156, 283, \dodoi{10.3847/1538-3881/aaebff}

\bibitem[{{Faedi} {et~al.}(2011){Faedi}, {Barros}, {Anderson}, {Brown}, {Collier Cameron}, {Pollacco}, {Boisse}, {H{\'e}brard}, {Lendl}, {Lister}, {Smalley}, {Street}, {Triaud}, {Bento}, {Bouchy}, {Butters}, {Enoch}, {Haswell}, {Hellier}, {Keenan}, {Miller}, {Moulds}, {Moutou}, {Norton}, {Queloz}, {Santerne}, {Simpson}, {Skillen}, {Smith}, {Udry}, {Watson}, {West}, \& {Wheatley}}]{faedi2011w39}
{Faedi}, F., {Barros}, S.~C.~C., {Anderson}, D.~R., {et~al.} 2011, \aap, 531, A40, \dodoi{10.1051/0004-6361/201116671}

\bibitem[{{Fairman} {et~al.}(2024){Fairman}, {Wakeford}, \& {MacDonald}}]{2024arXiv240307801F}
{Fairman}, C., {Wakeford}, H.~R., \& {MacDonald}, R.~J. 2024, arXiv e-prints, arXiv:2403.07801, \dodoi{10.48550/arXiv.2403.07801}

\bibitem[{{Feinstein} {et~al.}(2023){Feinstein}, {Radica}, {Welbanks}, {Murray}, {Ohno}, {Coulombe}, {Espinoza}, {Bean}, {Teske}, {Benneke}, {Line}, {Rustamkulov}, {Saba}, {Tsiaras}, {Barstow}, {Fortney}, {Gao}, {Knutson}, {MacDonald}, {Mikal-Evans}, {Rackham}, {Taylor}, {Parmentier}, {Batalha}, {Berta-Thompson}, {Carter}, {Changeat}, {dos Santos}, {Gibson}, {Goyal}, {Kreidberg}, {L{\'o}pez-Morales}, {Lothringer}, {Miguel}, {Molaverdikhani}, {Moran}, {Morello}, {Mukherjee}, {Sing}, {Stevenson}, {Wakeford}, {Ahrer}, {Alam}, {Alderson}, {Allen}, {Batalha}, {Bell}, {Blecic}, {Brande}, {Caceres}, {Casewell}, {Chubb}, {Crossfield}, {Crouzet}, {Cubillos}, {Decin}, {D{\'e}sert}, {Harrington}, {Heng}, {Henning}, {Iro}, {Kempton}, {Kendrew}, {Kirk}, {Krick}, {Lagage}, {Lendl}, {Mancini}, {Mansfield}, {May}, {Mayne}, {Nikolov}, {Palle}, {Petit dit de la Roche}, {Piaulet}, {Powell}, {Redfield}, {Rogers}, {Roman}, {Roy}, {Nixon}, {Schlawin}, {Tan}, {Tremblin}, {Turner}, {Venot}, {Waalkes}, {Wheatley}, \&
  {Zhang}}]{2023Natur.614..670F}
{Feinstein}, A.~D., {Radica}, M., {Welbanks}, L., {et~al.} 2023, \nat, 614, 670, \dodoi{10.1038/s41586-022-05674-1}

\bibitem[{Feroz {et~al.}(2009)Feroz, Hobson, \& Bridges}]{feroz2009multinest}
Feroz, F., Hobson, M., \& Bridges, M. 2009, Monthly Notices of the Royal Astronomical Society, 398, 1601

\bibitem[{{Fischer} {et~al.}(2016){Fischer}, {Knutson}, {Sing}, {Henry}, {Williamson}, {Fortney}, {Burrows}, {Kataria}, {Nikolov}, {Showman}, {Ballester}, {D{\'e}sert}, {Aigrain}, {Deming}, {Lecavelier des Etangs}, \& {Vidal-Madjar}}]{2016ApJ...827...19F}
{Fischer}, P.~D., {Knutson}, H.~A., {Sing}, D.~K., {et~al.} 2016, \apj, 827, 19, \dodoi{10.3847/0004-637X/827/1/19}

\bibitem[{{Fisher} \& {Heng}(2018)}]{2018MNRAS.481.4698F}
{Fisher}, C., \& {Heng}, K. 2018, \mnras, 481, 4698, \dodoi{10.1093/mnras/sty2550}

\bibitem[{Fletcher {et~al.}(2018)Fletcher, Gustafsson, \& Orton}]{Fletcher_2018}
Fletcher, L.~N., Gustafsson, M., \& Orton, G.~S. 2018, The Astrophysical Journal Supplement Series, 235, 24, \dodoi{10.3847/1538-4365/aaa07a}

\bibitem[{Foreman-Mackey(2016)}]{foreman2016corner}
Foreman-Mackey, D. 2016, The Journal of Open Source Software, 1, 24

\bibitem[{{Foreman-Mackey} {et~al.}(2017){Foreman-Mackey}, {Agol}, {Angus}, \& {Ambikasaran}}]{celerite}
{Foreman-Mackey}, D., {Agol}, E., {Angus}, R., \& {Ambikasaran}, S. 2017, ArXiv.
\newblock \url{https://arxiv.org/abs/1703.09710}

\bibitem[{Foreman-Mackey {et~al.}(2013)Foreman-Mackey, Hogg, Lang, \& Goodman}]{foreman2013emcee}
Foreman-Mackey, D., Hogg, D.~W., Lang, D., \& Goodman, J. 2013, Publications of the Astronomical Society of the Pacific, 125, 306

\bibitem[{{Fossati} {et~al.}(2023){Fossati}, {Pillitteri}, {Shaikhislamov}, {Bonfanti}, {Borsa}, {Carleo}, {Guilluy}, \& {Rumenskikh}}]{2023A&A...673A..37F}
{Fossati}, L., {Pillitteri}, I., {Shaikhislamov}, I.~F., {et~al.} 2023, \aap, 673, A37, \dodoi{10.1051/0004-6361/202245667}

\bibitem[{{Fossati} {et~al.}(2022){Fossati}, {Guilluy}, {Shaikhislamov}, {Carleo}, {Borsa}, {Bonomo}, {Giacobbe}, {Rainer}, {Cecchi-Pestellini}, {Khodachenko}, {Efimov}, {Rumenskikh}, {Miroshnichenko}, {Berezutsky}, {Nascimbeni}, {Brogi}, {Lanza}, {Mancini}, {Affer}, {Benatti}, {Biazzo}, {Bignamini}, {Carosati}, {Claudi}, {Cosentino}, {Covino}, {Desidera}, {Fiorenzano}, {Harutyunyan}, {Maggio}, {Malavolta}, {Maldonado}, {Micela}, {Molinari}, {Pagano}, {Pedani}, {Piotto}, {Poretti}, {Scandariato}, {Sozzetti}, \& {Stoev}}]{2022A&A...658A.136F}
{Fossati}, L., {Guilluy}, G., {Shaikhislamov}, I.~F., {et~al.} 2022, \aap, 658, A136, \dodoi{10.1051/0004-6361/202142336}

\bibitem[{{Fournier-Tondreau} {et~al.}(2024){Fournier-Tondreau}, {MacDonald}, {Radica}, {Lafreni{\`e}re}, {Welbanks}, {Piaulet}, {Coulombe}, {Allart}, {Morel}, {Artigau}, {Albert}, {Lim}, {Doyon}, {Benneke}, {Rowe}, {Darveau-Bernier}, {Cowan}, {Lewis}, {Cook}, {Flagg}, {Genest}, {Pelletier}, {Johnstone}, {Dang}, {Kaltenegger}, {Taylor}, \& {Turner}}]{2024MNRAS.528.3354F}
{Fournier-Tondreau}, M., {MacDonald}, R.~J., {Radica}, M., {et~al.} 2024, \mnras, 528, 3354, \dodoi{10.1093/mnras/stad3813}

\bibitem[{{Fraine} {et~al.}(2014){Fraine}, {Deming}, {Benneke}, {Knutson}, {Jord{\'a}n}, {Espinoza}, {Madhusudhan}, {Wilkins}, \& {Todorov}}]{2014Natur.513..526F}
{Fraine}, J., {Deming}, D., {Benneke}, B., {et~al.} 2014, \nat, 513, 526, \dodoi{10.1038/nature13785}

\bibitem[{{Fu} {et~al.}(2017){Fu}, {Deming}, {Knutson}, {Madhusudhan}, {Mandell}, \& {Fraine}}]{2017ApJ...847L..22F}
{Fu}, G., {Deming}, D., {Knutson}, H., {et~al.} 2017, \apjl, 847, L22, \dodoi{10.3847/2041-8213/aa8e40}

\bibitem[{{Fu} {et~al.}(2021){Fu}, {Deming}, {May}, {Stevenson}, {Sing}, {Lothringer}, {Wakeford}, {Nikolov}, {Mikal-Evans}, {Bourrier}, {dos Santos}, {Alam}, {Henry}, {Garc{\'\i}a Mu{\~n}oz}, \& {L{\'o}pez-Morales}}]{2021AJ....162..271F}
{Fu}, G., {Deming}, D., {May}, E., {et~al.} 2021, \aj, 162, 271, \dodoi{10.3847/1538-3881/ac3008}

\bibitem[{{Fu} {et~al.}(2022){Fu}, {Espinoza}, {Sing}, {Lothringer}, {Dos Santos}, {Rustamkulov}, {Deming}, {Kempton}, {Komacek}, {Knutson}, {Albert}, {Pontoppidan}, {Volk}, \& {Filippazzo}}]{2022ApJ...940L..35F}
{Fu}, G., {Espinoza}, N., {Sing}, D.~K., {et~al.} 2022, \apjl, 940, L35, \dodoi{10.3847/2041-8213/ac9977}

\bibitem[{{Fukui} {et~al.}(2013){Fukui}, {Narita}, {Kurosaki}, {Ikoma}, {Yanagisawa}, {Kuroda}, {Shimizu}, {Takahashi}, {Ohnuki}, {Onitsuka}, {Hirano}, {Suenaga}, {Kawauchi}, {Nagayama}, {Ohta}, {Yoshida}, {Kawai}, \& {Izumiura}}]{2013ApJ...770...95F}
{Fukui}, A., {Narita}, N., {Kurosaki}, K., {et~al.} 2013, \apj, 770, 95, \dodoi{10.1088/0004-637X/770/2/95}

\bibitem[{{Fukui} {et~al.}(2014){Fukui}, {Kawashima}, {Ikoma}, {Narita}, {Onitsuka}, {Ita}, {Onozato}, {Nishiyama}, {Baba}, {Ryu}, {Hirano}, {Hori}, {Kurosaki}, {Kawauchi}, {Takahashi}, {Nagayama}, {Tamura}, {Kawai}, {Kuroda}, {Nagayama}, {Ohta}, {Shimizu}, {Yanagisawa}, {Yoshida}, \& {Izumiura}}]{2014ApJ...790..108F}
{Fukui}, A., {Kawashima}, Y., {Ikoma}, M., {et~al.} 2014, \apj, 790, 108, \dodoi{10.1088/0004-637X/790/2/108}

\bibitem[{{Fulton} {et~al.}(2017){Fulton}, {Petigura}, {Howard}, {Isaacson}, {Marcy}, {Cargile}, {Hebb}, {Weiss}, {Johnson}, {Morton}, {Sinukoff}, {Crossfield}, \& {Hirsch}}]{2017AJ....154..109F}
{Fulton}, B.~J., {Petigura}, E.~A., {Howard}, A.~W., {et~al.} 2017, \aj, 154, 109, \dodoi{10.3847/1538-3881/aa80eb}

\bibitem[{{Gaia Collaboration} {et~al.}(2022){Gaia Collaboration}, {Creevey}, {Sarro}, {Lobel}, {Pancino}, {Andrae}, {Smart}, \& {et al.}}]{2022yCat..36740039G}
{Gaia Collaboration}, {Creevey}, O.~L., {Sarro}, L.~M., {et~al.} 2022

\bibitem[{{Gao} {et~al.}(2020){Gao}, {Thorngren}, {Lee}, {Fortney}, {Morley}, {Wakeford}, {Powell}, {Stevenson}, \& {Zhang}}]{2020NatAs...4..951G}
{Gao}, P., {Thorngren}, D.~P., {Lee}, E. K.~H., {et~al.} 2020, Nature Astronomy, 4, 951, \dodoi{10.1038/s41550-020-1114-3}

\bibitem[{Gharib-Nezhad {et~al.}(2021)Gharib-Nezhad, Iyer, Line, Freedman, Marley, \& Batalha}]{gharib2021exoplines}
Gharib-Nezhad, E., Iyer, A.~R., Line, M.~R., {et~al.} 2021, The Astrophysical Journal Supplement Series, 254, 34

\bibitem[{{Gibson} {et~al.}(2013){Gibson}, {Aigrain}, {Barstow}, {Evans}, {Fletcher}, \& {Irwin}}]{2013MNRAS.428.3680G}
{Gibson}, N.~P., {Aigrain}, S., {Barstow}, J.~K., {et~al.} 2013, \mnras, 428, 3680, \dodoi{10.1093/mnras/sts307}

\bibitem[{{Gibson} {et~al.}(2019){Gibson}, {de Mooij}, {Evans}, {Merritt}, {Nikolov}, {Sing}, \& {Watson}}]{2019MNRAS.482..606G}
{Gibson}, N.~P., {de Mooij}, E. J.~W., {Evans}, T.~M., {et~al.} 2019, \mnras, 482, 606, \dodoi{10.1093/mnras/sty2722}

\bibitem[{{Gibson} {et~al.}(2017){Gibson}, {Nikolov}, {Sing}, {Barstow}, {Evans}, {Kataria}, \& {Wilson}}]{2017MNRAS.467.4591G}
{Gibson}, N.~P., {Nikolov}, N., {Sing}, D.~K., {et~al.} 2017, \mnras, 467, 4591, \dodoi{10.1093/mnras/stx353}

\bibitem[{{Gibson} {et~al.}(2020){Gibson}, {Merritt}, {Nugroho}, {Cubillos}, {de Mooij}, {Mikal-Evans}, {Fossati}, {Lothringer}, {Nikolov}, {Sing}, {Spake}, {Watson}, \& {Wilson}}]{2020MNRAS.493.2215G}
{Gibson}, N.~P., {Merritt}, S., {Nugroho}, S.~K., {et~al.} 2020, \mnras, 493, 2215, \dodoi{10.1093/mnras/staa228}

\bibitem[{{Gillon} {et~al.}(2007){Gillon}, {Pont}, {Demory}, {Mallmann}, {Mayor}, {Mazeh}, {Queloz}, {Shporer}, {Udry}, \& {Vuissoz}}]{2007A&A...472L..13G}
{Gillon}, M., {Pont}, F., {Demory}, B.~O., {et~al.} 2007, \aap, 472, L13, \dodoi{10.1051/0004-6361:20077799}

\bibitem[{{Gillon} {et~al.}(2009){Gillon}, {Anderson}, {Triaud}, {Hellier}, {Maxted}, {Pollaco}, {Queloz}, {Smalley}, {West}, {Wilson}, {Bentley}, {Collier Cameron}, {Enoch}, {Hebb}, {Horne}, {Irwin}, {Joshi}, {Lister}, {Mayor}, {Pepe}, {Parley}, {Segransan}, {Udry}, \& {Wheatley}}]{2009Gillon}
{Gillon}, M., {Anderson}, D.~R., {Triaud}, A.~H.~M.~J., {et~al.} 2009, \aap, 501, 785, \dodoi{10.1051/0004-6361/200911749}

\bibitem[{{Gondoin}(2008)}]{gondoin2008}
{Gondoin}, P. 2008, \aap, 478, 883, \dodoi{10.1051/0004-6361:20078245}

\bibitem[{Gordon {et~al.}(2022)Gordon, Rothman, Hargreaves, Hashemi, Karlovets, Skinner, Conway, Hill, Kochanov, Tan, {et~al.}}]{gordon2022hitran2020}
Gordon, I.~E., Rothman, L.~S., Hargreaves, R., {et~al.} 2022, Journal of quantitative spectroscopy and radiative transfer, 277, 107949

\bibitem[{{Gordon} {et~al.}(2020){Gordon}, {Agol}, \& {Foreman-Mackey}}]{2020AJ....160..240G}
{Gordon}, T.~A., {Agol}, E., \& {Foreman-Mackey}, D. 2020, \aj, 160, 240, \dodoi{10.3847/1538-3881/abbc16}

\bibitem[{{Grant} {et~al.}(2023{\natexlab{a}}){Grant}, {Lothringer}, {Wakeford}, {Alam}, {Alderson}, {Bean}, {Benneke}, {D{\'e}sert}, {Daylan}, {Flagg}, {Hu}, {Inglis}, {Kirk}, {Kreidberg}, {L{\'o}pez-Morales}, {Mancini}, {Mikal-Evans}, {Molaverdikhani}, {Palle}, {Rackham}, {Redfield}, {Stevenson}, {Valenti}, {Wallack}, {Aggarwal}, {Ahrer}, {Crossfield}, {Crouzet}, {Iro}, {Nikolov}, {Wheatley}, \& {JWST Transiting Exoplanet Community ERS Team}}]{2023ApJ...949L..15G}
{Grant}, D., {Lothringer}, J.~D., {Wakeford}, H.~R., {et~al.} 2023{\natexlab{a}}, \apjl, 949, L15, \dodoi{10.3847/2041-8213/acd544}

\bibitem[{{Grant} {et~al.}(2023{\natexlab{b}}){Grant}, {Lewis}, {Wakeford}, {Batalha}, {Glidden}, {Goyal}, {Mullens}, {MacDonald}, {May}, {Seager}, {Stevenson}, {Valenti}, {Visscher}, {Alderson}, {Allen}, {Ca{\~n}as}, {Col{\'o}n}, {Clampin}, {Espinoza}, {Gressier}, {Huang}, {Lin}, {Long}, {Louie}, {Pe{\~n}a-Guerrero}, {Ranjan}, {Sotzen}, {Valentine}, {Anderson}, {Balmer}, {Bellini}, {Hoch}, {Kammerer}, {Libralato}, {Mountain}, {Perrin}, {Pueyo}, {Rickman}, {Rebollido}, {Sohn}, {van der Marel}, \& {Watkins}}]{2023ApJ...956L..29G}
{Grant}, D., {Lewis}, N.~K., {Wakeford}, H.~R., {et~al.} 2023{\natexlab{b}}, \apjl, 956, L29, \dodoi{10.3847/2041-8213/acfc3b}

\bibitem[{Gray {et~al.}(2001)Gray, Napier, \& Winkler}]{gray2001physical}
Gray, R., Napier, M., \& Winkler, L. 2001, The Astronomical Journal, 121, 2148

\bibitem[{{Guilluy} {et~al.}(2022){Guilluy}, {Giacobbe}, {Carleo}, {Cubillos}, {Sozzetti}, {Bonomo}, {Brogi}, {Gandhi}, {Fossati}, {Nascimbeni}, {Turrini}, {Schisano}, {Borsa}, {Lanza}, {Mancini}, {Maggio}, {Malavolta}, {Micela}, {Pino}, {Rainer}, {Bignamini}, {Claudi}, {Cosentino}, {Covino}, {Desidera}, {Fiorenzano}, {Harutyunyan}, {Lorenzi}, {Knapic}, {Molinari}, {Pacetti}, {Pagano}, {Pedani}, {Piotto}, \& {Poretti}}]{2022A&A...665A.104G}
{Guilluy}, G., {Giacobbe}, P., {Carleo}, I., {et~al.} 2022, \aap, 665, A104, \dodoi{10.1051/0004-6361/202243854}

\bibitem[{{Hartman} {et~al.}(2009){Hartman}, {Bakos}, {Torres}, {Kov{\'a}cs}, {Noyes}, {P{\'a}l}, {Latham}, {Sip{\H{o}}cz}, {Fischer}, {Johnson}, {Marcy}, {Butler}, {Howard}, {Esquerdo}, {Sasselov}, {Kov{\'a}cs}, {Stefanik}, {Fernandez}, {L{\'a}z{\'a}r}, {Papp}, \& {S{\'a}ri}}]{2009ApJ...706..785H}
{Hartman}, J.~D., {Bakos}, G.~{\'A}., {Torres}, G., {et~al.} 2009, \apj, 706, 785, \dodoi{10.1088/0004-637X/706/1/785}

\bibitem[{{Hartman} {et~al.}(2011{\natexlab{a}}){Hartman}, {Bakos}, {Kipping}, {Torres}, {Kov{\'a}cs}, {Noyes}, {Latham}, {Howard}, {Fischer}, {Johnson}, {Marcy}, {Isaacson}, {Quinn}, {Buchhave}, {B{\'e}ky}, {Sasselov}, {Stefanik}, {Esquerdo}, {Everett}, {Perumpilly}, {L{\'a}z{\'a}r}, {Papp}, \& {S{\'a}ri}}]{2011ApJ...728..138H}
{Hartman}, J.~D., {Bakos}, G.~{\'A}., {Kipping}, D.~M., {et~al.} 2011{\natexlab{a}}, \apj, 728, 138, \dodoi{10.1088/0004-637X/728/2/138}

\bibitem[{{Hartman} {et~al.}(2011{\natexlab{b}}){Hartman}, {Bakos}, {Sato}, {Torres}, {Noyes}, {Latham}, {Kov{\'a}cs}, {Fischer}, {Howard}, {Johnson}, {Marcy}, {Buchhave}, {F{\"u}resz}, {Perumpilly}, {B{\'e}ky}, {Stefanik}, {Sasselov}, {Esquerdo}, {Everett}, {Csubry}, {L{\'a}z{\'a}r}, {Papp}, \& {S{\'a}ri}}]{2011ApJ...726...52H}
{Hartman}, J.~D., {Bakos}, G.~{\'A}., {Sato}, B., {et~al.} 2011{\natexlab{b}}, \apj, 726, 52, \dodoi{10.1088/0004-637X/726/1/52}

\bibitem[{{Hawker} {et~al.}(2018){Hawker}, {Madhusudhan}, {Cabot}, \& {Gandhi}}]{2018ApJ...863L..11H}
{Hawker}, G.~A., {Madhusudhan}, N., {Cabot}, S. H.~C., \& {Gandhi}, S. 2018, \apjl, 863, L11, \dodoi{10.3847/2041-8213/aac49d}

\bibitem[{H{\'e}brard {et~al.}(2013)H{\'e}brard, Cameron, Brown, D{\'\i}az, Faedi, Smalley, Anderson, Armstrong, Barros, Bento, {et~al.}}]{hebrard2013wasp}
H{\'e}brard, G., Cameron, A.~C., Brown, D., {et~al.} 2013, Astronomy \& Astrophysics, 549, A134

\bibitem[{{Hellier} {et~al.}(2010){Hellier}, {Anderson}, {Collier Cameron}, {Gillon}, {Lendl}, {Maxted}, {Queloz}, {Smalley}, {Triaud}, {West}, {Brown}, {Enoch}, {Lister}, {Pepe}, {Pollacco}, {S{\'e}gransan}, \& {Udry}}]{2010ApJ...723L..60H}
{Hellier}, C., {Anderson}, D.~R., {Collier Cameron}, A., {et~al.} 2010, \apjl, 723, L60, \dodoi{10.1088/2041-8205/723/1/L60}

\bibitem[{{Hellier} {et~al.}(2014){Hellier}, {Anderson}, {Collier Cameron}, {Delrez}, {Gillon}, {Jehin}, {Lendl}, {Maxted}, {Pepe}, {Pollacco}, {Queloz}, {S{\'e}gransan}, {Smalley}, {Smith}, {Southworth}, {Triaud}, {Udry}, \& {West}}]{2014MNRAS.440.1982H}
---. 2014, \mnras, 440, 1982, \dodoi{10.1093/mnras/stu410}

\bibitem[{{Hellier} {et~al.}(2015){Hellier}, {Anderson}, {Collier Cameron}, {Delrez}, {Gillon}, {Jehin}, {Lendl}, {Maxted}, {Pepe}, {Pollacco}, {Queloz}, {S{\'e}gransan}, {Smalley}, {Smith}, {Southworth}, {Triaud}, {Turner}, {Udry}, \& {West}}]{2015AJ....150...18H}
---. 2015, \aj, 150, 18, \dodoi{10.1088/0004-6256/150/1/18}

\bibitem[{{Hirano} {et~al.}(2011){Hirano}, {Narita}, {Shporer}, {Sato}, {Aoki}, \& {Tamura}}]{2011PASJ...63S.531H}
{Hirano}, T., {Narita}, N., {Shporer}, A., {et~al.} 2011, \pasj, 63, 531, \dodoi{10.1093/pasj/63.sp2.S531}

\bibitem[{{Hodos{\'a}n} {et~al.}(2016){Hodos{\'a}n}, {Rimmer}, \& {Helling}}]{2016MNRAS.461.1222H}
{Hodos{\'a}n}, G., {Rimmer}, P.~B., \& {Helling}, C. 2016, \mnras, 461, 1222, \dodoi{10.1093/mnras/stw977}

\bibitem[{{Hoeijmakers} {et~al.}(2020){Hoeijmakers}, {Seidel}, {Pino}, {Kitzmann}, {Sindel}, {Ehrenreich}, {Oza}, {Bourrier}, {Allart}, {Gebek}, {Lovis}, {Yurchenko}, {Astudillo-Defru}, {Bayliss}, {Cegla}, {Lavie}, {Lendl}, {Melo}, {Murgas}, {Nascimbeni}, {Pepe}, {S{\'e}gransan}, {Udry}, {Wyttenbach}, \& {Heng}}]{2020A&A...641A.123H}
{Hoeijmakers}, H.~J., {Seidel}, J.~V., {Pino}, L., {et~al.} 2020, \aap, 641, A123, \dodoi{10.1051/0004-6361/202038365}

\bibitem[{{Holczer} {et~al.}(2015){Holczer}, {Shporer}, {Mazeh}, {Fabrycky}, {Nachmani}, {McQuillan}, {Sanchis-Ojeda}, {Orosz}, {Welsh}, {Ford}, \& {Jontof-Hutter}}]{2015ApJ...807..170H}
{Holczer}, T., {Shporer}, A., {Mazeh}, T., {et~al.} 2015, \apj, 807, 170, \dodoi{10.1088/0004-637X/807/2/170}

\bibitem[{{Howard} {et~al.}(2012){Howard}, {Marcy}, {Bryson}, {Jenkins}, {Rowe}, {Batalha}, {Borucki}, {Koch}, {Dunham}, {Gautier}, {Van Cleve}, {Cochran}, {Latham}, {Lissauer}, {Torres}, {Brown}, {Gilliland}, {Buchhave}, {Caldwell}, {Christensen-Dalsgaard}, {Ciardi}, {Fressin}, {Haas}, {Howell}, {Kjeldsen}, {Seager}, {Rogers}, {Sasselov}, {Steffen}, {Basri}, {Charbonneau}, {Christiansen}, {Clarke}, {Dupree}, {Fabrycky}, {Fischer}, {Ford}, {Fortney}, {Tarter}, {Girouard}, {Holman}, {Johnson}, {Klaus}, {Machalek}, {Moorhead}, {Morehead}, {Ragozzine}, {Tenenbaum}, {Twicken}, {Quinn}, {Isaacson}, {Shporer}, {Lucas}, {Walkowicz}, {Welsh}, {Boss}, {Devore}, {Gould}, {Smith}, {Morris}, {Prsa}, {Morton}, {Still}, {Thompson}, {Mullally}, {Endl}, \& {MacQueen}}]{2012ApJS..201...15H}
{Howard}, A.~W., {Marcy}, G.~W., {Bryson}, S.~T., {et~al.} 2012, \apjs, 201, 15, \dodoi{10.1088/0067-0049/201/2/15}

\bibitem[{{Huitson} {et~al.}(2013){Huitson}, {Sing}, {Pont}, {Fortney}, {Burrows}, {Wilson}, {Ballester}, {Nikolov}, {Gibson}, {Deming}, {Aigrain}, {Evans}, {Henry}, {Lecavelier des Etangs}, {Showman}, {Vidal-Madjar}, \& {Zahnle}}]{2013MNRAS.434.3252H}
{Huitson}, C.~M., {Sing}, D.~K., {Pont}, F., {et~al.} 2013, \mnras, 434, 3252, \dodoi{10.1093/mnras/stt1243}

\bibitem[{Hunter(2007)}]{hunter2007matplotlib}
Hunter, J.~D. 2007, Computing in science \& engineering, 9, 90

\bibitem[{Husser {et~al.}(2013)Husser, Wende-von Berg, Dreizler, Homeier, Reiners, Barman, \& Hauschildt}]{husser2013new}
Husser, T.-O., Wende-von Berg, S., Dreizler, S., {et~al.} 2013, Astronomy \& Astrophysics, 553, A6

\bibitem[{{Iyer} \& {Line}(2020)}]{2020ApJ...889...78I}
{Iyer}, A.~R., \& {Line}, M.~R. 2020, \apj, 889, 78, \dodoi{10.3847/1538-4357/ab612e}

\bibitem[{{Iyer} {et~al.}(2016){Iyer}, {Swain}, {Zellem}, {Line}, {Roudier}, {Rocha}, \& {Livingston}}]{2016ApJ...823..109I}
{Iyer}, A.~R., {Swain}, M.~R., {Zellem}, R.~T., {et~al.} 2016, \apj, 823, 109, \dodoi{10.3847/0004-637X/823/2/109}

\bibitem[{Jeffreys(1998)}]{jeffreys1998theory}
Jeffreys, H. 1998, The theory of probability (OuP Oxford)

\bibitem[{{Jiang} {et~al.}(2021){Jiang}, {Chen}, {Pall{\'e}}, {Murgas}, {Parviainen}, {Yan}, \& {Ma}}]{2021A&A...656A.114J}
{Jiang}, C., {Chen}, G., {Pall{\'e}}, E., {et~al.} 2021, \aap, 656, A114, \dodoi{10.1051/0004-6361/202141824}

\bibitem[{{Johnson} {et~al.}(2008){Johnson}, {Winn}, {Narita}, {Enya}, {Williams}, {Marcy}, {Sato}, {Ohta}, {Taruya}, {Suto}, {Turner}, {Bakos}, {Butler}, {Vogt}, {Aoki}, {Tamura}, {Yamada}, {Yoshii}, \& {Hidas}}]{johnson2008}
{Johnson}, J.~A., {Winn}, J.~N., {Narita}, N., {et~al.} 2008, \apj, 686, 649, \dodoi{10.1086/591078}

\bibitem[{{Jord{\'a}n} {et~al.}(2013){Jord{\'a}n}, {Espinoza}, {Rabus}, {Eyheramendy}, {Sing}, {D{\'e}sert}, {Bakos}, {Fortney}, {L{\'o}pez-Morales}, {Maxted}, {Triaud}, \& {Szentgyorgyi}}]{2013ApJ...778..184J}
{Jord{\'a}n}, A., {Espinoza}, N., {Rabus}, M., {et~al.} 2013, \apj, 778, 184, \dodoi{10.1088/0004-637X/778/2/184}

\bibitem[{{JWST Transiting Exoplanet Community Early Release Science Team} {et~al.}(2023){JWST Transiting Exoplanet Community Early Release Science Team}, {Ahrer}, {Alderson}, {Batalha}, {Batalha}, {Bean}, {Beatty}, {Bell}, {Benneke}, {Berta-Thompson}, {Carter}, {Crossfield}, {Espinoza}, {Feinstein}, {Fortney}, {Gibson}, {Goyal}, {Kempton}, {Kirk}, {Kreidberg}, {L{\'o}pez-Morales}, {Line}, {Lothringer}, {Moran}, {Mukherjee}, {Ohno}, {Parmentier}, {Piaulet}, {Rustamkulov}, {Schlawin}, {Sing}, {Stevenson}, {Wakeford}, {Allen}, {Birkmann}, {Brande}, {Crouzet}, {Cubillos}, {Damiano}, {D{\'e}sert}, {Gao}, {Harrington}, {Hu}, {Kendrew}, {Knutson}, {Lagage}, {Leconte}, {Lendl}, {MacDonald}, {May}, {Miguel}, {Molaverdikhani}, {Moses}, {Murray}, {Nehring}, {Nikolov}, {Petit dit de la Roche}, {Radica}, {Roy}, {Stassun}, {Taylor}, {Waalkes}, {Wachiraphan}, {Welbanks}, {Wheatley}, {Aggarwal}, {Alam}, {Banerjee}, {Barstow}, {Blecic}, {Casewell}, {Changeat}, {Chubb}, {Col{\'o}n}, {Coulombe}, {Daylan}, {de Val-Borro},
  {Decin}, {Dos Santos}, {Flagg}, {France}, {Fu}, {Garc{\'\i}a Mu{\~n}oz}, {Gizis}, {Glidden}, {Grant}, {Heng}, {Henning}, {Hong}, {Inglis}, {Iro}, {Kataria}, {Komacek}, {Krick}, {Lee}, {Lewis}, {Lillo-Box}, {Lustig-Yaeger}, {Mancini}, {Mandell}, {Mansfield}, {Marley}, {Mikal-Evans}, {Morello}, {Nixon}, {Ortiz Ceballos}, {Piette}, {Powell}, {Rackham}, {Ramos-Rosado}, {Rauscher}, {Redfield}, {Rogers}, {Roman}, {Roudier}, {Scarsdale}, {Shkolnik}, {Southworth}, {Spake}, {Steinrueck}, {Tan}, {Teske}, {Tremblin}, {Tsai}, {Tucker}, {Turner}, {Valenti}, {Venot}, {Waldmann}, {Wallack}, {Zhang}, \& {Zieba}}]{2023Natur.614..649J}
{JWST Transiting Exoplanet Community Early Release Science Team}, {Ahrer}, E.-M., {Alderson}, L., {et~al.} 2023, \nat, 614, 649, \dodoi{10.1038/s41586-022-05269-w}

\bibitem[{Kawashima \& Ikoma(2018)}]{kawashima2018theoretical}
Kawashima, Y., \& Ikoma, M. 2018, The Astrophysical Journal, 853, 7

\bibitem[{{Khalafinejad} {et~al.}(2018){Khalafinejad}, {Salz}, {Cubillos}, {Zhou}, {von Essen}, {Husser}, {Bayliss}, {L{\'o}pez-Morales}, {Dreizler}, {Schmitt}, \& {L{\"u}ftinger}}]{2018A&A...618A..98K}
{Khalafinejad}, S., {Salz}, M., {Cubillos}, P.~E., {et~al.} 2018, \aap, 618, A98, \dodoi{10.1051/0004-6361/201732029}

\bibitem[{{Khalafinejad} {et~al.}(2021){Khalafinejad}, {Molaverdikhani}, {Blecic}, {Mallonn}, {Nortmann}, {Caballero}, {Rahmati}, {Kaminski}, {Sadegi}, {Nagel}, {Carone}, {Amado}, {Azzaro}, {Bauer}, {Casasayas-Barris}, {Czesla}, {von Essen}, {Fossati}, {G{\"u}del}, {Henning}, {L{\'o}pez-Puertas}, {Lendl}, {L{\"u}ftinger}, {Montes}, {Oshagh}, {Pall{\'e}}, {Quirrenbach}, {Reffert}, {Reiners}, {Ribas}, {Stock}, {Yan}, {Zapatero Osorio}, \& {Zechmeister}}]{2021A&A...656A.142K}
{Khalafinejad}, S., {Molaverdikhani}, K., {Blecic}, J., {et~al.} 2021, \aap, 656, A142, \dodoi{10.1051/0004-6361/202141191}

\bibitem[{{Kirk} {et~al.}(2022){Kirk}, {Dos Santos}, {L{\'o}pez-Morales}, {Alam}, {Oklop{\v{c}}i{\'c}}, {MacLeod}, {Zeng}, \& {Zhou}}]{2022AJ....164...24K}
{Kirk}, J., {Dos Santos}, L.~A., {L{\'o}pez-Morales}, M., {et~al.} 2022, \aj, 164, 24, \dodoi{10.3847/1538-3881/ac722f}

\bibitem[{{Kirk} {et~al.}(2019){Kirk}, {L{\'o}pez-Morales}, {Wheatley}, {Weaver}, {Skillen}, {Louden}, {McCormac}, \& {Espinoza}}]{2019AJ....158..144K}
{Kirk}, J., {L{\'o}pez-Morales}, M., {Wheatley}, P.~J., {et~al.} 2019, \aj, 158, 144, \dodoi{10.3847/1538-3881/ab397d}

\bibitem[{{Kirk} {et~al.}(2017){Kirk}, {Wheatley}, {Louden}, {Doyle}, {Skillen}, {McCormac}, {Irwin}, \& {Karjalainen}}]{2017MNRAS.468.3907K}
{Kirk}, J., {Wheatley}, P.~J., {Louden}, T., {et~al.} 2017, \mnras, 468, 3907, \dodoi{10.1093/mnras/stx752}

\bibitem[{{Kirk} {et~al.}(2016){Kirk}, {Wheatley}, {Louden}, {Littlefair}, {Copperwheat}, {Armstrong}, {Marsh}, \& {Dhillon}}]{2016MNRAS.463.2922K}
---. 2016, \mnras, 463, 2922, \dodoi{10.1093/mnras/stw2205}

\bibitem[{{Kirk} {et~al.}(2018){Kirk}, {Wheatley}, {Louden}, {Skillen}, {King}, {McCormac}, \& {Irwin}}]{2018MNRAS.474..876K}
---. 2018, \mnras, 474, 876, \dodoi{10.1093/mnras/stx2826}

\bibitem[{{Knutson} {et~al.}(2014){Knutson}, {Benneke}, {Deming}, \& {Homeier}}]{2014Natur.505...66K}
{Knutson}, H.~A., {Benneke}, B., {Deming}, D., \& {Homeier}, D. 2014, \nat, 505, 66, \dodoi{10.1038/nature12887}

\bibitem[{{Knutson} {et~al.}(2007){Knutson}, {Charbonneau}, {Noyes}, {Brown}, \& {Gilliland}}]{2007ApJ...655..564K}
{Knutson}, H.~A., {Charbonneau}, D., {Noyes}, R.~W., {Brown}, T.~M., \& {Gilliland}, R.~L. 2007, \apj, 655, 564, \dodoi{10.1086/510111}

\bibitem[{{Knutson} {et~al.}(2010){Knutson}, {Howard}, \& {Isaacson}}]{Knutson2010}
{Knutson}, H.~A., {Howard}, A.~W., \& {Isaacson}, H. 2010, \apj, 720, 1569, \dodoi{10.1088/0004-637X/720/2/1569}

\bibitem[{{Knutson} {et~al.}(2011){Knutson}, {Madhusudhan}, {Cowan}, {Christiansen}, {Agol}, {Deming}, {D{\'e}sert}, {Charbonneau}, {Henry}, {Homeier}, {Langton}, {Laughlin}, \& {Seager}}]{2011ApJ...735...27K}
{Knutson}, H.~A., {Madhusudhan}, N., {Cowan}, N.~B., {et~al.} 2011, \apj, 735, 27, \dodoi{10.1088/0004-637X/735/1/27}

\bibitem[{Kokori {et~al.}(2021)Kokori, Tsiaras, Edwards, Rocchetto, Tinetti, Wünsche, Paschalis, Agnihotri, Bachschmidt, Bretton, \& et~al.}]{Kokori_2021}
Kokori, A., Tsiaras, A., Edwards, B., {et~al.} 2021, Experimental Astronomy, \dodoi{10.1007/s10686-020-09696-3}

\bibitem[{Kokori {et~al.}(2022)Kokori, Tsiaras, Edwards, Rocchetto, Tinetti, Bewersdorff, Jongen, Lekkas, Pantelidou, Poultourtzidis, {et~al.}}]{kokori2022exoclock}
---. 2022, The Astrophysical Journal Supplement Series, 258, 40

\bibitem[{Kokori {et~al.}(2023)Kokori, Tsiaras, Edwards, Jones, Pantelidou, Tinetti, Bewersdorff, Iliadou, Jongen, Lekkas, {et~al.}}]{kokori2023exoclock}
---. 2023, The Astrophysical Journal Supplement Series, 265, 4

\bibitem[{{Kosiarek} {et~al.}(2019){Kosiarek}, {Crossfield}, {Hardegree-Ullman}, {Livingston}, {Benneke}, {Henry}, {Howard}, {Berardo}, {Blunt}, {Fulton}, {Hirsch}, {Howard}, {Isaacson}, {Petigura}, {Sinukoff}, {Weiss}, {Bonfils}, {Dressing}, {Knutson}, {Schlieder}, {Werner}, {Gorjian}, {Krick}, {Morales}, {Astudillo-Defru}, {Almenara}, {Delfosse}, {Forveille}, {Lovis}, {Mayor}, {Murgas}, {Pepe}, {Santos}, {Udry}, {Corbett}, {Fors}, {Law}, {Ratzloff}, \& {del Ser}}]{2019AJ....157...97K}
{Kosiarek}, M.~R., {Crossfield}, I. J.~M., {Hardegree-Ullman}, K.~K., {et~al.} 2019, \aj, 157, 97, \dodoi{10.3847/1538-3881/aaf79c}

\bibitem[{Kramida(2013)}]{kramida2013critical}
Kramida, A. 2013, Fusion Science and Technology, 63, 313

\bibitem[{Kreidberg {et~al.}(2015)Kreidberg, Line, Bean, Stevenson, D{\'e}sert, Madhusudhan, Fortney, Barstow, Henry, Williamson, {et~al.}}]{kreidberg2015detection}
Kreidberg, L., Line, M.~R., Bean, J.~L., {et~al.} 2015, The Astrophysical Journal, 814, 66

\bibitem[{{Kreidberg} {et~al.}(2015){Kreidberg}, {Line}, {Bean}, {Stevenson}, {D{\'e}sert}, {Madhusudhan}, {Fortney}, {Barstow}, {Henry}, {Williamson}, \& {Showman}}]{2015ApJ...814...66K}
{Kreidberg}, L., {Line}, M.~R., {Bean}, J.~L., {et~al.} 2015, \apj, 814, 66, \dodoi{10.1088/0004-637X/814/1/66}

\bibitem[{{Lam} {et~al.}(2017{\natexlab{a}}){Lam}, {Faedi}, {Brown}, {Anderson}, {Delrez}, {Gillon}, {H{\'e}brard}, {Lendl}, {Mancini}, {Southworth}, {Smalley}, {Triaud}, {Turner}, {Hay}, {Armstrong}, {Barros}, {Bonomo}, {Bouchy}, {Boumis}, {Collier Cameron}, {Doyle}, {Hellier}, {Henning}, {Jehin}, {King}, {Kirk}, {Louden}, {Maxted}, {McCormac}, {Osborn}, {Palle}, {Pepe}, {Pollacco}, {Prieto-Arranz}, {Queloz}, {Rey}, {S{\'e}gransan}, {Udry}, {Walker}, {West}, \& {Wheatley}}]{lam2017}
{Lam}, K.~W.~F., {Faedi}, F., {Brown}, D.~J.~A., {et~al.} 2017{\natexlab{a}}, \aap, 599, A3, \dodoi{10.1051/0004-6361/201629403}

\bibitem[{{Lam} {et~al.}(2017{\natexlab{b}}){Lam}, {Faedi}, {Brown}, {Anderson}, {Delrez}, {Gillon}, {H{\'e}brard}, {Lendl}, {Mancini}, {Southworth}, {Smalley}, {Triaud}, {Turner}, {Hay}, {Armstrong}, {Barros}, {Bonomo}, {Bouchy}, {Boumis}, {Collier Cameron}, {Doyle}, {Hellier}, {Henning}, {Jehin}, {King}, {Kirk}, {Louden}, {Maxted}, {McCormac}, {Osborn}, {Palle}, {Pepe}, {Pollacco}, {Prieto-Arranz}, {Queloz}, {Rey}, {S{\'e}gransan}, {Udry}, {Walker}, {West}, \& {Wheatley}}]{2017A&A...599A...3L}
---. 2017{\natexlab{b}}, \aap, 599, A3, \dodoi{10.1051/0004-6361/201629403}

\bibitem[{{Lamp{\'o}n} {et~al.}(2023){Lamp{\'o}n}, {L{\'o}pez-Puertas}, {Sanz-Forcada}, {Czesla}, {Nortmann}, {Casasayas-Barris}, {Orell-Miquel}, {S{\'a}nchez-L{\'o}pez}, {Danielski}, {Pall{\'e}}, {Molaverdikhani}, {Henning}, {Caballero}, {Amado}, {Quirrenbach}, {Reiners}, \& {Ribas}}]{2023A&A...673A.140L}
{Lamp{\'o}n}, M., {L{\'o}pez-Puertas}, M., {Sanz-Forcada}, J., {et~al.} 2023, \aap, 673, A140, \dodoi{10.1051/0004-6361/202245649}

\bibitem[{{Lanotte} {et~al.}(2014){Lanotte}, {Gillon}, {Demory}, {Fortney}, {Astudillo}, {Bonfils}, {Magain}, {Delfosse}, {Forveille}, {Lovis}, {Mayor}, {Neves}, {Pepe}, {Queloz}, {Santos}, \& {Udry}}]{Lanotte2014}
{Lanotte}, A.~A., {Gillon}, M., {Demory}, B.~O., {et~al.} 2014, \aap, 572, A73, \dodoi{10.1051/0004-6361/201424373}

\bibitem[{{Lavie} {et~al.}(2017){Lavie}, {Ehrenreich}, {Bourrier}, {Lecavelier des Etangs}, {Vidal-Madjar}, {Delfosse}, {Gracia Berna}, {Heng}, {Thomas}, {Udry}, \& {Wheatley}}]{2017A&A...605L...7L}
{Lavie}, B., {Ehrenreich}, D., {Bourrier}, V., {et~al.} 2017, \aap, 605, L7, \dodoi{10.1051/0004-6361/201731340}

\bibitem[{{Lecavelier des Etangs} {et~al.}(2013){Lecavelier des Etangs}, {Sirothia}, {Gopal-Krishna}, \& {Zarka}}]{2013A&A...552A..65L}
{Lecavelier des Etangs}, A., {Sirothia}, S.~K., {Gopal-Krishna}, \& {Zarka}, P. 2013, \aap, 552, A65, \dodoi{10.1051/0004-6361/201219789}

\bibitem[{{Lecavelier Des Etangs} {et~al.}(2008){Lecavelier Des Etangs}, {Vidal-Madjar}, {D{\'e}sert}, \& {Sing}}]{2008A&A...485..865L}
{Lecavelier Des Etangs}, A., {Vidal-Madjar}, A., {D{\'e}sert}, J.~M., \& {Sing}, D. 2008, \aap, 485, 865, \dodoi{10.1051/0004-6361:200809704}

\bibitem[{Lee {et~al.}(2013)Lee, Heng, \& Irwin}]{lee2013atmospheric}
Lee, J.-M., Heng, K., \& Irwin, P.~G. 2013, The Astrophysical Journal, 778, 97

\bibitem[{L{\'e}pine \& Shara(2005)}]{lepine2005catalog}
L{\'e}pine, S., \& Shara, M.~M. 2005, The Astronomical Journal, 129, 1483

\bibitem[{{Libby-Roberts} {et~al.}(2022){Libby-Roberts}, {Berta-Thompson}, {Diamond-Lowe}, {Gully-Santiago}, {Irwin}, {Kempton}, {Rackham}, {Charbonneau}, {D{\'e}sert}, {Dittmann}, {Hofmann}, {Morley}, \& {Newton}}]{2022AJ....164...59L}
{Libby-Roberts}, J.~E., {Berta-Thompson}, Z.~K., {Diamond-Lowe}, H., {et~al.} 2022, \aj, 164, 59, \dodoi{10.3847/1538-3881/ac75de}

\bibitem[{{Line} {et~al.}(2013){Line}, {Knutson}, {Deming}, {Wilkins}, \& {Desert}}]{2013ApJ...778..183L}
{Line}, M.~R., {Knutson}, H., {Deming}, D., {Wilkins}, A., \& {Desert}, J.-M. 2013, \apj, 778, 183, \dodoi{10.1088/0004-637X/778/2/183}

\bibitem[{{Linsky} {et~al.}(2010){Linsky}, {Yang}, {France}, {Froning}, {Green}, {Stocke}, \& {Osterman}}]{2010ApJ...717.1291L}
{Linsky}, J.~L., {Yang}, H., {France}, K., {et~al.} 2010, \apj, 717, 1291, \dodoi{10.1088/0004-637X/717/2/1291}

\bibitem[{{Lothringer} {et~al.}(2018){Lothringer}, {Benneke}, {Crossfield}, {Henry}, {Morley}, {Dragomir}, {Barman}, {Knutson}, {Kempton}, {Fortney}, {McCullough}, \& {Howard}}]{2018AJ....155...66L}
{Lothringer}, J.~D., {Benneke}, B., {Crossfield}, I. J.~M., {et~al.} 2018, \aj, 155, 66, \dodoi{10.3847/1538-3881/aaa008}

\bibitem[{{Louden} {et~al.}(2017){Louden}, {Wheatley}, {Irwin}, {Kirk}, \& {Skillen}}]{2017MNRAS.470..742L}
{Louden}, T., {Wheatley}, P.~J., {Irwin}, P. G.~J., {Kirk}, J., \& {Skillen}, I. 2017, \mnras, 470, 742, \dodoi{10.1093/mnras/stx984}

\bibitem[{{Loyd} {et~al.}(2023){Loyd}, {Schneider}, {Jackman}, {France}, {Shkolnik}, {Arulanantham}, {Cauley}, {Llama}, \& {Schneider}}]{2023AJ....165..146L}
{Loyd}, R.~O.~P., {Schneider}, P.~C., {Jackman}, J. A.~G., {et~al.} 2023, \aj, 165, 146, \dodoi{10.3847/1538-3881/acbbc8}

\bibitem[{{Luque} \& {Pall{\'e}}(2022)}]{2022Sci...377.1211L}
{Luque}, R., \& {Pall{\'e}}, E. 2022, Science, 377, 1211, \dodoi{10.1126/science.abl7164}

\bibitem[{{Luque} {et~al.}(2020){Luque}, {Casasayas-Barris}, {Parviainen}, {Chen}, {Pall{\'e}}, {Livingston}, {B{\'e}jar}, {Crouzet}, {Esparza-Borges}, {Fukui}, {Hidalgo}, {Kawashima}, {Kawauchi}, {Klagyivik}, {Kurita}, {Kusakabe}, {de Leon}, {Madrigal-Aguado}, {Monta{\~n}{\'e}s-Rodr{\'\i}guez}, {Mori}, {Murgas}, {Narita}, {Nishiumi}, {Nowak}, {Oshagh}, {S{\'a}nchez-Benavente}, {Stangret}, {Tamura}, {Terada}, \& {Watanabe}}]{2020A&A...642A..50L}
{Luque}, R., {Casasayas-Barris}, N., {Parviainen}, H., {et~al.} 2020, \aap, 642, A50, \dodoi{10.1051/0004-6361/202038703}

\bibitem[{{Ma} {et~al.}(2023){Ma}, {Ito}, {Al-Refaie}, {Changeat}, {Edwards}, \& {Tinetti}}]{2023ApJ...957..104M}
{Ma}, S., {Ito}, Y., {Al-Refaie}, A.~F., {et~al.} 2023, \apj, 957, 104, \dodoi{10.3847/1538-4357/acf8ca}

\bibitem[{{MacDonald} \& {Batalha}(2023)}]{2023RNAAS...7...54M}
{MacDonald}, R.~J., \& {Batalha}, N.~E. 2023, Research Notes of the American Astronomical Society, 7, 54, \dodoi{10.3847/2515-5172/acc46a}

\bibitem[{{MacDonald} \& {Madhusudhan}(2017)}]{macdonald2017}
{MacDonald}, R.~J., \& {Madhusudhan}, N. 2017, \mnras, 469, 1979, \dodoi{10.1093/mnras/stx804}

\bibitem[{{MacDonald} \& {Madhusudhan}(2019)}]{2019MNRAS.486.1292M}
---. 2019, \mnras, 486, 1292, \dodoi{10.1093/mnras/stz789}

\bibitem[{{Maciejewski} {et~al.}(2016){Maciejewski}, {Dimitrov}, {Mancini}, {Southworth}, {Ciceri}, {D'Ago}, {Bruni}, {Raetz}, {Nowak}, {Ohlert}, {Puchalski}, {Saral}, {Derman}, {Petrucci}, {Jofre}, {Seeliger}, \& {Henning}}]{2016AcA....66...55M}
{Maciejewski}, G., {Dimitrov}, D., {Mancini}, L., {et~al.} 2016, \actaa, 66, 55, \dodoi{10.48550/arXiv.1603.03268}

\bibitem[{{Magrini} {et~al.}(2022){Magrini}, {Danielski}, {Bossini}, {Rainer}, {Turrini}, {Benatti}, {Brucalassi}, {Tsantaki}, {Delgado Mena}, {Sanna}, {Biazzo}, {Campante}, {Van der Swaelmen}, {Sousa}, {He{\l}miniak}, {Neitzel}, {Adibekyan}, {Bruno}, \& {Casali}}]{2022A&A...663A.161M}
{Magrini}, L., {Danielski}, C., {Bossini}, D., {et~al.} 2022, \aap, 663, A161, \dodoi{10.1051/0004-6361/202243405}

\bibitem[{{Mai} \& {Line}(2019)}]{2019ApJ...883..144M}
{Mai}, C., \& {Line}, M.~R. 2019, \apj, 883, 144, \dodoi{10.3847/1538-4357/ab3e6d}

\bibitem[{{Mallonn} {et~al.}(2015){Mallonn}, {Nascimbeni}, {Weingrill}, {von Essen}, {Strassmeier}, {Piotto}, {Pagano}, {Scandariato}, {Csizmadia}, {Herrero}, {Sada}, {Dhillon}, {Marsh}, {K{\"u}nstler}, {Bernt}, \& {Granzer}}]{2015A&A...583A.138M}
{Mallonn}, M., {Nascimbeni}, V., {Weingrill}, J., {et~al.} 2015, \aap, 583, A138, \dodoi{10.1051/0004-6361/201425395}

\bibitem[{{Mancini} {et~al.}(2014){Mancini}, {Southworth}, {Ciceri}, {Dominik}, {Henning}, {J{\o}rgensen}, {Lanza}, {Rabus}, {Snodgrass}, {Vilela}, {Alsubai}, {Bozza}, {Bramich}, {Calchi Novati}, {D'Ago}, {Figuera Jaimes}, {Galianni}, {Gu}, {Harps{\o}e}, {Hinse}, {Hundertmark}, {Juncher}, {Kains}, {Korhonen}, {Popovas}, {Rahvar}, {Skottfelt}, {Street}, {Surdej}, {Tsapras}, {Wang}, \& {Wertz}}]{2014A&A...562A.126M}
{Mancini}, L., {Southworth}, J., {Ciceri}, S., {et~al.} 2014, \aap, 562, A126, \dodoi{10.1051/0004-6361/201323265}

\bibitem[{{Mancini} {et~al.}(2018){Mancini}, {Esposito}, {Covino}, {Southworth}, {Biazzo}, {Bruni}, {Ciceri}, {Evans}, {Lanza}, {Poretti}, {Sarkis}, {Smith}, {Brogi}, {Affer}, {Benatti}, {Bignamini}, {Boccato}, {Bonomo}, {Borsa}, {Carleo}, {Claudi}, {Cosentino}, {Damasso}, {Desidera}, {Giacobbe}, {Gonz{\'a}lez-{\'A}lvarez}, {Gratton}, {Harutyunyan}, {Leto}, {Maggio}, {Malavolta}, {Maldonado}, {Martinez-Fiorenzano}, {Masiero}, {Micela}, {Molinari}, {Nascimbeni}, {Pagano}, {Pedani}, {Piotto}, {Rainer}, {Scandariato}, {Smareglia}, {Sozzetti}, {Andreuzzi}, \& {Henning}}]{Mancini2018}
{Mancini}, L., {Esposito}, M., {Covino}, E., {et~al.} 2018, \aap, 613, A41, \dodoi{10.1051/0004-6361/201732234}

\bibitem[{{Mancini} {et~al.}(2019){Mancini}, {Southworth}, {Molli{\`e}re}, {Tregloan-Reed}, {Juvan}, {Chen}, {Sarkis}, {Bruni}, {Ciceri}, {Andersen}, {Bozza}, {Bramich}, {Burgdorf}, {D'Ago}, {Dominik}, {Evans}, {Figuera Jaimes}, {Fossati}, {Henning}, {Hinse}, {Hundertmark}, {J{\o}rgensen}, {Kerins}, {Korhonen}, {K{\"u}ffmeier}, {Longa}, {Peixinho}, {Popovas}, {Rabus}, {Rahvar}, {Skottfelt}, {Snodgrass}, {Tronsgaard}, {Wang}, \& {Wertz}}]{2019MNRAS.485.5168M}
{Mancini}, L., {Southworth}, J., {Molli{\`e}re}, P., {et~al.} 2019, \mnras, 485, 5168, \dodoi{10.1093/mnras/stz661}

\bibitem[{{Mancini} {et~al.}(2022){Mancini}, {Esposito}, {Covino}, {Southworth}, {Poretti}, {Andreuzzi}, {Barbato}, {Biazzo}, {Borsato}, {Bruni}, {Damasso}, {Di Fabrizio}, {Evans}, {Granata}, {Lanza}, {Naponiello}, {Nascimbeni}, {Pinamonti}, {Sozzetti}, {Tregloan-Reed}, {Basilicata}, {Bignamini}, {Bonomo}, {Claudi}, {Cosentino}, {Desidera}, {Fiorenzano}, {Giacobbe}, {Harutyunyan}, {Henning}, {Knapic}, {Maggio}, {Micela}, {Molinari}, {Pagano}, {Pedani}, \& {Piotto}}]{2022A&A...664A.162M}
{Mancini}, L., {Esposito}, M., {Covino}, E., {et~al.} 2022, \aap, 664, A162, \dodoi{10.1051/0004-6361/202243742}

\bibitem[{Mandel \& Agol(2002)}]{mandel2002analytic}
Mandel, K., \& Agol, E. 2002, The Astrophysical Journal, 580, L171

\bibitem[{{Mansfield} {et~al.}(2018){Mansfield}, {Bean}, {Oklop{\v{c}}i{\'c}}, {Kreidberg}, {D{\'e}sert}, {Kempton}, {Line}, {Fortney}, {Henry}, {Mallonn}, {Stevenson}, {Dragomir}, {Allart}, \& {Bourrier}}]{2018ApJ...868L..34M}
{Mansfield}, M., {Bean}, J.~L., {Oklop{\v{c}}i{\'c}}, A., {et~al.} 2018, \apjl, 868, L34, \dodoi{10.3847/2041-8213/aaf166}

\bibitem[{{May} {et~al.}(2018){May}, {Zhao}, {Haidar}, {Rauscher}, \& {Monnier}}]{2018AJ....156..122M}
{May}, E.~M., {Zhao}, M., {Haidar}, M., {Rauscher}, E., \& {Monnier}, J.~D. 2018, \aj, 156, 122, \dodoi{10.3847/1538-3881/aad4a8}

\bibitem[{{Mayor} {et~al.}(2011){Mayor}, {Marmier}, {Lovis}, {Udry}, {S{\'e}gransan}, {Pepe}, {Benz}, {Bertaux}, {Bouchy}, {Dumusque}, {Lo Curto}, {Mordasini}, {Queloz}, \& {Santos}}]{2011arXiv1109.2497M}
{Mayor}, M., {Marmier}, M., {Lovis}, C., {et~al.} 2011, arXiv e-prints, arXiv:1109.2497, \dodoi{10.48550/arXiv.1109.2497}

\bibitem[{{McGruder} {et~al.}(2020){McGruder}, {L{\'o}pez-Morales}, {Espinoza}, {Rackham}, {Apai}, {Jord{\'a}n}, {Osip}, {Alam}, {Bixel}, {Fortney}, {Henry}, {Kirk}, {Lewis}, {Rodler}, \& {Weaver}}]{2020AJ....160..230M}
{McGruder}, C.~D., {L{\'o}pez-Morales}, M., {Espinoza}, N., {et~al.} 2020, \aj, 160, 230, \dodoi{10.3847/1538-3881/abb806}

\bibitem[{McKemmish {et~al.}(2019)McKemmish, Masseron, Hoeijmakers, P{\'e}rez-Mesa, Grimm, Yurchenko, \& Tennyson}]{mckemmish2019exomol}
McKemmish, L.~K., Masseron, T., Hoeijmakers, H.~J., {et~al.} 2019, Monthly Notices of the Royal Astronomical Society, 488, 2836

\bibitem[{McKemmish {et~al.}(2016)McKemmish, Yurchenko, \& Tennyson}]{mckemmish2016exomol}
McKemmish, L.~K., Yurchenko, S.~N., \& Tennyson, J. 2016, Monthly Notices of the Royal Astronomical Society, 463, 771

\bibitem[{{Merritt} {et~al.}(2021){Merritt}, {Gibson}, {Nugroho}, {de Mooij}, {Hooton}, {Lothringer}, {Matthews}, {Mikal-Evans}, {Nikolov}, {Sing}, \& {Watson}}]{2021MNRAS.506.3853M}
{Merritt}, S.~R., {Gibson}, N.~P., {Nugroho}, S.~K., {et~al.} 2021, \mnras, 506, 3853, \dodoi{10.1093/mnras/stab1878}

\bibitem[{{Mittag} {et~al.}(2013){Mittag}, {Schmitt}, \& {Schr{\"o}der}}]{Mittag2013}
{Mittag}, M., {Schmitt}, J.~H.~M.~M., \& {Schr{\"o}der}, K.~P. 2013, \aap, 549, A117, \dodoi{10.1051/0004-6361/201219868}

\bibitem[{{Montalto} {et~al.}(2015){Montalto}, {Iro}, {Santos}, {Desidera}, {Martins}, {Figueira}, \& {Alonso}}]{2015ApJ...811...55M}
{Montalto}, M., {Iro}, N., {Santos}, N.~C., {et~al.} 2015, \apj, 811, 55, \dodoi{10.1088/0004-637X/811/1/55}

\bibitem[{Morello {et~al.}(2020)Morello, Claret, Martin-Lagarde, Cossou, Tsiaras, \& Lagage}]{morello2020exotethys}
Morello, G., Claret, A., Martin-Lagarde, M., {et~al.} 2020, The Astronomical Journal, 159, 75

\bibitem[{{Morris} {et~al.}(2017{\natexlab{a}}){Morris}, {Hebb}, {Davenport}, {Rohn}, \& {Hawley}}]{2017ApJ...846...99M}
{Morris}, B.~M., {Hebb}, L., {Davenport}, J. R.~A., {Rohn}, G., \& {Hawley}, S.~L. 2017{\natexlab{a}}, \apj, 846, 99, \dodoi{10.3847/1538-4357/aa8555}

\bibitem[{{Morris} {et~al.}(2017{\natexlab{b}}){Morris}, {Hawley}, {Hebb}, {Sakari}, {Davenport}, {Isaacson}, {Howard}, {Montet}, \& {Agol}}]{2017ApJ...848...58M}
{Morris}, B.~M., {Hawley}, S.~L., {Hebb}, L., {et~al.} 2017{\natexlab{b}}, \apj, 848, 58, \dodoi{10.3847/1538-4357/aa8cca}

\bibitem[{{Mugnai} {et~al.}(2024){Mugnai}, {Swain}, {Estrela}, \& {Roudier}}]{2024MNRAS.531...35M}
{Mugnai}, L.~V., {Swain}, M.~R., {Estrela}, R., \& {Roudier}, G.~M. 2024, \mnras, 531, 35, \dodoi{10.1093/mnras/stae1073}

\bibitem[{{Murgas} {et~al.}(2020){Murgas}, {Chen}, {Nortmann}, {Palle}, \& {Nowak}}]{2020A&A...641A.158M}
{Murgas}, F., {Chen}, G., {Nortmann}, L., {Palle}, E., \& {Nowak}, G. 2020, \aap, 641, A158, \dodoi{10.1051/0004-6361/202038161}

\bibitem[{{Murgas} {et~al.}(2019){Murgas}, {Chen}, {Pall{\'e}}, {Nortmann}, \& {Nowak}}]{2019A&A...622A.172M}
{Murgas}, F., {Chen}, G., {Pall{\'e}}, E., {Nortmann}, L., \& {Nowak}, G. 2019, \aap, 622, A172, \dodoi{10.1051/0004-6361/201834063}

\bibitem[{Narita {et~al.}(2013)Narita, Fukui, Ikoma, Hori, Kurosaki, Kawashima, Nagayama, Onitsuka, Sukom, Nakajima, {et~al.}}]{narita2013multi}
Narita, N., Fukui, A., Ikoma, M., {et~al.} 2013, The Astrophysical Journal, 773, 144

\bibitem[{{Nielsen} {et~al.}(2019){Nielsen}, {De Rosa}, {Macintosh}, {Wang}, {Ruffio}, {Chiang}, {Marley}, {Saumon}, {Savransky}, {Ammons}, {Bailey}, {Barman}, {Blain}, {Bulger}, {Burrows}, {Chilcote}, {Cotten}, {Czekala}, {Doyon}, {Duch{\^e}ne}, {Esposito}, {Fabrycky}, {Fitzgerald}, {Follette}, {Fortney}, {Gerard}, {Goodsell}, {Graham}, {Greenbaum}, {Hibon}, {Hinkley}, {Hirsch}, {Hom}, {Hung}, {Dawson}, {Ingraham}, {Kalas}, {Konopacky}, {Larkin}, {Lee}, {Lin}, {Maire}, {Marchis}, {Marois}, {Metchev}, {Millar-Blanchaer}, {Morzinski}, {Oppenheimer}, {Palmer}, {Patience}, {Perrin}, {Poyneer}, {Pueyo}, {Rafikov}, {Rajan}, {Rameau}, {Rantakyr{\"o}}, {Ren}, {Schneider}, {Sivaramakrishnan}, {Song}, {Soummer}, {Tallis}, {Thomas}, {Ward-Duong}, \& {Wolff}}]{2019AJ....158...13N}
{Nielsen}, E.~L., {De Rosa}, R.~J., {Macintosh}, B., {et~al.} 2019, \aj, 158, 13, \dodoi{10.3847/1538-3881/ab16e9}

\bibitem[{{Nikolov} {et~al.}(2016){Nikolov}, {Sing}, {Gibson}, {Fortney}, {Evans}, {Barstow}, {Kataria}, \& {Wilson}}]{2016ApJ...832..191N}
{Nikolov}, N., {Sing}, D.~K., {Gibson}, N.~P., {et~al.} 2016, \apj, 832, 191, \dodoi{10.3847/0004-637X/832/2/191}

\bibitem[{{Nikolov} {et~al.}(2014){Nikolov}, {Sing}, {Pont}, {Burrows}, {Fortney}, {Ballester}, {Evans}, {Huitson}, {Wakeford}, {Wilson}, {Aigrain}, {Deming}, {Gibson}, {Henry}, {Knutson}, {Lecavelier des Etangs}, {Showman}, {Vidal-Madjar}, \& {Zahnle}}]{2014MNRAS.437...46N}
{Nikolov}, N., {Sing}, D.~K., {Pont}, F., {et~al.} 2014, \mnras, 437, 46, \dodoi{10.1093/mnras/stt1859}

\bibitem[{{Nikolov} {et~al.}(2015){Nikolov}, {Sing}, {Burrows}, {Fortney}, {Henry}, {Pont}, {Ballester}, {Aigrain}, {Wilson}, {Huitson}, {Gibson}, {D{\'e}sert}, {Lecavelier Des Etangs}, {Showman}, {Vidal-Madjar}, {Wakeford}, \& {Zahnle}}]{2015MNRAS.447..463N}
{Nikolov}, N., {Sing}, D.~K., {Burrows}, A.~S., {et~al.} 2015, \mnras, 447, 463, \dodoi{10.1093/mnras/stu2433}

\bibitem[{{Ninan} {et~al.}(2020){Ninan}, {Stefansson}, {Mahadevan}, {Bender}, {Robertson}, {Ramsey}, {Terrien}, {Wright}, {Diddams}, {Kanodia}, {Cochran}, {Endl}, {Ford}, {Fredrick}, {Halverson}, {Hearty}, {Jennings}, {Kaplan}, {Lubar}, {Metcalf}, {Monson}, {Nitroy}, {Roy}, \& {Schwab}}]{2020ApJ...894...97N}
{Ninan}, J.~P., {Stefansson}, G., {Mahadevan}, S., {et~al.} 2020, \apj, 894, 97, \dodoi{10.3847/1538-4357/ab8559}

\bibitem[{{Norris} {et~al.}(2023){Norris}, {Unruh}, {Witzke}, {Solanki}, {Krivova}, {Shapiro}, {Yeo}, {Cameron}, \& {Beeck}}]{norris2023}
{Norris}, C.~M., {Unruh}, Y.~C., {Witzke}, V., {et~al.} 2023, \mnras, 524, 1139, \dodoi{10.1093/mnras/stad1738}

\bibitem[{{Nortmann} {et~al.}(2018){Nortmann}, {Pall{\'e}}, {Salz}, {Sanz-Forcada}, {Nagel}, {Alonso-Floriano}, {Czesla}, {Yan}, {Chen}, {Snellen}, {Zechmeister}, {Schmitt}, {L{\'o}pez-Puertas}, {Casasayas-Barris}, {Bauer}, {Amado}, {Caballero}, {Dreizler}, {Henning}, {Lamp{\'o}n}, {Montes}, {Molaverdikhani}, {Quirrenbach}, {Reiners}, {Ribas}, {S{\'a}nchez-L{\'o}pez}, {Schneider}, \& {Zapatero Osorio}}]{2018Sci...362.1388N}
{Nortmann}, L., {Pall{\'e}}, E., {Salz}, M., {et~al.} 2018, Science, 362, 1388, \dodoi{10.1126/science.aat5348}

\bibitem[{{Noyes} {et~al.}(1984){Noyes}, {Hartmann}, {Baliunas}, {Duncan}, \& {Vaughan}}]{Noyes1984}
{Noyes}, R.~W., {Hartmann}, L.~W., {Baliunas}, S.~L., {Duncan}, D.~K., \& {Vaughan}, A.~H. 1984, \apj, 279, 763, \dodoi{10.1086/161945}

\bibitem[{Oliphant(2006)}]{oliphant2006guide}
Oliphant, T.~E. 2006, A guide to NumPy, Vol.~1 (Trelgol Publishing USA)

\bibitem[{{Ouyang} {et~al.}(2023){Ouyang}, {Wang}, {Zhai}, {Chen}, {Rojo}, {Liu}, {Zhao}, {Huang}, \& {Zhao}}]{2023MNRAS.521.5860O}
{Ouyang}, Q., {Wang}, W., {Zhai}, M., {et~al.} 2023, \mnras, 521, 5860, \dodoi{10.1093/mnras/stad893}

\bibitem[{{Palle} {et~al.}(2017){Palle}, {Chen}, {Prieto-Arranz}, {Nowak}, {Murgas}, {Nortmann}, {Pollacco}, {Lam}, {Montanes-Rodriguez}, {Parviainen}, \& {Casasayas-Barris}}]{2017A&A...602L..15P}
{Palle}, E., {Chen}, G., {Prieto-Arranz}, J., {et~al.} 2017, \aap, 602, L15, \dodoi{10.1051/0004-6361/201731018}

\bibitem[{{Panja} {et~al.}(2020){Panja}, {Cameron}, \& {Solanki}}]{panja2020}
{Panja}, M., {Cameron}, R., \& {Solanki}, S.~K. 2020, \apj, 893, 113, \dodoi{10.3847/1538-4357/ab8230}

\bibitem[{{Paragas} {et~al.}(2021){Paragas}, {Vissapragada}, {Knutson}, {Oklop{\v{c}}i{\'c}}, {Chachan}, {Greklek-McKeon}, {Dai}, {Tinyanont}, \& {Vasisht}}]{2021ApJ...909L..10P}
{Paragas}, K., {Vissapragada}, S., {Knutson}, H.~A., {et~al.} 2021, \apjl, 909, L10, \dodoi{10.3847/2041-8213/abe706}

\bibitem[{{Parviainen} {et~al.}(2018){Parviainen}, {Pall{\'e}}, {Chen}, {Nortmann}, {Murgas}, {Nowak}, {Aigrain}, {Booth}, {Abazorius}, \& {Iro}}]{2018A&A...609A..33P}
{Parviainen}, H., {Pall{\'e}}, E., {Chen}, G., {et~al.} 2018, \aap, 609, A33, \dodoi{10.1051/0004-6361/201731113}

\bibitem[{Patrascu {et~al.}(2015)Patrascu, Yurchenko, \& Tennyson}]{patrascu2015exomol}
Patrascu, A.~T., Yurchenko, S.~N., \& Tennyson, J. 2015, Monthly Notices of the Royal Astronomical Society, 449, 3613

\bibitem[{Peek {et~al.}(2019)Peek, Desai, White, D'Abrusco, Mazzarella, Grant, Novacescu, Scire, \& Winkelman}]{peek2019robust}
Peek, J.~E., Desai, V., White, R.~L., {et~al.} 2019, arXiv preprint arXiv:1907.06234

\bibitem[{{Pinhas} {et~al.}(2018){Pinhas}, {Rackham}, {Madhusudhan}, \& {Apai}}]{2018MNRAS.480.5314P}
{Pinhas}, A., {Rackham}, B.~V., {Madhusudhan}, N., \& {Apai}, D. 2018, \mnras, 480, 5314, \dodoi{10.1093/mnras/sty2209}

\bibitem[{Polyansky {et~al.}(2018)Polyansky, Kyuberis, Zobov, Tennyson, Yurchenko, \& Lodi}]{10.1093/mnras/sty1877}
Polyansky, O.~L., Kyuberis, A.~A., Zobov, N.~F., {et~al.} 2018, Monthly Notices of the Royal Astronomical Society, 480, 2597, \dodoi{10.1093/mnras/sty1877}

\bibitem[{{Pont} {et~al.}(2009){Pont}, {Gilliland}, {Knutson}, {Holman}, \& {Charbonneau}}]{2009MNRAS.393L...6P}
{Pont}, F., {Gilliland}, R.~L., {Knutson}, H., {Holman}, M., \& {Charbonneau}, D. 2009, \mnras, 393, L6, \dodoi{10.1111/j.1745-3933.2008.00582.x}

\bibitem[{{Powell} {et~al.}(2024){Powell}, {Feinstein}, {Lee}, {Zhang}, {Tsai}, {Taylor}, {Kirk}, {Bell}, {Barstow}, {Gao}, {Bean}, {Blecic}, {Chubb}, {Crossfield}, {Jordan}, {Kitzmann}, {Moran}, {Morello}, {Moses}, {Welbanks}, {Yang}, {Zhang}, {Ahrer}, {Bello-Arufe}, {Brande}, {Casewell}, {Crouzet}, {Cubillos}, {Demory}, {Dyrek}, {Flagg}, {Hu}, {Inglis}, {Jones}, {Kreidberg}, {L{\'o}pez-Morales}, {Lagage}, {Meier Vald{\'e}s}, {Miguel}, {Parmentier}, {Piette}, {Rackham}, {Radica}, {Redfield}, {Stevenson}, {Wakeford}, {Aggarwal}, {Alam}, {Batalha}, {Batalha}, {Benneke}, {Berta-Thompson}, {Brady}, {Caceres}, {Carter}, {D{\'e}sert}, {Harrington}, {Iro}, {Line}, {Lothringer}, {MacDonald}, {Mancini}, {Molaverdikhani}, {Mukherjee}, {Nixon}, {Oza}, {Palle}, {Rustamkulov}, {Sing}, {Steinrueck}, {Venot}, {Wheatley}, \& {Yurchenko}}]{2024Natur.626..979P}
{Powell}, D., {Feinstein}, A.~D., {Lee}, E. K.~H., {et~al.} 2024, \nat, 626, 979, \dodoi{10.1038/s41586-024-07040-9}

\bibitem[{Price-Whelan {et~al.}(2018)Price-Whelan, Sip{\H{o}}cz, G{\"u}nther, Lim, Crawford, Conseil, Shupe, Craig, Dencheva, Ginsburg, {et~al.}}]{price2018astropy}
Price-Whelan, A.~M., Sip{\H{o}}cz, B., G{\"u}nther, H., {et~al.} 2018, The Astronomical Journal, 156, 123

\bibitem[{Rackham {et~al.}(2018)Rackham, Apai, \& Giampapa}]{rackham2018transit}
Rackham, B.~V., Apai, D., \& Giampapa, M.~S. 2018, The Astrophysical Journal, 853, 122

\bibitem[{Rackham {et~al.}(2019)Rackham, Apai, \& Giampapa}]{rackham2019transit}
---. 2019, The Astronomical Journal, 157, 96

\bibitem[{{Rathcke} {et~al.}(2023{\natexlab{a}}){Rathcke}, {Buchhave}, {Mendon{\c{c}}a}, {Sing}, {L{\'o}pez-Morales}, {Alam}, {Henry}, {Nikolov}, {Garc{\'\i}a Mu{\~n}oz}, {Mikal-Evans}, {Wakeford}, {Dos Santos}, \& {Rajpaul}}]{2023MNRAS.522..582R}
{Rathcke}, A.~D., {Buchhave}, L.~A., {Mendon{\c{c}}a}, J.~M., {et~al.} 2023{\natexlab{a}}, \mnras, 522, 582, \dodoi{10.1093/mnras/stad1010}

\bibitem[{{Rathcke} {et~al.}(2023{\natexlab{b}}){Rathcke}, {Buchhave}, {Mendon{\c{c}}a}, {Sing}, {L{\'o}pez-Morales}, {Alam}, {Henry}, {Nikolov}, {Garc{\'\i}a Mu{\~n}oz}, {Mikal-Evans}, {Wakeford}, {Dos Santos}, \& {Rajpaul}}]{Rathcke2023}
---. 2023{\natexlab{b}}, \mnras, 522, 582, \dodoi{10.1093/mnras/stad1010}

\bibitem[{{Ribas} {et~al.}(2008){Ribas}, {Font-Ribera}, \& {Beaulieu}}]{2008ApJ...677L..59R}
{Ribas}, I., {Font-Ribera}, A., \& {Beaulieu}, J.-P. 2008, \apjl, 677, L59, \dodoi{10.1086/587961}

\bibitem[{Robinson \& Catling(2014)}]{robinson2014common}
Robinson, T.~D., \& Catling, D.~C. 2014, Nature Geoscience, 7, 12

\bibitem[{Rothman \& Gordon(2014)}]{rothman2014status}
Rothman, L.~S., \& Gordon, I.~E. 2014, in 13th International HITRAN Conference, 49

\bibitem[{Rothman {et~al.}(1987)Rothman, Gamache, Goldman, Brown, Toth, Pickett, Poynter, Flaud, Camy-Peyret, Barbe, {et~al.}}]{rothman1987hitran}
Rothman, L.~S., Gamache, R.~R., Goldman, A., {et~al.} 1987, Applied optics, 26, 4058

\bibitem[{Rothman {et~al.}(2010)Rothman, Gordon, Barber, Dothe, Gamache, Goldman, Perevalov, Tashkun, \& Tennyson}]{rothman2010hitemp}
Rothman, L.~S., Gordon, I., Barber, R., {et~al.} 2010, Journal of Quantitative Spectroscopy and Radiative Transfer, 111, 2139

\bibitem[{{Roy} {et~al.}(2023){Roy}, {Benneke}, {Piaulet}, {Gully-Santiago}, {Crossfield}, {Morley}, {Kreidberg}, {Mikal-Evans}, {Brande}, {Delisle}, {Greene}, {Hardegree-Ullman}, {Barman}, {Christiansen}, {Dragomir}, {Fortney}, {Howard}, {Kosiarek}, \& {Lothringer}}]{2023ApJ...954L..52R}
{Roy}, P.-A., {Benneke}, B., {Piaulet}, C., {et~al.} 2023, \apjl, 954, L52, \dodoi{10.3847/2041-8213/acebf0}

\bibitem[{{Rustamkulov} {et~al.}(2023){Rustamkulov}, {Sing}, {Mukherjee}, {May}, {Kirk}, {Schlawin}, {Line}, {Piaulet}, {Carter}, {Batalha}, {Goyal}, {L{\'o}pez-Morales}, {Lothringer}, {MacDonald}, {Moran}, {Stevenson}, {Wakeford}, {Espinoza}, {Bean}, {Batalha}, {Benneke}, {Berta-Thompson}, {Crossfield}, {Gao}, {Kreidberg}, {Powell}, {Cubillos}, {Gibson}, {Leconte}, {Molaverdikhani}, {Nikolov}, {Parmentier}, {Roy}, {Taylor}, {Turner}, {Wheatley}, {Aggarwal}, {Ahrer}, {Alam}, {Alderson}, {Allen}, {Banerjee}, {Barat}, {Barrado}, {Barstow}, {Bell}, {Blecic}, {Brande}, {Casewell}, {Changeat}, {Chubb}, {Crouzet}, {Daylan}, {Decin}, {D{\'e}sert}, {Mikal-Evans}, {Feinstein}, {Flagg}, {Fortney}, {Harrington}, {Heng}, {Hong}, {Hu}, {Iro}, {Kataria}, {Kempton}, {Krick}, {Lendl}, {Lillo-Box}, {Louca}, {Lustig-Yaeger}, {Mancini}, {Mansfield}, {Mayne}, {Miguel}, {Morello}, {Ohno}, {Palle}, {Petit dit de la Roche}, {Rackham}, {Radica}, {Ramos-Rosado}, {Redfield}, {Rogers}, {Shkolnik}, {Southworth}, {Teske}, {Tremblin},
  {Tucker}, {Venot}, {Waalkes}, {Welbanks}, {Zhang}, \& {Zieba}}]{2023Natur.614..659R}
{Rustamkulov}, Z., {Sing}, D.~K., {Mukherjee}, S., {et~al.} 2023, \nat, 614, 659, \dodoi{10.1038/s41586-022-05677-y}

\bibitem[{{Saba} {et~al.}(2022){Saba}, {Tsiaras}, {Morvan}, {Thompson}, {Changeat}, {Edwards}, {Jolly}, {Waldmann}, \& {Tinetti}}]{sabatransmission}
{Saba}, A., {Tsiaras}, A., {Morvan}, M., {et~al.} 2022, \aj, 164, 2, \dodoi{10.3847/1538-3881/ac6c01}

\bibitem[{{Salz} {et~al.}(2015){Salz}, {Schneider}, {Czesla}, \& {Schmitt}}]{2015Salz}
{Salz}, M., {Schneider}, P.~C., {Czesla}, S., \& {Schmitt}, J.~H.~M.~M. 2015, \aap, 576, A42, \dodoi{10.1051/0004-6361/201425243}

\bibitem[{{Sedaghati} {et~al.}(2017){Sedaghati}, {Boffin}, {Delrez}, {Gillon}, {Csizmadia}, {Smith}, \& {Rauer}}]{2017MNRAS.468.3123S}
{Sedaghati}, E., {Boffin}, H. M.~J., {Delrez}, L., {et~al.} 2017, \mnras, 468, 3123, \dodoi{10.1093/mnras/stx646}

\bibitem[{{Sedaghati} {et~al.}(2016){Sedaghati}, {Boffin}, {Je{\v{r}}abkov{\'a}}, {Garc{\'\i}a Mu{\~n}oz}, {Grenfell}, {Smette}, {Ivanov}, {Csizmadia}, {Cabrera}, {Kabath}, {Rocchetto}, \& {Rauer}}]{2016A&A...596A..47S}
{Sedaghati}, E., {Boffin}, H.~M.~J., {Je{\v{r}}abkov{\'a}}, T., {et~al.} 2016, \aap, 596, A47, \dodoi{10.1051/0004-6361/201629090}

\bibitem[{{Seidel} {et~al.}(2020){Seidel}, {Lendl}, {Bourrier}, {Ehrenreich}, {Allart}, {Sousa}, {Cegla}, {Bonfils}, {Conod}, {Grandjean}, {Wyttenbach}, {Astudillo-Defru}, {Bayliss}, {Heng}, {Lavie}, {Lovis}, {Melo}, {Pepe}, {S{\'e}gransan}, \& {Udry}}]{2020A&A...643A..45S}
{Seidel}, J.~V., {Lendl}, M., {Bourrier}, V., {et~al.} 2020, \aap, 643, A45, \dodoi{10.1051/0004-6361/202039058}

\bibitem[{{Simpson} {et~al.}(2023){Simpson}, {Fetherolf}, {Kane}, {Pepper}, {Mo{\v{c}}nik}, \& {Dalba}}]{2023AJ....166...72S}
{Simpson}, E.~R., {Fetherolf}, T., {Kane}, S.~R., {et~al.} 2023, \aj, 166, 72, \dodoi{10.3847/1538-3881/acda26}

\bibitem[{{Sing} {et~al.}(2008){Sing}, {Vidal-Madjar}, {D{\'e}sert}, {Lecavelier des Etangs}, \& {Ballester}}]{2008ApJ...686..658S}
{Sing}, D.~K., {Vidal-Madjar}, A., {D{\'e}sert}, J.~M., {Lecavelier des Etangs}, A., \& {Ballester}, G. 2008, \apj, 686, 658, \dodoi{10.1086/590075}

\bibitem[{{Sing} {et~al.}(2011){Sing}, {Pont}, {Aigrain}, {Charbonneau}, {D{\'e}sert}, {Gibson}, {Gilliland}, {Hayek}, {Henry}, {Knutson}, {Lecavelier Des Etangs}, {Mazeh}, \& {Shporer}}]{2011MNRAS.416.1443S}
{Sing}, D.~K., {Pont}, F., {Aigrain}, S., {et~al.} 2011, \mnras, 416, 1443, \dodoi{10.1111/j.1365-2966.2011.19142.x}

\bibitem[{{Sing} {et~al.}(2013){Sing}, {Lecavelier des Etangs}, {Fortney}, {Burrows}, {Pont}, {Wakeford}, {Ballester}, {Nikolov}, {Henry}, {Aigrain}, {Deming}, {Evans}, {Gibson}, {Huitson}, {Knutson}, {Showman}, {Vidal-Madjar}, {Wilson}, {Williamson}, \& {Zahnle}}]{2013MNRAS.436.2956S}
{Sing}, D.~K., {Lecavelier des Etangs}, A., {Fortney}, J.~J., {et~al.} 2013, \mnras, 436, 2956, \dodoi{10.1093/mnras/stt1782}

\bibitem[{{Sing} {et~al.}(2015){Sing}, {Wakeford}, {Showman}, {Nikolov}, {Fortney}, {Burrows}, {Ballester}, {Deming}, {Aigrain}, {D{\'e}sert}, {Gibson}, {Henry}, {Knutson}, {Lecavelier des Etangs}, {Pont}, {Vidal-Madjar}, {Williamson}, \& {Wilson}}]{2015MNRAS.446.2428S}
{Sing}, D.~K., {Wakeford}, H.~R., {Showman}, A.~P., {et~al.} 2015, \mnras, 446, 2428, \dodoi{10.1093/mnras/stu2279}

\bibitem[{Sing {et~al.}(2016)Sing, Fortney, Nikolov, Wakeford, Kataria, Evans, Aigrain, Ballester, Burrows, Deming, {et~al.}}]{sing2016continuum}
Sing, D.~K., Fortney, J.~J., Nikolov, N., {et~al.} 2016, Nature, 529, 59

\bibitem[{{Sing} {et~al.}(2019){Sing}, {Lavvas}, {Ballester}, {Lecavelier des Etangs}, {Marley}, {Nikolov}, {Ben-Jaffel}, {Bourrier}, {Buchhave}, {Deming}, {Ehrenreich}, {Mikal-Evans}, {Kataria}, {Lewis}, {L{\'o}pez-Morales}, {Garc{\'\i}a Mu{\~n}oz}, {Henry}, {Sanz-Forcada}, {Spake}, {Wakeford}, \& {PanCET Collaboration}}]{2019AJ....158...91S}
{Sing}, D.~K., {Lavvas}, P., {Ballester}, G.~E., {et~al.} 2019, \aj, 158, 91, \dodoi{10.3847/1538-3881/ab2986}

\bibitem[{Skaf {et~al.}(2020)Skaf, Bieger, Edwards, Changeat, Morvan, Kiefer, Blain, Zingales, Poveda, Al-Refaie, {et~al.}}]{skaf2020ares}
Skaf, N., Bieger, M.~F., Edwards, B., {et~al.} 2020, The Astronomical Journal, 160, 109

\bibitem[{Skilling(2006)}]{skilling2006nested}
Skilling, J. 2006, Bayesian Analysis, 1, 833, \dodoi{10.1214/06-BA127}

\bibitem[{{Skinner} \& {Cho}(2022)}]{skinner2021}
{Skinner}, J.~W., \& {Cho}, J.~Y.~K. 2022, \mnras, 511, 3584, \dodoi{10.1093/mnras/stab2809}

\bibitem[{{Snellen} {et~al.}(2008){Snellen}, {Albrecht}, {de Mooij}, \& {Le Poole}}]{2008A&A...487..357S}
{Snellen}, I.~A.~G., {Albrecht}, S., {de Mooij}, E.~J.~W., \& {Le Poole}, R.~S. 2008, \aap, 487, 357, \dodoi{10.1051/0004-6361:200809762}

\bibitem[{{Southworth} {et~al.}(2012){Southworth}, {Hinse}, {Dominik}, {Fang}, {Harps{\o}e}, {J{\o}rgensen}, {Kerins}, {Liebig}, {Mancini}, {Skottfelt}, {Anderson}, {Smalley}, {Tregloan-Reed}, {Wertz}, {Alsubai}, {Bozza}, {Calchi Novati}, {Dreizler}, {Gu}, {Hundertmark}, {Jessen-Hansen}, {Kains}, {Kjeldsen}, {Lund}, {Lundkvist}, {Mathiasen}, {Penny}, {Rahvar}, {Ricci}, {Scarpetta}, {Snodgrass}, \& {Surdej}}]{southworth2012w17}
{Southworth}, J., {Hinse}, T.~C., {Dominik}, M., {et~al.} 2012, \mnras, 426, 1338, \dodoi{10.1111/j.1365-2966.2012.21781.x}

\bibitem[{{Spake} {et~al.}(2021){Spake}, {Sing}, {Wakeford}, {Nikolov}, {Mikal-Evans}, {Deming}, {Barstow}, {Anderson}, {Carter}, {Gillon}, {Goyal}, {Hebrard}, {Hellier}, {Kataria}, {Lam}, {Triaud}, \& {Wheatley}}]{2021MNRAS.500.4042S}
{Spake}, J.~J., {Sing}, D.~K., {Wakeford}, H.~R., {et~al.} 2021, \mnras, 500, 4042, \dodoi{10.1093/mnras/staa3116}

\bibitem[{{Speagle}(2020)}]{2020MNRAS.493.3132S}
{Speagle}, J.~S. 2020, \mnras, 493, 3132, \dodoi{10.1093/mnras/staa278}

\bibitem[{{Spyratos} {et~al.}(2023){Spyratos}, {Nikolov}, {Constantinou}, {Southworth}, {Madhusudhan}, {Sedaghati}, {Ehrenreich}, \& {Mancini}}]{2023MNRAS.521.2163S}
{Spyratos}, P., {Nikolov}, N.~K., {Constantinou}, S., {et~al.} 2023, \mnras, 521, 2163, \dodoi{10.1093/mnras/stad637}

\bibitem[{{Sreejith} {et~al.}(2020){Sreejith}, {Fossati}, {Youngblood}, {France}, \& {Ambily}}]{Sreejith2020}
{Sreejith}, A.~G., {Fossati}, L., {Youngblood}, A., {France}, K., \& {Ambily}, S. 2020, \aap, 644, A67, \dodoi{10.1051/0004-6361/202039167}

\bibitem[{{Stassun} {et~al.}(2017){Stassun}, {Collins}, \& {Gaudi}}]{2017AJ....153..136S}
{Stassun}, K.~G., {Collins}, K.~A., \& {Gaudi}, B.~S. 2017, \aj, 153, 136, \dodoi{10.3847/1538-3881/aa5df3}

\bibitem[{{Stassun} {et~al.}(2019){Stassun}, {Oelkers}, {Paegert}, {Torres}, {Pepper}, {De Lee}, {Collins}, {Latham}, {Muirhead}, {Chittidi}, {Rojas-Ayala}, {Fleming}, {Rose}, {Tenenbaum}, {Ting}, {Kane}, {Barclay}, {Bean}, {Brassuer}, {Charbonneau}, {Ge}, {Lissauer}, {Mann}, {McLean}, {Mullally}, {Narita}, {Plavchan}, {Ricker}, {Sasselov}, {Seager}, {Sharma}, {Shiao}, {Sozzetti}, {Stello}, {Vanderspek}, {Wallace}, \& {Winn}}]{2019AJ....158..138S}
{Stassun}, K.~G., {Oelkers}, R.~J., {Paegert}, M., {et~al.} 2019, \aj, 158, 138, \dodoi{10.3847/1538-3881/ab3467}

\bibitem[{{Stef{\`a}nsson} {et~al.}(2022){Stef{\`a}nsson}, {Mahadevan}, {Petrovich}, {Winn}, {Kanodia}, {Millholland}, {Maney}, {Ca{\~n}as}, {Wisniewski}, {Robertson}, {Ninan}, {Ford}, {Bender}, {Blake}, {Cegla}, {Cochran}, {Diddams}, {Dong}, {Endl}, {Fredrick}, {Halverson}, {Hearty}, {Hebb}, {Hirano}, {Lin}, {Logsdon}, {Lubar}, {McElwain}, {Metcalf}, {Monson}, {Rajagopal}, {Ramsey}, {Roy}, {Schwab}, {Schweiker}, {Terrien}, \& {Wright}}]{2022ApJ...931L..15S}
{Stef{\`a}nsson}, G., {Mahadevan}, S., {Petrovich}, C., {et~al.} 2022, \apjl, 931, L15, \dodoi{10.3847/2041-8213/ac6e3c}

\bibitem[{{Stevenson} {et~al.}(2016){Stevenson}, {Bean}, {Seifahrt}, {Gilbert}, {Line}, {D{\'e}sert}, \& {Fortney}}]{2016ApJ...817..141S}
{Stevenson}, K.~B., {Bean}, J.~L., {Seifahrt}, A., {et~al.} 2016, \apj, 817, 141, \dodoi{10.3847/0004-637X/817/2/141}

\bibitem[{Stotesbury {et~al.}(2022)Stotesbury, Edwards, Lavigne, Pesquita, Veilleux, Windred, Al-Refaie, Bradley, Ma, Savini, {et~al.}}]{stotesbury2022twinkle}
Stotesbury, I., Edwards, B., Lavigne, J.-F., {et~al.} 2022, in Space Telescopes and Instrumentation 2022: Optical, Infrared, and Millimeter Wave, Vol. 12180, SPIE, 1117--1130

\bibitem[{{Swain} {et~al.}(2009){Swain}, {Vasisht}, {Tinetti}, {Bouwman}, {Chen}, {Yung}, {Deming}, \& {Deroo}}]{2009ApJ...690L.114S}
{Swain}, M.~R., {Vasisht}, G., {Tinetti}, G., {et~al.} 2009, \apjl, 690, L114, \dodoi{10.1088/0004-637X/690/2/L114}

\bibitem[{{Tennyson} {et~al.}(2016){Tennyson}, {Yurchenko}, {Al-Refaie}, {Barton}, {Chubb}, {Coles}, {Diamantopoulou}, {Gorman}, {Hill}, {Lam}, {Lodi}, {McKemmish}, {Na}, {Owens}, {Polyansky}, {Rivlin}, {Sousa-Silva}, {Underwood}, {Yachmenev}, \& {Zak}}]{2016JMoSp.327...73T}
{Tennyson}, J., {Yurchenko}, S.~N., {Al-Refaie}, A.~F., {et~al.} 2016, Journal of Molecular Spectroscopy, 327, 73, \dodoi{10.1016/j.jms.2016.05.002}

\bibitem[{{Thompson} {et~al.}(2024){Thompson}, {Biagini}, {Cracchiolo}, {Petralia}, {Changeat}, {Saba}, {Morello}, {Morvan}, {Micela}, \& {Tinetti}}]{2024ApJ...960..107T}
{Thompson}, A., {Biagini}, A., {Cracchiolo}, G., {et~al.} 2024, \apj, 960, 107, \dodoi{10.3847/1538-4357/ad0369}

\bibitem[{Tinetti {et~al.}(2018)Tinetti, Drossart, Eccleston, Hartogh, Heske, Leconte, Micela, Ollivier, Pilbratt, Puig, {et~al.}}]{tinetti2018chemical}
Tinetti, G., Drossart, P., Eccleston, P., {et~al.} 2018, Experimental Astronomy, 46, 135

\bibitem[{{Tinetti} {et~al.}(2021){Tinetti}, {Eccleston}, {Haswell}, {Lagage}, {Leconte}, {L{\"u}ftinger}, {Micela}, {Min}, {Pilbratt}, {Puig}, {Swain}, {Testi}, {Turrini}, {Vandenbussche}, {Rosa Zapatero Osorio}, {Aret}, {Beaulieu}, {Buchhave}, {Ferus}, {Griffin}, {Guedel}, {Hartogh}, {Machado}, {Malaguti}, {Pall{\'e}}, {Rataj}, {Ray}, {Ribas}, {Szab{\'o}}, {Tan}, {Werner}, {Ratti}, {Scharmberg}, {Salvignol}, {Boudin}, {Halain}, {Haag}, {Crouzet}, {Kohley}, {Symonds}, {Renk}, {Caldwell}, {Abreu}, {Alonso}, {Amiaux}, {Berth{\'e}}, {Bishop}, {Bowles}, {Carmona}, {Coffey}, {Colom{\'e}}, {Crook}, {D{\'e}sjonqueres}, {D{\'\i}az}, {Drummond}, {Focardi}, {G{\'o}mez}, {Holmes}, {Krijger}, {Kovacs}, {Hunt}, {Machado}, {Morgante}, {Ollivier}, {Ottensamer}, {Pace}, {Pagano}, {Pascale}, {Pearson}, {M{\o}ller Pedersen}, {Pniel}, {Roose}, {Savini}, {Stamper}, {Szirovicza}, {Szoke}, {Tosh}, {Vilardell}, {Barstow}, {Borsato}, {Casewell}, {Changeat}, {Charnay}, {Civi{\v{s}}}, {Coud{\'e} du Foresto}, {Coustenis}, {Cowan},
  {Danielski}, {Demangeon}, {Drossart}, {Edwards}, {Gilli}, {Encrenaz}, {Kiss}, {Kokori}, {Ikoma}, {Morales}, {Mendon{\c{c}}a}, {Moneti}, {Mugnai}, {Garc{\'\i}a Mu{\~n}oz}, {Helled}, {Kama}, {Miguel}, {Nikolaou}, {Pagano}, {Panic}, {Rengel}, {Rickman}, {Rocchetto}, {Sarkar}, {Selsis}, {Tennyson}, {Tsiaras}, {Venot}, {Vida}, {Waldmann}, {Yurchenko}, {Szab{\'o}}, {Zellem}, {Al-Refaie}, {Perez Alvarez}, {Anisman}, {Arhancet}, {Ateca}, {Baeyens}, {Barnes}, {Bell}, {Benatti}, {Biazzo}, {B{\l}{\k{e}}cka}, {Bonomo}, {Bosch}, {Bossini}, {Bourgalais}, {Brienza}, {Brucalassi}, {Bruno}, {Caines}, {Calcutt}, {Campante}, {Canestrari}, {Cann}, {Casali}, {Casas}, {Cassone}, {Cara}, {Carmona}, {Carone}, {Carrasco}, {Changeat}, {Chioetto}, {Cortecchia}, {Czupalla}, {Chubb}, {Ciaravella}, {Claret}, {Claudi}, {Codella}, {Garcia Comas}, {Cracchiolo}, {Cubillos}, {Da Peppo}, {Decin}, {Dejabrun}, {Delgado-Mena}, {Di Giorgio}, {Diolaiti}, {Dorn}, {Doublier}, {Doumayrou}, {Dransfield}, {Dumaye}, {Dunford}, {Jimenez Escobar}, {Van
  Eylen}, {Farina}, {Fedele}, {Fern{\'a}ndez}, {Fleury}, {Fonte}, {Fontignie}, {Fossati}, {Funke}, {Galy}, {Garai}, {Garc{\'\i}a}, {Garc{\'\i}a-Rigo}, {Garufi}, {Germano Sacco}, {Giacobbe}, {G{\'o}mez}, {Gonzalez}, {Gonzalez-Galindo}, {Grassi}, {Griffith}, {Guarcello}, {Goujon}, {Gressier}, {Grzegorczyk}, {Guillot}, {Guilluy}, {Hargrave}, {Hellin}, {Herrero}, {Hills}, {Horeau}, {Ito}, {Jessen}, {Kabath}, {K{\'a}lm{\'a}n}, {Kawashima}, {Kimura}, {Kn{\'\i}{\v{z}}ek}, {Kreidberg}, {Kruid}, {Kruijssen}, {Kubel{\'\i}k}, {Lara}, {Lebonnois}, {Lee}, {Lefevre}, {Lichtenberg}, {Locci}, {Lombini}, {Sanchez Lopez}, {Lorenzani}, {MacDonald}, {Magrini}, {Maldonado}, {Marcq}, {Migliorini}, {Modirrousta-Galian}, {Molaverdikhani}, {Molinari}, {Molli{\`e}re}, {Moreau}, {Morello}, {Morinaud}, {Morvan}, {Moses}, {Mouzali}, {Nakhjiri}, {Naponiello}, {Narita}, {Nascimbeni}, {Nikolaou}, {Noce}, {Oliva}, {Palladino}, {Papageorgiou}, {Parmentier}, {Peres}, {P{\'e}rez}, {Perez-Hoyos}, {Perger}, {Cecchi Pestellini}, {Petralia},
  {Philippon}, {Piccialli}, {Pignatari}, {Piotto}, {Podio}, {Polenta}, {Preti}, {Pribulla}, {Lopez Puertas}, {Rainer}, {Reess}, {Rimmer}, {Robert}, {Rosich}, {Rossi}, {Rust}, {Saleh}, {Sanna}, {Schisano}, {Schreiber}, {Schwartz}, {Scippa}, {Seli}, {Shibata}, {Simpson}, {Shorttle}, {Skaf}, {Skup}, {Sobiecki}, {Sousa}, {Sozzetti}, {{\v{S}}poner}, {Steiger}, {Tanga}, {Tackley}, {Taylor}, {Tecza}, {Terenzi}, {Tremblin}, {Tozzi}, {Triaud}, {Trompet}, {Tsai}, {Tsantaki}, {Valencia}, {Carine Vandaele}, {Van der Swaelmen}, {Adibekyan}, {Vasisht}, {Vazan}, {Del Vecchio}, {Waltham}, {Wawer}, {Widemann}, {Wolkenberg}, {Hou Yip}, {Yung}, {Zilinskas}, {Zingales}, \& {Zuppella}}]{2021arXiv210404824T}
{Tinetti}, G., {Eccleston}, P., {Haswell}, C., {et~al.} 2021, arXiv e-prints, arXiv:2104.04824, \dodoi{10.48550/arXiv.2104.04824}

\bibitem[{{Torres} {et~al.}(2008){Torres}, {Winn}, \& {Holman}}]{2008ApJ...677.1324T}
{Torres}, G., {Winn}, J.~N., \& {Holman}, M.~J. 2008, \apj, 677, 1324, \dodoi{10.1086/529429}

\bibitem[{{Triaud} {et~al.}(2010){Triaud}, {Collier Cameron}, {Queloz}, {Anderson}, {Gillon}, {Hebb}, {Hellier}, {Loeillet}, {Maxted}, {Mayor}, {Pepe}, {Pollacco}, {S{\'e}gransan}, {Smalley}, {Udry}, {West}, \& {Wheatley}}]{triaud2010}
{Triaud}, A.~H.~M.~J., {Collier Cameron}, A., {Queloz}, D., {et~al.} 2010, \aap, 524, A25, \dodoi{10.1051/0004-6361/201014525}

\bibitem[{{Triaud} {et~al.}(2013){Triaud}, {Anderson}, {Collier Cameron}, {Doyle}, {Fumel}, {Gillon}, {Hellier}, {Jehin}, {Lendl}, {Lovis}, {Maxted}, {Pepe}, {Pollacco}, {Queloz}, {S{\'e}gransan}, {Smalley}, {Smith}, {Udry}, {West}, \& {Wheatley}}]{triaud2013}
{Triaud}, A.~H.~M.~J., {Anderson}, D.~R., {Collier Cameron}, A., {et~al.} 2013, \aap, 551, A80, \dodoi{10.1051/0004-6361/201220900}

\bibitem[{{Triaud} {et~al.}(2015{\natexlab{a}}){Triaud}, {Gillon}, {Ehrenreich}, {Herrero}, {Lendl}, {Anderson}, {Collier Cameron}, {Delrez}, {Demory}, {Hellier}, {Heng}, {Jehin}, {Maxted}, {Pollacco}, {Queloz}, {Ribas}, {Smalley}, {Smith}, \& {Udry}}]{triaud2015dayside}
{Triaud}, A. H.~M.~J., {Gillon}, M., {Ehrenreich}, D., {et~al.} 2015{\natexlab{a}}, \mnras, 450, 2279, \dodoi{10.1093/mnras/stv706}

\bibitem[{{Triaud} {et~al.}(2015{\natexlab{b}}){Triaud}, {Gillon}, {Ehrenreich}, {Herrero}, {Lendl}, {Anderson}, {Collier Cameron}, {Delrez}, {Demory}, {Hellier}, {Heng}, {Jehin}, {Maxted}, {Pollacco}, {Queloz}, {Ribas}, {Smalley}, {Smith}, \& {Udry}}]{2015MNRAS.450.2279T}
---. 2015{\natexlab{b}}, \mnras, 450, 2279, \dodoi{10.1093/mnras/stv706}

\bibitem[{Trotta(2008)}]{trotta2008bayes}
Trotta, R. 2008, Contemporary Physics, 49, 71

\bibitem[{{Tsai} {et~al.}(2023){Tsai}, {Lee}, {Powell}, {Gao}, {Zhang}, {Moses}, {H{\'e}brard}, {Venot}, {Parmentier}, {Jordan}, {Hu}, {Alam}, {Alderson}, {Batalha}, {Bean}, {Benneke}, {Bierson}, {Brady}, {Carone}, {Carter}, {Chubb}, {Inglis}, {Leconte}, {Line}, {L{\'o}pez-Morales}, {Miguel}, {Molaverdikhani}, {Rustamkulov}, {Sing}, {Stevenson}, {Wakeford}, {Yang}, {Aggarwal}, {Baeyens}, {Barat}, {de Val-Borro}, {Daylan}, {Fortney}, {France}, {Goyal}, {Grant}, {Kirk}, {Kreidberg}, {Louca}, {Moran}, {Mukherjee}, {Nasedkin}, {Ohno}, {Rackham}, {Redfield}, {Taylor}, {Tremblin}, {Visscher}, {Wallack}, {Welbanks}, {Youngblood}, {Ahrer}, {Batalha}, {Behr}, {Berta-Thompson}, {Blecic}, {Casewell}, {Crossfield}, {Crouzet}, {Cubillos}, {Decin}, {D{\'e}sert}, {Feinstein}, {Gibson}, {Harrington}, {Heng}, {Henning}, {Kempton}, {Krick}, {Lagage}, {Lendl}, {Lothringer}, {Mansfield}, {Mayne}, {Mikal-Evans}, {Palle}, {Schlawin}, {Shorttle}, {Wheatley}, \& {Yurchenko}}]{2023Natur.617..483T}
{Tsai}, S.-M., {Lee}, E. K.~H., {Powell}, D., {et~al.} 2023, \nat, 617, 483, \dodoi{10.1038/s41586-023-05902-2}

\bibitem[{Tsiaras {et~al.}(2016{\natexlab{a}})Tsiaras, Waldmann, Rocchetto, Varley, Morello, Damiano, \& Tinetti}]{tsiaras2016new}
Tsiaras, A., Waldmann, I., Rocchetto, M., {et~al.} 2016{\natexlab{a}}, The Astrophysical Journal, 832, 202

\bibitem[{Tsiaras {et~al.}(2019)Tsiaras, Waldmann, Tinetti, Tennyson, \& Yurchenko}]{tsiaras2019water}
Tsiaras, A., Waldmann, I.~P., Tinetti, G., Tennyson, J., \& Yurchenko, S.~N. 2019, Nature Astronomy, 3, 1086

\bibitem[{Tsiaras {et~al.}(2016{\natexlab{b}})Tsiaras, Rocchetto, Waldmann, Venot, Varley, Morello, Damiano, Tinetti, Barton, Yurchenko, {et~al.}}]{tsiaras2016detection}
Tsiaras, A., Rocchetto, M., Waldmann, I., {et~al.} 2016{\natexlab{b}}, The Astrophysical Journal, 820, 99

\bibitem[{Tsiaras {et~al.}(2018)Tsiaras, Waldmann, Zingales, Rocchetto, Morello, Damiano, Karpouzas, Tinetti, McKemmish, Tennyson, {et~al.}}]{tsiaras2018population}
Tsiaras, A., Waldmann, I., Zingales, T., {et~al.} 2018, The Astronomical Journal, 155, 156

\bibitem[{{Turner} {et~al.}(2017){Turner}, {Leiter}, {Biddle}, {Pearson}, {Hardegree-Ullman}, {Thompson}, {Teske}, {Cates}, {Cook}, {Berube}, {Nieberding}, {Jones}, {Raphael}, {Wallace}, {Watson}, \& {Johnson}}]{2017MNRAS.472.3871T}
{Turner}, J.~D., {Leiter}, R.~M., {Biddle}, L.~I., {et~al.} 2017, \mnras, 472, 3871, \dodoi{10.1093/mnras/stx2221}

\bibitem[{{Vidal-Madjar} {et~al.}(2003){Vidal-Madjar}, {Lecavelier des Etangs}, {D{\'e}sert}, {Ballester}, {Ferlet}, {H{\'e}brard}, \& {Mayor}}]{2003Natur.422..143V}
{Vidal-Madjar}, A., {Lecavelier des Etangs}, A., {D{\'e}sert}, J.~M., {et~al.} 2003, \nat, 422, 143, \dodoi{10.1038/nature01448}

\bibitem[{{Vidal-Madjar} {et~al.}(2004){Vidal-Madjar}, {D{\'e}sert}, {Lecavelier des Etangs}, {H{\'e}brard}, {Ballester}, {Ehrenreich}, {Ferlet}, {McConnell}, {Mayor}, \& {Parkinson}}]{2004ApJ...604L..69V}
{Vidal-Madjar}, A., {D{\'e}sert}, J.~M., {Lecavelier des Etangs}, A., {et~al.} 2004, \apjl, 604, L69, \dodoi{10.1086/383347}

\bibitem[{{Villanueva} {et~al.}(2024){Villanueva}, {Fauchez}, {Kofman}, {Alei}, {Lee}, {Janin}, {Himes}, {Leconte}, {Leung}, {Faggi}, {Mak}, {Sergeev}, {Kozakis}, {Manners}, {Mayne}, {Schwieterman}, {Howe}, \& {Batalha}}]{2024PSJ.....5...64V}
{Villanueva}, G.~L., {Fauchez}, T.~J., {Kofman}, V., {et~al.} 2024, \psj, 5, 64, \dodoi{10.3847/PSJ/ad2681}

\bibitem[{{Vissapragada} {et~al.}(2020){Vissapragada}, {Knutson}, {Jovanovic}, {Harada}, {Oklop{\v{c}}i{\'c}}, {Eriksen}, {Mawet}, {Millar-Blanchaer}, {Tinyanont}, \& {Vasisht}}]{2020AJ....159..278V}
{Vissapragada}, S., {Knutson}, H.~A., {Jovanovic}, N., {et~al.} 2020, \aj, 159, 278, \dodoi{10.3847/1538-3881/ab8e34}

\bibitem[{{von Essen} {et~al.}(2019){von Essen}, {Wedemeyer}, {Sosa}, {Hjorth}, {Parkash}, {Freudenthal}, {Mallonn}, {Micul{\'a}n}, {Zibecchi}, {Cellone}, \& {Torres}}]{2019A&A...628A.116V}
{von Essen}, C., {Wedemeyer}, S., {Sosa}, M.~S., {et~al.} 2019, \aap, 628, A116, \dodoi{10.1051/0004-6361/201731966}

\bibitem[{{{\v{Z}}{\'a}k} {et~al.}(2019){{\v{Z}}{\'a}k}, {Kab{\'a}th}, {Boffin}, {Ivanov}, \& {Skarka}}]{2019AJ....158..120Z}
{{\v{Z}}{\'a}k}, J., {Kab{\'a}th}, P., {Boffin}, H. M.~J., {Ivanov}, V.~D., \& {Skarka}, M. 2019, \aj, 158, 120, \dodoi{10.3847/1538-3881/ab32ec}

\bibitem[{{Wakeford} {et~al.}(2013){Wakeford}, {Sing}, {Deming}, {Gibson}, {Fortney}, {Burrows}, {Ballester}, {Nikolov}, {Aigrain}, {Henry}, {Knutson}, {Lecavelier des Etangs}, {Pont}, {Showman}, {Vidal-Madjar}, \& {Zahnle}}]{2013MNRAS.435.3481W}
{Wakeford}, H.~R., {Sing}, D.~K., {Deming}, D., {et~al.} 2013, \mnras, 435, 3481, \dodoi{10.1093/mnras/stt1536}

\bibitem[{{Wakeford} {et~al.}(2017{\natexlab{a}}){Wakeford}, {Sing}, {Kataria}, {Deming}, {Nikolov}, {Lopez}, {Tremblin}, {Amundsen}, {Lewis}, {Mandell}, {Fortney}, {Knutson}, {Benneke}, \& {Evans}}]{2017Sci...356..628W}
{Wakeford}, H.~R., {Sing}, D.~K., {Kataria}, T., {et~al.} 2017{\natexlab{a}}, Science, 356, 628, \dodoi{10.1126/science.aah4668}

\bibitem[{{Wakeford} {et~al.}(2017{\natexlab{b}}){Wakeford}, {Stevenson}, {Lewis}, {Sing}, {L{\'o}pez-Morales}, {Marley}, {Kataria}, {Mandell}, {Ballester}, {Barstow}, {Ben-Jaffel}, {Bourrier}, {Buchhave}, {Ehrenreich}, {Evans}, {Garc{\'\i}a Mu{\~n}oz}, {Henry}, {Knutson}, {Lavvas}, {Lecavelier des Etangs}, {Nikolov}, \& {Sanz-Forcada}}]{2017ApJ...835L..12W}
{Wakeford}, H.~R., {Stevenson}, K.~B., {Lewis}, N.~K., {et~al.} 2017{\natexlab{b}}, \apjl, 835, L12, \dodoi{10.3847/2041-8213/835/1/L12}

\bibitem[{{Wakeford} {et~al.}(2018){Wakeford}, {Sing}, {Deming}, {Lewis}, {Goyal}, {Wilson}, {Barstow}, {Kataria}, {Drummond}, {Evans}, {Carter}, {Nikolov}, {Knutson}, {Ballester}, \& {Mandell}}]{2018AJ....155...29W}
{Wakeford}, H.~R., {Sing}, D.~K., {Deming}, D., {et~al.} 2018, \aj, 155, 29, \dodoi{10.3847/1538-3881/aa9e4e}

\bibitem[{{Welbanks} {et~al.}(2019){Welbanks}, {Madhusudhan}, {Allard}, {Hubeny}, {Spiegelman}, \& {Leininger}}]{2019ApJ...887L..20W}
{Welbanks}, L., {Madhusudhan}, N., {Allard}, N.~F., {et~al.} 2019, \apjl, 887, L20, \dodoi{10.3847/2041-8213/ab5a89}

\bibitem[{{Wilson} {et~al.}(2020){Wilson}, {Gibson}, {Nikolov}, {Constantinou}, {Madhusudhan}, {Goyal}, {Barstow}, {Carter}, {de Mooij}, {Drummond}, {Mikal-Evans}, {Helling}, {Mayne}, \& {Sing}}]{2020MNRAS.497.5155W}
{Wilson}, J., {Gibson}, N.~P., {Nikolov}, N., {et~al.} 2020, \mnras, 497, 5155, \dodoi{10.1093/mnras/staa2307}

\bibitem[{{Wilson} {et~al.}(2015){Wilson}, {Sing}, {Nikolov}, {Lecavelier des Etangs}, {Pont}, {Fortney}, {Ballester}, {L{\'o}pez-Morales}, {D{\'e}sert}, \& {Vidal-Madjar}}]{2015MNRAS.450..192W}
{Wilson}, P.~A., {Sing}, D.~K., {Nikolov}, N., {et~al.} 2015, \mnras, 450, 192, \dodoi{10.1093/mnras/stv642}

\bibitem[{{Winn} {et~al.}(2010){Winn}, {Johnson}, {Howard}, {Marcy}, {Isaacson}, {Shporer}, {Bakos}, {Hartman}, \& {Albrecht}}]{2010ApJ...723L.223W}
{Winn}, J.~N., {Johnson}, J.~A., {Howard}, A.~W., {et~al.} 2010, \apjl, 723, L223, \dodoi{10.1088/2041-8205/723/2/L223}

\bibitem[{{Wong} {et~al.}(2020){Wong}, {Benneke}, {Gao}, {Knutson}, {Chachan}, {Henry}, {Deming}, {Kataria}, {Lee}, {Nikolov}, {Sing}, {Ballester}, {Baskin}, {Wakeford}, \& {Williamson}}]{2020AJ....159..234W}
{Wong}, I., {Benneke}, B., {Gao}, P., {et~al.} 2020, \aj, 159, 234, \dodoi{10.3847/1538-3881/ab880d}

\bibitem[{{Wong} {et~al.}(2022){Wong}, {Chachan}, {Knutson}, {Henry}, {Adams}, {Kataria}, {Benneke}, {Gao}, {Deming}, {L{\'o}pez-Morales}, {Sing}, {Alam}, {Ballester}, {Barstow}, {Buchhave}, {dos Santos}, {Fu}, {Garc{\'\i}a Mu{\~n}oz}, {MacDonald}, {Mikal-Evans}, {Sanz-Forcada}, \& {Wakeford}}]{2022AJ....164...30W}
{Wong}, I., {Chachan}, Y., {Knutson}, H.~A., {et~al.} 2022, \aj, 164, 30, \dodoi{10.3847/1538-3881/ac7234}

\bibitem[{{Wood} {et~al.}(2011){Wood}, {Maxted}, {Smalley}, \& {Iro}}]{2011MNRAS.412.2376W}
{Wood}, P.~L., {Maxted}, P.~F.~L., {Smalley}, B., \& {Iro}, N. 2011, \mnras, 412, 2376, \dodoi{10.1111/j.1365-2966.2010.18061.x}

\bibitem[{{Yan} {et~al.}(2020){Yan}, {Espinoza}, {Molaverdikhani}, {Henning}, {Mancini}, {Mallonn}, {Rackham}, {Apai}, {Jord{\'a}n}, {Molli{\`e}re}, {Chen}, {Carone}, \& {Reiners}}]{2020A&A...642A..98Y}
{Yan}, F., {Espinoza}, N., {Molaverdikhani}, K., {et~al.} 2020, \aap, 642, A98, \dodoi{10.1051/0004-6361/201937265}

\bibitem[{{Yang} {et~al.}(2017){Yang}, {Liu}, {Gao}, {Fang}, {Guo}, {Zhang}, {Hou}, {Wang}, \& {Cao}}]{2017ApJ...849...36Y}
{Yang}, H., {Liu}, J., {Gao}, Q., {et~al.} 2017, \apj, 849, 36, \dodoi{10.3847/1538-4357/aa8ea2}

\bibitem[{{Yee} {et~al.}(2018){Yee}, {Petigura}, {Fulton}, {Knutson}, {Batygin}, {Bakos}, {Hartman}, {Hirsch}, {Howard}, {Isaacson}, {Kosiarek}, {Sinukoff}, \& {Weiss}}]{2018AJ....155..255Y}
{Yee}, S.~W., {Petigura}, E.~A., {Fulton}, B.~J., {et~al.} 2018, \aj, 155, 255, \dodoi{10.3847/1538-3881/aabfec}

\bibitem[{{Yip} {et~al.}(2024){Yip}, {Changeat}, {Al-Refaie}, \& {Waldmann}}]{2024ApJ...961...30Y}
{Yip}, K.~H., {Changeat}, Q., {Al-Refaie}, A., \& {Waldmann}, I.~P. 2024, \apj, 961, 30, \dodoi{10.3847/1538-4357/ad063f}

\bibitem[{Yip {et~al.}(2020)Yip, Changeat, Edwards, Morvan, Chubb, Tsiaras, Waldmann, \& Tinetti}]{yip2020compatibility}
Yip, K.~H., Changeat, Q., Edwards, B., {et~al.} 2020, The Astronomical Journal, 161, 4

\bibitem[{{Zhou} \& {Bayliss}(2012)}]{2012MNRAS.426.2483Z}
{Zhou}, G., \& {Bayliss}, D.~D.~R. 2012, \mnras, 426, 2483, \dodoi{10.1111/j.1365-2966.2012.21817.x}

\end{thebibliography}
\bibliographystyle{aasjournal}
\end{document}